# Impact of interim analyses on Bayesian and frequentist operating characteristics of Bayesian clinical trials

Clément R. Massonnaud,[1,2] Lucas Manns,[1] Cédric Laouénan,[1,2]* André Gillibert[3]*

1. Université Paris Cité and Université Sorbonne Paris Nord, Inserm, IAME, F-75018 Paris, France
2. Département d'Épidémiologie, Biostatistique, Recherche Clinique, AP-HP, Hôpital Bichat Claude-Bernard, F-75018 Paris, France
3. Department of Biostatistics, CHU Rouen, F-76000, Rouen, France

*co-last authors

**Corresponding author:**

Clément R. Massonnaud

ORCID : 0000-0003-0292-9668

AP-HP, Hôpital Bichat, Département d'Épidémiologie, Biostatistique, Recherche Clinique

46 rue Henri Huchard, 75018 Paris, France

Clement.massonnaud@aphp.fr

**Declaration of conflicting interest:** no conflict of interest.

**Funding statement**: no funding.

**Ethical approval and informed consent statements**: not applicable.

**Data availability statement** : all the data and code used in this work are fully and freely available at https://github.com/ClementMassonnaud/BayesianInterim

# Abstract

While Bayesian methods are increasingly used in clinical research, confusion persists as to whether Bayesian designs are affected by repeated interim analyses, and how such effects should be evaluated. We aimed to clarify this question by evaluating both frequentist and Bayesian operating characteristics of group-sequential trials. We conducted simulation studies with normally distributed outcomes, examining designs with repeated analyses, with and without futility stopping rules and multiplicity adjustments. We show that repeated interim analyses alter both frequentist and Bayesian operating characteristics in group-sequential trials, regardless of the inferential framework adopted. Without proper adjustment for multiplicity, the Type I error rate increases with the number of analyses. Bayesian operating characteristics such as the risk of erroneous conclusions and the informative value of an efficacy conclusion are meaningful alternatives to classical frequentist metrics, but they are sensitive to prior divergence between stakeholders, and this sensitivity grows with the number of analyses. Even under full prior agreement, the informative value of an efficacy conclusion is reduced. While it is possible to control frequentist operating characteristics of Bayesian trials with appropriate methods, it is not possible to guarantee such control for Bayesian operating characteristics because they are prior-dependent and different stakeholders may adopt different priors. Contrary to claims that Bayesian inference is immune to multiplicity, our results show that Bayesian clinical trials are no less affected by repeated analyses than frequentist ones, regardless of the framework used to evaluate them.

# Keywords

## 1. Introduction

Group-sequential clinical trial designs extend the classical fixed-sample framework by allowing interim analyses as data accumulates, according to pre-specified stopping rules, thereby permitting decisions before the planned maximum sample size is reached. These designs can improve efficiency, reduce costs, and limit patient exposure to inferior treatments. Their use has expanded in recent years with the development of adaptive and platform trial designs, in which multiple interventions may be evaluated simultaneously and trial features, such as randomisation probabilities or treatment arms, can be modified over time. These designs rely heavily on repeated interim analyses to guide adaptations, making the control and interpretation of their operating characteristics increasingly complex.

In the traditional frequentist Neyman-Pearson hypothesis testing framework, clinical trials are designed to ensure adequate power, that is, a high probability of detecting an effect if it is equal to a predefined non-null value, such as a plausible or clinically meaningful effect, while controlling the probability of rejecting the null hypothesis if it is true (Type I error), at a prespecified level.[1] Interim analyses allow early stopping for efficacy, but without appropriate correction for multiple testing, inflate the Type I error rate, an issue long recognised and studied in the statistical literature.[2–8]

The Bayesian approach differs from the frequentist Neyman-Pearson hypothesis testing framework in that inference targets the posterior distribution of the parameter of interest, given the observed data and a prior distribution, rather than a dichotomized probability on the observed data under a null hypothesis. This is often regarded as more flexible, informative, and intuitive than the frequentist approach,[9–12] and in part explains the growing use of Bayesian methods in clinical trials.[13–16]

Since the posterior distribution depends only on (i) the prior distribution, (ii) the actual observed data, and (iii) a likelihood function, it is invariant to the sampling plan and the timing or number of interim analyses, a property related to the likelihood principle.[17–19] Consequently, repeated analyses of the data do not, in themselves, invalidate this Bayesian inference, provided the model is correctly specified. This contrasts with the frequentist setting, which requires explicit control of the Type I error rate to account for multiple analyses of the data, typically by adjusting decision boundaries.

Although several works have consistently highlighted that repeated interim analyses in Bayesian designs will alter their frequentist operating characteristics,[19–22] several authors argue that Bayesian methods do not suffer from the multiple-testing issues encountered in the frequentist framework. For example, Berry stated that "*the reason for stopping a trial affects frequentist measures but not Bayesian inferences*" and that "*There is no penalty for continual data analysis with the possibility of stopping*",[9] and Heuts and colleagues, wrote more recently that *"because Bayesian inference depends solely on observed and not unobserved data, no statistical penalty is required for multiple looks at the data".* [23] These statements are generally intended to mean that the inflation of the frequentist Type I error rate induced by repeated testing is a consequence of the frequentist framework and is therefore not directly relevant to Bayesian inference. However, the distinction between absence of Type I error inflation and whether Type I error is relevant or not is subtle and can easily be misunderstood, particularly by non-statisticians. Several authors have suggested considering alternative, Bayesian-consistent metrics when evaluating a trial design, a consideration which was also discussed in the recently released Food and Drug Administration (FDA) guidance for the use of Bayesian designs in clinical trials.[24]

However, the impact of repeated interim analyses on these alternative metrics has received comparatively little attention in the literature. The aim of this work is threefold: (i) to confirm that repeated interim analyses in Bayesian designs alter frequentist operating characteristics, (ii) to assess the impact of repeated interim analyses on alternative, Bayesian-consistent metrics, and (iii) to explore whether it is needed, possible, and meaningful to adjust the Bayesian designs to account for repeated interim analyses.

Section 2 introduces the rationale and definitions of frequentist and Bayesian operating characteristics, the statistical framework, and the simulation scenarios. In Section 3 we study the frequentist properties of both frequentist and Bayesian group-sequential designs. In Section 4, we analyse the Bayesian operating characteristics of Bayesian group-sequential designs. In Section 5, we extend the analysis to methods that aim to control frequentist operating characteristics of Bayesian trials, and investigate their relevance for the control of the Bayesian operating characteristics. Finally, Section 6 presents a comprehensive discussion of the results, their implications for current practice, and broader considerations regarding the appropriate frameworks for evaluating trial designs.

# 2. Methods

## 2.1. Rationale

In a fixed design with a single final analysis, one can discuss whether the primary goal is to make a binary decision of efficacy or non-efficacy of the treatment (decision only), to provide the most precise and accurate estimate of $\mu$ (statistical inference only), or both. In the case of a group-sequential design, however, binary decisions must be made at each interim analysis to decide whether the trial should continue. While operating characteristics such as bias, mean squared error, and the coverage of confidence or credible intervals are important when precise estimation of $\mu$ is the primary goal, we focus here on decision-related metrics. These metrics constitute the first step in interpreting trials that stop early and represent the most debated aspect of Bayesian interim analyses.

## 2.2. Operating characteristics

We consider a randomized controlled trial in which data are collected to estimate a continuous measure of a treatment effect, denoted $\mu$. The objective is to assess whether the treatment effect is positive, defined by $\mu > 0$. The event $V$ is defined as the binary event that a trial concludes that the treatment effect is positive (this conclusion can be reached at any analysis). The complementary event $\bar{V}$ is the absence of an efficacy conclusion (which can be defined as the absence of a conclusion of efficacy or a conclusion of futility).

Two decision-related frequentist operating characteristics are considered. The Type I error rate ($\alpha$) is the probability, calculated under the null hypothesis $\mu = 0$, of concluding efficacy. The power ($1 - \beta$), for a pre-specified value $\mu_1 > 0$, is the probability, calculated under the alternative hypothesis that $\mu = \mu_1$, of concluding efficacy.

$$\alpha = \Pr_{\mu=0}(V)$$

$$1 - \beta = \Pr_{\mu=\mu_1}(V)$$

A more Bayesian-consistent approach for computing decision-related operating characteristics is to evaluate them across all possible values of μ, rather than at fixed values, and to average the results with respect to a prior distribution on μ. In practice, this can be implemented via Monte Carlo simulation by generating trial data under parameter values μ randomly drawn from a prior distribution. We refer to this prior as the *reviewer's prior* ($\pi_{reviewer}$) as it represents the perspective of an external reviewer seeking to assess the properties of the trial design in light of their own prior beliefs about μ. Importantly, this prior may or may not be identical to the prior used for the analyses, which was chosen by the investigators responsible for designing the trial, and which we refer to as the *investigator's prior* ($\pi_{investigator}$).

Following this approach, two Bayesian operating characteristics analogous to the frequentist ones can be computed: the probability of concluding for efficacy conditional to $\mu \leq 0$ (analogous to Type I error rate, sometimes referred to as the Bayesian Type I error, which we call $A$) and the probability of concluding efficacy conditional to μ > 0 (analogous to the power, sometimes referred to as the Bayesian power or Bayesian conditional power, which we call $1 - B$).

$$A = \Pr_{\mu \sim \pi_{reviewer}} (V | \mu \leq 0)$$

$$1 - B = \Pr_{\mu \sim \pi_{reviewer}} (V | \mu > 0)$$

These probabilities are akin to the sensitivity and specificity of a diagnostic test where the binary conclusion of efficacy in the trial represents the test and the actual positivity of μ represents the reference standard. Therefore, similarly to the positive predictive value of a diagnostic test and following the definition proposed by Storey,[25] we can define the positive false discovery rate (pFDR) and the positive true discovery rate (pTDR), as:

$$pTDR = \Pr_{\mu \sim \pi_{reviewer}} (\mu > 0 | V)$$

$$pFDR = 1 - pTDR = \Pr_{\mu \sim \pi_{reviewer}} (\mu \leq 0 | \mathrm{V})$$

These are perhaps more aligned with the Bayesian paradigm, as they represent posterior beliefs about the sign of $\mu$, after the trial has concluded efficacy. Note that this differs from the usual calculation of the posterior belief about the sign of μ which is conditioned on the observed data.

Using the odds formulation of Bayes' theorem and defining $\mathrm{Odds}_{\mu \sim \pi}(E) = \frac{\Pr_{\mu \sim \pi}(E)}{1 - \Pr_{\mu \sim \pi}(E)}$ for any event $E$ and distribution $\pi$, we have the equivalence:

$$\mathrm{Odds}_{\mu \sim \pi_{reviewer}} (\mu > 0 | V) = \mathrm{Odds}_{\mu \sim \pi_{reviewer}} (\mu > 0) \cdot \frac{\Pr_{\mu \sim \pi_{reviewer}} (V | \mu > 0)}{\Pr_{\mu \sim \pi_{reviewer}} (V | \mu \leq 0)}$$

Where:

$$\frac{\Pr_{\mu \sim \pi_{reviewer}} (V | \mu > 0)}{\Pr_{\mu \sim \pi_{reviewer}} (V | \mu \leq 0)} = \frac{1 - B}{A}$$

is the Bayes factor of the event $V$ on the assertion that $\mu > 0$.

Moreover, we define the positive true discovery odds $pTDO = \underset{\mu \sim \pi_{reviewer}}{\text{Odds}}(\mu > 0|V)$, which is by definition of odds:

$$\text{pTDO} = \underset{\mu \sim \pi_{reviewer}}{\text{Odds}}(\mu > 0|V) = \frac{\underset{\mu \sim \pi_{reviewer}}{\text{Pr}}(\mu > 0|V)}{\underset{\mu \sim \pi_{reviewer}}{\text{Pr}}(\mu \leq 0|\text{V})} = \frac{\text{pTDR}}{\text{pFDR}}$$

In practice, this ratio represents the informative value of an efficacy conclusion of a trial. This is an interesting operating characteristic to consider because it has a clear and intuitive practical interpretation, and because it is inherently Bayesian. It derives from Bayes' theorem, and in the special case where $\pi_{reviewer}$ is symmetric and centered on zero, we have $\underset{\mu \sim \pi_{reviewer}}{\text{Odds}}(\mu > 0) = 1$, and the $pTDO$ becomes strictly equivalent to the Bayes factor of the event $V$ on the assertion that $\mu > 0$.

$$pTDO = \frac{\underset{\mu \sim \pi_{reviewer}}{\text{Pr}}(\mu > 0|V)}{\underset{\mu \sim \pi_{reviewer}}{\text{Pr}}(\mu \leq 0|\text{V})} = \frac{\underset{\mu \sim \pi_{reviewer}}{\text{Pr}}(V|\mu > 0)}{\underset{\mu \sim \pi_{reviewer}}{\text{Pr}}(V|\mu \leq 0)}$$

$$\text{pTDO} = \frac{pTDR}{pFDR} = \frac{1-B}{A}$$

Table 1 summarises the operating characteristics considered in this work.

**Table 1. Definition and description of the operating characteristics considered. $\mu$ is the efficacy parameter of interest, $V$ is the binary event that a trial concludes efficacy, $\pi_{reviewer}$ is a distribution which represents the prior belief about $\mu$ of an external reviewer seeking to assess the operating characteristics of a trial design.**

| **Operating characteristics** | **Description** |
|---|---|
| $\alpha = \underset{\mu=0}{\text{Pr}}(V)$ | The frequentist Type I error rate |
| $1-\beta = \underset{\mu=\mu_1}{\text{Pr}}(V)$ | The frequentist power |
| $A = \underset{\mu \sim \pi_{reviewer}}{\text{Pr}}(V|\mu \leq 0)$ | Probability that a trial concludes efficacy, given that the treatment is not efficacious |
| $1-B = \underset{\mu \sim \pi_{reviewer}}{\text{Pr}}(V|\mu > 0)$ | Probability that a trial concludes efficacy, given that the treatment is efficacious |
| $pTDR = \underset{\mu \sim \pi_{reviewer}}{\text{Pr}}(\mu > 0|V)$ | Probability that the treatment is efficacious, given that the trial concluded efficacy (positive true discovery rate) |
| $pFDR = 1 - pTDR = \underset{\mu \sim \pi_{reviewer}}{\text{Pr}}(\mu \leq 0|\text{V})$ | Probability that the treatment is not efficacious, given that the trial concluded efficacy (positive false discovery rate) |
| $\text{pTDO} = \frac{pTDR}{pFDR}$ | Positive true discovery odds, the informative value of an efficacy conclusion of a trial.<br>In all simulated scenarios it is actually equal to the Bayesian factor of the conclusion of efficacy on $\mu > 0$, since the prior odds of $\mu > 0$ is always assumed to be one. |

## 2.3. Statistical framework

We consider a trial with sequential enrolment. Let $X_1, X_2, \dots, X_n$ denote $n$ independent efficacy observations. Depending on the trial design, $X_i$ may represent an individual outcome (single-arm trial) or a treatment-effect estimate derived from a matched pair or another comparison unit (two-arm trial). We assume that the $X_i$ are independent and identically distributed random variables following a $\mathcal{N}(\mu, 1)$ distribution, where $\mu$ is the parameter of interest. By convention, results can be transposed to any known variance $\sigma^2$. We assume that larger absolute values of μ indicate greater effects, with positive values representing a beneficial effect.

Through data accrual, analyses (interim or final) are planned when the sample size reaches $(k_1, \dots, k_m)$ with $k_1 < \cdots < k_{m-1} < k_m = n$ and all $k_i \in \mathbb{N}$ for $i \in 1, \dots, m$. Each interim analysis is a ternary process that can conclude *efficacy*, *futility* or *ongoing inclusions*. The final analysis is a binary process that can conclude *efficacy* or *futility* only. Analyses start at $k = k_1$ up to the first interim analysis concluding *efficacy* or *futility,* or at the final analysis at $k_m = n$, if all interim analyses concluded to *ongoing inclusions*. In this process, we define the growing sample $X_{1,\dots,k} = (X_1, \dots, X_k)$.

For any frequentist analysis number $i$ at sample size $k = k_i$, we test the hypotheses:

$$H_0\colon \mu = 0 \text{ vs. } H_1\colon \mu > 0$$

The z-value, following a $\mathcal{N}(0,1)$ distribution under $H_0$ is given by:

$$Z_k = \frac{\overline{X_{1,\dots,k}} - \mu}{\sigma/\sqrt{k}} = \frac{\overline{X_{1,\dots,k}}\sqrt{k}}{\sigma} = \overline{X_{1,\dots,k}}\sqrt{k}$$

where $\overline{X_{1,\dots,k}} = \frac{1}{k}\sum_{i=1}^{k} X_i$.

Then, we compute a one-sided P-value $p_{value} = \Phi(-Z_k)$, where $\Phi$ denotes the cumulative distribution function of the standard normal distribution, and conclude *efficacy* if $p_{value}$ is below an efficacy threshold $e_i \in ]0,1[$, *futility* if $p_{value}$ is above a futility threshold $f_i \in ]0,1]$, and *ongoing inclusions* if $p_{value} \in [e_i, f_i]$. The futility threshold is set to one in scenarios with no futility interim analysis.

For the Bayesian interim/final analyses, we assume a normal prior $\pi_{investigator}$:

$$\mu \sim \mathcal{N}\left(\mu_{investigator}, \sigma^2_{investigator}\right)$$

Since both the prior and the likelihood are normally distributed, the posterior distribution is also normal (i.e., the prior and likelihood are conjugate):

$$\mu \mid X_{1,\dots,k} \sim \mathcal{N}\left(\mu_{posterior}, \sigma^2_{posterior}\right)$$

This property gives us a closed-form expression for the parameters of the posterior distribution of the investigator:

$$\mu_{posterior} = \frac{\tau_0 \mu_{investigator} + \tau \overline{X_{1,\dots,k}}}{\tau_0 + \tau} \tag{1}$$

$$\sigma^2_{posterior} = \frac{1}{\tau_0 + \tau} \tag{2}$$

where $\tau_0 = 1/\sigma^2_{investigator}$ and $\tau = k/\sigma^2 = k$ denote the information of the prior and the data, respectively.

Then, the posterior probability $\Pr\left(\mu > 0 \mid X_{1,\ldots,k}\right) = \Phi\left(\frac{\mu_{posterior}}{\sigma_{posterior}}\right)$ is calculated and the investigator concludes *efficacy* if this probability exceeds $1 - e_i$, *futility* if this probability is less than $1 - f_i$, and *ongoing inclusions* if this probability is in $[1 - f_i, 1 - e_i]$.

In the limit case where $\sigma^2_{investigator} \to +\infty$, we have $\tau_0 \to 0$, and:

$$\mu_{posterior} = \overline{X_{1,\ldots,k}}$$

$$\sigma^2_{posterior} = \frac{\sigma^2}{k}$$

So the posterior $Pr\left(\mu > 0 \mid X_{1,\ldots,k}\right) = \Phi\left(\frac{\overline{X_{1,\ldots,k}}\sqrt{k}}{\sigma}\right) = \Phi(Z_k) = 1 - p_{value}$ is exactly equal to one minus the frequentist one-sided *P*-value. Therefore, when $\sigma^2_{investigator} \to +\infty$ (non-informative prior), the frequentist and Bayesian decision rules are identical, leading to the same inference.

Table 2 summarises the main mathematical notations used in this work.

**Table 2. Summary of the main notations.**

| | |
|---|---|
| $n$ | Maximum sample size (if the trial does not stops early) |
| $X_1, \ldots, X_n$ | Random variables of the efficacy measures |
| $\mu$ | The efficacy parameter of interest. |
| $\pi_{investigator} = \mathcal{N}\left(0, \sigma^2_{investigator}\right)$ | The prior distribution of $\mu$, used for the analyses (interim or final).<br>Represents the prior belief of the investigators designing the trial. |
| $\pi_{reviewer} = \mathcal{N}\left(0, \sigma^2_{reviewer}\right)$ | The prior distribution of $\mu$, used to generate the data in the simulations in order to assess the operating characteristics of the trial.<br>Represents the prior belief of an external reviewer assessing the properties of the trial design. |
| $V$ | The binary event that a trial concludes that the treatment is efficacious (conclusion of efficacy at any analysis). |
| $m$ | The number of analyses (interim and final) |
| $k_i$ | Sample size at analysis $i \in \{1, \ldots m\}$. |
| $e_i$ | Efficacy significance threshold at analysis $i \in \{1, \ldots m\}$.<br>The frequentist analysis concludes efficacy if $p_{value} < e_i$.<br>The Bayesian analysis concludes efficacy if $Pr\left(\mu > 0 \middle| X_{1,\ldots,k_i}\right) > 1 - e_i$ |
| $f_i$ | Futility significance threshold at interim analysis $i$.<br>The frequentist analysis concludes futility if $p_{value} > f_i$.<br>The Bayesian analysis concludes futility if $\Pr\left(\mu > 0 \middle| X_{1,\ldots,k_i}\right) < 1 - f_i$ |

## 2.4. Simulations

We consider trials with sequential enrolment and interim analyses at prespecified sample sizes $k_1, \dots, k_m$, with a final analysis performed after $n = 500$ efficacy measures have been observed. We consider both a single final analysis and group-sequential designs, with $m \in \{1, 2, 5, 10, 20, 50, 100, 250, 500\}$ the number of analyses. The interim analyses are equally spaced in terms of information, with $m = 500$ corresponding to the extreme case of one analysis after each newly observed efficacy measure. Formally, given the data $X = (X_1, X_2, \dots, X_{500})$, we perform frequentist and Bayesian analyses on cumulative subsets $X_{1,\dots,k_i}$ where $k_i = \frac{500\,i}{m}$ for $i \in \{1, \dots, m\}$.

To assess the Type I error rates of frequentist and Bayesian designs, we simulate trials under the null hypothesis $X_i \sim \mathcal{N}(\mu, 1)$ with $\mu = 0$ for all $i \in \{1, \dots, 500\}$. The Type I error rate is estimated as the proportion of trials that declare efficacy at any analysis.

To assess power, we simulate trials under an alternative hypothesis with $\mu = 0.1112$. This value was chosen such that a design with a single frequentist analysis at $n = 500$ achieves 80% power at a 5% significance level. This is given by:

$$\Delta = \sqrt{\frac{\left(\Phi^{-1}(0.95) + \Phi^{-1}(0.80)\right)^2}{500}}$$

Power is then estimated as the proportion of trials that declare efficacy at any analysis.

To evaluate the Bayesian operating characteristics, we simulate trials such as:

$$X_i \sim \mathcal{N}(\mu, 1)$$

$$\mu \sim \mathcal{N}\left(0, \sigma^2_{reviewer}\right)$$

We then compute the proportion of trials that concluded efficacy given that $\mu \leq 0$ (i.e. $A$), and the proportion that concluded efficacy given that $\mu > 0$ (i.e. $1 - B$). Conversely, among trials that concluded efficacy, we compute the proportion of trials which were simulated under $\mu > 0$ (i.e. pTDR), and under $\mu \leq 0$ (i.e. pFDR). If no trial concluded efficacy in the simulations, pTDR and pFDR are undefined. In cases where pFDR was equal to zero, we reported $pTDO = pTDR/pFDR$ as the lower bound value for the limit of the number of simulations.

Table 3 shows the six scenarios considered. For the first scenario, we defined all significance thresholds $e_i = 0.05$ for $i \in \{1, \dots, m\}$ and $f_i = 1$ for $i \in \{1, \dots, m-1\}$, *i.e.* no correction for multiple testing and no futility interim analyses. For the second, third and fourth scenarios, we studied various futility stopping rules $f_i = 0.95$, $f_i = 0.90$ and $f_i = 0.50$, $i \in \{1, \dots, m-1\}$, respectively, maintaining $e_i = 0.05$, $i \in \{1, \dots, m\}$. For the fifth scenario we chose $f_i = 1$ and computed efficacy thresholds $e_i$ according to Pocock's significance boundaries and the sample size $n$ to control the one-sided Type I error rate $\alpha$ at 0.05 and the power $1 - \beta$ at 0.80. For the sixth scenario we chose $f_i = 1$ and computed efficacy thresholds $e_i$ and sample size $n$ by simulation to control the Bayesian operating characteristics $A$ at 0.05 and $1 - B$ at 0.80. For each scenario, 100 000 trials were simulated. The core procedures for the simulations and analyses were implemented in the R programming language,[26] and the code, together with pseudocode explaining the underlying algorithms, is provided in the supplementary methods. All simulations, analyses and presentation of results were performed using R, version 4.5.3, and the complete set of programs required to reproduce the results presented here is freely available online at https://github.com/ClementMassonnaud/BayesianInterim.

**Table 3. Summary of the simulation scenarios considered. See Table 2 for a description of the notations used.**

| Trial design | Stopping rules scenarios | Operating characteristics evaluated |
|---|---|---|
| $X_i \sim \mathcal{N}(\mu, 1)$<br>$\mu = 0$ and $\mu = \mu_1 = 0.1112$<br>$\sigma_{investigator} \in \{0.05, 0.1, 0.25, 1, 10\}$ | **For all scenarios:**<br>$e_i = 0.05$ for $i \in \{1, \ldots, m\}$<br><br>**Scenario 1:** $f_i = 1.00$<br>**Scenario 2 :** $f_i = 0.95$<br>**Scenario 3 :** $f_i = 0.90$<br>**Scenario 4 :** $f_i = 0.50$<br><br>for $i \in \{1, \ldots, m-1\}$ | $\alpha = \Pr_{\mu=0}(V)$<br>$1 - \beta = \Pr_{\mu=\mu_1}(V)$ |
| $X_i \sim \mathcal{N}(\mu, 1)$<br>$\mu \sim \mathcal{N}(0, \sigma_{reviewer}^2)$<br>$\sigma_{reviewer} \in \{0.001, 0.01, 0.05, 0.1, 0.25, 1, 10, 100\}$<br>$\sigma_{investigator} \in \{0.001, 0.01, 0.05, 0.1, 0.25, 1, 10, 100\}$ | | $A = \Pr_{\mu \sim \pi_{reviewer}}(V \mid \mu \leq 0)$<br>$1 - B = \Pr_{\mu \sim \pi_{reviewer}}(V \mid \mu > 0)$<br>$pTDR = \Pr_{\mu \sim \pi_{reviewer}}(\mu > 0 \mid V)$<br>$pFDR = \Pr_{\mu \sim \pi_{reviewer}}(\mu \leq 0 \mid V)$<br>$pTDO = pTDR/pFDR$ |
| $X_i \sim \mathcal{N}(\mu, 1)$<br>$\mu = 0$ and $\mu = \mu_1 = 0.1112$<br>$\sigma_{investigator} = 10$ | **Scenario 5**: $e_i, e_m$ and $n$ computed according to Pocock's boundaries to achieve $\alpha = 0.05$ and $1 - \beta = 0.80$<br><br>$f_i = 1.00$<br>$e_m = f_m$<br><br>$i \in \{1, \ldots, m-1\}$ | $\alpha = \Pr_{\mu=0}(V)$<br>$1 - \beta = \Pr_{\mu=\mu_1}(V)$<br>Expected sample size |
| $X_i \sim \mathcal{N}(\mu, 1)$<br>$\mu \sim \mathcal{N}(0, \sigma_{reviewer}^2)$<br>$\sigma_{reviewer} \in \{0.001, 0.01, 0.05, 0.1, 0.25, 1, 10, 100\}$<br>$\sigma_{investigator} = 0.25$ | **Scenario 6**: $e_i, e_m$ and $n$ derived by simulation to achieve $A = 0.05$ and $1 - B = 0.80$ under $\mu \sim \mathcal{N}(0, 0.25^2)$<br><br>$f_i = 1.00$<br>$e_m = f_m$<br><br>$i \in \{1, \ldots, m-1\}$ | $A = \Pr_{\mu \sim \pi_{reviewer}}(V \mid \mu \leq 0)$<br>$1 - B = \Pr_{\mu \sim \pi_{reviewer}}(V \mid \mu > 0)$<br>$pTDR = \Pr_{\mu \sim \pi_{reviewer}}(\mu > 0 \mid V)$<br>$pFDR = \Pr_{\mu \sim \pi_{reviewer}}(\mu \leq 0 \mid V)$<br>$pTDO = pTDR/pFDR$ |

# 3. Frequentist operating characteristics of frequentist and Bayesian designs

## 3.1. Allowing early stopping for efficacy only

Figure 1 shows the Type I error rate $(\alpha)$ as a function of the number of analyses (panel A) and the power $(1-\beta)$ as a function of the Type I error rate (panel B), for both frequentist and Bayesian decision rules ($p_{value} < e_i$ and $\Pr_{\mu \sim \pi_{investigator}} \left(\mu > 0 \middle| X_{1,\dots,k_i}\right) > 1 - e_i, i \in \{1, \dots, m\}$, respectively) in scenario number one ($e_1 = \cdots = e_m = 0.05, f_1 = \cdots = f_{m-1} = 1$).

As expected, the Type I error rate for the frequentist is approximately 5% with a single analysis and increases to around 40% when analyses are performed after each new observation. For Bayesian analyses, the Type I error rate depends on the prior used for analyses. With a weakly informative prior (e.g. $\sigma^2_{investigator} = 10$), the error rates approximate those of the frequentist analyses. This is consistent with the fact that, as $\sigma^2_{investigator} \to \infty$, $\Pr_{\mu \sim \pi_{investigator}} \left(\mu > 0 \middle| X_{1,\dots,k_i}\right)$ tends towards $1 - p_{value}$ so that the decision rule of the Bayesian and frequentist design converges (see section 2.2. for derivation).

Also, the Type I error rate increases much faster than the power with the number of interim analyses, as is reflected by the ratio $(1-\beta)/\alpha$, which drops significantly from 16 for one analysis to 2 for 500 analyses in the frequentist trials and Bayesian trials with the weakly informative prior $\mathcal{N}(0,10^2)$, and from 44 to 20 in the Bayesian trial with the strongly informative prior $\mathcal{N}(0,0.05^2)$ (Table 4).

Although it is possible to control the Type I error rate for a given number of analyses by choosing a sufficiently informative prior (i.e. reducing $\sigma^2_{investigator}$), this comes at the cost of reduced power. As shown in Figure 1.B., for fixed efficacy significance thresholds $e_i$ and sample size $n$, it is not possible to increase the number of interim analyses while both controlling the Type I error rate and maintaining high power. Only in the case of a single analysis with a weakly informative prior can we achieve both targets of 5% Type I error rate and 80% power. Supplementary table S1 shows $\frac{1-\beta}{\alpha}$ ratios for other values of $\sigma_{investigator}$.

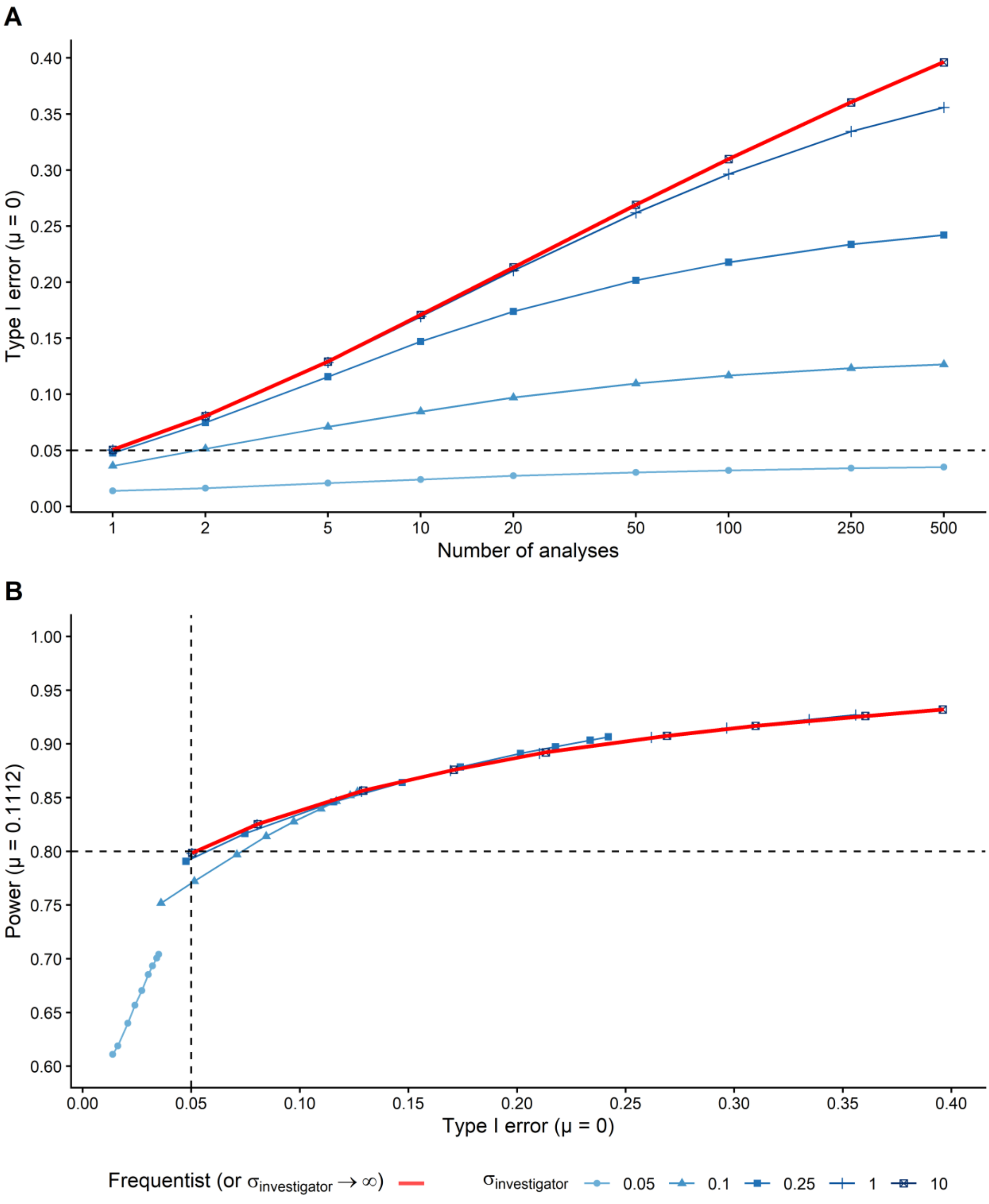

**Figure 1. Type I error rate and power of frequentist and Bayesian trial designs with efficacy stopping rules (frequentist: one-sided $p_{value} < 0.05$, Bayesian: $Pr(\mu > 0|X_{1,\dots,k_i}) > 0.95, i \in \{1, \dots, m\}$), no futility stopping rules, and no correction for multiplicity of analyses. $\sigma_{investigator}$ is the standard deviation of the prior $\mathcal{N}(0, \sigma^2_{investigator})$ used for the Bayesian analyses. Panel A: Type I error rate (y-axis) as a function of the number of equally-spaced analyses (x-axis logarithm scale) and $\sigma_{investigator}$. Panel B: Power (y-axis) as a function of Type I error (x-axis), according to the number of equally-spaced analyses and $\sigma_{investigator}$, where each point represents a design with varying number of analyses (1 to 500), increasing from left to right.**

**Table 4. Type I error rate (α), power $(1-\beta)$, and ratio $(1-\beta)/\alpha$ of frequentist and Bayesian trial designs with efficacy stopping rules (frequentist: one-sided $p_{value} < 0.05$, Bayesian: $Pr(\mu > 0 | X_{1,\dots,k_i}) > 0.95, i \in \{1, \dots, m\}$), no futility stopping rules, and no correction for multiplicity of analyses. $\sigma_{investigator}$ is the standard deviation of the prior $\mathcal{N}(0, \sigma^2_{investigator})$ used for the Bayesian analyses.**

| Number of analyses | Frequentist trials | | | Bayesian trials | | | | | |
|---|---|---|---|---|---|---|---|---|---|
| | | | | $\sigma_{investigator} = 10$ | | | $\sigma_{investigator} = 0.05$ | | |
| | Type I error rate (α) | Power $(1-\beta)$ | $\frac{1-\beta}{\alpha}$ | Type I error rate (α) | Power $(1-\beta)$ | $\frac{1-\beta}{\alpha}$ | Type I error rate (α) | Power $(1-\beta)$ | $\frac{1-\beta}{\alpha}$ |
| 1 | 0.05 | 0.80 | 16 | 0.05 | 0.80 | 16 | 0.01 | 0.61 | 44 |
| 2 | 0.08 | 0.83 | 10 | 0.08 | 0.83 | 10 | 0.02 | 0.62 | 38 |
| 5 | 0.13 | 0.86 | 7 | 0.13 | 0.86 | 7 | 0.02 | 0.64 | 31 |
| 10 | 0.17 | 0.88 | 5 | 0.17 | 0.88 | 5 | 0.02 | 0.66 | 27 |
| 20 | 0.21 | 0.89 | 4 | 0.21 | 0.89 | 4 | 0.03 | 0.67 | 25 |
| 50 | 0.27 | 0.91 | 3 | 0.27 | 0.91 | 3 | 0.03 | 0.69 | 23 |
| 100 | 0.31 | 0.92 | 3 | 0.31 | 0.92 | 3 | 0.03 | 0.69 | 22 |
| 250 | 0.36 | 0.93 | 3 | 0.36 | 0.93 | 3 | 0.03 | 0.70 | 21 |
| 500 | 0.40 | 0.93 | 2 | 0.40 | 0.93 | 2 | 0.04 | 0.70 | 20 |

## 3.2. Allowing early stopping for efficacy or futility

We repeated the simulations with the added interim futility stopping rules $f_1 = \cdots = f_{m-1} = 0.95$ (scenario 2), $f_1 = \cdots = f_{m-1} = 0.90$ (scenario 3) and $f_1 = \cdots = f_{m-1} = 0.50$ (scenario 4), while keeping $e_1 = \cdots = e_m = 0.05$. Introducing a futility stopping rule reduced the Type I error rate but it remained well above 5% in most cases when performing multiple analyses, while power was systematically reduced (Figure 2 and supplementary table S2). Decreasing the futility threshold to 0.50 further decreased the Type I error rate, but at the cost of additional loss of power. Moreover, in contrast to scenarios allowing early stopping for efficacy only, a large number of interim analyses could decrease the power in addition to increase the Type I error rate.

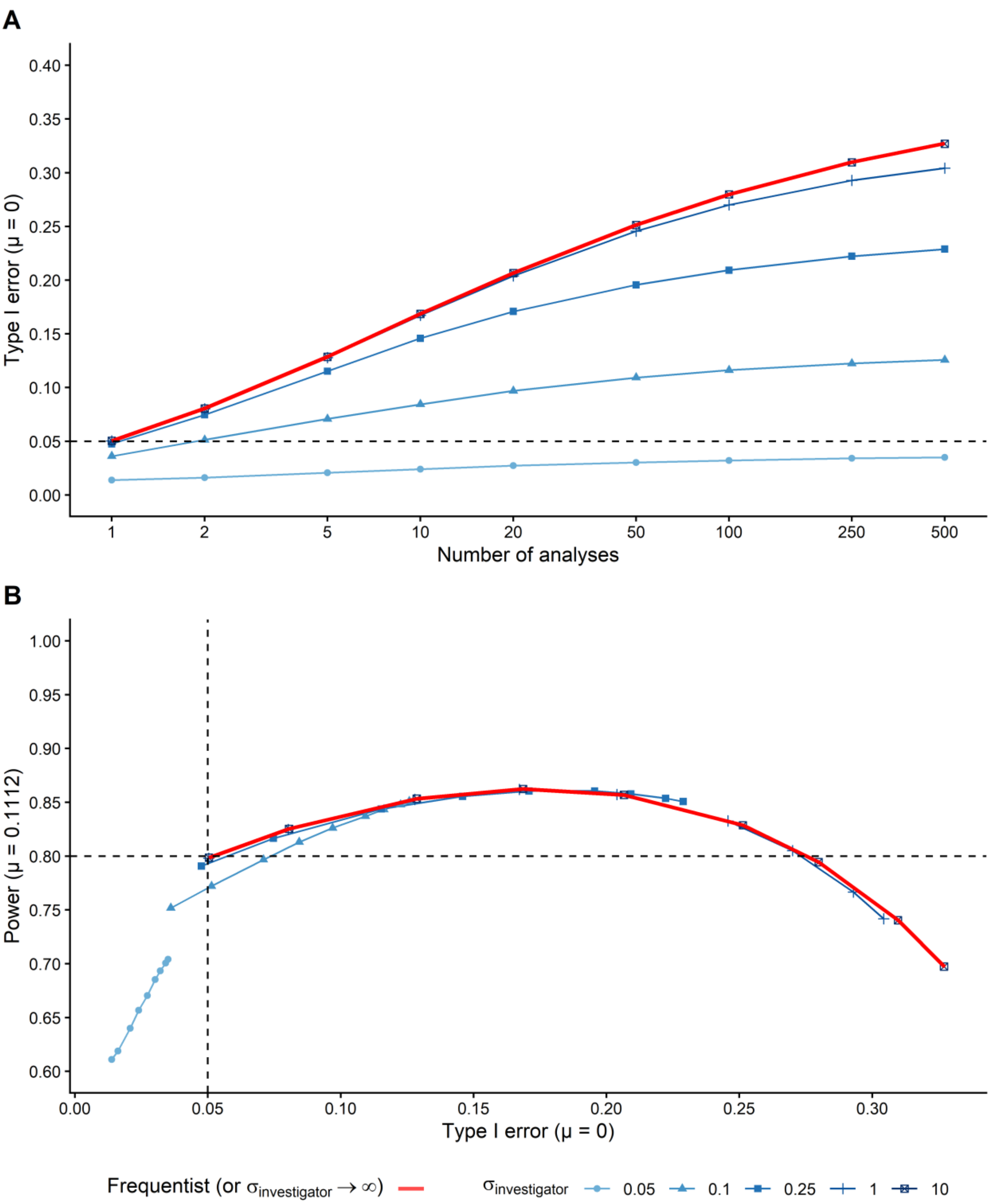

**Figure 2. Type I error rate and power of frequentist and Bayesian trial designs with efficacy stopping rules (frequentist: one-sided $p_{value} < 0.05$, Bayesian: $Pr(\mu > 0|X_{1,\ldots,k_i}) > 0.95, i \in \{1, \ldots, m\}$), and added futility stopping rules (frequentist: one-sided $p_{value} > 0.90$, Bayesian: $Pr(\mu > 0|X_{1,\ldots,k_i}) < 0.10, i \in \{1, \ldots, m\}$), with no correction for multiplicity of analyses. $\sigma_{investigator}$ is the standard deviation of the prior $\mathcal{N}(0, \sigma^2_{investigator})$ used for the Bayesian analyses. Panel A: Type I error rate (y-axis) as a function of the number of equally-spaced analyses (x-axis, logarithmic scale) and $\sigma_{investigator}$. Panel B: Power (y-axis) as a function of Type I error rate (x-axis), according to the number of equally-spaced analyses and $\sigma_{investigator}$, where each point represents a design with varying number of analyses (1 to 500), increasing from left to right.**

# 4. Bayesian properties of Bayesian trial designs

## 4.1. Allowing early stopping for efficacy only

Figure 3 shows the two operating characteristics $1 - \mathrm{B} = \Pr_{\mu \sim \pi_{reviewer}}(V|\mu > 0)$ and $A = \Pr_{\mu \sim \pi_{reviewer}}(V|\mu \leq 0)$ computed under $\pi_{reviewer} = \mathcal{N}(0, \sigma^2_{reviewer})$ of Bayesian trials designed with various investigator priors $\mathcal{N}(0, \sigma^2_{investigator})$.

We see that in the special case where $\sigma_{reviewer} = \sigma_{investigator}$ the risk $A$ is always below 5%, even when performing one analysis after each new observation (Figure 3.D.). For non-informative priors, such as $\mathcal{N}(0,100^2)$, this risk approaches zero and the probability that the trial concludes efficacy conditional to $\mu > 0$ approaches 100%. This is explained by the fact that for $\sigma_{reviewer} = 100$, we simulate trials for which the treatment effect $\mu$ is often very large, positive or negative. In these cases where $\mu \gg 0$, it is easy to conclude efficacy (usually early in the trial), and it is unlikely to conclude efficacy in cases where $\mu \ll 0$. Conversely, for strongly informative priors, such as $\mathcal{N}(0,0.01^2)$, we generate trials with smaller effects and the posterior distribution of $\mu$ is strongly pulled towards zero. Therefore, it becomes unlikely to conclude efficacy, whether $\mu$ was in fact slightly above or below zero. The point $\sigma_{reviewer} = \sigma_{investigator} = 0.25$ is inbetween and has the highest risk $A$ among the cases where the external reviewer and investigator agree on the prior.

However, we see that the operating characteristics vary when $\sigma_{reviewer} \neq \sigma_{investigator}$. Typically, if an external reviewer believes that $\mu$ is more likely to be closer to zero than what was assumed by the investigators designing the trial ($\sigma_{reviewer} < \sigma_{investigator}$), the operating characteristics deteriorate quickly and increasingly with the number of interim analyses. We also confirm that when using a non-informative prior for the analyses, the risks $A$ converge to the frequentist Type I error rates when $\sigma_{reviewer}$ approaches zero. The results for different number of interim analyses are shown in appendix (supplementary figures S1 to S9).

Table 5 shows the informative value of an efficacy conclusion of a trial, expressed as the positive true discovery odds $pTDO = \mathrm{pTDR}/\mathrm{pFDR}$. These odds are positively correlated with $\sigma_{reviewer}$ and negatively correlated with $\sigma_{investigator}$. This is explained by the fact that if an external reviewer believes strong effects are likely ($\sigma_{reviewer}$ is large), they will think that an error in the sign of μ is very unlikely and would therefore consider that they would gain strong information from a trial that concluded efficacy, especially if the prior used for the analysis was informative around zero. Conversely, if the reviewer is more sceptical than the investigator ($\sigma_{reviewer} < \sigma_{investigator}$), the informative value of an efficacy conclusion can be very low (e.g. $pTDO = 2.8$ for $\sigma_{reviewer} = 0.05$, $\sigma_{investigator} = 10$, and $m = 100$ analyses).

Critically, we see that even in the special case $\sigma_{reviewer} = \sigma_{investigator}$, the pTDO decreases with the number of interim analyses. This means that even if there is a perfect alignment between the investigator's and reviewer's priors, the informative value of an efficacy conclusion of a trial deteriorates with the number of interim analyses. The pTDO for other priors and the individual values pTDR and pFDR for different number of analyses and priors are shown in appendix (supplementary table S3 and figures S10 to S18).

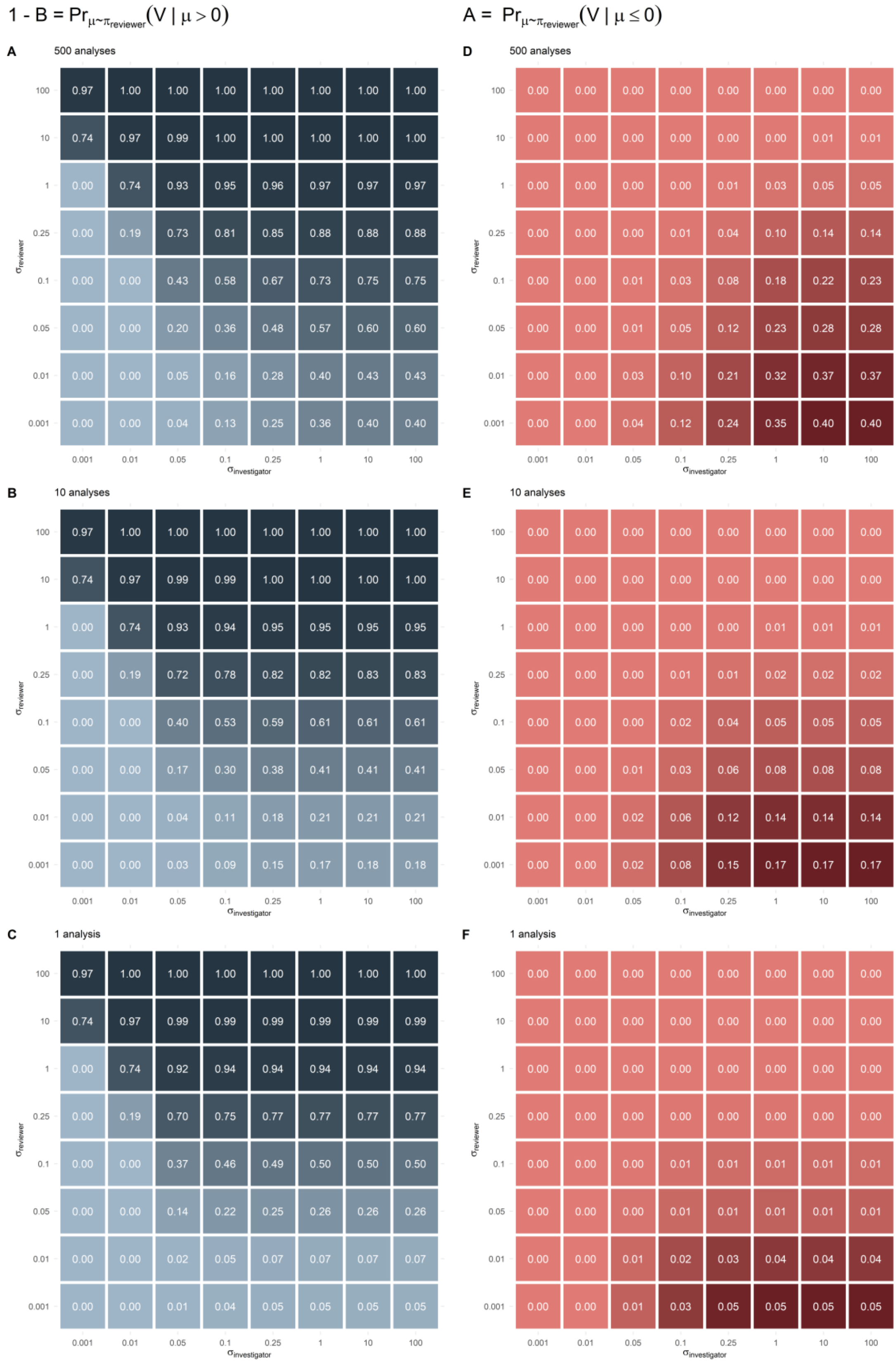

**Figure 3. Probability of concluding efficacy (binary event $V$) conditional to $\mu > 0$ (panels A, B, C) and $\mu \leq 0$ (panels D, E, F), according the number of analyses, in Bayesian trials with efficacy stopping rules $Pr(\mu > 0|X_{1,\dots,k_i}) > 0.95, i \in \{1,\dots,m\}$, no futility stopping rules, and no correction for multiplicity of analyses. $\sigma_{investigator}$ is the standard deviation of the prior $\pi_{investigator} = \mathcal{N}(0, \sigma^2_{investigator})$ used for the analyses (interim and final), and $\sigma_{reviewer}$ is the standard deviation of the prior $\pi_{reviewer} = \mathcal{N}(0, \sigma^2_{reviewer})$ used to compute the operating characteristics of the trial.**

**Table 5. Informative value of an efficacy conclusion of a trial expressed as the positive true discovery odds $pTDO = pTDR/pFDR$, in Bayesian trials with efficacy stopping rules $Pr(\mu > 0|X_{1,\dots,k_i}) > 0.95, i \in \{1, \dots, m\}$, no futility stopping rules, and no correction for multiplicity of analyses. $\sigma_{investigator}$ is the standard deviation of the prior $\mathcal{N}(0, \sigma^2_{investigator})$ used for the analyses, and $\sigma_{reviewer}$ is the standard deviation of the prior $\mathcal{N}(0, \sigma^2_{reviewer})$ used to compute the operating characteristics of the trial.**

| Number of analyses | $\sigma_{reviewer}$ | $\text{pTDO} = \frac{pTDR}{pFDR} = \frac{Pr_{\mu\sim\pi_{reviewer}}(\mu > 0\|V)}{Pr_{\mu\sim\pi_{reviewer}}(\mu \leq 0\|V)}$ | | | | |
|---|---|---|---|---|---|---|
| | | $\sigma_{investigator}$ | $\sigma_{investigator}$ | $\sigma_{investigator}$ | $\sigma_{investigator}$ | $\sigma_{investigator}$ |
| | | **0.05** | **0.1** | **0.25** | **1** | **10** |
| **1** | **0.05** | 47 | 24 | 20 | 19 | 19 |
| **2** | **0.05** | 36 | 16 | 12 | 12 | 12 |
| **5** | **0.05** | 29 | 12 | 7.9 | 7.2 | 7.2 |
| **10** | **0.05** | 27 | 10 | 6.3 | 5.4 | 5.4 |
| **20** | **0.05** | 24 | 9.3 | 5.4 | 4.3 | 4.2 |
| **50** | **0.05** | 21 | 8.4 | 4.6 | 3.4 | 3.3 |
| **100** | **0.05** | 21 | 8.1 | 4.3 | 3.0 | 2.8 |
| **250** | **0.05** | 20 | 7.8 | 4.1 | 2.6 | 2.4 |
| **500** | **0.05** | 19 | 7.6 | 3.9 | 2.5 | 2.2 |
| **1** | **0.1** | 249 | 90 | 70 | 66 | 66 |
| **2** | **0.1** | 166 | 52 | 39 | 36 | 35 |
| **5** | **0.1** | 142 | 37 | 21 | 18 | 18 |
| **10** | **0.1** | 110 | 31 | 15 | 13 | 12 |
| **20** | **0.1** | 99 | 27 | 12 | 9.2 | 8.9 |
| **50** | **0.1** | 82 | 23 | 10 | 6.6 | 6.2 |
| **100** | **0.1** | 76 | 21 | 9.2 | 5.5 | 5.0 |
| **250** | **0.1** | 76 | 20 | 8.4 | 4.4 | 3.9 |
| **500** | **0.1** | 70 | 19 | 8.0 | 4.0 | 3.3 |
| **1** | **0.25** | 999 | 499 | 332 | 332 | 332 |
| **2** | **0.25** | 999 | 249 | 166 | 142 | 142 |
| **5** | **0.25** | 999 | 166 | 82 | 70 | 66 |
| **10** | **0.25** | 999 | 124 | 55 | 42 | 41 |
| **20** | **0.25** | 499 | 110 | 41 | 27 | 26 |
| **50** | **0.25** | 499 | 90 | 31 | 17 | 16 |
| **100** | **0.25** | 499 | 82 | 27 | 13 | 12 |
| **250** | **0.25** | 499 | 76 | 23 | 9.9 | 8.0 |
| **500** | **0.25** | 332 | 76 | 22 | 8.5 | 6.4 |
| **1** | **1** | ≥ 1000 | 999 | 999 | 999 | 999 |
| **2** | **1** | ≥ 1000 | 999 | 499 | 499 | 499 |
| **5** | **1** | ≥ 1000 | 499 | 332 | 249 | 249 |
| **10** | **1** | ≥ 1000 | 499 | 249 | 199 | 199 |
| **20** | **1** | ≥ 1000 | 499 | 199 | 124 | 124 |
| **50** | **1** | 999 | 332 | 142 | 76 | 70 |
| **100** | **1** | 999 | 332 | 124 | 55 | 47 |
| **250** | **1** | 999 | 332 | 110 | 39 | 29 |
| **500** | **1** | 999 | 332 | 99 | 32 | 21 |
| **1** | **10** | ≥ 1000 | ≥ 1000 | ≥ 1000 | ≥ 1000 | ≥ 1000 |
| **2** | **10** | ≥ 1000 | ≥ 1000 | ≥ 1000 | ≥ 1000 | ≥ 1000 |
| **5** | **10** | ≥ 1000 | ≥ 1000 | ≥ 1000 | ≥ 1000 | ≥ 1000 |
| **10** | **10** | ≥ 1000 | ≥ 1000 | 999 | 999 | 999 |
| **20** | **10** | ≥ 1000 | ≥ 1000 | 999 | 999 | 999 |
| **50** | **10** | ≥ 1000 | ≥ 1000 | 999 | 499 | 499 |
| **100** | **10** | ≥ 1000 | ≥ 1000 | 999 | 499 | 499 |
| **250** | **10** | ≥ 1000 | ≥ 1000 | 999 | 332 | 249 |
| **500** | **10** | ≥ 1000 | ≥ 1000 | 999 | 332 | 199 |

## 4.2. Allowing early stopping for efficacy or futility

We repeated the simulations with the added interim futility stopping rules $f_1 = \cdots = f_{m-1} = 0.95$ (scenario 2), $f_1 = \cdots = f_{m-1} = 0.90$ (scenario 3) and $f_1 = \cdots = f_{m-1} = 0.50$ (scenario 4), while keeping $e_1 = \cdots = e_m = 0.05$ for efficacy stopping rules. The operating characteristics $A$, $1 - B$, pTDR, pFDR, and the pTDO computed under these scenarios are shown in supplementary figures S19 to S72. Similarly to the frequentist operating characteristics, introducing futility stopping rules slightly reduced the risks $A$ but also reduced $1 - B$. The pFDR and pTDR were also reduced, but much less than $A$ and $1 - B$, as these are probabilities conditional on the fact that the trial has concluded efficacy. The pTDO behaves similarly to the scenario with only efficacy stopping rules, that is it decreases with the number of interim analyses, even when there is perfect alignment between the investigator's and reviewer's priors. The pTDO was even worse than without a futility stopping rule when the number of analyses was high and the investigator had a weakly informative prior while the external reviewer had a strongly informative prior (supplementary tables S4 to S6). For instance, for $\sigma_{investigator} = 10$, $\sigma_{reviewer} = 0.05$, and $m = 50$ analyses the $pTDO$ was 3.3 without a futility stopping rule and was 2.4 with $f_i = 0.50$ (supplementary table S6). This could be explained by the fact that $1 - B$ was reduced more strongly than $A$ by the futility rule, so that $pTDO = (1 - B)/A$ decreased. In that case, the futility rule reduces the chance of concluding efficacy while decreasing the informative value of an efficacy conclusion.

# 5. Controlling the operating characteristics of frequentist and Bayesian designs

## 5.1. From the frequentist perspective

The methods to control the Type I error rate inflation due to multiple testing in frequentist group-sequential trials are well known.[27–31] It is equally possible to control the frequentist Type I error rate inflation in Bayesian group-sequential trials by carefully planning the trial and calibrating the decision rules. In fact, under simple conditions, it is possible to directly transcribe frequentists control methods to Bayesian equivalents. For instance, in a frequentist hypothesis-testing framework that rejects $H_0$ when any interim statistic $p_{value}$ is less than a common boundary $e_1 = \cdots = e_m$, the method proposed by Pocock is to decrease $e_i$ such that the overall Type I error rate is controlled at the desired level, say 5%. Although the maximum sample size $n$ must be increased accordingly to maintain adequate power, e.g. 80%, the expected sample size under an alternative hypothesis $H_1$ of a non-null treatment effect should decrease, since the trial as a chance to demonstrate efficacy at an interim analysis and terminate early.

Similarly, in a Bayesian framework, with an efficacy decision rule $\Pr(\mu > 0 | X_{1,\dots,k_i}) > 1 - \mathrm{e}_i$, the efficacy thresholds $e_i$ can be set to ensure that the overall Type I error is controlled at the desired level. Likewise, the sample size has to be increased accordingly to maintain adequate power. As demonstrated in section 2.2., under the simple scenario of a normal-normal conjugate prior and model, with a non-informative prior, there is an equivalence between the frequentist boundary $e_i$ and the Bayesian efficacy threshold $1 - e_i$, since $\lim_{\sigma_{investigator} \to +\infty} \left(\Pr(\mu > 0 | X_{1,\dots,k_i})\right) = 1 - p_{value}$.

Therefore, the Bayesian and frequentist decision rules are equivalent for non-informative priors so the same methods can be used to control the Type I error rate.

We can show that the asymptotic equivalence is already very good with a weakly informative prior $\mathcal{N}(0,10^2)$. We simulated trials under $H_0: \mu = 0$ and $H_1: \mu = 0.1112$, and for each trial, we performed a single final analysis, or 2, 5, 10, 20, 50, 100, 250, or 500 equally-spaced analyses. We concluded efficacy if $p_{value} < e_i$ for frequentist analyses, and if $\Pr(\mu > 0 | X_{1,\dots,k_i}) > 1 - e_i$ for Bayesian analyses, with $e_i$ adjusted for the number of analyses. For a single analysis, the sample size that ensures 80% power is $n = 500$. The Pocock boundary $e_i$ and sample size $n$ that ensures 5% overall Type I error and 80% power according to the number of analyses can be easily computed numerically, or found using pre-existing statistical software. Examples of such computations are shown in appendix (section 1.2).

Table 6 shows the operating characteristics of the simulated trials. We see that they are appropriately controlled for all scenarios and are equivalent for both the frequentist and Bayesian designs. Interestingly, although the maximum sample size increases monotonically with the number of analyses, the expected sample size decreases only up to 10 analyses, then increases again. In these scenarios, performing 20 analyses or more no longer provides any benefit. Due to the equivalence between frequentist and Bayesian analyses with weakly informative priors, other methods to control the multiplicity (i.e. O'Brien-Fleming or alpha-spending functions) could also be applied to the Bayesian analyses, and would retain their advantages and limits. Using more informative priors for the Bayesian analyses would lose the direct equivalence with frequentist inference. However, as the posterior distribution is a monotonic transformation of the $p_{value}$ in the normal-normal conjugate prior-model case for a given sample size $k_i$, an equivalence could still be found between the frequentist significance threshold and the Bayesian decision rule, although with varying $e_i$. In other cases, the Bayesian thresholds $1 - e_i$ can be estimated numerically to obtain the same alpha spending function as in a frequentist design.

**Table 6. Frequentist operating characteristics of frequentist and Bayesian trials, controlled for multiplicity of analyses. For frequentist trials, the boundary $e_i$ for one-sided tests $p_{value} < e_i$ was derived following Pocock's method, and the Bayesian decision rule $Pr(\mu > 0 \mid X_{1,\dots,k_i}) > 1 - e_i$ was used with the same $e_i$ values with $\pi_{investigator} = \mathcal{N}(0, 10^2)$. Frequentist Type I error rates were computed under $H_0: \mu = 0$, and the power and expected sample sizes were computed under $H_1: \mu = 0.1112$.**

| Number of analyses | $e_i$ | $1 - e_i$ | Maximum sample size ($n$) | Frequentist analyses | | | Bayesian analyses | | |
|---|---|---|---|---|---|---|---|---|---|
| | | | | Type I error rate | Power | Expected sample size | Type I error rate | Power | Expected sample size |
| 1 | 0.050 | 0.950 | 500 | 0.05 | 0.8 | 500 | 0.05 | 0.8 | 500 |
| 2 | 0.030 | 0.970 | 561 | 0.05 | 0.8 | 423 | 0.05 | 0.8 | 423 |
| 5 | 0.017 | 0.983 | 627 | 0.05 | 0.8 | 394 | 0.05 | 0.8 | 394 |
| 10 | 0.012 | 0.988 | 669 | 0.05 | 0.8 | 393 | 0.05 | 0.8 | 393 |
| 20 | 0.008 | 0.992 | 705 | 0.05 | 0.8 | 399 | 0.05 | 0.8 | 399 |
| 50 | 0.006 | 0.994 | 748 | 0.05 | 0.8 | 413 | 0.05 | 0.8 | 413 |
| 100 | 0.005 | 0.995 | 775 | 0.05 | 0.8 | 423 | 0.05 | 0.8 | 423 |
| 250 | 0.004 | 0.996 | 813 | 0.05 | 0.8 | 442 | 0.05 | 0.8 | 442 |
| 500 | 0.003 | 0.997 | 840 | 0.05 | 0.8 | 456 | 0.05 | 0.8 | 457 |

## 5.2. From the Bayesian perspective

Let's consider a Bayesian design with a reasonably informative investigator prior $\mathcal{N}(0{,}0.25^2)$, which yields about 33% probability that $\mu > 0.1112$, the effect assumed for $H_1$ in the frequentist scenarios. If we evaluate the probability $1 - B$ of concluding efficacy conditional on $\mu > 0$ and the probability $A$ of concluding efficacy conditional on $\mu \leq 0$ under the same reviewer prior $\mathcal{N}(0{,}0.25^2)$ without correction for multiple decision, we found that $A$ is actually well below 5% in most scenarios while $1 - B$ might not be sufficiently high (Figure 3, section 4.1.). These designs might therefore be considered suboptimal and one might want to adjust the efficacy thresholds $e_i$ and the maximum sample size $n$ to control $A$ and $1 - B$ at desired levels, say 5% and 80%.

In the frequentist framework, for given number of analyses $m$, fraction of information $k_i/n$ at which analyses are performed, and for given decision thresholds $e_i$, the type I error rate is independent on the maximum sample size $n$. This means it is possible to first compute $k_i/n$ and $e_i$ to control the Type I error rate for a given number of analyses $m$, and then increase the sample size $n$ to control the desired power.

However, this is no longer the case for operating characteristics such $A$ and $1 - B$ as they are averaged of the prior distribution of $\mu$. The chance of early stopping for efficacy for any given $\mu_t < 0$ decreases with $k_i$, since a larger sample size at the time of analysis leads to a lower risk of error on the sign of $\mu_t$. Therefore, for given $m$, $e_i$, and $k_i/n$, the risk $A$ of concluding efficacy averaged on all negative $\mu_t$ values decreases with $n$. Conversely, $1 - B$ increases as $n$ grows. Since, for given $m$, $e_i$, and $k_i/n$, $A$ and $1 - B$ are both dependent on $n$, all parameters $(n, e_i)$ must be estimated simultaneously to produce the desired Bayesian operating characteristics. This is an optimisation problem that can be solved numerically through simulations. Under simple conditions of normal prior and model, and a single decision threshold $e_1 = \cdots = e_m = e$, this can be done relatively easily with conventional optimisation algorithms (appendix section 1.3).

Table 7 shows the parameters that ensure $\mathrm{A} = 0.05$ and $1 - \mathrm{B} = 0.80$ for $\pi_{reviewer} = \pi_{investigator} = \mathcal{N}(0{,}0.25^2)$ for varying number of analyses. In other words, the investigator in charge of the design seeks to control the operating characteristics at pre-specified levels, computed under the same prior as the one used for the analyses. The expected sample size was computed as the mean of the number of observations at the time of efficacy conclusion, in trials with $\mu > 0$ and with $\mu \leq 0$.

**Table 7. Bayesian operating characteristics of Bayesian trials with $\pi_{investigator} = \mathcal{N}(0, 0.25^2)$. The maximum sample size $n$ and the efficacy threshold $1 - e_i$ for the decision rule $Pr(\mu > 0 \mid X_{1,\dots,k_i}) > 1 - e_i$ were derived through numerical optimisation to ensure $Pr_{\mu\sim\mathcal{N}(0{,}0.25^2)}(V \mid \mu \leq 0) = 0.05$ and $Pr_{\mu\sim\mathcal{N}(0{,}0.25^2)}(V \mid \mu > 0) = 0.80$, that is the operating characteristics are controlled for $\pi_{reviewer} = \pi_{investigator}$.**

| Number of analyses | $1 - e_i$ | Maximum sample size | $\underset{\mu\sim\mathcal{N}(0{,}0.25^2)}{Pr}(V \mid \mu \leq 0)$ | $\underset{\mu\sim\mathcal{N}(0{,}0.25^2)}{Pr}(V \mid \mu > 0)$ | Expected sample size in trials with $\mu > 0$ | Expected sample size in trials with $\mu \leq 0$ |
|---|---|---|---|---|---|---|
| 1 | 0.701 | 121 | 0.05 | 0.8 | 121 | 121 |
| 2 | 0.787 | 145 | 0.05 | 0.8 | 97 | 142 |
| 5 | 0.860 | 178 | 0.05 | 0.8 | 87 | 173 |
| 10 | 0.890 | 199 | 0.05 | 0.8 | 88 | 192 |
| 20 | 0.904 | 207 | 0.05 | 0.8 | 86 | 200 |
| 50 | 0.919 | 220 | 0.05 | 0.8 | 88 | 212 |
| 100 | 0.925 | 225 | 0.05 | 0.8 | 89 | 217 |

Although it is possible to control the operating characteristics at pre-specified levels for a given number of analyses and a given prior, an external reviewer might hold different beliefs about $\mu$, and might therefore wish to assess the operating characteristics under alternative priors. Figure 4 shows the operating characteristics $A$ and $1 - B$ of the trial designs in Table 7, but computed under different reviewer priors $\pi_{reviewer} = \mathcal{N}\left(0, \sigma_{reviewer}^2\right)$. We see that if an external reviewer believes that μ is likely to be closer to zero, that is $\sigma_{reviewer} < 0.25$, then the operating characteristics are no longer at the levels estimated by the investigator: the probability $A$ increases above 5% and the probability $1 - B$ decreases below 80%. Supplementary figure S73 shows that the positive true and false discovery rates (pTDR and pFDR) behave similarly, meaning that the informative value of an efficacy conclusion of trial is reduced, reaching near-zero informative value ($pTDO < 1.25$) when $\sigma_{reviewer} \leq 0.01$.

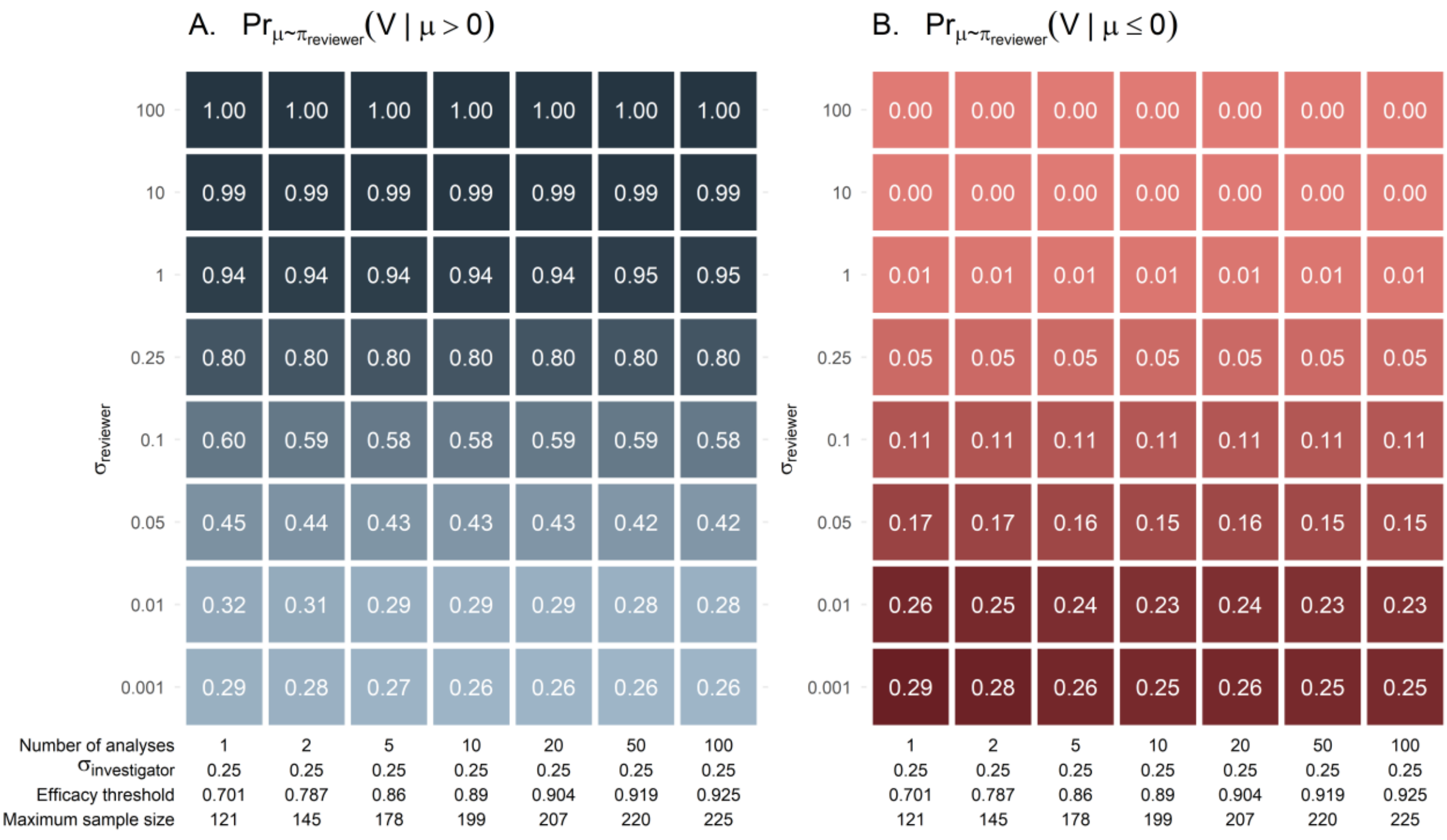

**Figure 4. Bayesian operating characteristics $Pr_{\mu\sim\mathcal{N}(0,\sigma^2_{\text{reviewer}})}(V \mid \mu \leq 0)$ and $Pr_{\mu\sim\mathcal{N}(0,\sigma^2_{reviewer})}(V \mid \mu > 0)$ computed under various $\sigma_{reviewer}$. The efficacy thresholds $1 - e_i$ for the decision rule $Pr(\mu > 0 \mid X_{1,\dots,k_i}) > 1 - e_i$, and the maximum sample size $n$ were calibrated according to the number of analyses to achieve $Pr_{\mu\sim N(0,0.25^2)}(V \mid \mu \leq 0) = 0.05$ and $Pr_{\mu\sim N(0,0.25^2)}(V \mid \mu > 0) = 0.80$.**

# 6. Discussion

In this work, we show that repeated interim analyses of accumulating data in clinical trials, without adjustment for multiple testing, fundamentally alter their operating characteristics, regardless of whether a frequentist or Bayesian framework is adopted for the design of the interim analyses, the interpretation of the operating characteristics, or both. We discuss here the interpretation of these results and their implications for current and future practice.

## 6.1. Frequentist operating characteristics

We demonstrate that when early stopping for efficacy is permitted at interim analyses, without adjustment for multiple testing, the frequentist operating characteristics of Bayesian trials are no longer controlled, leading to inevitable Type I error inflation similar to that observed in frequentist trials. When non-informative priors are used in Bayesian designs, the inference is equivalent to frequentist inference, and the Type I error rates of both designs are identical. Although the use of priors that place greater mass on the null value $\mu = 0$ can mitigate Type I error inflation in Bayesian trials, this comes at the cost of reduced power, and the Type I error rate will still tend towards 1 in perpetual trials when $n \to \infty$. Neither the Type I error rate nor the power can be maintained at their desired levels unless the efficacy thresholds and maximum sample size are adjusted to account for the number of interim analyses.

## 6.2. Bayesian operating characteristics

We can derive Bayesian operating characteristics analogous to the frequentist ones by considering $\mu$ to be a random variable rather than a fixed parameter. Because the choice of prior is subjective, Bayesian operating characteristics are not universal. They might differ between designers of a trial, an external observer interpreting the results of the trial, and a regulator evaluating the evidence provided by the trial. This implies considering two prior distributions for the parameter of interest: one used to estimate the posterior distribution at the interim or final analyses ($\pi_{investigator}$) and another used to compute the operating characteristics ($\pi_{reviewer}$).

The use of non-informative priors in the analyses is usually motivated by the fact that they are quickly overwhelmed by the data (hence the term “non-informative”). However, they are far from neutral and are implausible for continuous variables. For instance, a prior $\mathcal{N}(0,100^2)$ implies that there is a 96% probability that the treatment effect is greater or lower than five standard deviations (an extremely large effect never seen in practice), and only 1.6% probability that the effect is between −2 and +2 standard deviations. An investigator choosing such a prior would evaluate Bayesian operating characteristics for their own design with inevitably very low risks to conclude in the wrong direction, even for decision rules that would appear very liberal to an external reviewer. Even choosing only weakly informative priors, the investigator would evaluate very low risks of erroneous conclusions for their design.

However, any sensible external reviewer who disagrees with such implausible priors would consider that the risk of incorrect conclusions is high, which we observed when $\sigma_{reviewer} < \sigma_{investigator}$. We show that in such cases this risk inflates, increasingly with the number of analyses and the magnitude of the prior divergence between the reviewer and investigator. This increase is less severe when $\pi_{investigator}$ is more informative, but it persists as long as $\sigma_{reviewer} < \sigma_{investigator}$. Although it is technically possible for an investigator to choose a non-informative prior for the analyses and to evaluate the Bayesian operating characteristic under a more informative prior, this is a conceptual paradox as it essentially means that the investigator believes simultaneously in two alternative distributions for the parameter. In any case, it does not solve the issue that a reviewer would evaluate higher risks of erroneous efficacy conclusions if they consider a more sceptical prior.

Considering other operating characteristics such as the positive true discovery rate (i.e. the probability that $\mu > 0$ given that the trial has concluded efficacy) also reveals that repeating interim analyses without correction for multiplicity of analyses is not without consequences. Importantly, we propose a new operating characteristic, the positive true discovery odds, defined as $Pr(\mu > 0 \mid V)/Pr(\mu \leq 0 \mid V)$. This measure is interesting for the evaluation of Bayesian designs for at least two reasons: (i) it is inherently Bayesian, and in the special case where the parameter distribution is centered at zero ($Pr(\mu > 0) = 0.5$), it is exactly equivalent to the Bayes factor of an efficacy conclusion on the assertion that μ is positive, (ii) it provides an intuitive measure of the informative value of an efficacy conclusion of a trial. In fact, it is analogous to the odds of a positive predictive value of a diagnostic test.

We show that, without correction for multiplicity of analyses, even in the special case where $\pi_{reviewer} = \pi_{investigator}$, the informative value of an efficacy conclusion decreases as the number of analyses increases. In practice, this means that even if the reviewer agrees with the investigator's choice of prior, increasing the number of analyses reduces the informative value of an efficacy conclusion, all else being equal. This reduction in informative value is exacerbated when the reviewer prior is more sceptical than the investigator's prior.

## 6.3. Measures to control operating characteristics

In all cases, futility stopping rules mitigate the risks of incorrectly concluding efficacy, regardless of how they are computed, but do not, by themselves, guarantee control of these risks at a given level, regardless of the number of analyses. Moreover, they systematically lead to reduced probabilities of correctly concluding efficacy.

As with the well-known frequentist methods to control the Type I error inflation due to multiple testing, it is possible to adjust the posterior decision thresholds of Bayesian designs to control their frequentist operating characteristics. Methods to derive such adjusted thresholds have been proposed in the literature.[32,33]

It is also possible to control Bayesian operating characteristics. Whereas adjusted thresholds can be derived analytically to control frequentist operating characteristics, doing so for Bayesian operating characteristics is generally more difficult. It is usually achieved through numerical optimisation and can become computationally challenging for complex models.

However, the consequences of such adjustments in the Bayesian setting must be carefully examined. When repeating interim analyses without adjustments, an investigator evaluating the Bayesian operating characteristics of the trial under the same prior as the one used for the analyses would estimate a risk $A$ of erroneously concluding efficacy well below the traditional benchmark of 5%. Therefore, if the investigator wants to adjust the decision thresholds to control $A$ at say, 5%, they would select less stringent thresholds. However, this choice would worsen all Bayesian operating characteristics from the perspective of a more sceptical reviewer ($\sigma_{reviewer} < \sigma_{investigator}$) compared to the investigator having not adjusted for multiplicity of analyses at all.

## 6.4. Going beyond binary hypothesis-testing

Errors in conclusions arise from the use of statistical tests to support binary decisions. Ryan and colleagues have argued that, rather than relying on a statistical test alone, decisions should be based on the full posterior distribution, interpreted in relation to clinically meaningful values.[19] Similarly, Amrhein, Greenland, and McShane wrote that "for the choices often required in regulatory, policy and business environments, decisions based on the costs, benefits and likelihoods of all potential consequences always beat those made based solely on statistical significance".[34] These ideas are reflected in Bayesian decision-theoretic approaches, in which decisions are made by maximizing expected utility across the posterior distribution.

However, interim analyses imply binary conclusions. We can show that when the number of interim analyses increases, the posterior probability of $\mu > 0$ observed in case of an efficacy conclusion approaches the value of the efficacy decision threshold itself (see supplementary figures S74 to S78). In the limit case of continuous analysis of accumulating data, with an efficacy decision threshold identical for all analyses, the posterior probability of $\mu > 0$ observed in case of an efficacy conclusion becomes deterministic and is actually a methodological choice (i.e. the efficacy decision threshold). Then, the posterior simply acts as a binary hypothesis-testing tool, losing most or all of its quantitative interpretation. This limits the possibility of interpretation by external observers in disagreement with some of the design choices. For instance, if many interim analyses are performed and the study is terminated whenever $\Pr_{\mu \sim \mathcal{N}(0,1)}(\mu > 0 | X_{1,\dots,k}) \geq 0.95$, the lower bound of the 95% one-sided credible interval will only be slightly above zero. This will preclude any meaningful conclusion by an external reviewer who would want to assess whether $\mu$ is greater than a minimal clinically important difference (MCID), or a regulator requiring 97.5% or 99% certainty before making a decision, or a reviewer disagreeing with a prior belief that there is a 15.9% probability that the average treatment effect is greater than one standard deviation. Any interpretation requiring more conservative assumptions than those of the investigators who designed the trial is bound to be inconclusive. It should be noted that the frequentist Neyman-Pearson paradigm is subject to the same constraints on quantitative interpretation when repeated interim analyses are performed.

## 6.5. Summary of findings and implications for current practice

The results of this work highlight six elements to consider in the design and analysis of group-sequential trials:

(i) Bayesian inference is affected by the repetition of interim analyses. This is an inherent consequence of repeated decision-making and not a feature exclusive to frequentist inference. Even considering Bayesian operating characteristics, Bayesian trials should not be viewed as inherently immune to multiple testing issues. Therefore, the issue of multiple testing should be taken into account in the design of all trials with interim analyses.

(ii) Bayesian operating characteristics are worth considering as alternatives to the frequentist Type I error rate and power, especially for Bayesian trials. However, these operating characteristics are inherently sensitive to divergence in prior belief about the parameter of interest, and increasingly so as the number of analyses grows. Even when there is consensus on the prior, increasing the number of interim analyses without adjusting for multiplicity reduces the informative value of an efficacy conclusion. While it is possible to calibrate the decision thresholds to control these Bayesian operating characteristics at desired levels, this calibration further increases their sensitivity to prior divergence.

(iii) Whether considering frequentist or Bayesian operating characteristics, and whether or not multiplicity adjustments are applied, the posterior probability of a treatment effect conditional on the observed data loses most, if not all, of its quantitative interpretability when analyses are repeated. In such cases, only binary conclusions can be drawn and the posterior simply acts as a binary hypothesis-testing tool, much like a frequentist p-value used solely for significance testing. Therefore, when evaluating a positive trial with repeated interim analyses, reviewers should no longer rely on the posterior reported by the authors but consider other measures such as the positive true discovery rate or the positive true discovery odds, in light of their own prior.

(iv) A null hypothesis, such as $H_0: \mu = 0$, is actually the most sceptical prior (null effect and zero variance) and the only one that can reach consensus as it will yield upper bounds for the estimated risks of erroneous efficacy conclusions. Therefore, although rooted in frequentist methodology and often viewed as irrelevant from a Bayesian standpoint, the Type I error rate remains a key operating characteristic for Bayesian trials with repeated interim analyses.

(v) The design and evaluation of operating characteristics of Bayesian trials often rely on simulations. Ensuring the transparency and reproducibility of these simulations is critical for assessing the robustness of the designs, and is particularly challenging in adaptive designs, where prospective modifications to the protocol may alter the operating characteristics of the trial and necessitate updated simulations. Moreover, reporting standards remain insufficient. For instance, a systematic review found that detailed simulation procedures and results were publicly available for only 62% (24/58) of Bayesian platform trials, and fully reproducible (i.e., with publicly accessible simulation code) for only two trials.[14] Therefore, efforts should be made to make the detailed description and source code used for the simulations publicly available to allow for reproducible and transparent assessment of operating characteristics of Bayesian designs.

(vi) In recent years, so-called “perpetual” trials have emerged.[14] These Bayesian designs do not pre-specify a final analysis at a fixed sample size but instead perform continuous interim analyses until a decision is reached or the trial is stopped for operational reasons (e.g., funding limitations or recruitment challenges). It should be clear from the results presented here and their implications that such designs cannot control the risks of erroneous efficacy conclusions indefinitely, and that the informative value of an efficacy conclusion will approach zero as the trial progresses. In practice, perpetual trials seek to control operating characteristics only up to a finite, very large, maximum sample size. However, the robustness of their design is difficult to assess given the limited availability and reproducibility of the underlying simulation studies.[14]

### 6.6. Strengths and limitations

The analyses were conducted under a deliberately simplified setting. The purpose of this work was to propose a coherent framework for studying the impact of repeated interim analyses on both frequentist and Bayesian inferences, regardless of the actual specification of the trial design. We did not seek to precisely quantify the alteration of the operating characteristics of trials with specific designs, outcomes, interim analysis plans, models, or testing hypotheses. Although this choice may limit the direct quantitative generalisability of the results, the qualitative findings are not tied to a specific model or design, but rather reflect the fundamental issues caused by repeated decision-making in the interpretation of both frequentist and Bayesian inferences. The choice of a normal model is motivated by the fact that, under fairly general conditions, most importantly that successive observations are independent and identically distributed, most statistics are asymptotically normal by the central limit theorem. These approximations are accurate enough that statistical software apply them by default to common estimators such as the log odds ratio, log hazard ratio, or difference of proportions. Consequently, the results presented here extend to most study designs and statistics, with quantitative but no qualitative differences. Even when observations are not fully independent, independence typically holds at a higher level of aggregation (e.g. cluster-randomised trials), and the normal approximation remains adequate. Our framework already accounts for partial correlation between statistics computed at successive interim analyses. The exact degree of correlation may vary across more complex designs, but this does not affect the qualitative conclusions of our work. Choosing the simplest framework that allows a clear and complete exploration of our questions enables analytical tractability and facilitates a clear interpretation of the results. Importantly, all simulations are explicitly described, and the entire code and results are made freely and publicly available, providing complete transparency and reproducibility to allow independent verification and reuse.

## 7. Conclusion

Consistently with the current literature, we confirmed that group-sequential Bayesian trials are not immune to Type I error rate inflation. Moreover, we showed that Bayesian operating characteristics such as the risk of erroneous conclusions and the informative value of an efficacy conclusion degrade as the number of interim analyses increases and are sensitive to prior divergence between stakeholders. While it is possible to control the frequentist operating characteristics of Bayesian trials with appropriate methods, it is not possible to guarantee such control for Bayesian operating characteristics because they are prior-dependent and different stakeholders may reasonably adopt different priors.

Supplementary material for the article

# *Impact of interim analyses on Bayesian and frequentist operating characteristics of Bayesian clinical trials*

Clément R. Massonnaud et al.

# 1 Supplementary methods

## 1.1 Simulation procedures and computations

### 1.1.1 Pseudo-code

```
BEGIN

SIMULATE TRIALS:
        set n_trials : number of trials
        set n_obs : number of observations per trial
        set SD_data : standard deviation of the data-generating distribution N(mean=0,
sd=SD_data), i.e. $\mathcal{N}(0, \sigma^2_{reviewer})$.
        set theta :
                if theta is fixed : use provided value (e.g. 0 for H0, $\mu_1$ for H1)
                if theta is random : draw n_trials values from N(mean=0, sd=SD_data)

        for each trial i in 1 … n_trials :
                draw n_obs values from N(mean=theta_i, sd=1)
                compute cumulative means xbar_j = mean(x_1..x_j) for j in 1..n_obs

COMPUTE FREQUENTIST AND BAYESIAN ANALYSES:
        set SD_prior : standard deviation of the prior N(mean=0, sd=SD_prior) for Bayesian
analyses (i.e. $\mathcal{N}(0, \sigma^2_{investigator})$)
        set K : number of analyses
        compute analysis times : subset of indices in 1, …, n_obs (e.g. {250, 500})

        for each trial and for each analysis time:

                // Frequentist : compute one-sided p-value
                // Bayesian : compute posterior probabilities (conjugate normal update)

COMPUTE TRIAL CONCLUSIONS:
        set boundary for the frequentist efficacy Z test
        set efficacy thresholds for Bayesian efficacy analyses
        set futility thresholds for frequentist and Bayesian futility analyses

        compute conclusions for each trial and return the first analysis time for which a
decision rule is met

END
```

### 1.1.2 Core R functions

#### *1.1.2.1 Packages*

```
library(data.table)
```

#### *1.1.2.2 Simulate N trials of n observations*

```
#' Function to simulate n_trials of n_obs observations each
#'
#' @param n_trials numeric scalar. The number of trials.
#' @param n_obs numeric scalar. The number of observations for each trial
#' @param theta numeric scalar. The value of the parameter (mean) for data generation. If N
ULL (default), theta is drawn from a normal distribution N(mu = 0, sd = SD_data)
#' @param SD_data numeric scalar. The standard deviation of the normal distribution of mean
0, from which the parameter (mean) for data generation is drawn.
#'
#' @returns a data.table with n_trials * n_obs rows
#' @export
#'
#' @examples


simulate_trials <- function(n_trials,
                            n_obs,
                            theta = NULL,
                            SD_data = NULL)


{
  # ------------- #
  # Simulate data #
  # ------------- #

  # Draw parameter value
  if (is.null(theta)) theta = rnorm(n_trials, mean = 0, sd = SD_data)

  # Matrix of dimension n_trials * n_obs, one row per simulated trial
  # (n_trials) and one column per observation in a trial (n_obs), with
  # repeated values of theta
  theta_matrix = matrix(rep(theta, n_obs),
                        nrow = n_trials,
                        ncol = n_obs,
                        byrow = FALSE,
                        dimnames = list(1:n_trials, 1:n_obs))

  # Matrix of dimension n_obs * n_trials, one column per simulated trial
  # (n_trials) and one row per observation in a trial (n_obs), with
  # simulated value for each observation from rnorm
  data_matrix = t(apply(theta_matrix,
                        MARGIN = 2,
                        FUN = function(t) {
                          rnorm(n_trials, mean = theta, sd = 1)
                        }))
  dimnames(data_matrix)[[2]] = 1:n_trials

  # Compute observed means for each additional observation
  xbar_matrix = apply(data_matrix,
                      MARGIN = 2,
                      FUN = function(x) cumsum(x) / seq_along(x))
```

```r
  # Store results in a data.table
  res_dt = as.data.table(xbar_matrix,
                         keep.rownames = "analysis.idx") |>
    melt(id.vars = "analysis.idx",
         variable.name = "ID.trial",
         value.name = "xbar")
  res_dt = merge(res_dt,
                 data.table(ID.trial = factor(1:n_trials),
                            theta = theta))
  res_dt[, analysis.idx := as.numeric(analysis.idx)]
  res_dt[, SD.data := ifelse(is.null(SD_data), NA, SD_data)]

  return(res_dt)
}
```

#### 1.1.2.3 Get indices of equally spaced analyses of the trial

```r
#' Get indices of equally spaced analyses of the trial (including the final analysis)
#'
#' @param n_obs numeric scalar. The number of observations per trial.
#' @param n_analyses numeric scalar. The number of analyses.
#' @param spacing_analyses numeric scalar. The number of observations between each analysis
. Defaults to 'NULL'. If not NULL, first_analysis must be specified and n_analyses must be
set to NULL.
#' @param first_analysis numeric scalar. The index of the first analysis when spacing_analy
ses is not NULL. Defaults to NULL.
#'
#' @returns a numeric vector of size n_analyses. The indices of the analyses.
#' @export
#'
#' @examples
#'
get_analyses_idx = function(n_obs, n_analyses = NULL, spacing_analyses = NULL, first_analys
is = NULL) {

    if (!is.null(spacing_analyses)) {
      stopifnot(is.null(n_analyses))
      if(is.null(first_analysis)) stop("If spacing_analyses is not null, first_analysis mus
t be provided")

      return(seq(from = first_analysis, to = n_obs, by = spacing_analyses))

    } else if(!is.null(n_analyses)) {
      stopifnot(is.null(spacing_analyses) & is.null(first_analysis))

      return(floor(seq(from = 1, to = n_obs, length.out = n_analyses + 1))[-1])
    }
}
```

#### 1.1.2.4 Compute normal posterior distribution

```r
#' Compute normal posterior distribution
#'
#' @param mu_prior numeric scalar. Mean of prior normal distribution
#' @param sigma_prior numeric scalar. Standard deviation of the prior normal distribution
#' @param xbar numeric scalar. The observed mean for the sample X of size n_obs
#' @param sigma_obs numeric scalar. Known observed standard deviation.
#' @param n_obs numeric scalar. The number of observations in sample X
#'
#' @returns A list with posterior distribution parameters mu_post and sigma_post
#' @export
#'
#' @examples
get_normal_post = function(mu_prior, sigma_prior, xbar, sigma_obs, n_obs) {

  # --- Posterior (analytic conjugate update) ---
  # prior precision = 1/tau, data precision = n_obs/sigma_obs^2
  tau = sigma_prior^2
  prec = 1 / tau
  prec_data = n_obs / (sigma_obs^2)
  prec_post = prec + prec_data

  tau_post = 1 / prec_post       # posterior variance
  sigma_post = sqrt(tau_post)    # posterior sd

  mu_post = (prec * mu_prior + prec_data * xbar) / prec_post

  return(list(mu.post = mu_post,
              sigma.post = sigma_post))
}
```

#### 1.1.2.5 Compute frequentist and Bayesian statistical tests at each time for all trials

```r
#' Function that computes the statistical tests at each time for all trials
#'
#' @param dt data.table. The results of simulate_trials(), a data table with n_trials * n_o
bs rows
#' @param prior list. A list with names 'mu' and 'sigma', the parameters of the normal prio
r
#' @param q_eff numeric scalar. The quantile for the frequentist, one-sided, superiority Z-
test, and the Bayesian posterior probability efficacy test P(mu > q_eff | X) > a
#' @param q_fut numeric scalar. The quantile for the Bayesian posterior probability futilit
y test P(mu > q_fut | X) < b if fut_rule == "higher", or P(mu < q_fut | X) > b if fut_rule
== "lower"
#' @param fut_rule character. The futility stopping rule, one of "higher", "lower", or "non
e"
#'
#' @returns a data.table with n_trials * n_obs rows
#' @export
#'
#' @examples


compute_analyses <- function(dt,
                             prior,
                             q_eff,
```

```
                             q_fut = NULL,
                             fut_rule = "none")
{

  if (!is.null(q_fut) & fut_rule == "none") stop()
  if (is.null(q_fut) & fut_rule %in% c("lower", "higher")) stop()

  # ----------------------------- #
  # Compute frequentist analysis #
  # ----------------------------- #
  dt[, z := (xbar - q_eff) * sqrt(analysis.idx)]
  dt[, prob.freq := pnorm(z, mean = 0, sd = 1, lower.tail = FALSE)]

  # ----------------------------- #
  #   Compute Bayesian analysis  #
  # ----------------------------- #


  # Compute posterior
  dt = cbind(dt,
             dt[, get_normal_post(mu_prior = prior$mu,
                                  sigma_prior = prior$sigma,
                                  xbar,
                                  sigma_obs = 1,
                                  n_obs = analysis.idx)])

  # Compute probabilities

  # Efficacy
  dt[, prob.bay.eff := pnorm(q_eff, mean = mu.post, sd = sigma.post, lower.tail = FALSE)]

  # Futility
  if (fut_rule == "lower") {
    # P(mu < mu1) > a
    dt[, prob.bay.fut := pnorm(q_fut, mean = mu.post, sd = sigma.post)]
  } else if (fut_rule == "higher") {
    # P(mu > 0) < b
    dt[, prob.bay.fut := pnorm(q_fut, mean = mu.post, sd = sigma.post, lower.tail = FALSE)]
  } else {
    dt[, prob.bay.fut := NA]
  }

  dt$SD.prior = factor(prior$sigma)
  return(dt)

}
```

### *1.1.2.6 Compute trial conclusions based on a set of decision rules*

```r
#' Function to get the conclusion for the trial based on a set of decision rules
#'
#' @param dt data.table. The results of compute_analyses(), a data table n_trials * n_obs r
ows
#' @param n_obs numeric scalar. The number of observations per trial.
#' @param z_boundary numeric scalar. The boundary for the Z test.
#' @param bay_threshold_eff numeric scalar. The threshold for the Bayesian posterior probab
ility efficacy test P(mu > 0 | X) > bay_threshold_eff.
#' @param bay_threshold_fut numeric scalar. The threshold for the Bayesian posterior probab
ility futility test P(mu > 0 | X) < bay_threshold_fut if fut_rule == "higher", P(mu < mu1 |
X) > bay_threshold_fut if fut_rule == "lower".
#' @param fut_rule character. The futility decision rule, one of "higher", "lower", or "non
e"
#'
#' @returns a data.table with one row per trial.
#' @export
#'
#' @examples


get_trial_conclusion <- function(dt,
                                 n_obs,
                                 z_boundary,
                                 bay_threshold_eff,
                                 bay_threshold_fut = NULL,
                                 fut_rule = "none") {

  if (!is.null(bay_threshold_fut) & fut_rule == "none") stop()
  if (is.null(bay_threshold_fut) & fut_rule %in% c("lower", "higher")) stop()

  #----------------------------#
  # COMPUTE TRIALS CONCLUSIONS #
  #----------------------------#
  setkey(dt, analysis.idx)

  # Frequentist efficacy
  dt_all = merge(unique(dt[, .(SD.prior, SD.data, ID.trial, theta)]),
                 dt[z > z_boundary,
                    .SD[1, ],
                    by = ID.trial][, .(SD.prior,
                                       SD.data,
                                       ID.trial,
                                       time.freq = analysis.idx)],
                 by = c("SD.prior", "SD.data", "ID.trial"),
                 all.x = T)


  # Bayesian efficacy
  dt_all = merge(dt_all,
                 dt[prob.bay.eff > bay_threshold_eff,
                    .SD[1, ],
                    by = ID.trial][, .(SD.prior,
                                       SD.data,
                                       ID.trial,
                                       time.bay.eff = analysis.idx)],
                 by = c("SD.prior", "SD.data", "ID.trial"),
                 all.x = T)
```

```r
  # Bayesian futility
  if(fut_rule == "lower") {
    dt_all = merge(dt_all,
                   dt[prob.bay.fut > bay_threshold_fut,
                      .SD[1, ],
                      by = ID.trial][, .(SD.prior,
                                         SD.data,
                                         ID.trial,
                                         time.bay.fut = analysis.idx)],
                   by = c("SD.prior", "SD.data", "ID.trial"),
                   all.x = T)
  } else if (fut_rule == "higher") {
    dt_all = merge(dt_all,
                   dt[prob.bay.fut < bay_threshold_fut,
                      .SD[1, ],
                      by = ID.trial][, .(SD.prior,
                                         SD.data,
                                         ID.trial,
                                         time.bay.fut = analysis.idx)],
                   by = c("SD.prior", "SD.data", "ID.trial"),
                   all.x = T)
  } else {
    dt_all$time.bay.fut = NA_real_
  }

  # Free memory
  rm(dt)
  gc()

  #--------------------------#
  # WRITE TRIALS CONCLUSIONS #
  #--------------------------#
  dt_all[, res.freq := ifelse(is.na(time.freq), "comp", "eff")]
  dt_all[res.freq == "comp", time.freq := n_obs]

  dt_all[is.na(time.bay.eff), time.bay.eff := n_obs + 1]
  dt_all[is.na(time.bay.fut), time.bay.fut := n_obs + 1]

  dt_all[time.bay.eff < time.bay.fut, `:=`(res.bay = "eff", time.bay = time.bay.eff)]
  dt_all[time.bay.fut < time.bay.eff, `:=`(res.bay = "fut", time.bay = time.bay.fut)]
  dt_all[time.bay.fut == time.bay.eff, `:=`(res.bay = "both", time.bay = time.bay.eff)]
  dt_all[time.bay.eff == time.bay.fut & time.bay.eff == n_obs + 1, `:=`(res.bay = "comp",
                                                                         time.bay = n_obs)]

  return(dt_all[, .(SD.prior, SD.data, ID.trial, res.freq, time.freq, res.bay, time.bay, th
eta)])

}
```

#### 1.1.2.7 Examples

```
# Examples ----------------------------------------------------------------

## Efficacy stopping rules only -------------------------------------------

# n_trials = 1e5
# n_obs = 500
# SD_data = 0.001
# prior = list(mu = 0, sigma = 100)
# q_eff = 0
# z_boundary = qnorm(0.05, lower.tail = FALSE)
# bay_threshold_eff = 0.95
#
# trials = simulate_trials(n_trials = n_trials,
#                              n_obs = n_obs,
#                              SD_data = SD_data)
#
# # Look only at analyses after 250 and 500 observations
# sims = compute_analyses(dt = trials[analysis.idx %in% c(250, 500), ],
#                          prior = prior,
#                          q_eff = q_eff)
#
# conc = get_trial_conclusion(sims,
#                              n_obs = n_obs,
#                              z_boundary = z_boundary,
#                              bay_threshold_eff = bay_threshold_eff)

## Efficacy and futility stopping rules -----------------------------------
#
# n_trials = 1e2
# n_obs = 500
# SD_data = 0.01
# prior = list(mu = 0, sigma = 100)
# q_eff = 0
# q_fut = 0
# fut_rule = "higher"
# z_boundary = qnorm(0.05, lower.tail = FALSE)
# bay_threshold_eff = 0.95
# bay_threshold_fut = 0.05
#
# trials_dt = simulate_trials(n_trials = n_trials,
#                              n_obs = n_obs,
#                              SD_data = SD_data)
#
# sims = compute_analyses(dt = trials_dt,
#                          prior = prior,
#                          q_eff = q_eff,
#                          q_fut = q_fut,
#                          fut_rule = fut_rule)
#
# # Look only at analyses after 250 and 500 observations
# dt = sims[analysis.idx %in% c(250, 500), ]
#
# conc = get_trial_conclusion(dt,
#                              n_obs = n_obs,
#                              z_boundary = z_boundary,
#                              bay_threshold_eff = bay_threshold_eff,
#                              bay_threshold_fut = bay_threshold_fut,
#                              fut_rule = fut_rule)
```

## 1.2 Numerical computation of Pocock boundaries

### 1.2.1 Finding the Z-test boundary c

The objective is to find a value $c$ that ensures an overall Type I error rate controlled at the 5% level when performing $K$ analyses in trials of sample size $n$. For one-sided tests, the overall Type I error rate is the probability, under $H_0: \mu = 0$, that any statistic $Z_k$ computed on a sample of observations $X_{1,\dots,k}$ exceeds the threshold $c$. It is more easily computed as the complement of the probability that none of the statistics $Z_k$ exceed $c$:

$$\mathrm{P}\left(\bigcup_{k\in\{k_1,\dots,k_m\}} Z_k > c\right) = 1 - \mathrm{P}\left(\bigcap_{k\in\{k_1,\dots,k_m\}} Z_k \le c\right)$$

All statistics $Z_k$ are correlated. For $K$ analyses, we have the correlation matrix:

$$R = \begin{pmatrix} \mathrm{Cor}(Z_1, Z_1) & \mathrm{Cor}(Z_1, Z_2) & \cdots & \mathrm{Cor}(Z_1, Z_K) \\ \mathrm{Cor}(Z_2, Z_1) & \mathrm{Cor}(Z_2, Z_2) & \cdots & \mathrm{Cor}(Z_2, Z_K) \\ \vdots & \vdots & \ddots & \vdots \\ \mathrm{Cor}(Z_K, Z_1) & \mathrm{Cor}(Z_K, Z_2) & \cdots & \mathrm{Cor}(Z_K, Z_K) \end{pmatrix}$$

Where each element is:

$$\mathrm{Cor}(\mathrm{Z}_i, Z_j) = \sqrt{\frac{\min(t_i, t_j)}{\max(t_i, t_j)}}$$

With $t_i$ the information fraction at analysis $i$, given by $t_i = \frac{n_i}{n_{max}}$.

Therefore, the vector of statistics $Z_k$ can be modelled under $H_0$ as a multivariate normal distribution of mean zero and covariance matrix $R$. Indeed, under $H_0$ and even under an alternative hypothesis $H_1$, $VAR(Z_i) = 1$ for all $i$ and the covariance matrix is equal to the correlation matrix.

$$\begin{pmatrix} Z_1 \\ Z_2 \\ \vdots \\ Z_K \end{pmatrix} \sim \mathcal{N}_{\mathcal{K}}\left(\begin{pmatrix} 0 \\ 0 \\ \vdots \\ 0 \end{pmatrix}, R\right)$$

This can be implemented in R like so:

```
# Compute the correlation matrix for analyses at various sample sizes
corr_matrix <- function(analyses_idx) {

  # Convert to information fraction
  t = analyses_idx / max(analyses_idx)
  # Correlation matrix
  R = outer(t, t, function(ti, tj) sqrt(pmin(ti, tj) / pmax(ti, tj)))

}

# Compute the multivariate normal probability that none of Z statistic exceeds c under H0
prob_no_crossing <- function(c, R, mean) {

  mvtnorm::pmvnorm(lower = rep(-Inf, nrow(R)),
                   upper = rep(c, nrow(R)),
                   mean  = mean,
                   sigma = R)[1]
}
```

Then the objective is to find the value $c$ that gives:

$$P\left(\bigcup_{k\in\{k_1,\dots,k_m\}} Z_k > c\right) = 1 - P\left(\bigcap_{k\in\{k_1,\dots,k_m\}} Z_k \leq c\right) = 0.05$$

or:

$$P\left(\bigcap_{k\in\{k_1,\dots,k_m\}} Z_k \leq c\right) - 0.95 = 0$$

This is easily done in R like so:

```
# Define the objective function
# P(Z_1 < c, ..., Z_k < c) - 0.95 = 0
objective_error <- function(c, R, alpha = 0.05) {
  prob_no_crossing(c = c,
                   R = R,
                   mean = rep(0, nrow(R))) - (1 - alpha)
}

# Find c for Type 1 error = 5 %
n_analyses = c(1, 2, 5, 10, 20, 50, 100, 250, 500)

c_thresholds = sapply(n_analyses,
                      function (K) {
  print(K)
  # Define K analyses at various times in trials of sample size 500
  idx = get_analyses_idx(n_obs = 500, n_analyses = K)
  # Compute the correlation matrix
  R = corr_matrix(idx)

  # Find c
  c_threshold = uniroot(objective_error,
                        interval = c(1, 4),
                        R = R,
                        maxiter = 100)$root
})
```

### 1.2.2 Finding the sample size to obtain adequate power

Conversely, for one-sided tests, the power error is the probability, under $H_1: \mu = \mu_1$, that any statistic $Z_k$ computed on a sample of observations $X_{1,\dots,k}$ exceeds the threshold $c$. As this threshold has been increased to control the overall Type I error rate under $H_0$, the sample size has to be increased to maintain the desired power. Therefore, to achieve 80% power, the objective is to find the sample size $n$ that gives under $H_1$:

$$\mathrm{P}\left(\bigcup_{k\in\{k_1,\dots,k_m\}} Z_k > c\right) = 1 - \mathrm{P}\left(\bigcap_{k\in\{k_1,\dots,k_m\}} Z_k \le c\right) = 0.80$$

That is:

$$1 - \mathrm{P}\left(\bigcap_{k\in\{k_1,\dots,k_m\}} Z_k \le c\right) - 0.80 = 0$$

With:

$$\begin{pmatrix} Z_1 \\ Z_2 \\ \vdots \\ Z_K \end{pmatrix} \sim \mathcal{N}_{\mathcal{K}}\left(\begin{pmatrix} \mu_1 \\ \mu_1 \\ \vdots \\ \mu_1 \end{pmatrix}, R\right)$$

This is easily done in R like so:

```
# Find sample size n that increases power back to 80%
objective_power <- function(n, K, c, power = 0.80) {

  # Define K analyses at various times in trials of sample size n
  idx = get_analyses_idx(n_obs = n, n_analyses = K)
  # Compute the correlation matrix
  R = corr_matrix(idx)

  # Under H1
  delta = qnorm(0.95, 0, 1) + qnorm(0.8, 0, 1)
  theta_H1 = 1 / sqrt(500) * delta
  # Vector of Z statistics
  z_vec = theta_H1 * sqrt(idx)
  # Compute the objective
  # 1 - P(Z_1 < c, ..., Z_k < c) - 0.80 = 0
  1 - prob_no_crossing(c = c,
                       R = R,
                       mean = z_vec) - power
}

sample_sizes = sapply(c(1, 2, 5, 10, 20, 50, 100, 250, 500), function (K) {
  print(K)
  set.seed(1234)
  round(uniroot(objective_power,
                interval = c(499, 850),
                K = K,
                c = c_thresholds[[paste0("N", K)]],
                power = 0.80,
                maxiter = 100)$root)
})
```

## 1.3 Numerical computation of Bayesian decision thresholds to control the Bayesian operating characteristics

Our objective is to find the parameters $(n, e)$ that produce 5% of risk $A$ of incorrectly concluding efficacy conditional to $\mu < 0$ and 80% of probability $1 - B$ of correctly concluding efficacy conditional to $\mu \geq 0$ with a common decision threshold $e = e_1, \dots, e_m$ for all interim analyses, computed under the common prior $\mathcal{N}\left(0, \sigma^2_{reviewer}\right) = \mathcal{N}\left(0, \sigma^2_{investigator}\right)$ for $m$ equal-spaced interim analyses. We therefore define an objective function to be minimized:

$$t = (A - 0.05)^2 + (B - 0.20)^2$$

We used non-linear optimisation algorithms from the NLopt library, through the R package *nloptr* (version 2.2.1), specifically the derivative-free Nelder–Mead simplex algorithm, which is specifically suitable for non-smooth objective functions and does not require gradient information. The optimisation was conducted with a maximum of 100 function evaluations. Convergence was assessed using an absolute tolerance on the parameters of $(0.1, 0.001)$ and a relative tolerance on the objective function of $10^{-5}$. The complete set of programs required to reproduce the results presented here is freely available online at https://github.com/ClementMassonnaud/BayesianInterim.

# 2 Supplementary results

## 2.1 Frequentist operating characteristics

### 2.1.1 Stopping for efficacy only

**Table S1. Type I error rate (α), power $(1-\beta)$, and ratio $(1-\beta)/\alpha$ of frequentist and Bayesian trial designs with efficacy stopping rules (frequentist: one-sided $p_{value} < 0.05$, Bayesian: $Pr(\mu > 0 | X_{1,\ldots,k_i}) > 0.95, i \in \{1, \ldots, m\}$), no futility stopping rules, and no correction for multiplicity of analyses. $\sigma_{investigator}$ is the standard deviation of the prior $\mathcal{N}(0, \sigma^2_{investigator})$ used for the Bayesian analyses.**

| Number of analyses | Bayesian trials | | | | | | | | | | | | | | |
|---|---|---|---|---|---|---|---|---|---|---|---|---|---|---|---|
| | $\sigma_{investigator} = 10$ | | | $\sigma_{investigator} = 1$ | | | $\sigma_{investigator} = 0.25$ | | | $\sigma_{investigator} = 0.1$ | | | $\sigma_{investigator} = 0.05$ | | |
| | Type I error rate (α) | Power $(1-\beta)$ | $\frac{1-\beta}{\alpha}$ | Type I error rate (α) | Power $(1-\beta)$ | $\frac{1-\beta}{\alpha}$ | Type I error rate (α) | Power $(1-\beta)$ | $\frac{1-\beta}{\alpha}$ | Type I error rate (α) | Power $(1-\beta)$ | $\frac{1-\beta}{\alpha}$ | Type I error rate (α) | Power $(1-\beta)$ | $\frac{1-\beta}{\alpha}$ |
| 1 | 0.05 | 0.8 | 16 | 0.05 | 0.8 | 16 | 0.05 | 0.79 | 17 | 0.04 | 0.75 | 21 | 0.01 | 0.61 | 44 |
| 2 | 0.08 | 0.83 | 10 | 0.08 | 0.82 | 10 | 0.07 | 0.82 | 11 | 0.05 | 0.77 | 15 | 0.02 | 0.62 | 38 |
| 5 | 0.13 | 0.86 | 7 | 0.13 | 0.86 | 7 | 0.12 | 0.85 | 7 | 0.07 | 0.8 | 11 | 0.02 | 0.64 | 31 |
| 10 | 0.17 | 0.88 | 5 | 0.17 | 0.88 | 5 | 0.15 | 0.86 | 6 | 0.08 | 0.81 | 10 | 0.02 | 0.66 | 27 |
| 20 | 0.21 | 0.89 | 4 | 0.21 | 0.89 | 4 | 0.17 | 0.88 | 5 | 0.1 | 0.83 | 9 | 0.03 | 0.67 | 25 |
| 50 | 0.27 | 0.91 | 3 | 0.26 | 0.91 | 3 | 0.2 | 0.89 | 4 | 0.11 | 0.84 | 8 | 0.03 | 0.69 | 23 |
| 100 | 0.31 | 0.92 | 3 | 0.3 | 0.91 | 3 | 0.22 | 0.9 | 4 | 0.12 | 0.85 | 7 | 0.03 | 0.69 | 22 |
| 250 | 0.36 | 0.93 | 3 | 0.33 | 0.92 | 3 | 0.23 | 0.9 | 4 | 0.12 | 0.85 | 7 | 0.03 | 0.7 | 21 |
| 500 | 0.4 | 0.93 | 2 | 0.36 | 0.93 | 3 | 0.24 | 0.91 | 4 | 0.13 | 0.86 | 7 | 0.04 | 0.7 | 20 |

### 2.1.2 Stopping for efficacy or futility

**Table S2. Type I error rate and power of frequentist and Bayesian trial designs with efficacy stopping rules (frequentist: one-sided $p_{value} < 0.05$, Bayesian: $Pr(\mu > 0 | X_{1,\ldots,k_i}) > 0.95, i \in \{1, \ldots, m\}$), and added futility stopping rules (frequentist: one-sided $p_{value} > \mathrm{f_i}$, Bayesian: $Pr(\mu > 0 | X_{1,\ldots,k_i}) < 1 - \mathrm{f_i}, i \in \{1, \ldots, m-1\}, f_m = 0.05$), with no correction for multiplicity of analyses. $\sigma_{investigator}$ is the standard deviation of the prior $\mathcal{N}(0, \sigma^2_{investigator})$ used for the Bayesian analyses.**

| Futility threshold $(f_i)$ | Number of analyses | Frequentist analyses | | Bayesian analyses | | | | | | | | | |
|---|---|---|---|---|---|---|---|---|---|---|---|---|---|
| | | | | $\sigma_{prior} = 10$ | | $\sigma_{prior} = 1$ | | $\sigma_{prior} = 0.25$ | | $\sigma_{prior} = 0.1$ | | $\sigma_{prior} = 0.05$ | |
| | | Type I error rate | Power | Type I error rate | Power | Type I error rate | Power | Type I error rate | Power | Type I error rate | Power | Type I error rate | Power |
| 0.05 | 1 | 0.05 | 0.8 | 0.05 | 0.8 | 0.05 | 0.8 | 0.05 | 0.79 | 0.04 | 0.75 | 0.01 | 0.61 |
| | 2 | 0.08 | 0.83 | 0.08 | 0.83 | 0.08 | 0.82 | 0.07 | 0.82 | 0.05 | 0.77 | 0.02 | 0.62 |
| | 5 | 0.13 | 0.86 | 0.13 | 0.86 | 0.13 | 0.86 | 0.12 | 0.85 | 0.07 | 0.8 | 0.02 | 0.64 |
| | 10 | 0.17 | 0.87 | 0.17 | 0.87 | 0.17 | 0.87 | 0.15 | 0.86 | 0.08 | 0.81 | 0.02 | 0.66 |
| | 20 | 0.21 | 0.88 | 0.21 | 0.88 | 0.21 | 0.88 | 0.17 | 0.87 | 0.1 | 0.83 | 0.03 | 0.67 |
| | 50 | 0.26 | 0.87 | 0.26 | 0.87 | 0.26 | 0.87 | 0.2 | 0.88 | 0.11 | 0.84 | 0.03 | 0.69 |
| | 100 | 0.3 | 0.86 | 0.3 | 0.86 | 0.28 | 0.87 | 0.22 | 0.89 | 0.12 | 0.85 | 0.03 | 0.69 |
| | 250 | 0.33 | 0.83 | 0.33 | 0.83 | 0.32 | 0.85 | 0.23 | 0.89 | 0.12 | 0.85 | 0.03 | 0.7 |
| | 500 | 0.36 | 0.8 | 0.36 | 0.8 | 0.33 | 0.84 | 0.24 | 0.89 | 0.13 | 0.85 | 0.04 | 0.7 |
| 0.1 | 1 | 0.05 | 0.8 | 0.05 | 0.8 | 0.05 | 0.8 | 0.05 | 0.79 | 0.04 | 0.75 | 0.01 | 0.61 |
| | 2 | 0.08 | 0.83 | 0.08 | 0.83 | 0.08 | 0.82 | 0.07 | 0.82 | 0.05 | 0.77 | 0.02 | 0.62 |
| | 5 | 0.13 | 0.85 | 0.13 | 0.85 | 0.13 | 0.85 | 0.12 | 0.84 | 0.07 | 0.8 | 0.02 | 0.64 |
| | 10 | 0.17 | 0.86 | 0.17 | 0.86 | 0.17 | 0.86 | 0.15 | 0.86 | 0.08 | 0.81 | 0.02 | 0.66 |
| | 20 | 0.21 | 0.86 | 0.21 | 0.86 | 0.2 | 0.86 | 0.17 | 0.86 | 0.1 | 0.83 | 0.03 | 0.67 |
| | 50 | 0.25 | 0.83 | 0.25 | 0.83 | 0.25 | 0.83 | 0.2 | 0.86 | 0.11 | 0.84 | 0.03 | 0.69 |
| | 100 | 0.28 | 0.79 | 0.28 | 0.79 | 0.27 | 0.81 | 0.21 | 0.86 | 0.12 | 0.84 | 0.03 | 0.69 |
| | 250 | 0.31 | 0.74 | 0.31 | 0.74 | 0.29 | 0.77 | 0.22 | 0.85 | 0.12 | 0.85 | 0.03 | 0.7 |
| | 500 | 0.33 | 0.7 | 0.33 | 0.7 | 0.3 | 0.74 | 0.23 | 0.85 | 0.13 | 0.85 | 0.04 | 0.7 |
| 0.5 | 1 | 0.05 | 0.8 | 0.05 | 0.8 | 0.05 | 0.8 | 0.05 | 0.79 | 0.04 | 0.75 | 0.01 | 0.61 |
| | 2 | 0.08 | 0.82 | 0.08 | 0.82 | 0.08 | 0.82 | 0.07 | 0.81 | 0.05 | 0.77 | 0.02 | 0.62 |
| | 5 | 0.12 | 0.77 | 0.12 | 0.77 | 0.11 | 0.77 | 0.1 | 0.76 | 0.06 | 0.72 | 0.02 | 0.59 |
| | 10 | 0.14 | 0.68 | 0.14 | 0.68 | 0.13 | 0.68 | 0.11 | 0.67 | 0.06 | 0.63 | 0.02 | 0.53 |
| | 20 | 0.15 | 0.56 | 0.15 | 0.56 | 0.14 | 0.56 | 0.11 | 0.55 | 0.06 | 0.52 | 0.02 | 0.45 |
| | 50 | 0.15 | 0.43 | 0.15 | 0.43 | 0.15 | 0.43 | 0.1 | 0.41 | 0.05 | 0.38 | 0.01 | 0.33 |
| | 100 | 0.16 | 0.35 | 0.16 | 0.35 | 0.14 | 0.34 | 0.08 | 0.31 | 0.04 | 0.29 | 0.01 | 0.25 |
| | 250 | 0.16 | 0.27 | 0.16 | 0.27 | 0.13 | 0.25 | 0.06 | 0.21 | 0.03 | 0.19 | 0.01 | 0.17 |
| | 500 | 0.16 | 0.24 | 0.16 | 0.24 | 0.12 | 0.2 | 0.05 | 0.15 | 0.02 | 0.14 | 0.01 | 0.12 |

## 2.2 Bayesian operating characteristics

### 2.2.1 Stopping for efficacy only

#### *2.2.1.1 A = Pr(V | μ > 0) and 1 – B = Pr(V | μ ≤ 0)*

**Figure S1 to S9. Probability of concluding efficacy (binary event $V$) conditional on $\mu > 0$ (panel A) and $\mu \leq 0$ (panel B) in Bayesian trials with efficacy stopping rules $Pr(\mu > 0 | X_{1,\ldots,k_i}) > 0.95, i \in \{1, \ldots, m\}$, no futility stopping rules, and no correction for multiplicity of analyses. $\sigma_{investigator}$ is the standard deviation of the prior $\pi_{investigator} = \mathcal{N}(0, \sigma^2_{investigator})$ used for the analyses, and $\sigma_{reviewer}$ is the standard deviation of the prior $\pi_{reviewer} = \mathcal{N}(0, \sigma^2_{reviewer})$ used to compute the operating characteristics of the trial.**

**Figure S1. For one analysis.**

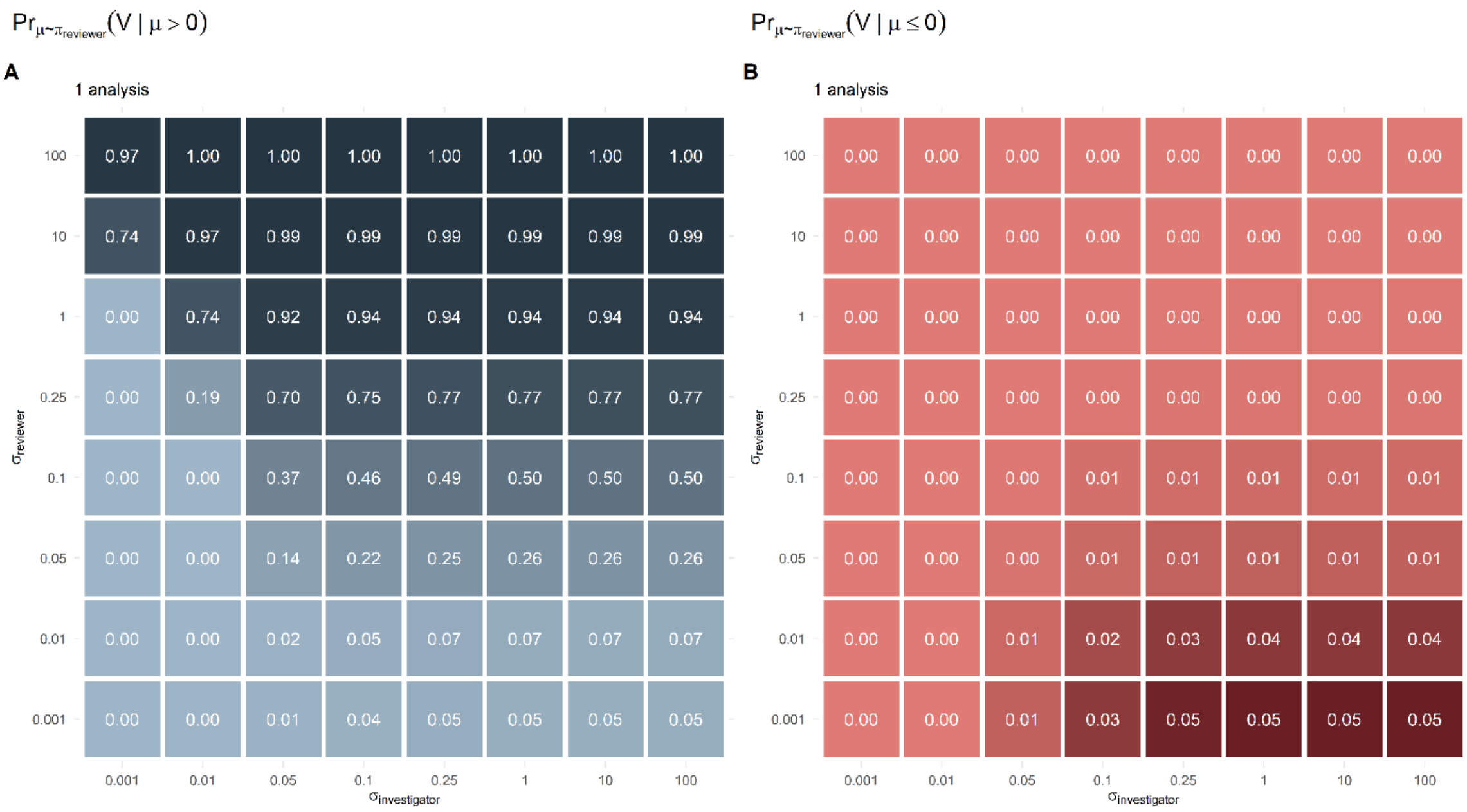

| $\sigma_{reviewer}$ \ $\sigma_{investigator}$ (A) | 0.001 | 0.01 | 0.05 | 0.1 | 0.25 | 1 | 10 | 100 |
|---|---|---|---|---|---|---|---|---|
| 100 | 0.97 | 1.00 | 1.00 | 1.00 | 1.00 | 1.00 | 1.00 | 1.00 |
| 10 | 0.74 | 0.97 | 0.99 | 0.99 | 0.99 | 0.99 | 0.99 | 0.99 |
| 1 | 0.00 | 0.74 | 0.92 | 0.94 | 0.94 | 0.94 | 0.94 | 0.94 |
| 0.25 | 0.00 | 0.19 | 0.70 | 0.75 | 0.77 | 0.77 | 0.77 | 0.77 |
| 0.1 | 0.00 | 0.00 | 0.37 | 0.46 | 0.49 | 0.50 | 0.50 | 0.50 |
| 0.05 | 0.00 | 0.00 | 0.14 | 0.22 | 0.25 | 0.26 | 0.26 | 0.26 |
| 0.01 | 0.00 | 0.00 | 0.02 | 0.05 | 0.07 | 0.07 | 0.07 | 0.07 |
| 0.001 | 0.00 | 0.00 | 0.01 | 0.04 | 0.05 | 0.05 | 0.05 | 0.05 |

| $\sigma_{reviewer}$ \ $\sigma_{investigator}$ (B) | 0.001 | 0.01 | 0.05 | 0.1 | 0.25 | 1 | 10 | 100 |
|---|---|---|---|---|---|---|---|---|
| 100 | 0.00 | 0.00 | 0.00 | 0.00 | 0.00 | 0.00 | 0.00 | 0.00 |
| 10 | 0.00 | 0.00 | 0.00 | 0.00 | 0.00 | 0.00 | 0.00 | 0.00 |
| 1 | 0.00 | 0.00 | 0.00 | 0.00 | 0.00 | 0.00 | 0.00 | 0.00 |
| 0.25 | 0.00 | 0.00 | 0.00 | 0.00 | 0.00 | 0.00 | 0.00 | 0.00 |
| 0.1 | 0.00 | 0.00 | 0.00 | 0.01 | 0.01 | 0.01 | 0.01 | 0.01 |
| 0.05 | 0.00 | 0.00 | 0.00 | 0.01 | 0.01 | 0.01 | 0.01 | 0.01 |
| 0.01 | 0.00 | 0.00 | 0.01 | 0.02 | 0.03 | 0.04 | 0.04 | 0.04 |
| 0.001 | 0.00 | 0.00 | 0.01 | 0.03 | 0.05 | 0.05 | 0.05 | 0.05 |

**Figure S2. For two analyses.**

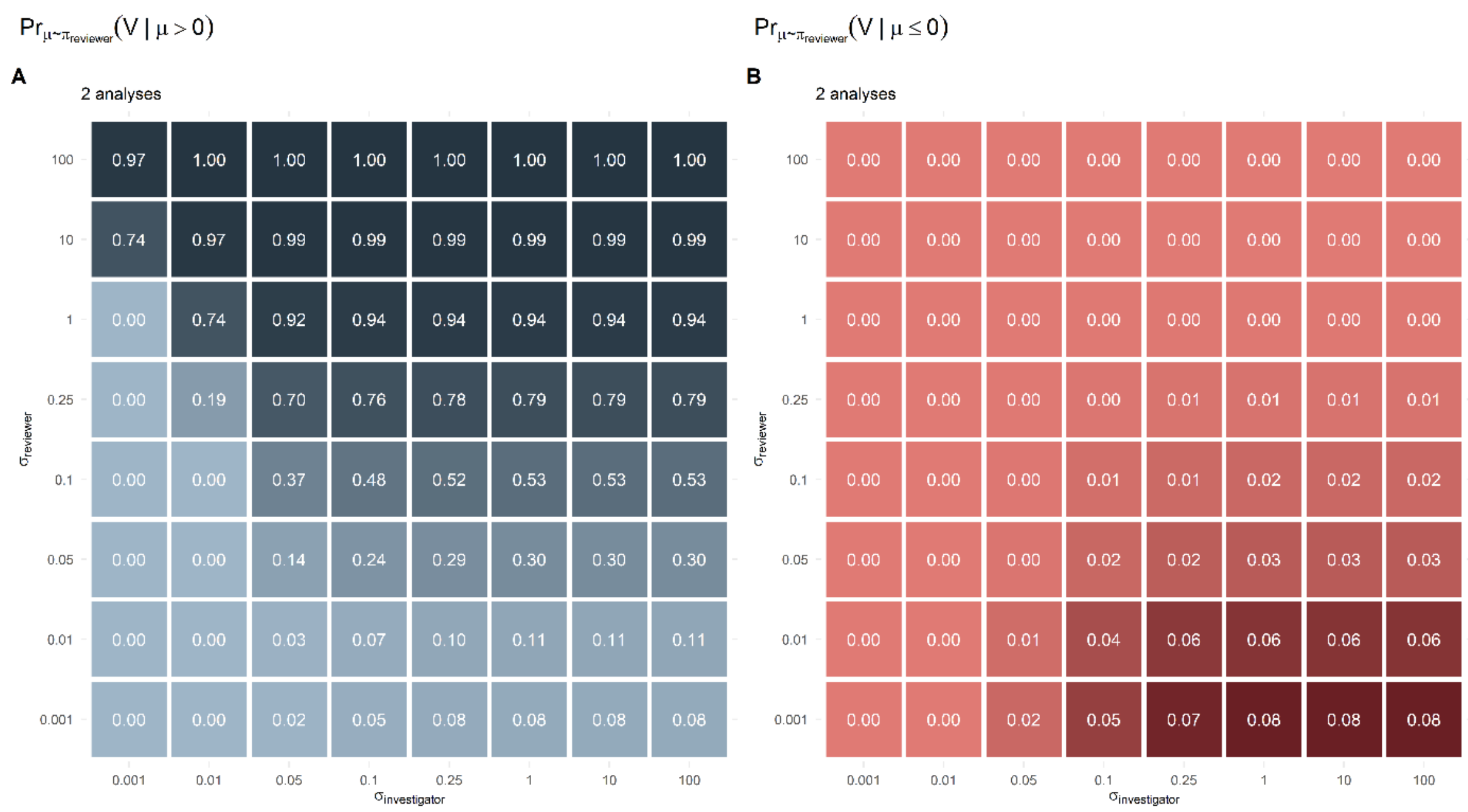

| $\sigma_{reviewer}$ \ $\sigma_{investigator}$ (A) | 0.001 | 0.01 | 0.05 | 0.1 | 0.25 | 1 | 10 | 100 |
|---|---|---|---|---|---|---|---|---|
| 100 | 0.97 | 1.00 | 1.00 | 1.00 | 1.00 | 1.00 | 1.00 | 1.00 |
| 10 | 0.74 | 0.97 | 0.99 | 0.99 | 0.99 | 0.99 | 0.99 | 0.99 |
| 1 | 0.00 | 0.74 | 0.92 | 0.94 | 0.94 | 0.94 | 0.94 | 0.94 |
| 0.25 | 0.00 | 0.19 | 0.70 | 0.76 | 0.78 | 0.79 | 0.79 | 0.79 |
| 0.1 | 0.00 | 0.00 | 0.37 | 0.48 | 0.52 | 0.53 | 0.53 | 0.53 |
| 0.05 | 0.00 | 0.00 | 0.14 | 0.24 | 0.29 | 0.30 | 0.30 | 0.30 |
| 0.01 | 0.00 | 0.00 | 0.03 | 0.07 | 0.10 | 0.11 | 0.11 | 0.11 |
| 0.001 | 0.00 | 0.00 | 0.02 | 0.05 | 0.08 | 0.08 | 0.08 | 0.08 |

| $\sigma_{reviewer}$ \ $\sigma_{investigator}$ (B) | 0.001 | 0.01 | 0.05 | 0.1 | 0.25 | 1 | 10 | 100 |
|---|---|---|---|---|---|---|---|---|
| 100 | 0.00 | 0.00 | 0.00 | 0.00 | 0.00 | 0.00 | 0.00 | 0.00 |
| 10 | 0.00 | 0.00 | 0.00 | 0.00 | 0.00 | 0.00 | 0.00 | 0.00 |
| 1 | 0.00 | 0.00 | 0.00 | 0.00 | 0.00 | 0.00 | 0.00 | 0.00 |
| 0.25 | 0.00 | 0.00 | 0.00 | 0.00 | 0.01 | 0.01 | 0.01 | 0.01 |
| 0.1 | 0.00 | 0.00 | 0.00 | 0.01 | 0.01 | 0.02 | 0.02 | 0.02 |
| 0.05 | 0.00 | 0.00 | 0.00 | 0.02 | 0.02 | 0.03 | 0.03 | 0.03 |
| 0.01 | 0.00 | 0.00 | 0.01 | 0.04 | 0.06 | 0.06 | 0.06 | 0.06 |
| 0.001 | 0.00 | 0.00 | 0.02 | 0.05 | 0.07 | 0.08 | 0.08 | 0.08 |

**Figure S3. For five analyses.**

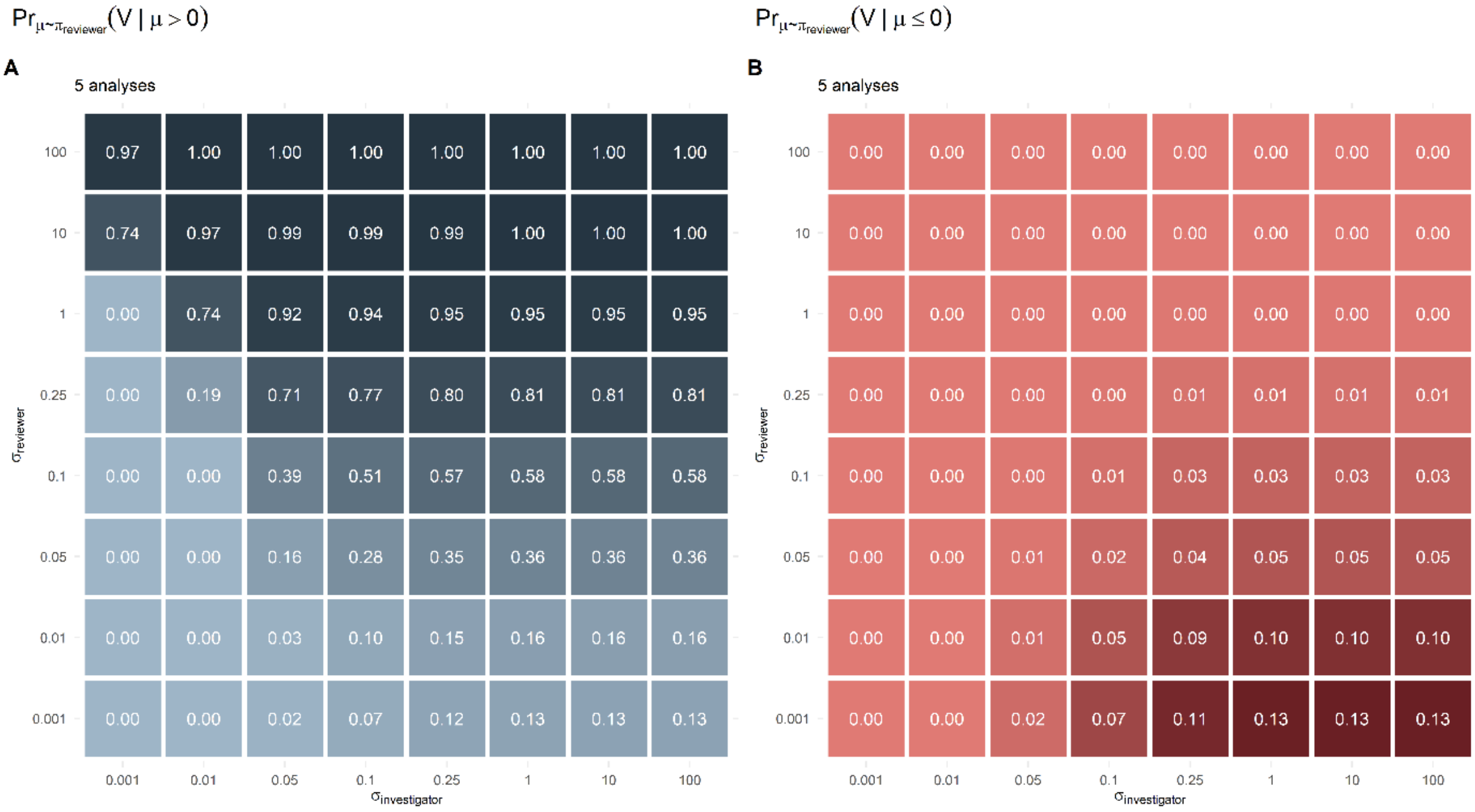

**Figure S4. For 10 analyses.**

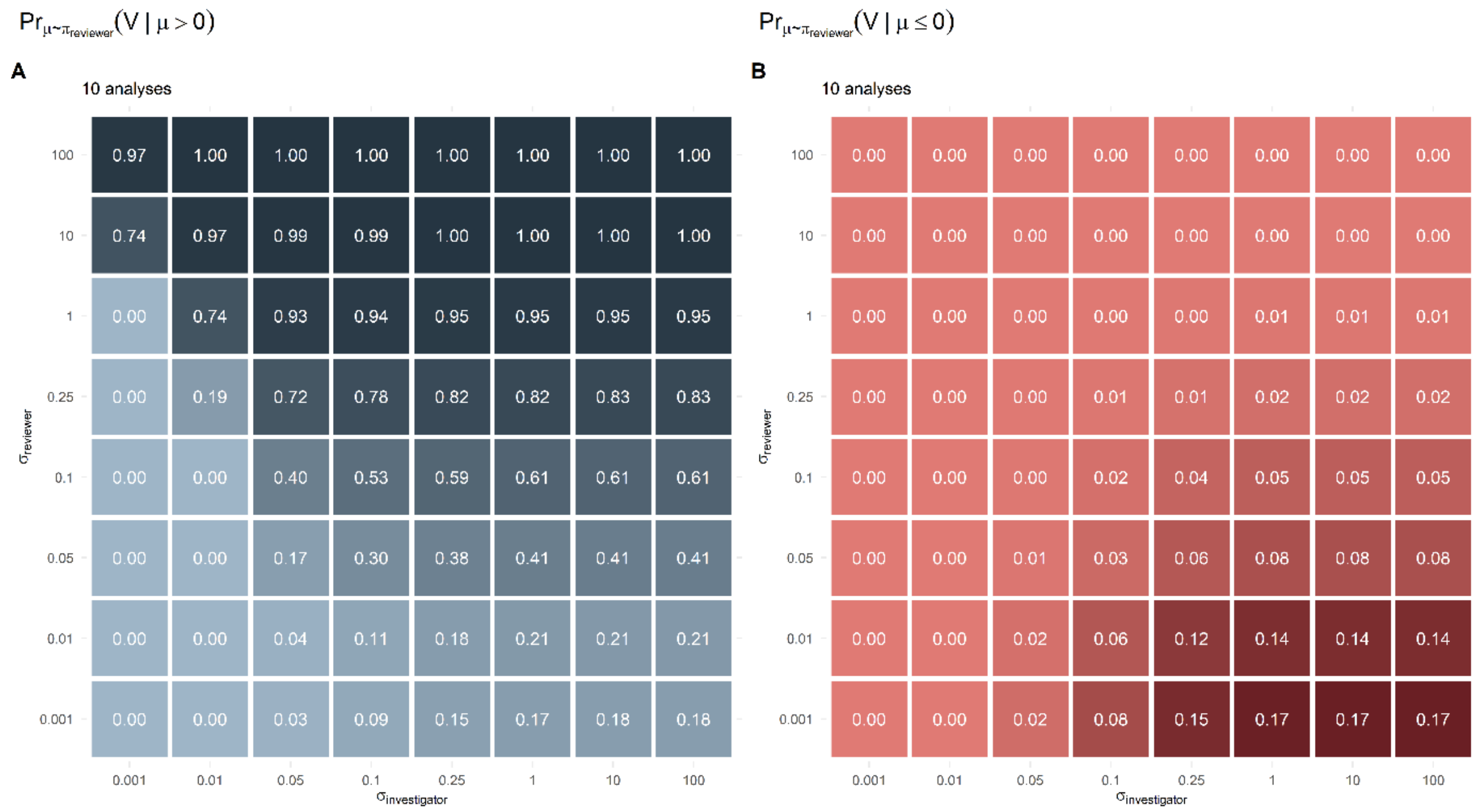

**Figure S5. For 20 analyses.**

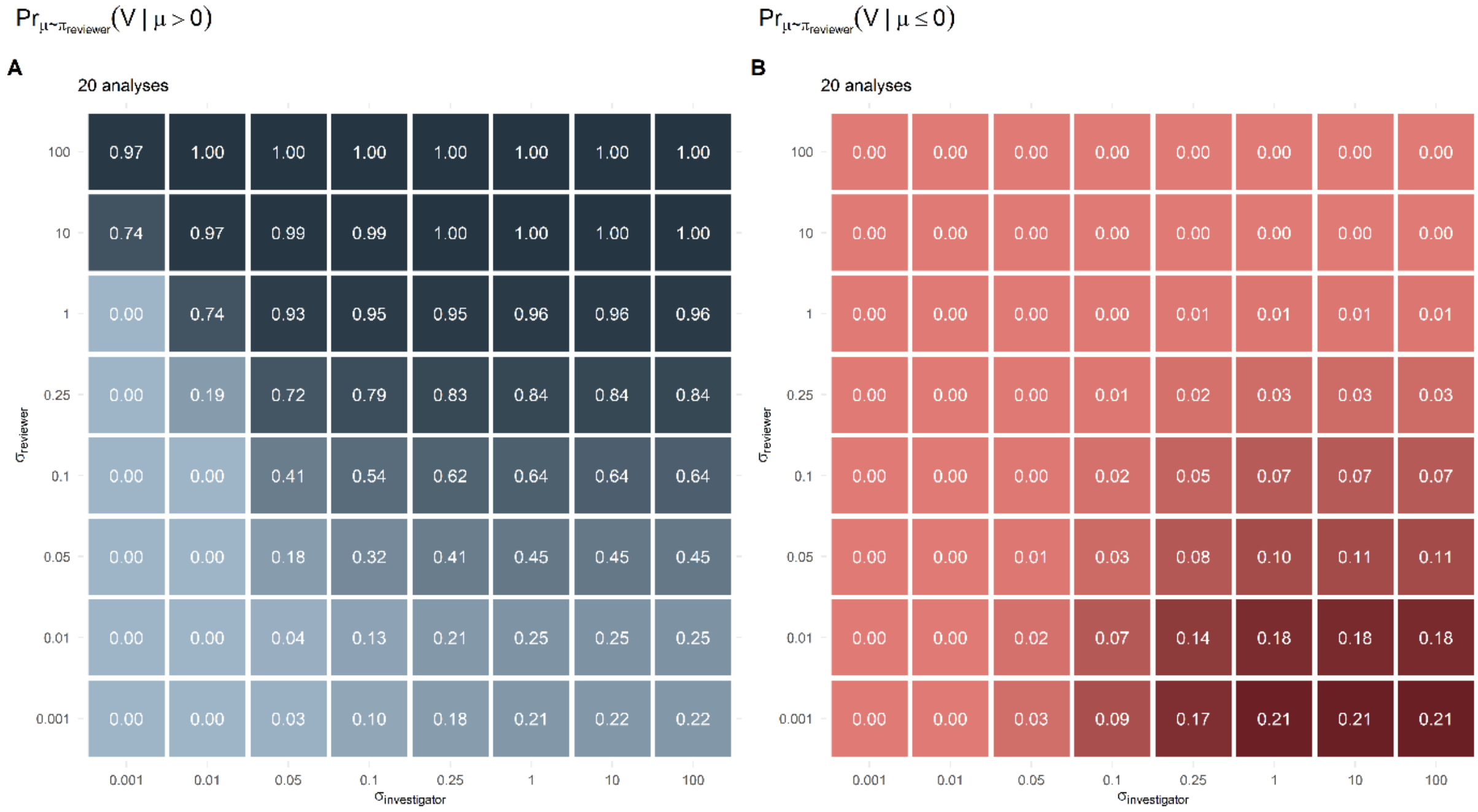

**Figure S6. For 50 analyses.**

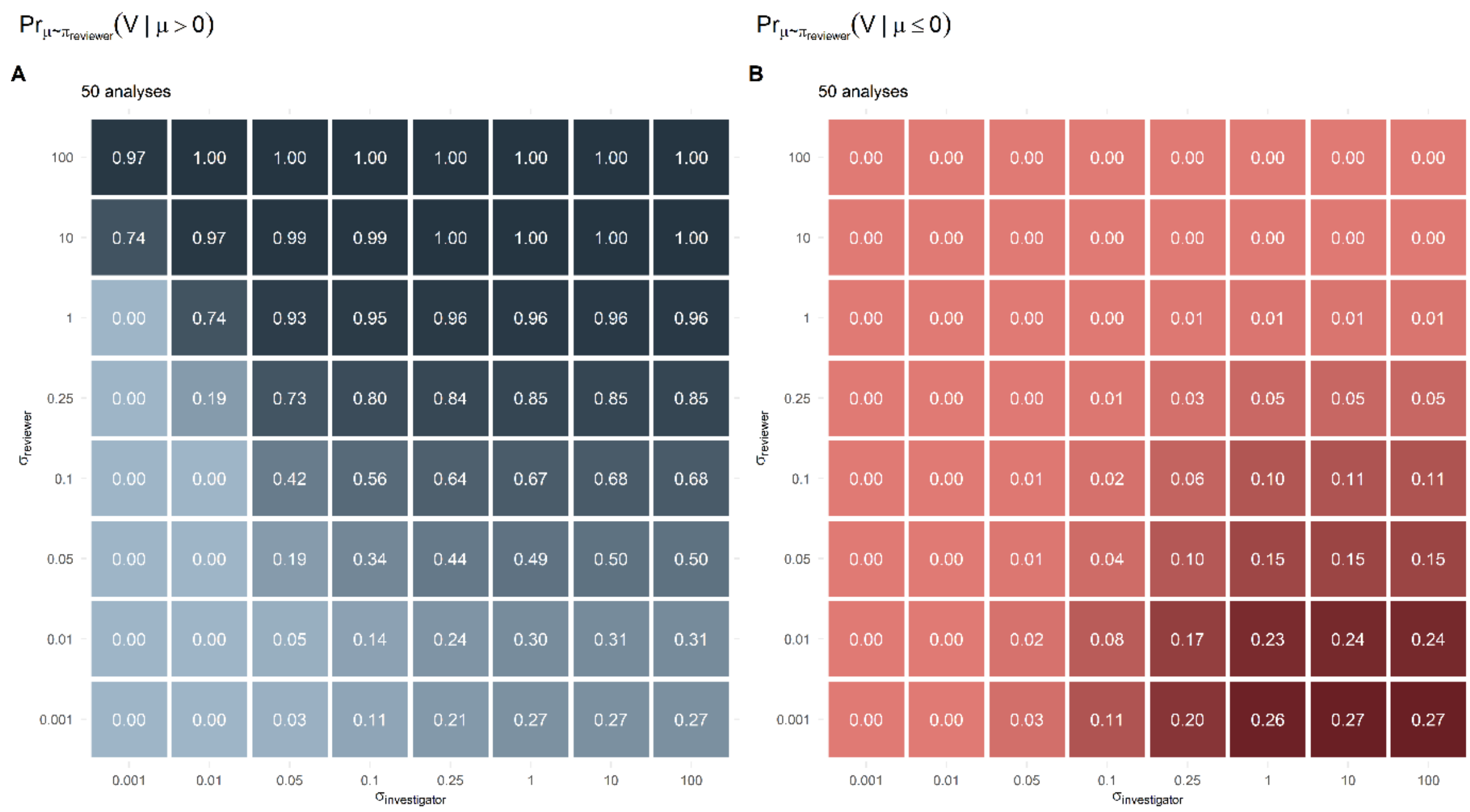

**Figure S7. For 100 analyses.**

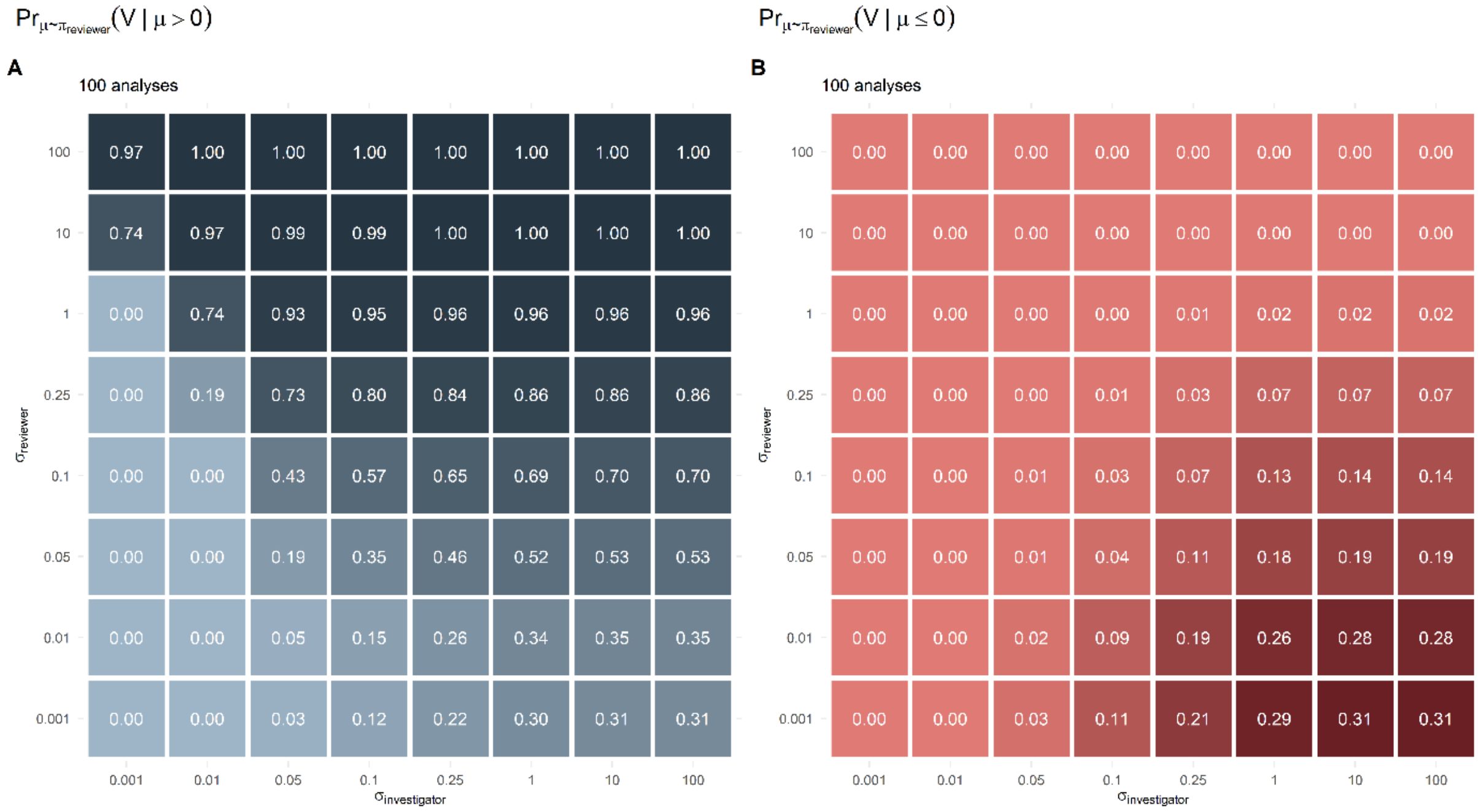

**Figure S8. For 250 analyses.**

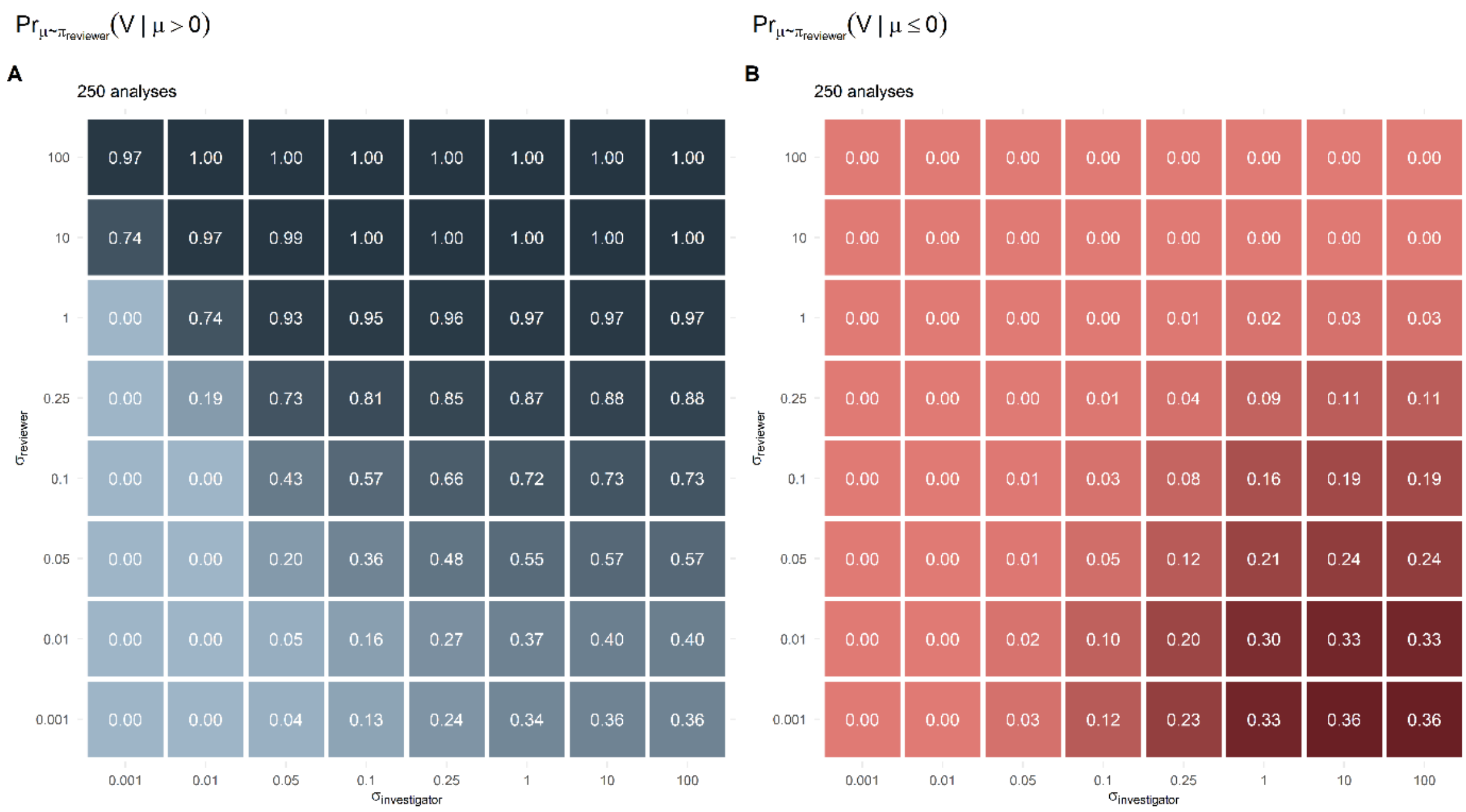

**Figure S9. For 500 analyses.**

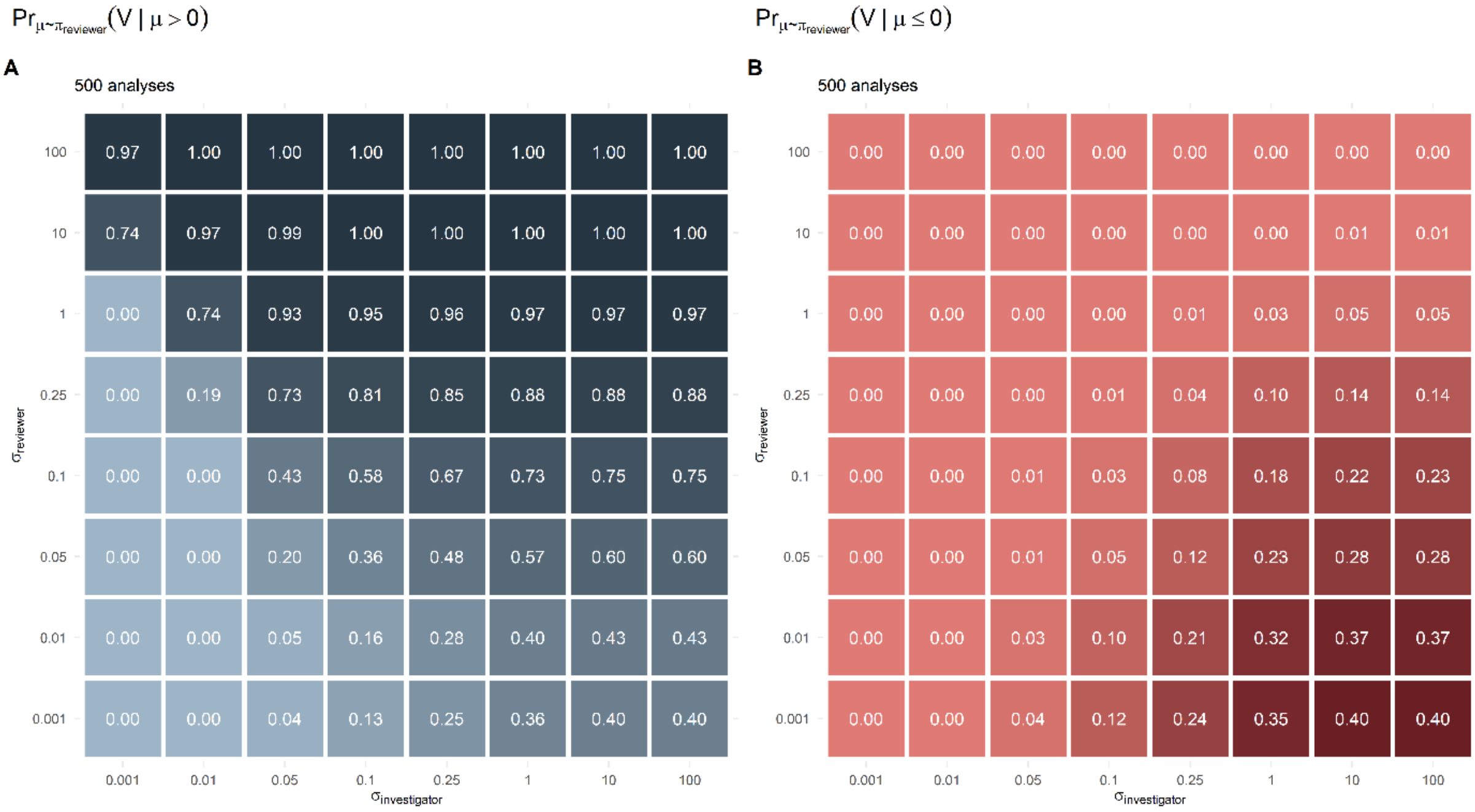

### 2.2.1.2 pTDR = Pr(μ > 0 | V) and pFDR = Pr(μ ≤ 0 | V)

**Figure S10 to S18. Probability of $\mu > 0$ (panel A) and $\mu \leq 0$ (panel B) conditional on having concluded efficacy (binary event $V$) in Bayesian trials with efficacy stopping rules $Pr\left(\mu > 0 \middle| X_{1,\dots,k_i}\right) > 0.95, i \in \{1, \dots, m\}$, no futility stopping rules, and no correction for multiplicity of analyses. $\sigma_{investigator}$ is the standard deviation of the prior $\pi_{investigator} = \mathcal{N}\left(0, \sigma^2_{investigator}\right)$ used for the analyses, and $\sigma_{reviewer}$ is the standard deviation of the prior $\pi_{reviewer} = \mathcal{N}\left(0, \sigma^2_{reviewer}\right)$ used to compute the operating characteristics of the trial. Grey cells are not computable because no event of efficacy conclusion occurred in the simulations.**

**Figure S10. For one analysis.**

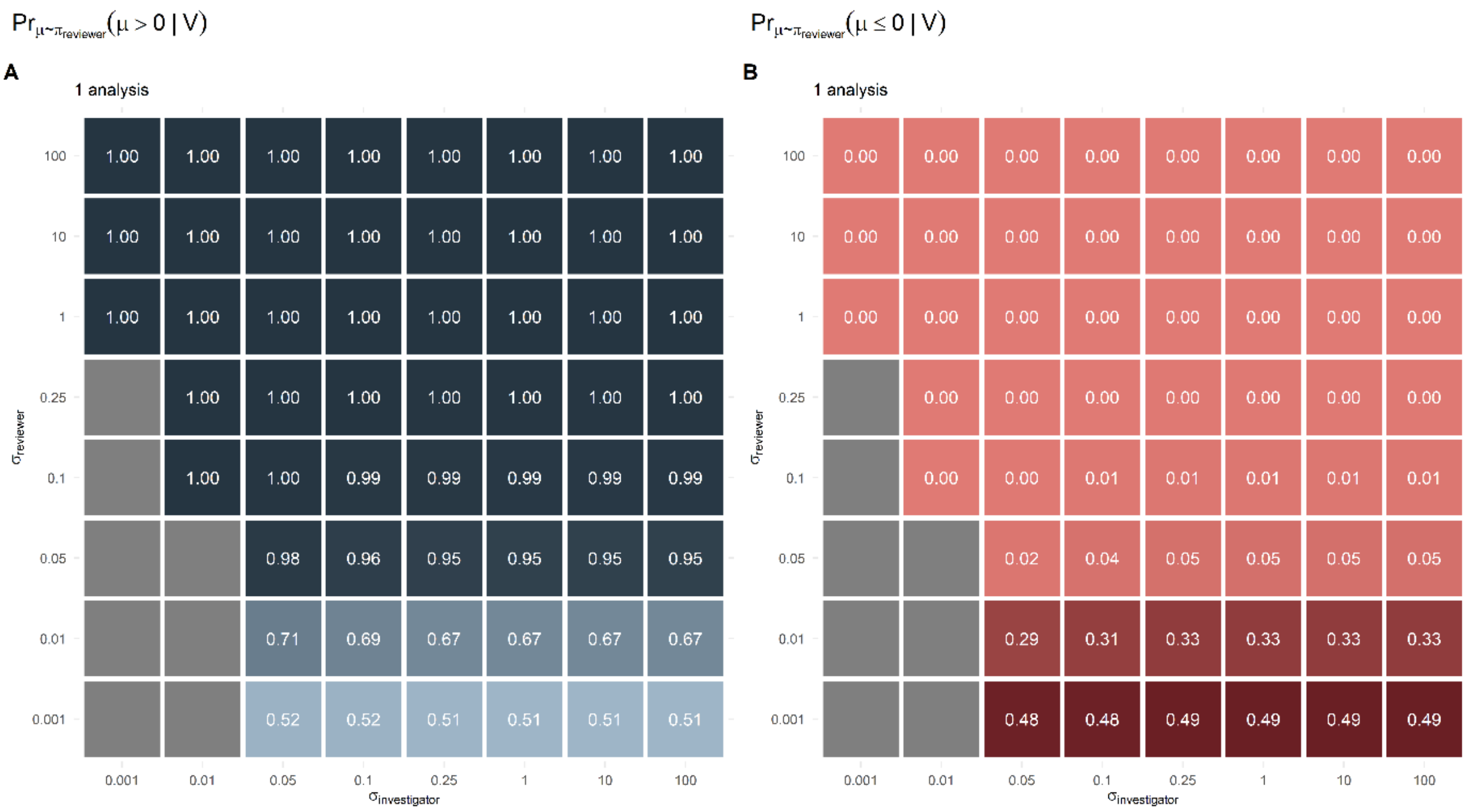

**Figure S11. For two analyses.**

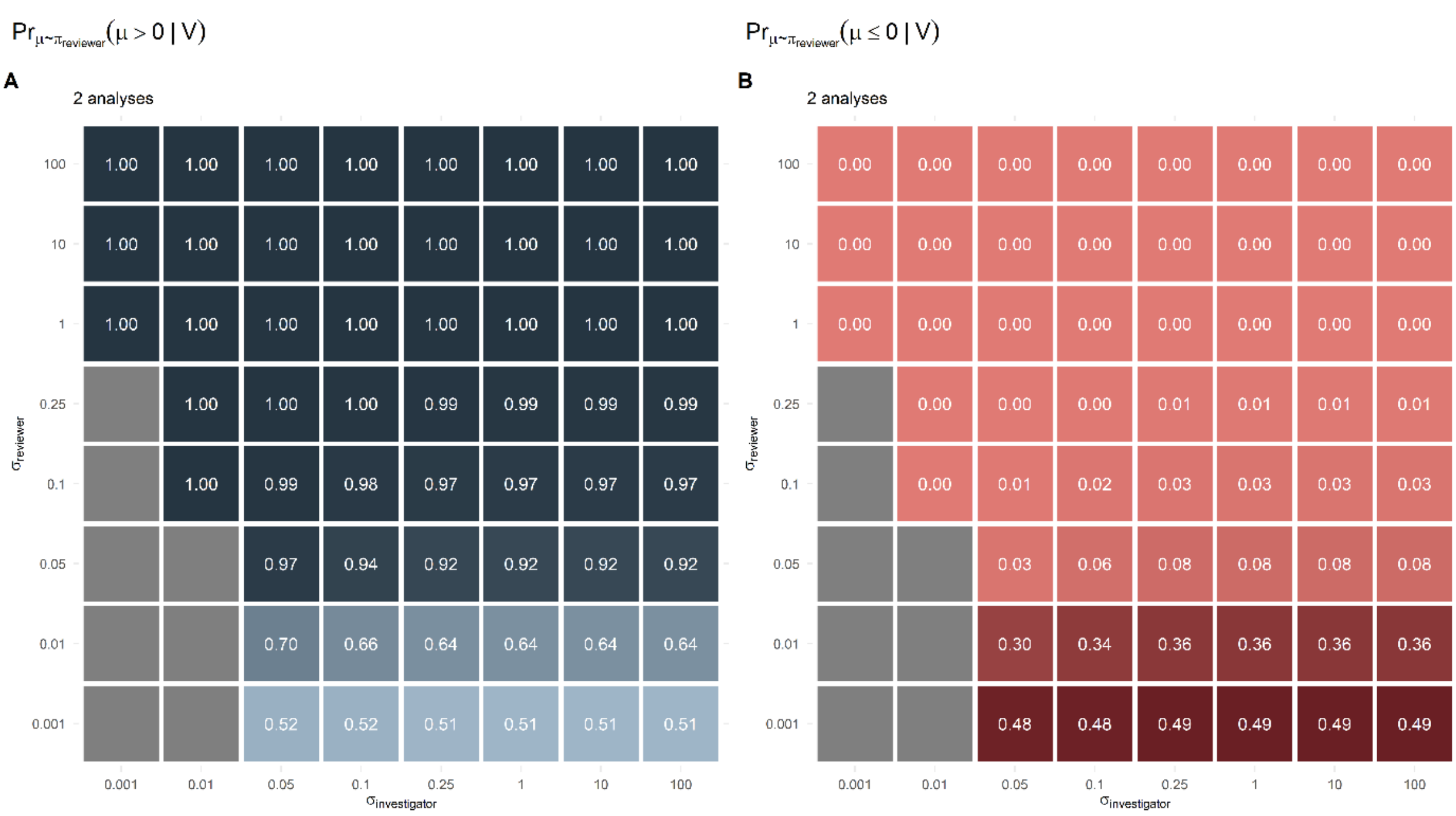

**Figure S12. For five analyses.**

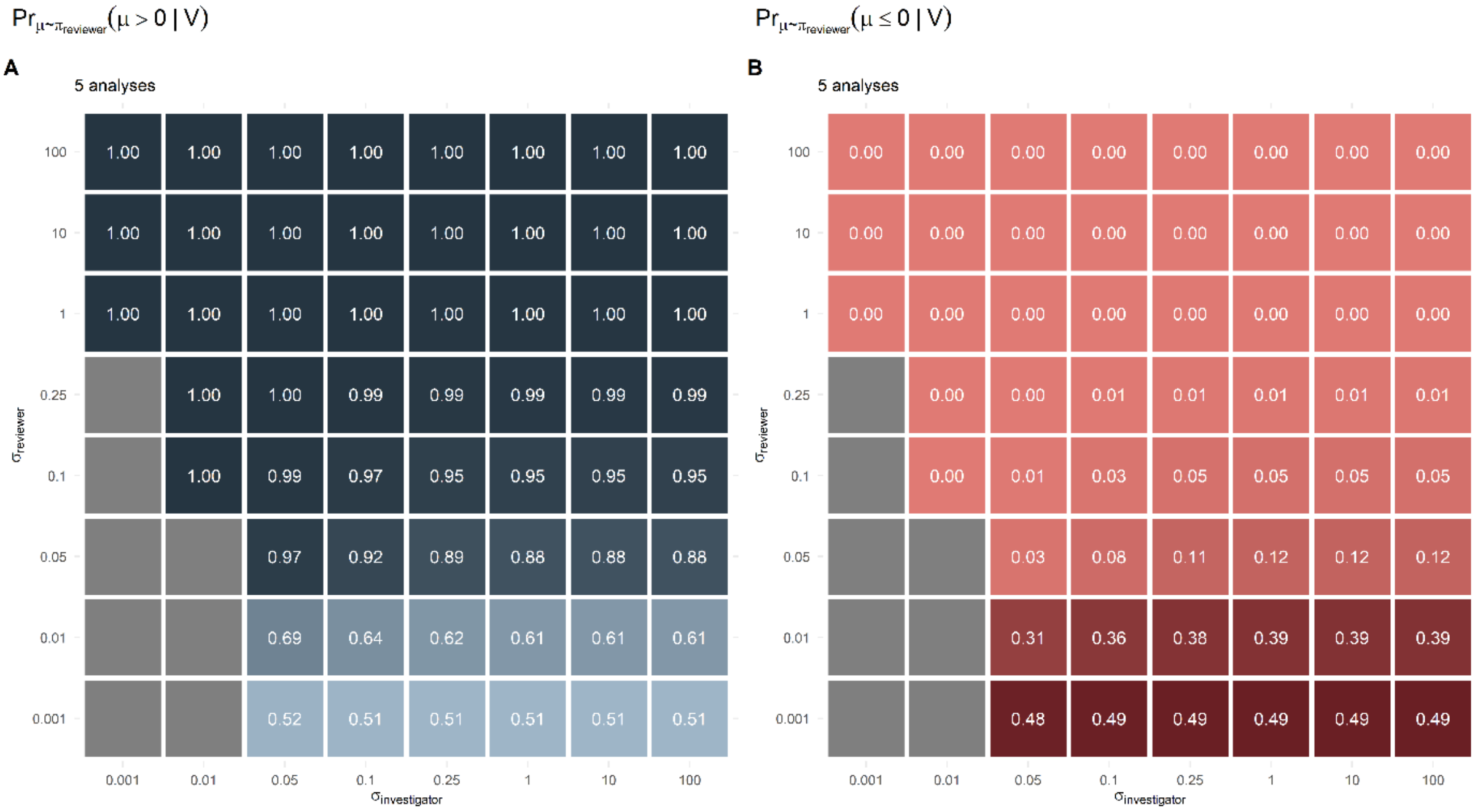

**Figure S13. For 10 analyses.**

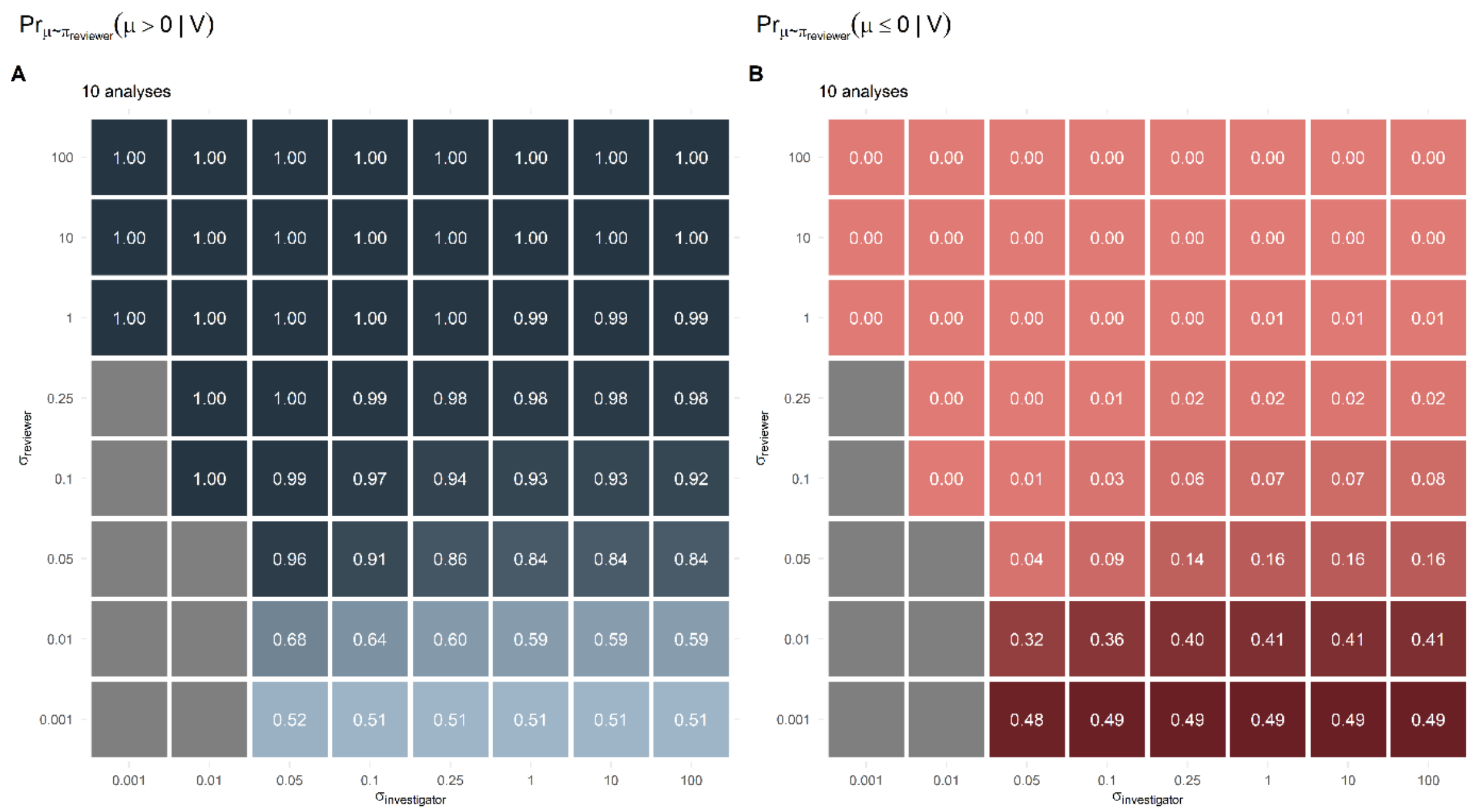

**Figure S14. For 20 analyses.**

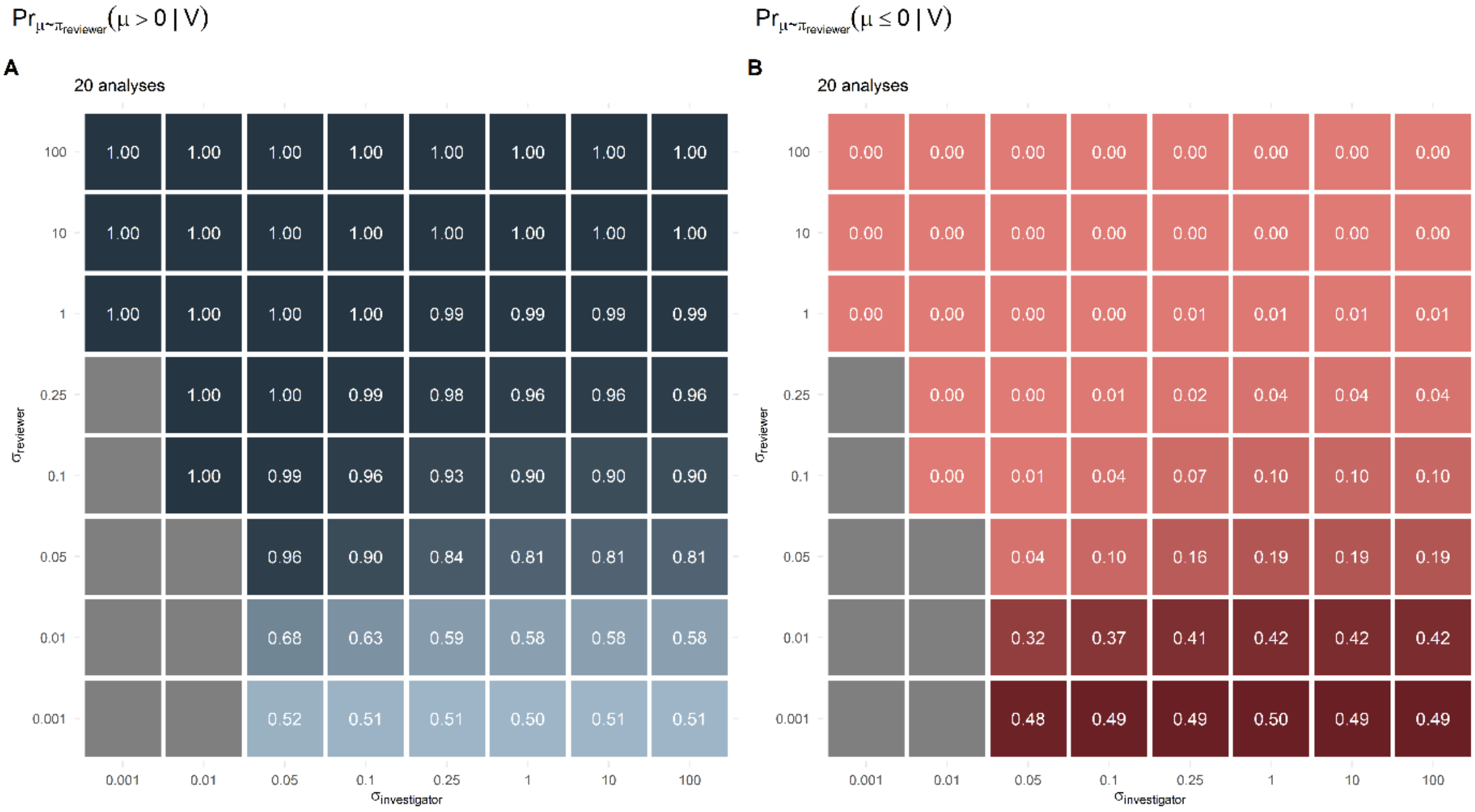

**Figure S15. For 50 analyses.**

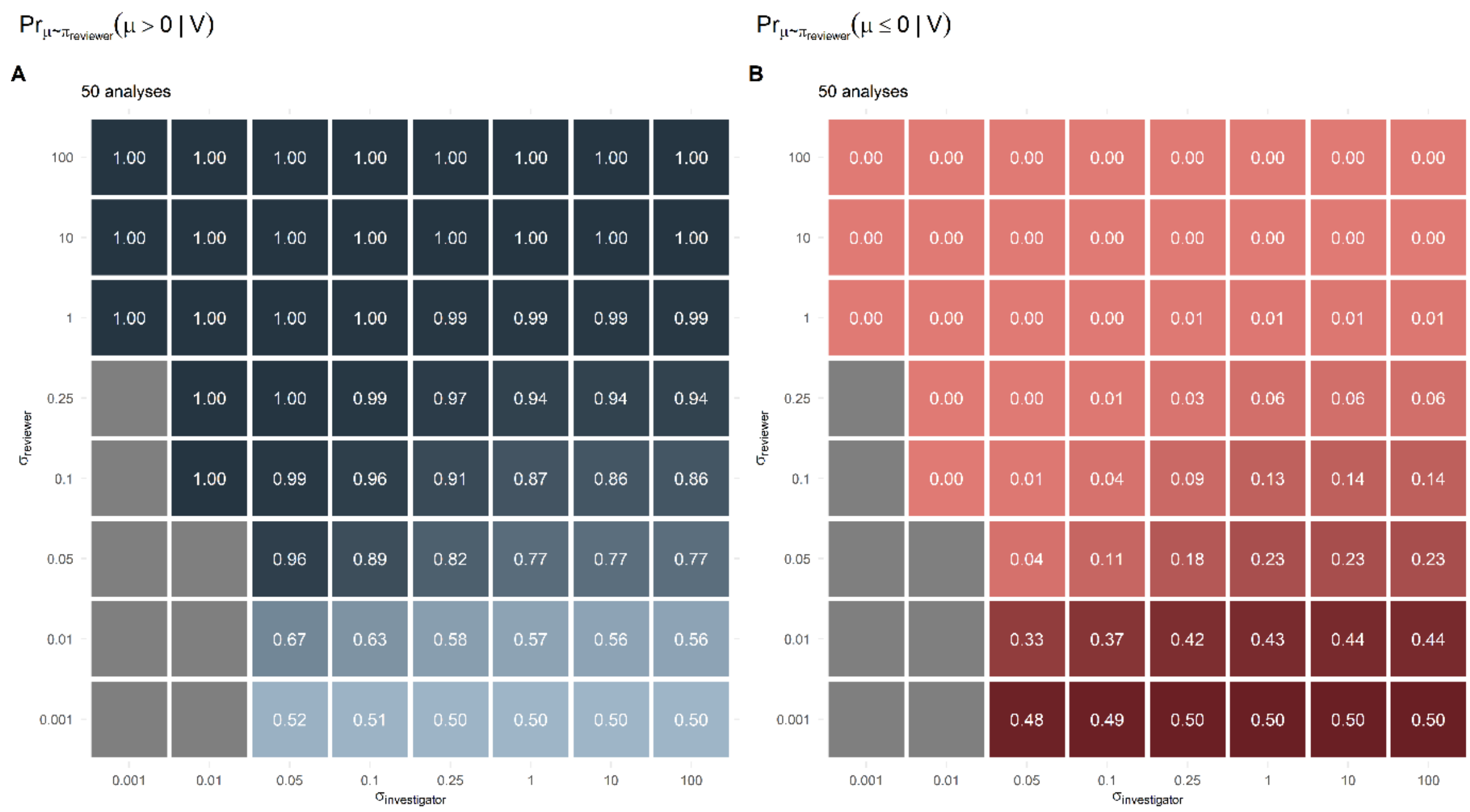

**Figure S16. For 100 analyses.**

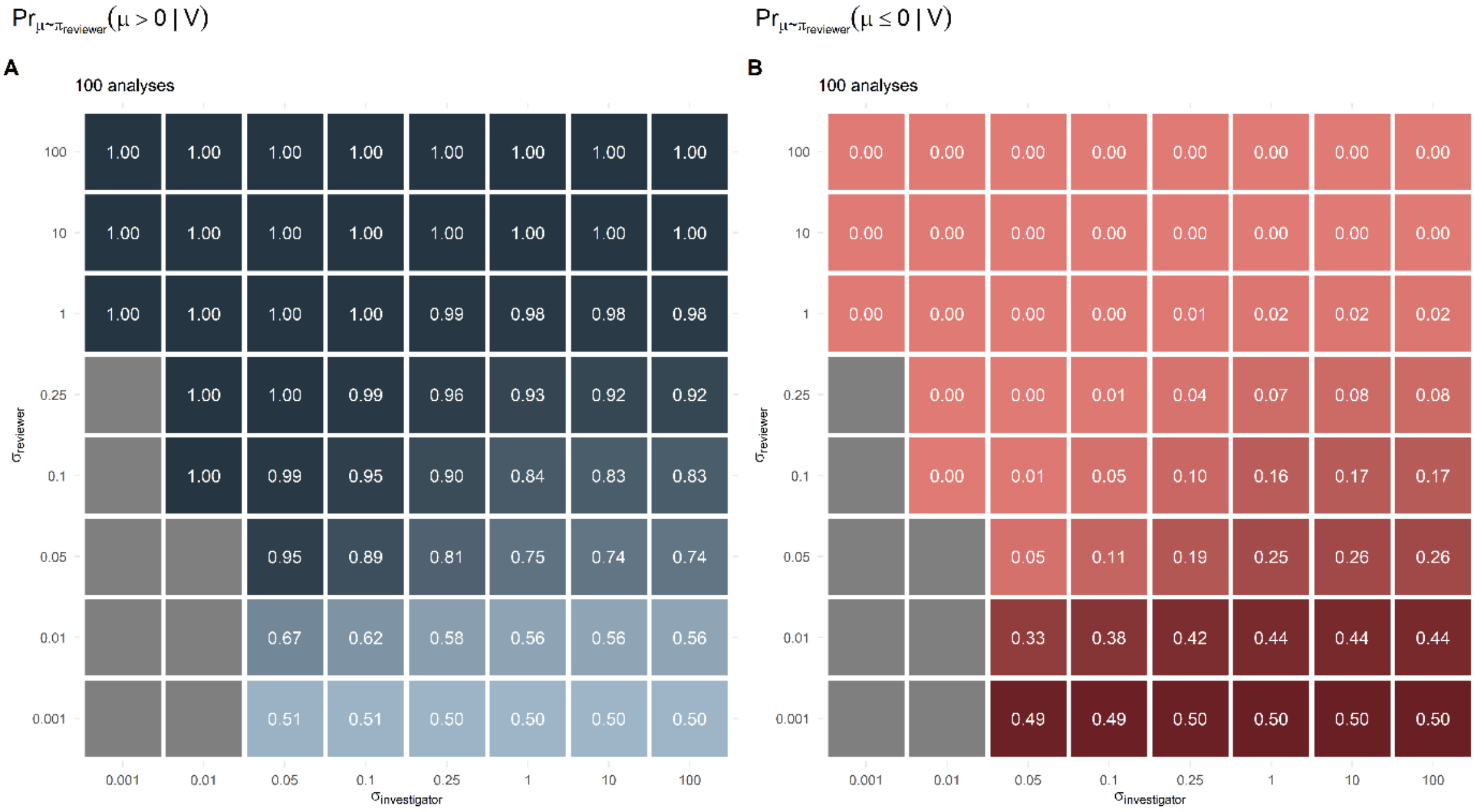

**Figure S17. For 250 analyses.**

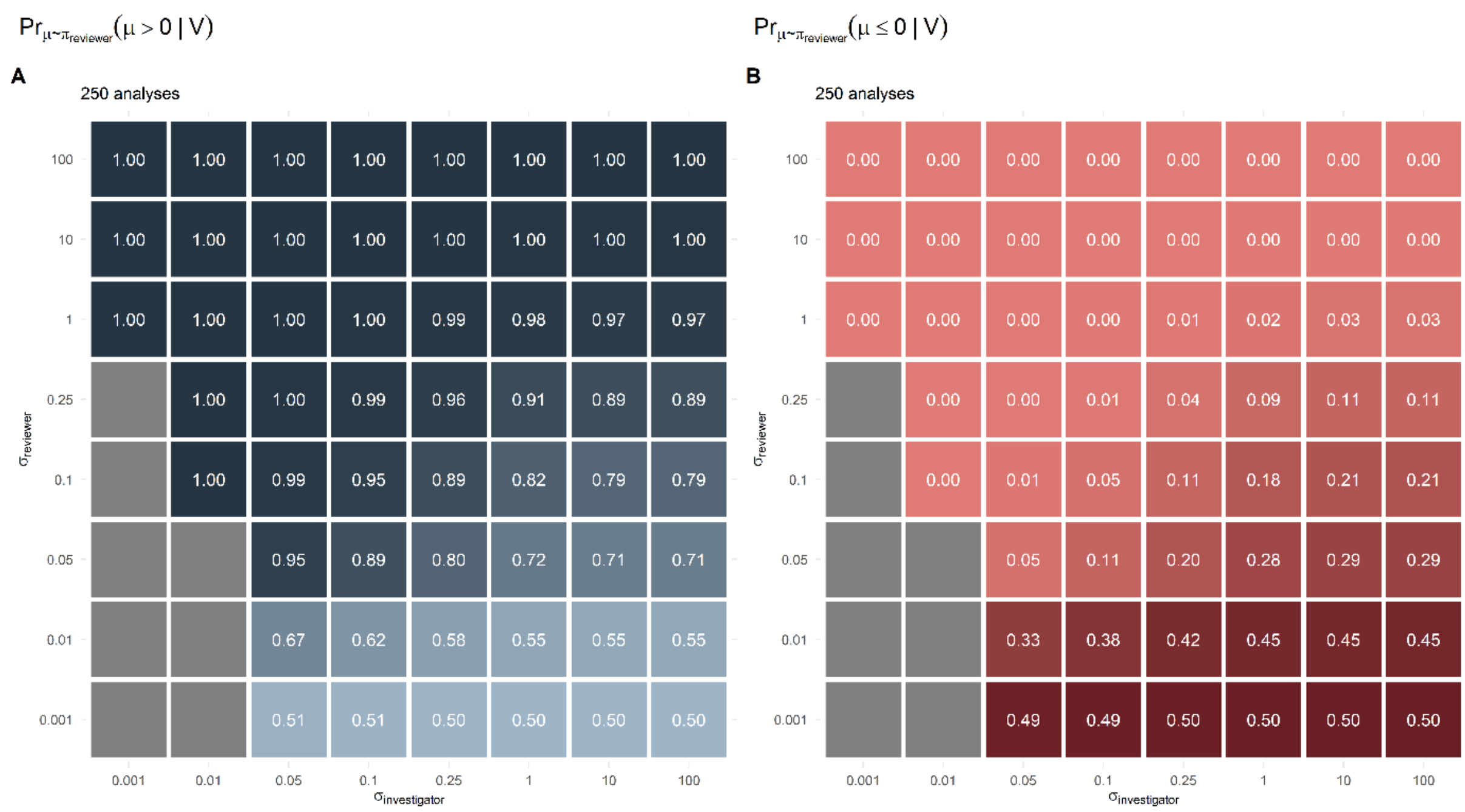

**Figure S18. For 500 analyses.**

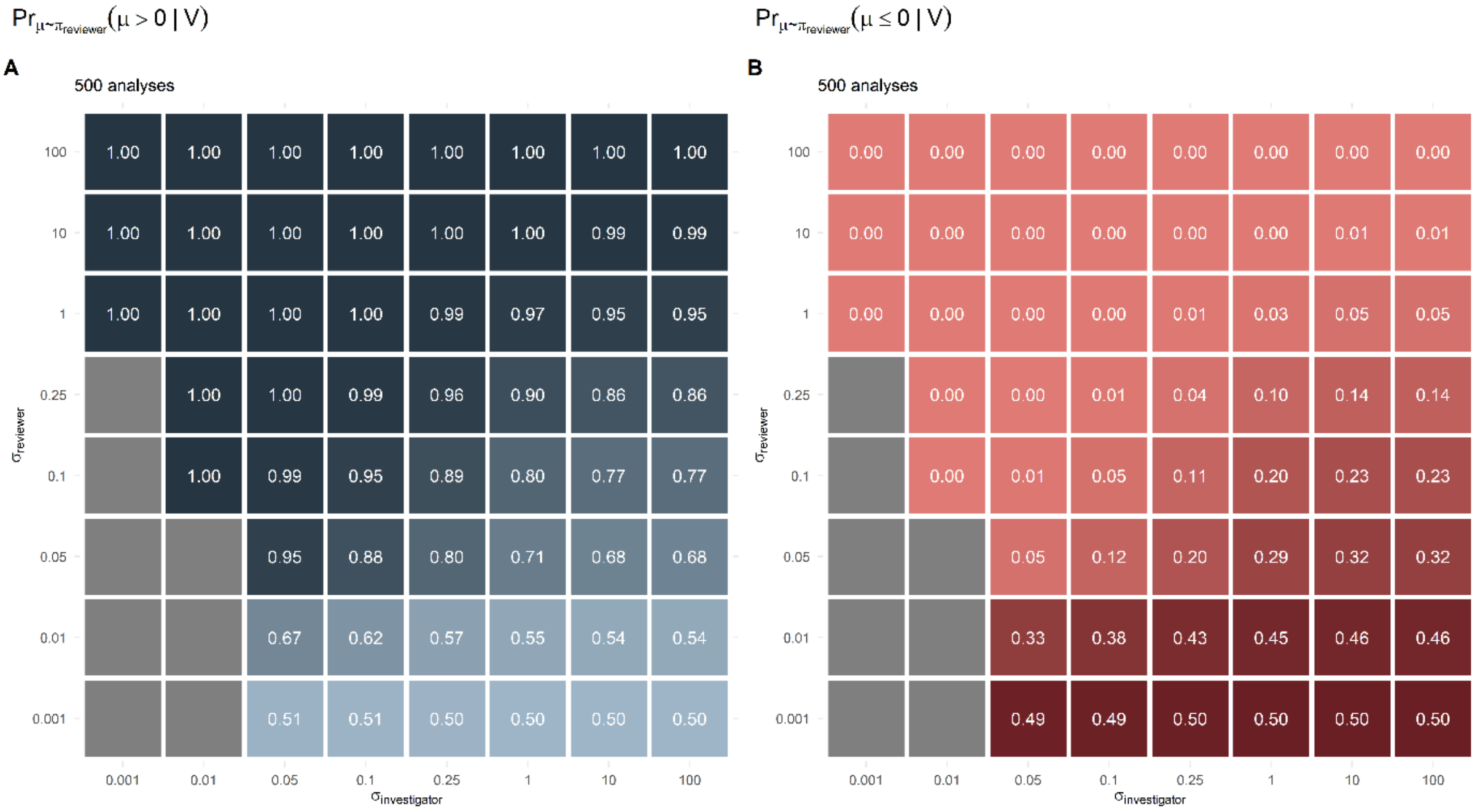

**Table S3. Informative value of a trial concluding efficacy expressed as $pTDO = pTDR/pFDR$, in Bayesian trials with efficacy stopping rules $Pr(\mu > 0|X_{1,\ldots,k_i}) > 0.95, i \in \{1, \ldots, m\}$, no futility stopping rules, and no correction for multiplicity of analyses. $\sigma_{investigator}$ is the standard deviation of the prior $\mathcal{N}(0, \sigma^2_{investigator})$ used for the analyses, and $\sigma_{reviewer}$ is the standard deviation of the prior $\mathcal{N}(0, \sigma^2_{reviewer})$ used to compute the operating characteristics of the trial. Empty cells are not computable because no event of efficacy conclusion occurred in the simulations.**

| Number of analyses | $\sigma_{reviewer}$ | $\text{pTDO} = \frac{pTDR}{pFDR} = \frac{Pr_{\mu \sim \pi_{reviewer}}(\mu > 0 \mid V)}{Pr_{\mu \sim \pi_{reviewer}}(\mu \leq 0 \mid V)}$ | | | | | | | |
|---|---|---|---|---|---|---|---|---|---|
| | | $\sigma_{investigator}$ **0.001** | $\sigma_{investigator}$ **0.01** | $\sigma_{investigator}$ **0.05** | $\sigma_{investigator}$ **0.1** | $\sigma_{investigator}$ **0.25** | $\sigma_{investigator}$ **1** | $\sigma_{investigator}$ **10** | $\sigma_{investigator}$ **100** |
| 1 | 0.001 | | | 1.1 | 1.1 | 1.1 | 1.0 | 1.0 | 1.0 |
| 2 | 0.001 | | | 1.1 | 1.1 | 1.0 | 1.0 | 1.0 | 1.0 |
| 5 | 0.001 | | | 1.1 | 1.0 | 1.0 | 1.0 | 1.0 | 1.0 |
| 10 | 0.001 | | | 1.1 | 1.0 | 1.0 | 1.0 | 1.0 | 1.0 |
| 20 | 0.001 | | | 1.1 | 1.0 | 1.0 | 1.0 | 1.0 | 1.0 |
| 50 | 0.001 | | | 1.1 | 1.0 | 1.0 | 1.0 | 1.0 | 1.0 |
| 100 | 0.001 | | | 1.0 | 1.0 | 1.0 | 1.0 | 1.0 | 1.0 |
| 250 | 0.001 | | | 1.0 | 1.0 | 1.0 | 1.0 | 1.0 | 1.0 |
| 500 | 0.001 | | | 1.0 | 1.0 | 1.0 | 1.0 | 1.0 | 1.0 |
| 1 | 0.01 | | | 2.5 | 2.2 | 2.1 | 2.1 | 2.1 | 2.1 |
| 2 | 0.01 | | | 2.3 | 1.9 | 1.8 | 1.8 | 1.8 | 1.8 |
| 5 | 0.01 | | | 2.2 | 1.8 | 1.6 | 1.6 | 1.6 | 1.6 |
| 10 | 0.01 | | | 2.2 | 1.7 | 1.5 | 1.5 | 1.5 | 1.5 |
| 20 | 0.01 | | | 2.1 | 1.7 | 1.5 | 1.4 | 1.4 | 1.4 |
| 50 | 0.01 | | | 2.0 | 1.7 | 1.4 | 1.3 | 1.3 | 1.3 |
| 100 | 0.01 | | | 2.0 | 1.7 | 1.4 | 1.3 | 1.3 | 1.3 |
| 250 | 0.01 | | | 2.0 | 1.6 | 1.4 | 1.2 | 1.2 | 1.2 |
| 500 | 0.01 | | | 2.0 | 1.6 | 1.3 | 1.2 | 1.2 | 1.2 |
| 1 | 0.05 | | | 47 | 24 | 20 | 19 | 19 | 19 |
| 2 | 0.05 | | | 36 | 16 | 12 | 12 | 12 | 12 |
| 5 | 0.05 | | | 29 | 12 | 7.9 | 7.2 | 7.2 | 7.2 |
| 10 | 0.05 | | | 27 | 10 | 6.3 | 5.4 | 5.4 | 5.4 |
| 20 | 0.05 | | | 24 | 9.3 | 5.4 | 4.3 | 4.2 | 4.2 |
| 50 | 0.05 | | | 21 | 8.4 | 4.6 | 3.4 | 3.3 | 3.3 |
| 100 | 0.05 | | | 21 | 8.1 | 4.3 | 3.0 | 2.8 | 2.8 |
| 250 | 0.05 | | | 20 | 7.8 | 4.1 | 2.6 | 2.4 | 2.4 |
| 500 | 0.05 | | | 19 | 7.6 | 3.9 | 2.5 | 2.2 | 2.2 |
| 1 | 0.1 | | ≥ 1000 | 249 | 90 | 70 | 66 | 66 | 66 |
| 2 | 0.1 | | ≥ 1000 | 166 | 52 | 39 | 36 | 35 | 35 |
| 5 | 0.1 | | ≥ 1000 | 142 | 37 | 21 | 18 | 18 | 18 |
| 10 | 0.1 | | ≥ 1000 | 110 | 31 | 15 | 13 | 12 | 12 |
| 20 | 0.1 | | ≥ 1000 | 99 | 27 | 12 | 9.2 | 8.9 | 8.9 |
| 50 | 0.1 | | ≥ 1000 | 82 | 23 | 10 | 6.6 | 6.2 | 6.2 |
| 100 | 0.1 | | ≥ 1000 | 76 | 21 | 9.2 | 5.5 | 5.0 | 5.0 |
| 250 | 0.1 | | ≥ 1000 | 76 | 20 | 8.4 | 4.4 | 3.9 | 3.9 |
| 500 | 0.1 | | ≥ 1000 | 70 | 19 | 8.0 | 4.0 | 3.3 | 3.3 |
| 1 | 0.25 | | ≥ 1000 | 999 | 499 | 332 | 332 | 332 | 332 |
| 2 | 0.25 | | ≥ 1000 | 999 | 249 | 166 | 142 | 142 | 142 |
| 5 | 0.25 | | ≥ 1000 | 999 | 166 | 82 | 70 | 66 | 66 |
| 10 | 0.25 | | ≥ 1000 | 999 | 124 | 55 | 42 | 41 | 41 |
| 20 | 0.25 | | ≥ 1000 | 499 | 110 | 41 | 27 | 26 | 26 |
| 50 | 0.25 | | ≥ 1000 | 499 | 90 | 31 | 17 | 16 | 16 |
| 100 | 0.25 | | ≥ 1000 | 499 | 82 | 27 | 13 | 12 | 12 |
| 250 | 0.25 | | ≥ 1000 | 499 | 76 | 23 | 9.9 | 8.0 | 7.9 |
| 500 | 0.25 | | ≥ 1000 | 332 | 76 | 22 | 8.5 | 6.4 | 6.4 |
| 1 | 1 | ≥ 1000 | ≥ 1000 | ≥ 1000 | 999 | 999 | 999 | 999 | 999 |
| 2 | 1 | ≥ 1000 | ≥ 1000 | ≥ 1000 | 999 | 499 | 499 | 499 | 499 |
| 5 | 1 | ≥ 1000 | ≥ 1000 | ≥ 1000 | 499 | 332 | 249 | 249 | 249 |
| 10 | 1 | ≥ 1000 | ≥ 1000 | ≥ 1000 | 499 | 249 | 199 | 199 | 199 |
| 20 | 1 | ≥ 1000 | ≥ 1000 | ≥ 1000 | 499 | 199 | 124 | 124 | 124 |
| 50 | 1 | ≥ 1000 | ≥ 1000 | 999 | 332 | 142 | 76 | 70 | 70 |
| 100 | 1 | ≥ 1000 | ≥ 1000 | 999 | 332 | 124 | 55 | 47 | 47 |
| 250 | 1 | ≥ 1000 | ≥ 1000 | 999 | 332 | 110 | 39 | 29 | 29 |
| 500 | 1 | ≥ 1000 | ≥ 1000 | 999 | 332 | 99 | 32 | 21 | 20 |
| 1 | 10 | ≥ 1000 | ≥ 1000 | ≥ 1000 | ≥ 1000 | ≥ 1000 | ≥ 1000 | ≥ 1000 | ≥ 1000 |
| 2 | 10 | ≥ 1000 | ≥ 1000 | ≥ 1000 | ≥ 1000 | ≥ 1000 | ≥ 1000 | ≥ 1000 | ≥ 1000 |
| 5 | 10 | ≥ 1000 | ≥ 1000 | ≥ 1000 | ≥ 1000 | ≥ 1000 | ≥ 1000 | ≥ 1000 | ≥ 1000 |
| 10 | 10 | ≥ 1000 | ≥ 1000 | ≥ 1000 | ≥ 1000 | 999 | 999 | 999 | 999 |
| 20 | 10 | ≥ 1000 | ≥ 1000 | ≥ 1000 | ≥ 1000 | 999 | 999 | 999 | 999 |
| 50 | 10 | ≥ 1000 | ≥ 1000 | ≥ 1000 | ≥ 1000 | 999 | 499 | 499 | 499 |
| 100 | 10 | ≥ 1000 | ≥ 1000 | ≥ 1000 | ≥ 1000 | 999 | 499 | 499 | 499 |
| 250 | 10 | ≥ 1000 | ≥ 1000 | ≥ 1000 | ≥ 1000 | 999 | 332 | 249 | 249 |
| 500 | 10 | ≥ 1000 | ≥ 1000 | ≥ 1000 | ≥ 1000 | 999 | 332 | 199 | 199 |
| 1 | 100 | ≥ 1000 | ≥ 1000 | ≥ 1000 | ≥ 1000 | ≥ 1000 | ≥ 1000 | ≥ 1000 | ≥ 1000 |
| 2 | 100 | ≥ 1000 | ≥ 1000 | ≥ 1000 | ≥ 1000 | ≥ 1000 | ≥ 1000 | ≥ 1000 | ≥ 1000 |
| 5 | 100 | ≥ 1000 | ≥ 1000 | ≥ 1000 | ≥ 1000 | ≥ 1000 | ≥ 1000 | ≥ 1000 | ≥ 1000 |
| 10 | 100 | ≥ 1000 | ≥ 1000 | ≥ 1000 | ≥ 1000 | ≥ 1000 | ≥ 1000 | ≥ 1000 | ≥ 1000 |
| 20 | 100 | ≥ 1000 | ≥ 1000 | ≥ 1000 | ≥ 1000 | ≥ 1000 | ≥ 1000 | ≥ 1000 | ≥ 1000 |
| 50 | 100 | ≥ 1000 | ≥ 1000 | ≥ 1000 | ≥ 1000 | ≥ 1000 | ≥ 1000 | ≥ 1000 | ≥ 1000 |
| 100 | 100 | ≥ 1000 | ≥ 1000 | ≥ 1000 | ≥ 1000 | ≥ 1000 | ≥ 1000 | ≥ 1000 | ≥ 1000 |
| 250 | 100 | ≥ 1000 | ≥ 1000 | ≥ 1000 | ≥ 1000 | ≥ 1000 | ≥ 1000 | ≥ 1000 | ≥ 1000 |
| 500 | 100 | ≥ 1000 | ≥ 1000 | ≥ 1000 | ≥ 1000 | ≥ 1000 | ≥ 1000 | ≥ 1000 | ≥ 1000 |

### 2.2.2 Stopping for efficacy or futility

#### *2.2.2.1 Futility threshold = 0.05*

***A = Pr(V | μ > 0) and 1 – B = Pr(V | μ ≤ 0)***

**Figure S19 to S27. Probability of concluding efficacy (binary event $V$) conditional on $\mu > 0$ (panel A) and $\mu \leq 0$ (panel B) in Bayesian trials with efficacy stopping rules $Pr(\mu > 0|X_{1,\ldots,k_i}) > 0.95, i \in \{1, \ldots, m\}$, futility stopping rules $Pr(\mu > 0|X_{1,\ldots,k_i}) < 0.05, i \in \{1, \ldots, m-1\}$, and no correction for multiplicity of analyses. $\sigma_{investigator}$ is the standard deviation of the prior $\pi_{investigator} = \mathcal{N}(0, \sigma^2_{investigator})$ used for the analyses, and $\sigma_{reviewer}$ is the standard deviation of the prior $\pi_{reviewer} = \mathcal{N}(0, \sigma^2_{reviewer})$ used to compute the operating characteristics of the trial.**

**Figure S19. For one analysis.**

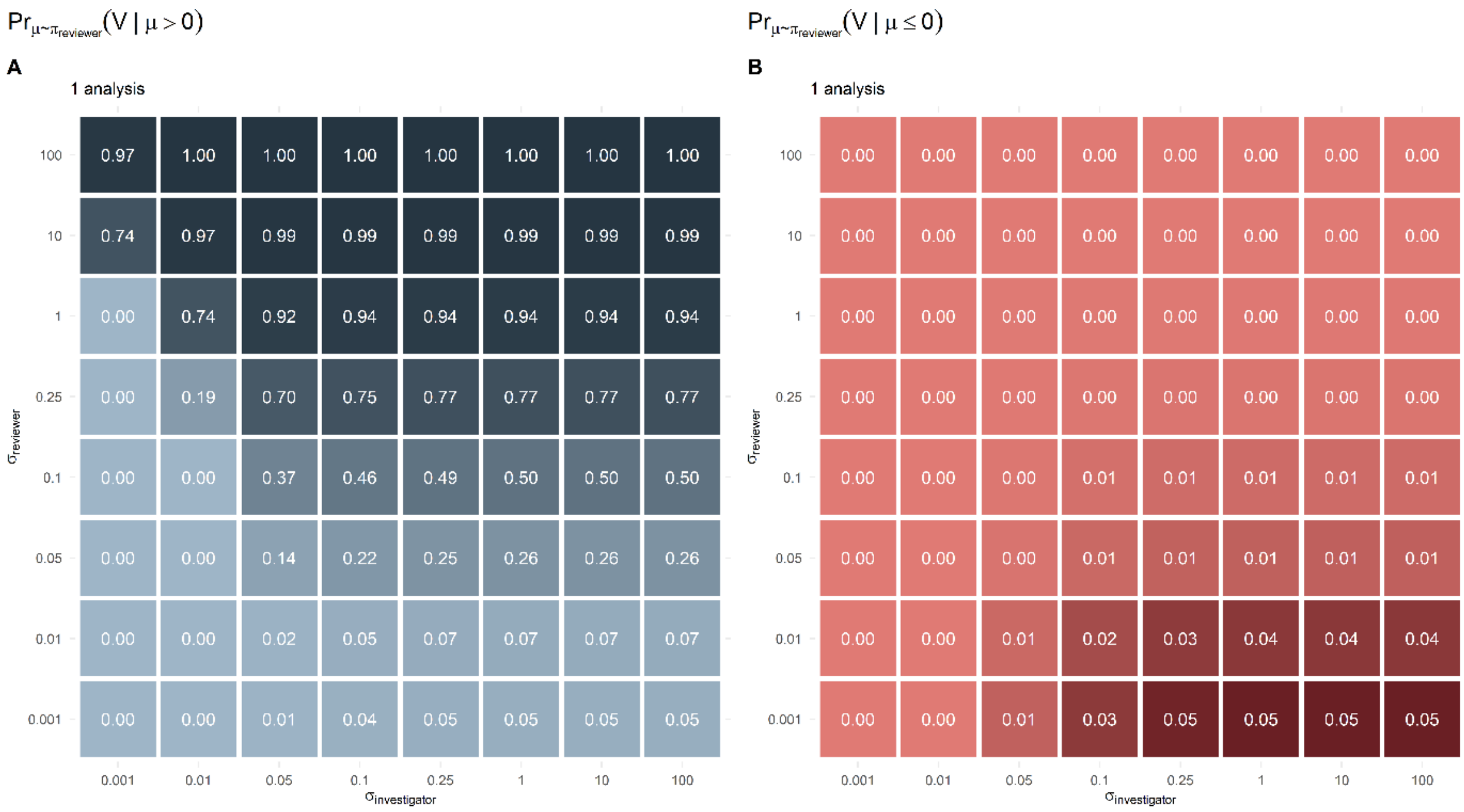

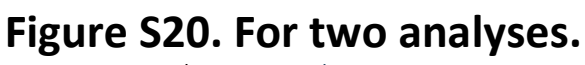

**Figure S20. For two analyses.**

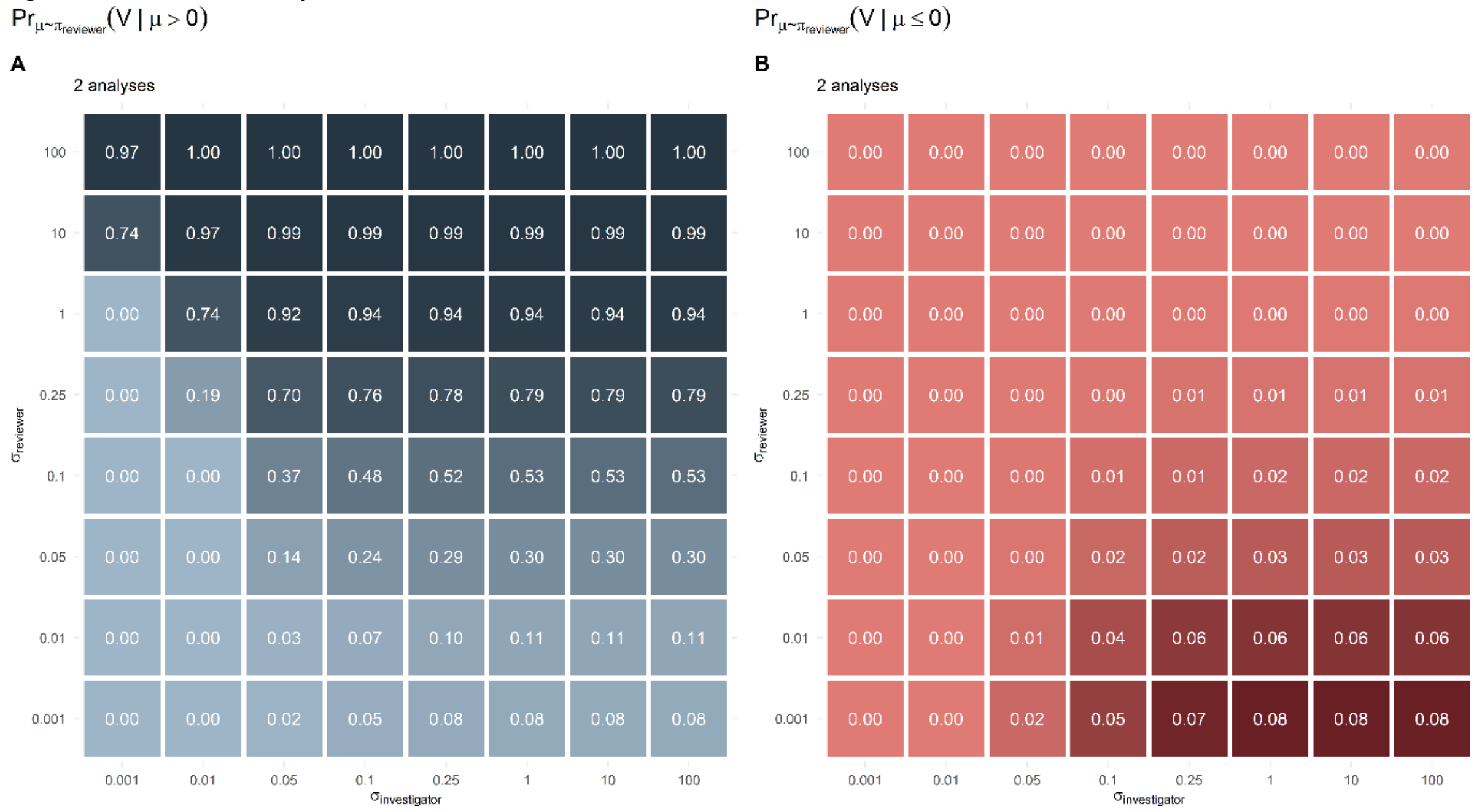

**Figure S21. For five analyses.**

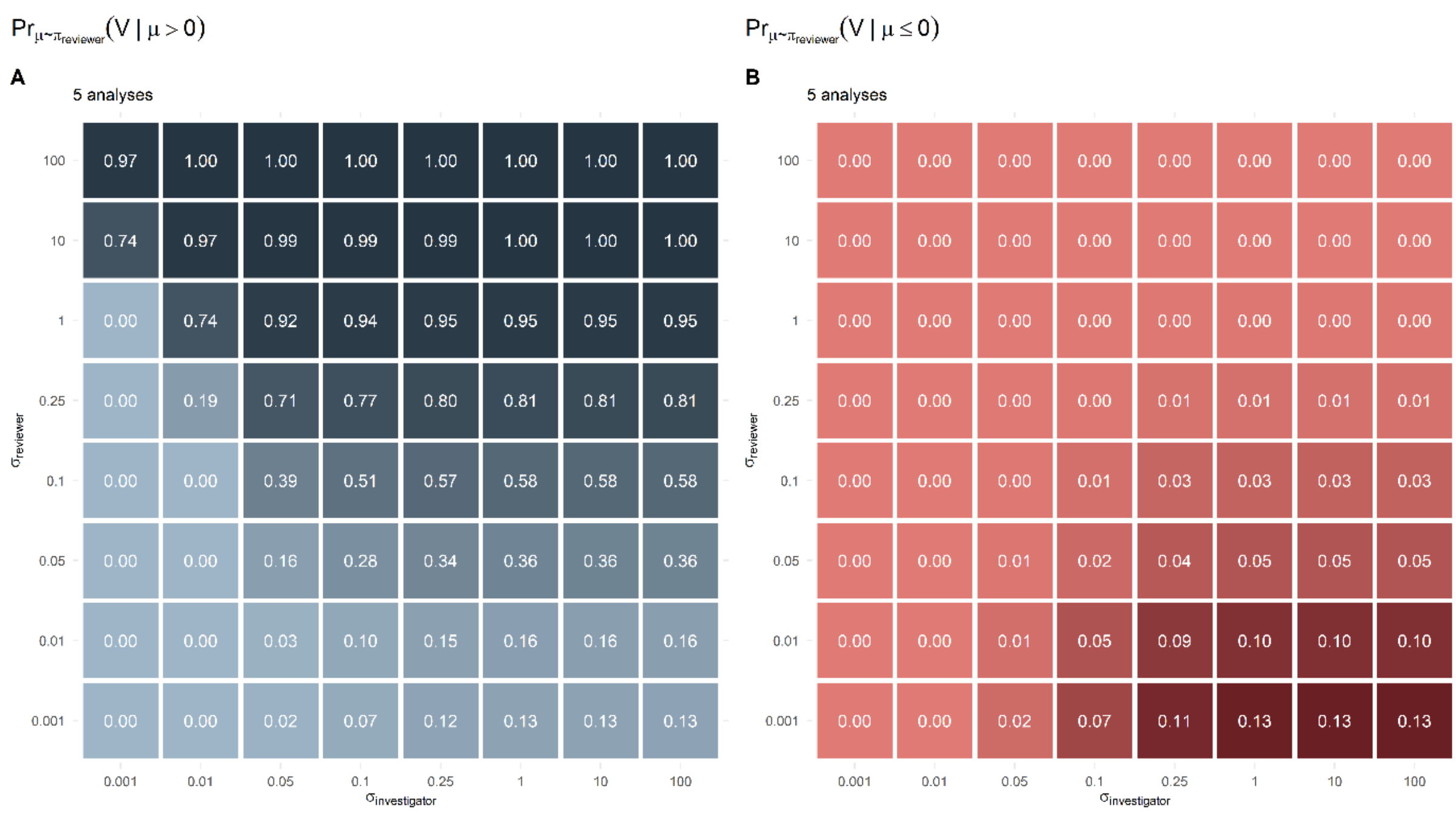

**Figure S22. For 10 analyses.**

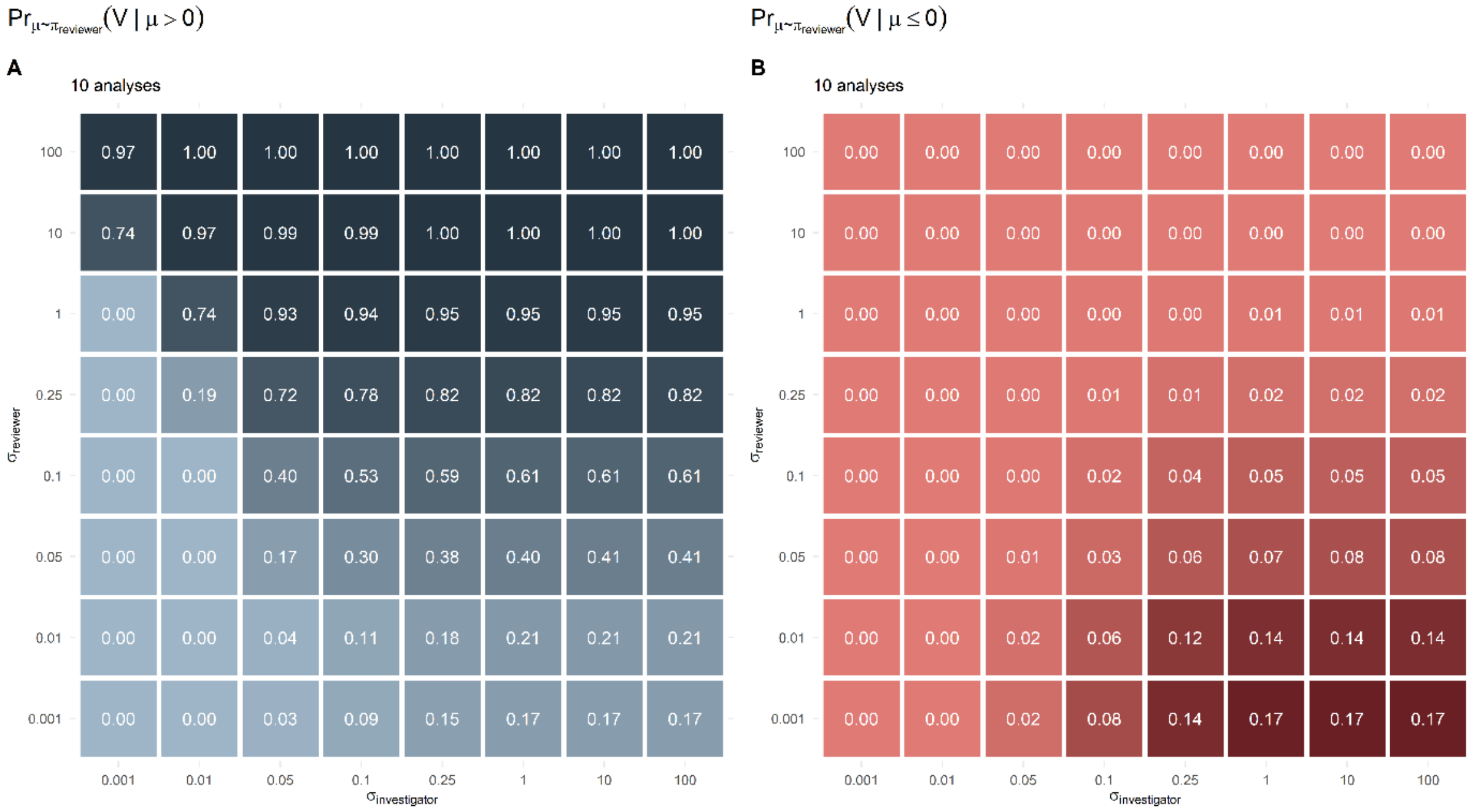

**Figure S23. For 20 analyses.**

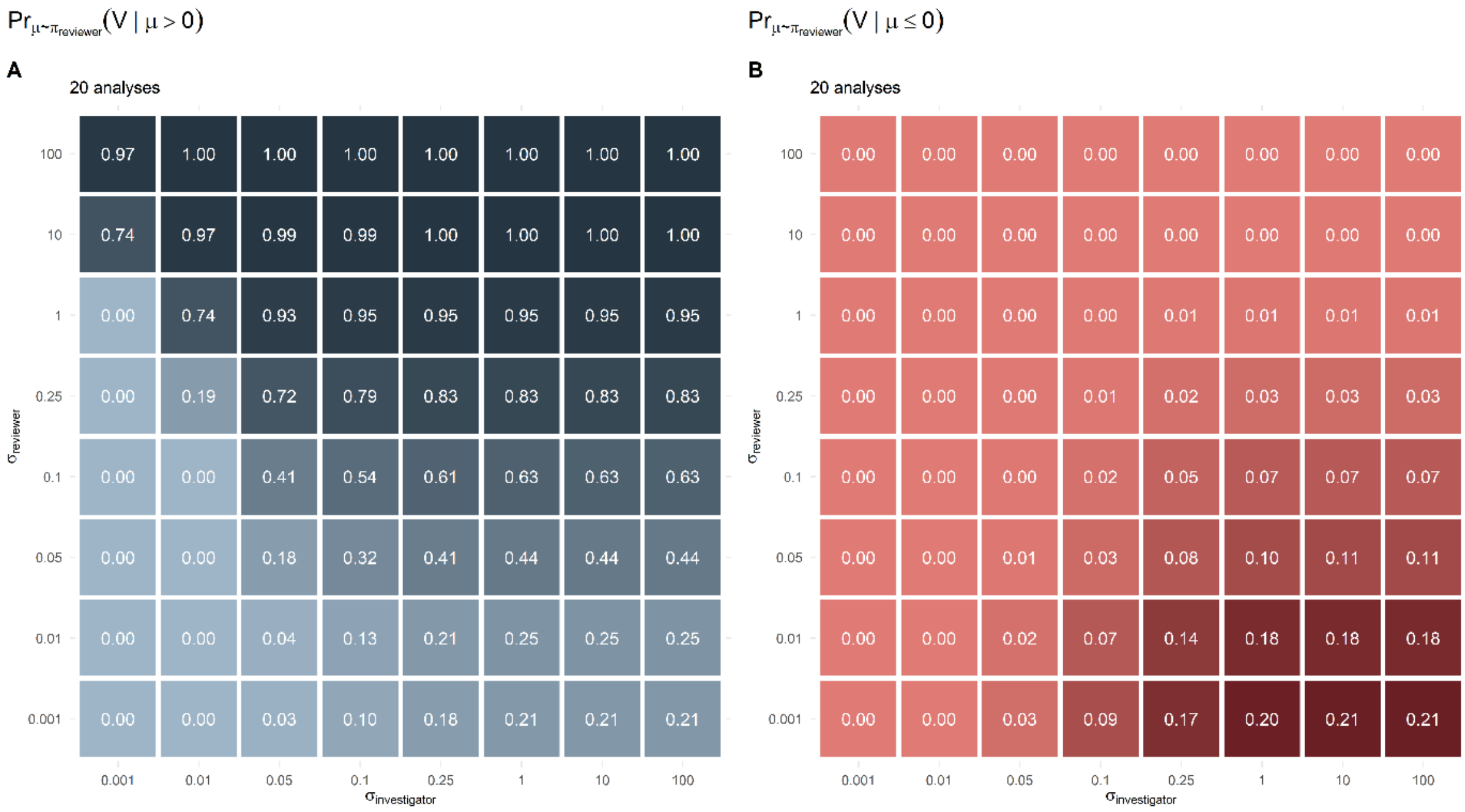

**Figure S24. For 50 analyses.**

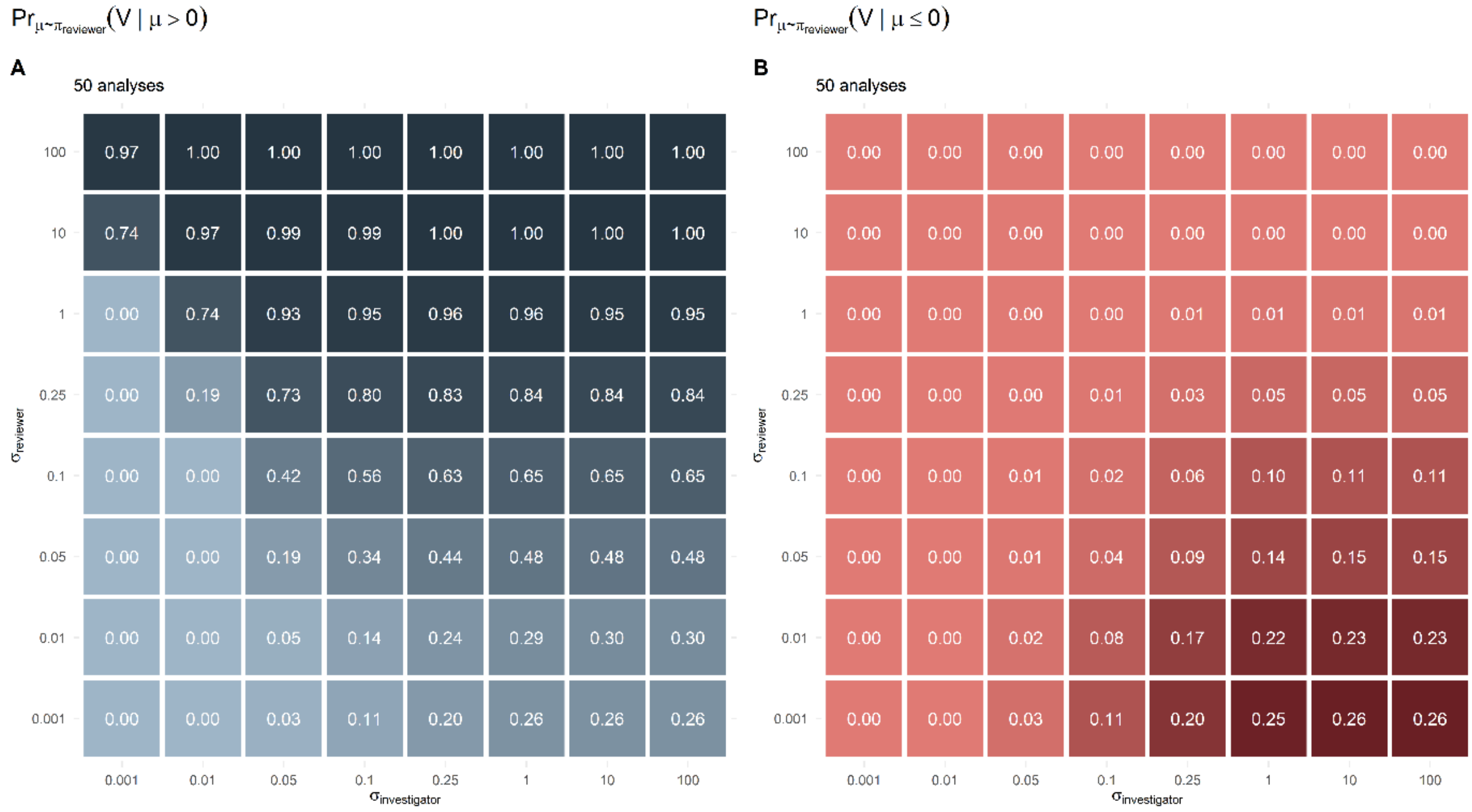

**Figure S25. For 100 analyses.**

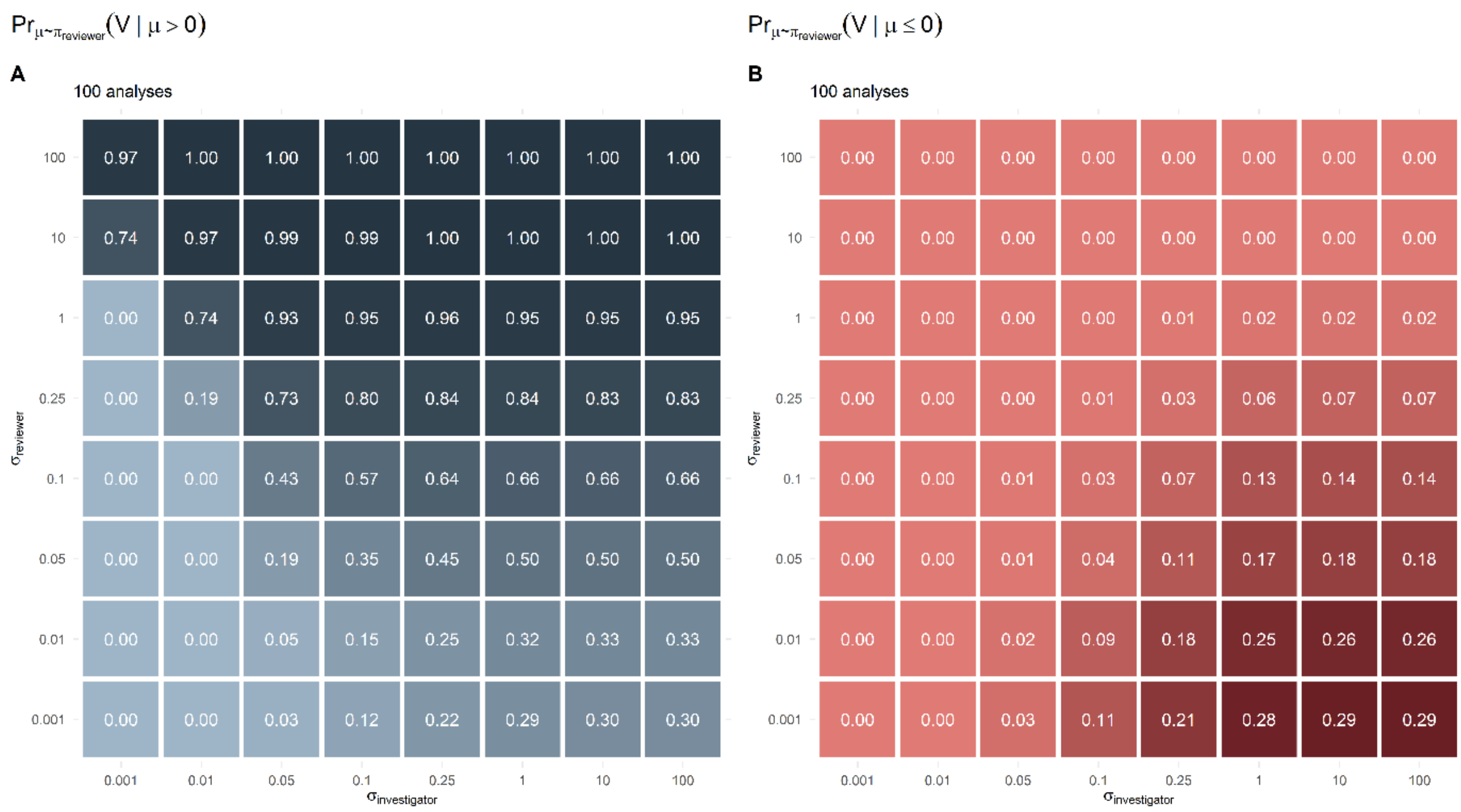

**Figure S26. For 250 analyses.**

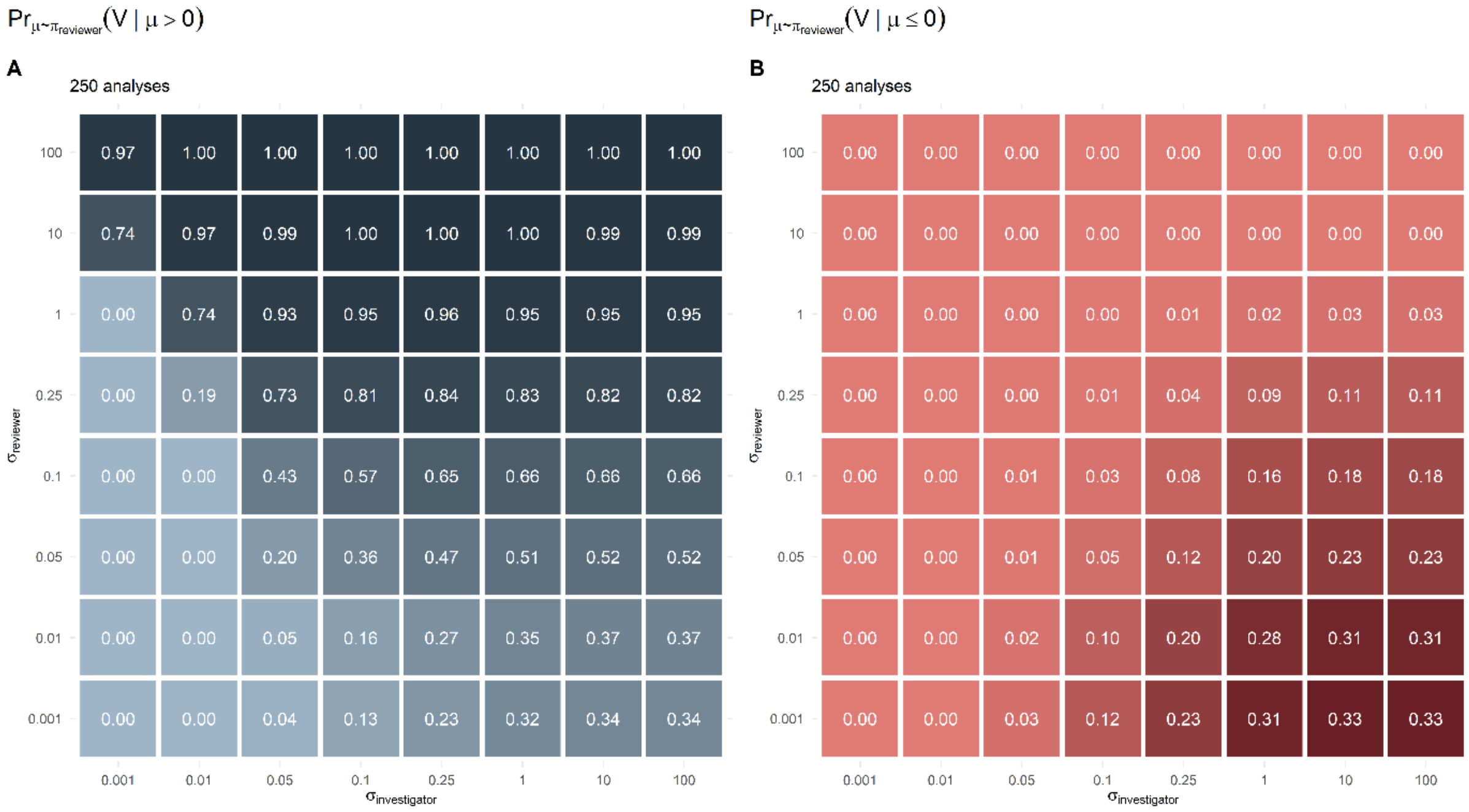

**Figure S27. For 500 analyses.**

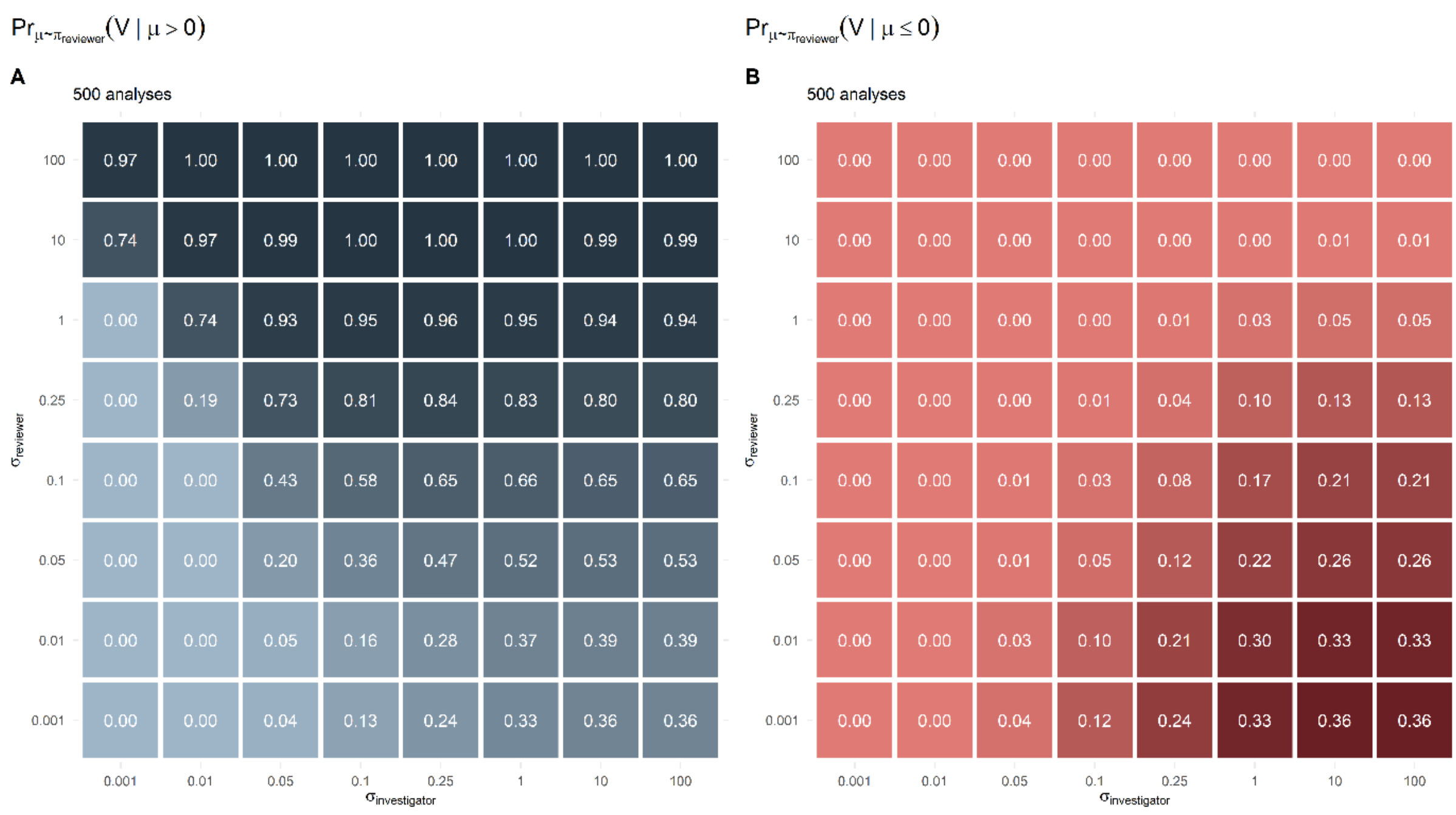

***pTDR = Pr(μ > 0 | V) and pFDR = Pr(μ ≤ 0 | V)***

**Figure S28 to S36. Probability of $\mu > 0$ (panel A) and $\mu \leq 0$ (panel B) conditional on having concluded efficacy (binary event $V$) in Bayesian trials with efficacy stopping rules $Pr(\mu > 0|X_{1,\dots,k_i}) > 0.95, i \in \{1,\dots,m\}$, futility stopping rules $Pr(\mu > 0|X_{1,\dots,k_i}) < 0.05, i \in \{1,\dots,m-1\}$, and no correction for multiplicity of analyses. $\sigma_{investigator}$ is the standard deviation of the prior $\pi_{investigator} = \mathcal{N}(0, \sigma^2_{investigator})$ used for the analyses, and $\sigma_{reviewer}$ is the standard deviation of the prior $\pi_{reviewer} = \mathcal{N}(0, \sigma^2_{reviewer})$ used to compute the operating characteristics of the trial. Grey cells are not computable because no event of efficacy conclusion occurred in the simulations.**

**Figure S28. For one analysis.**

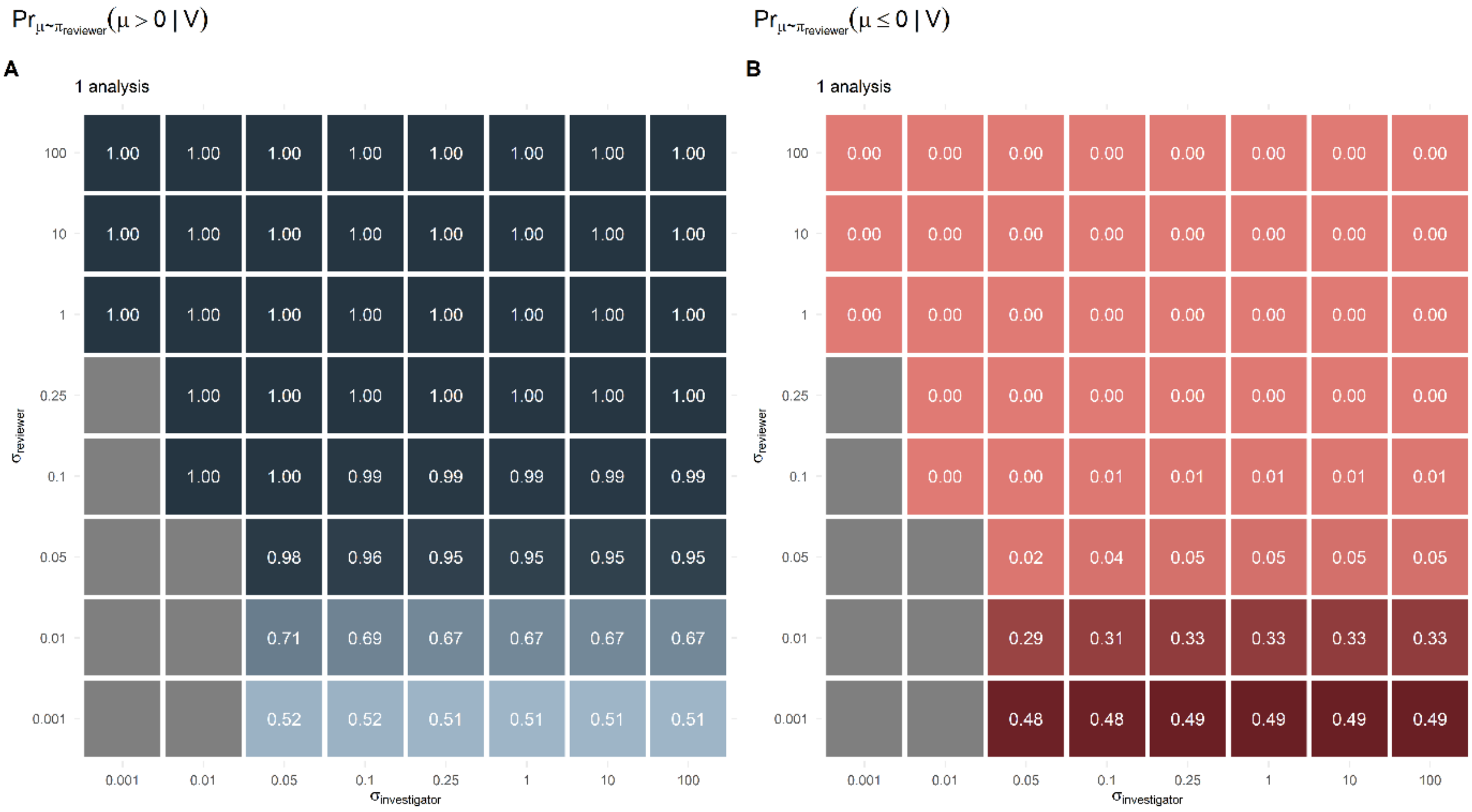

**Figure S29. For two analyses.**

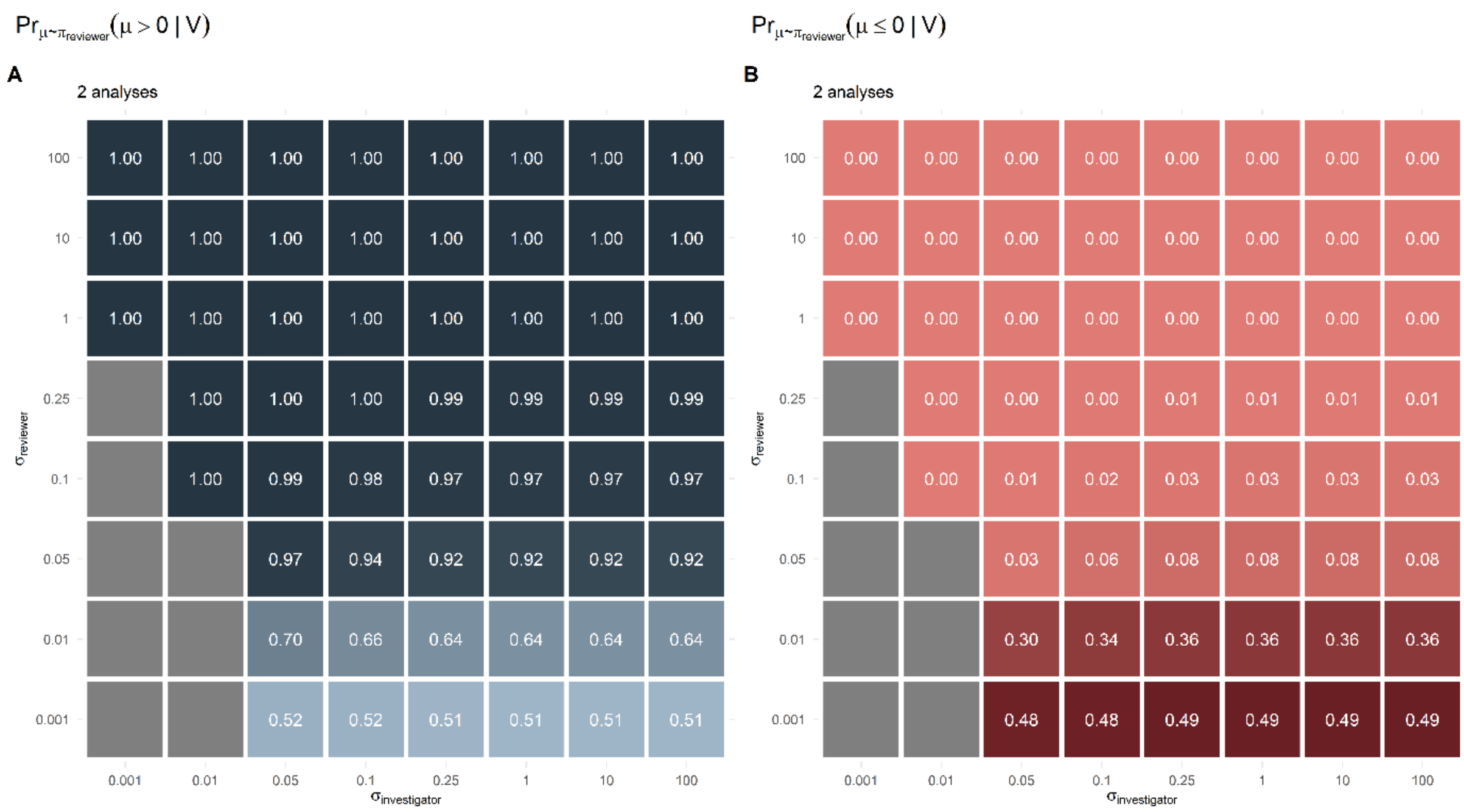

**Figure S30. For five analyses.**

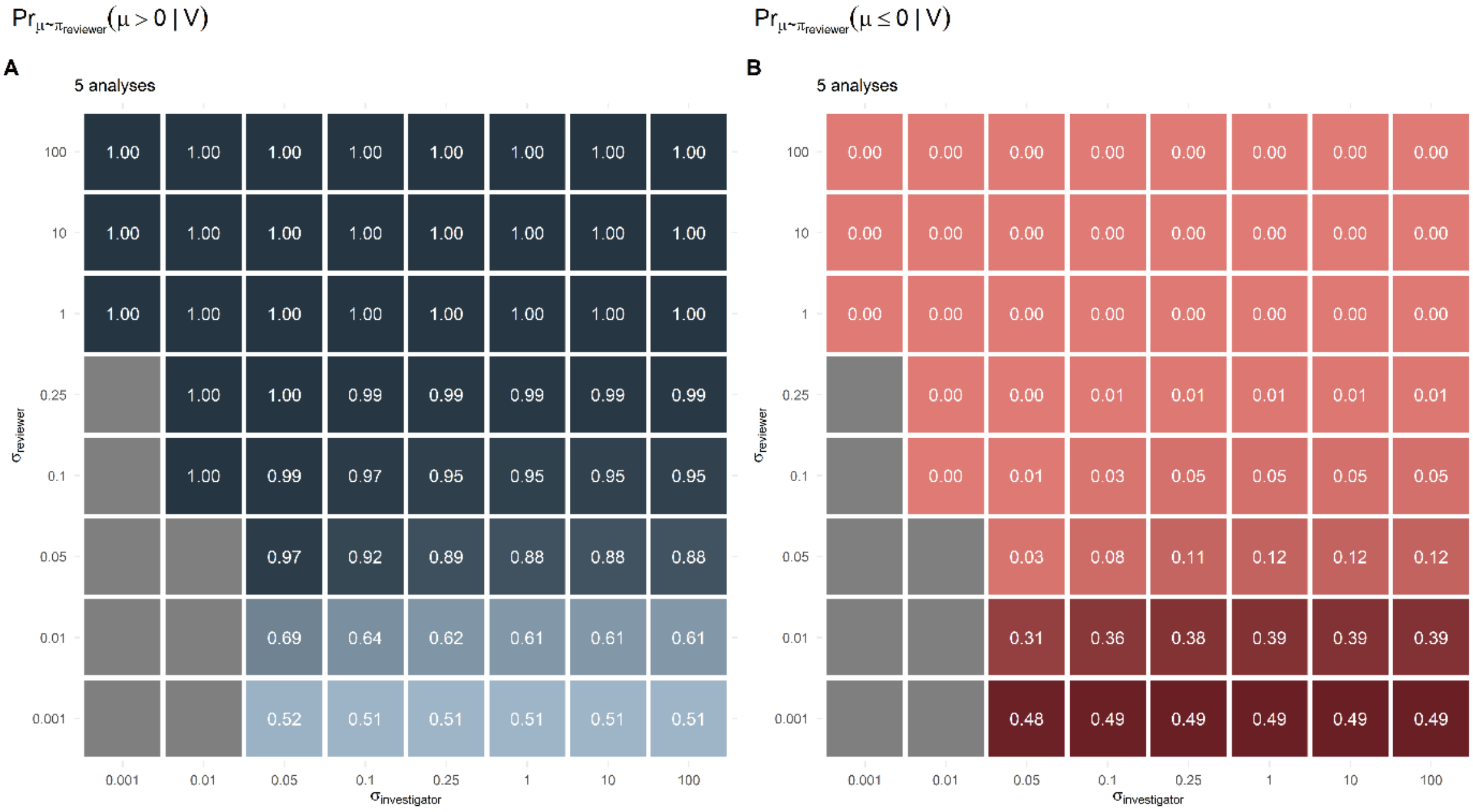

**Figure S31. For 10 analyses.**

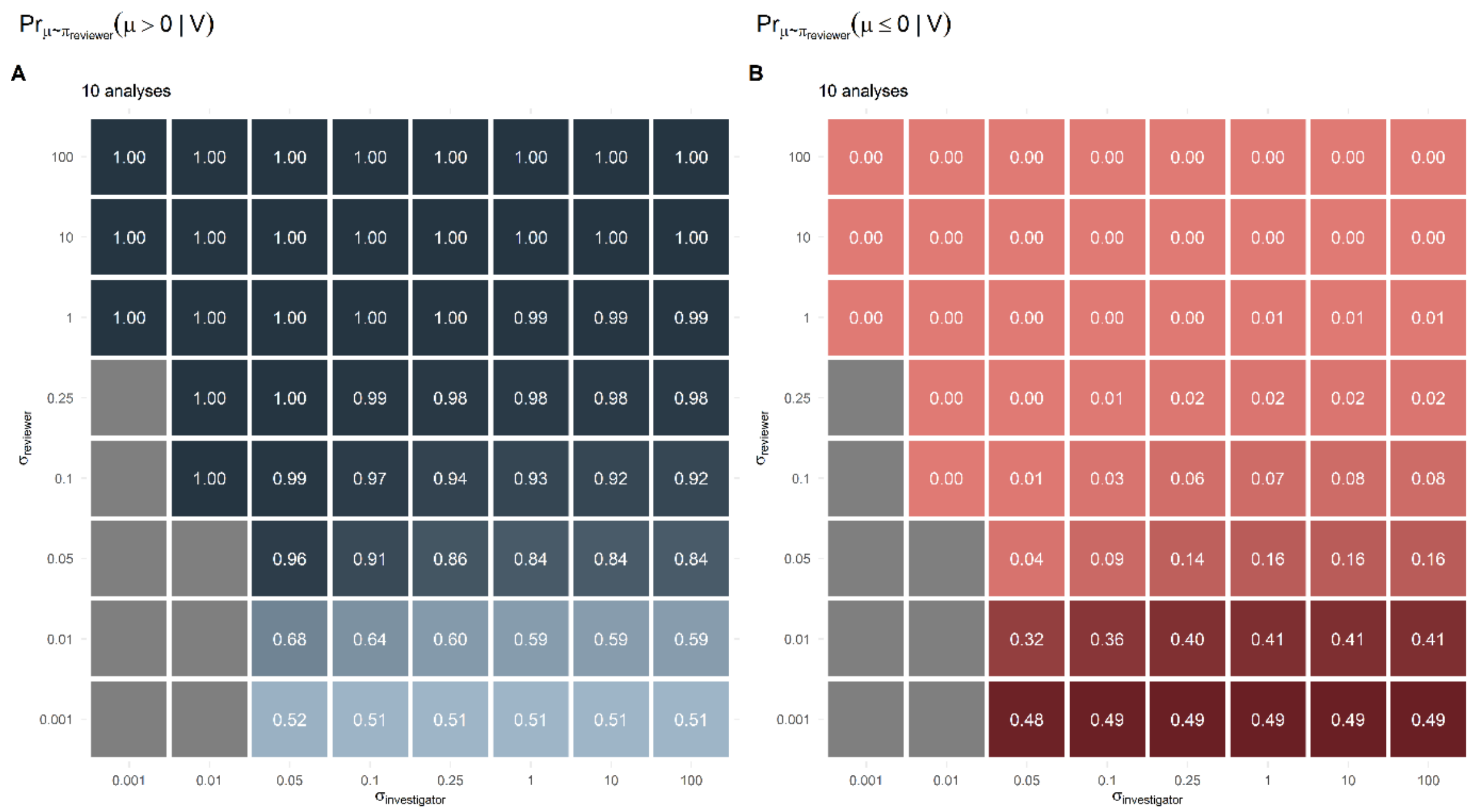

**Figure S32. For 20 analyses.**

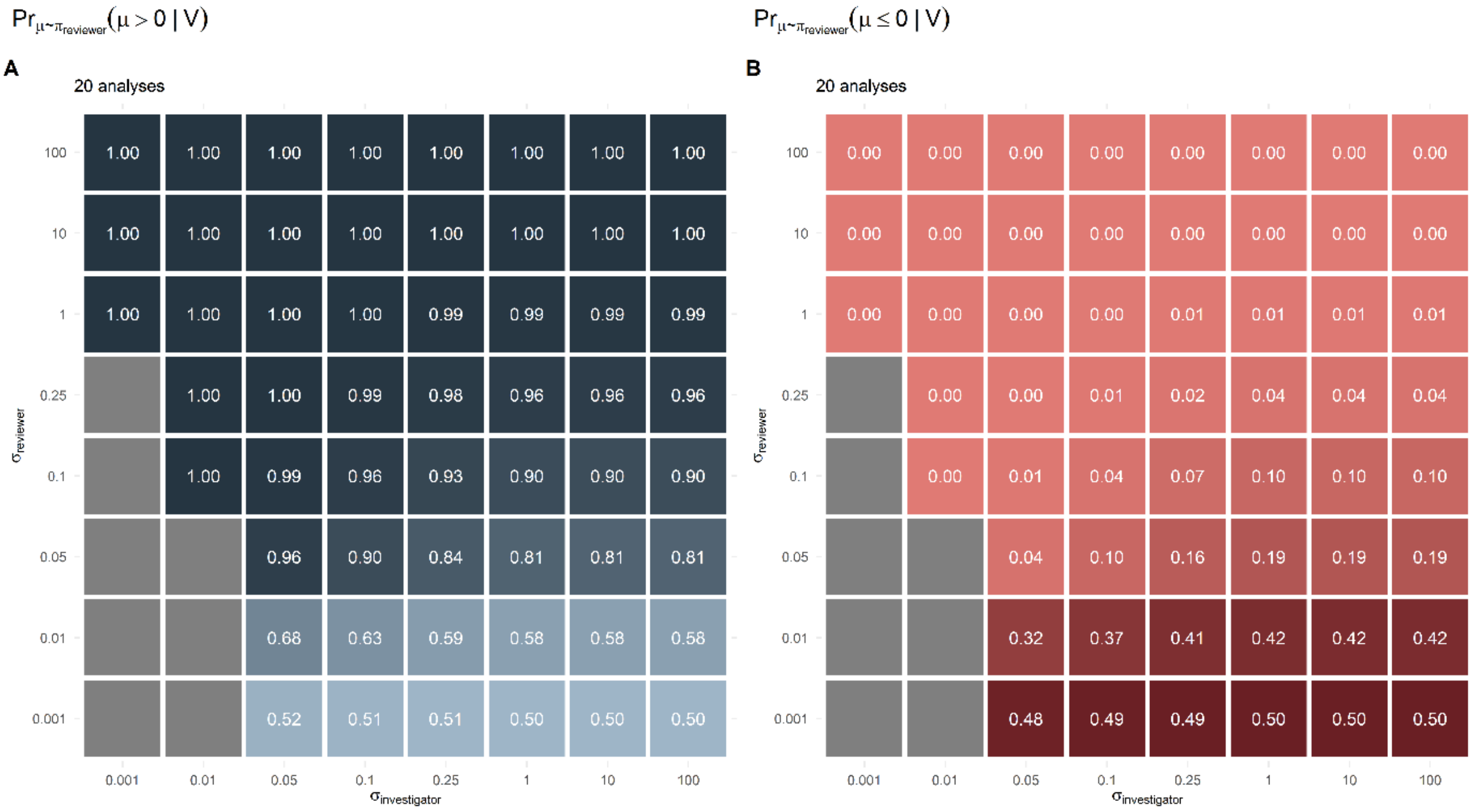

**Figure S33. For 50 analyses.**

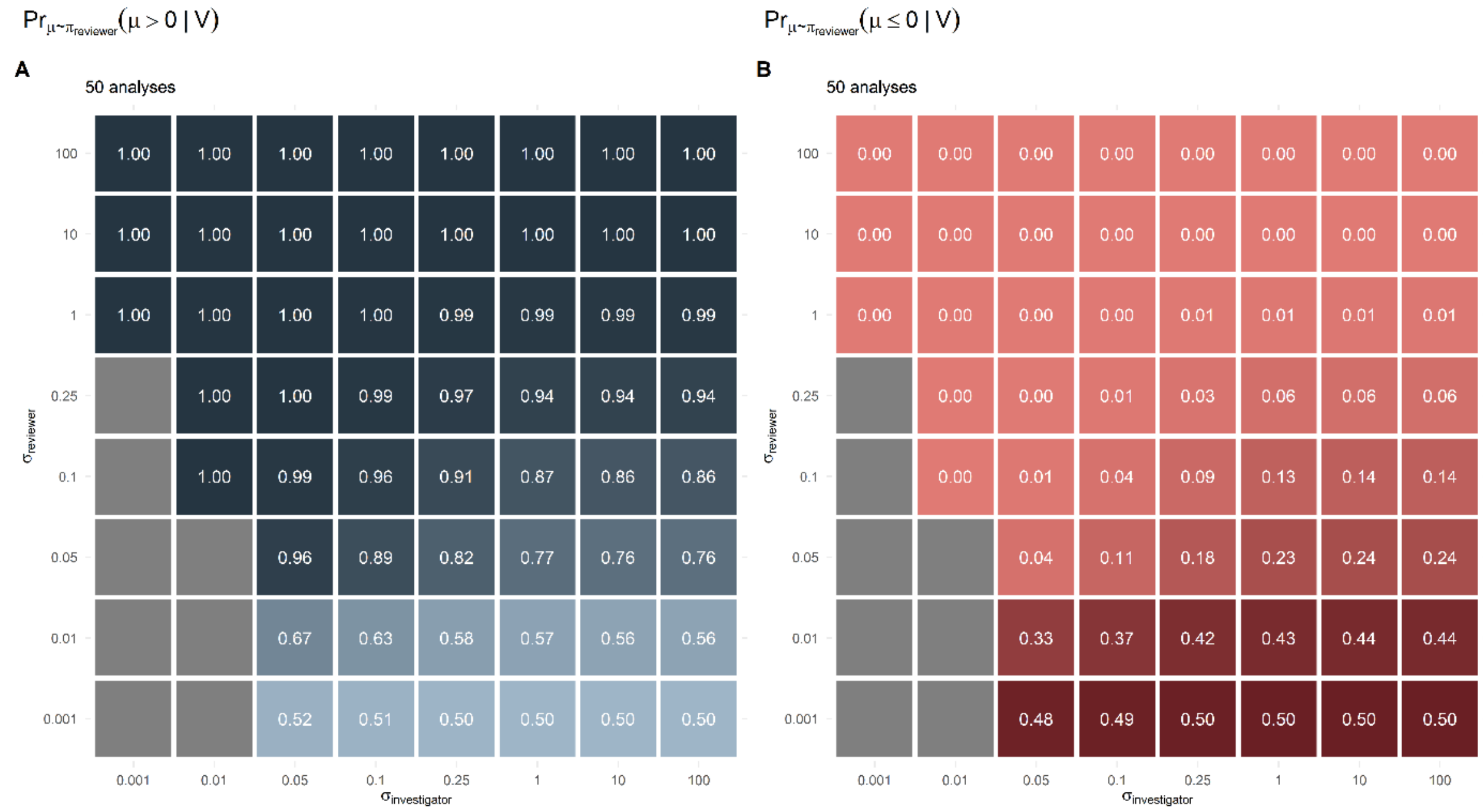

**Figure S34. For 100 analyses.**

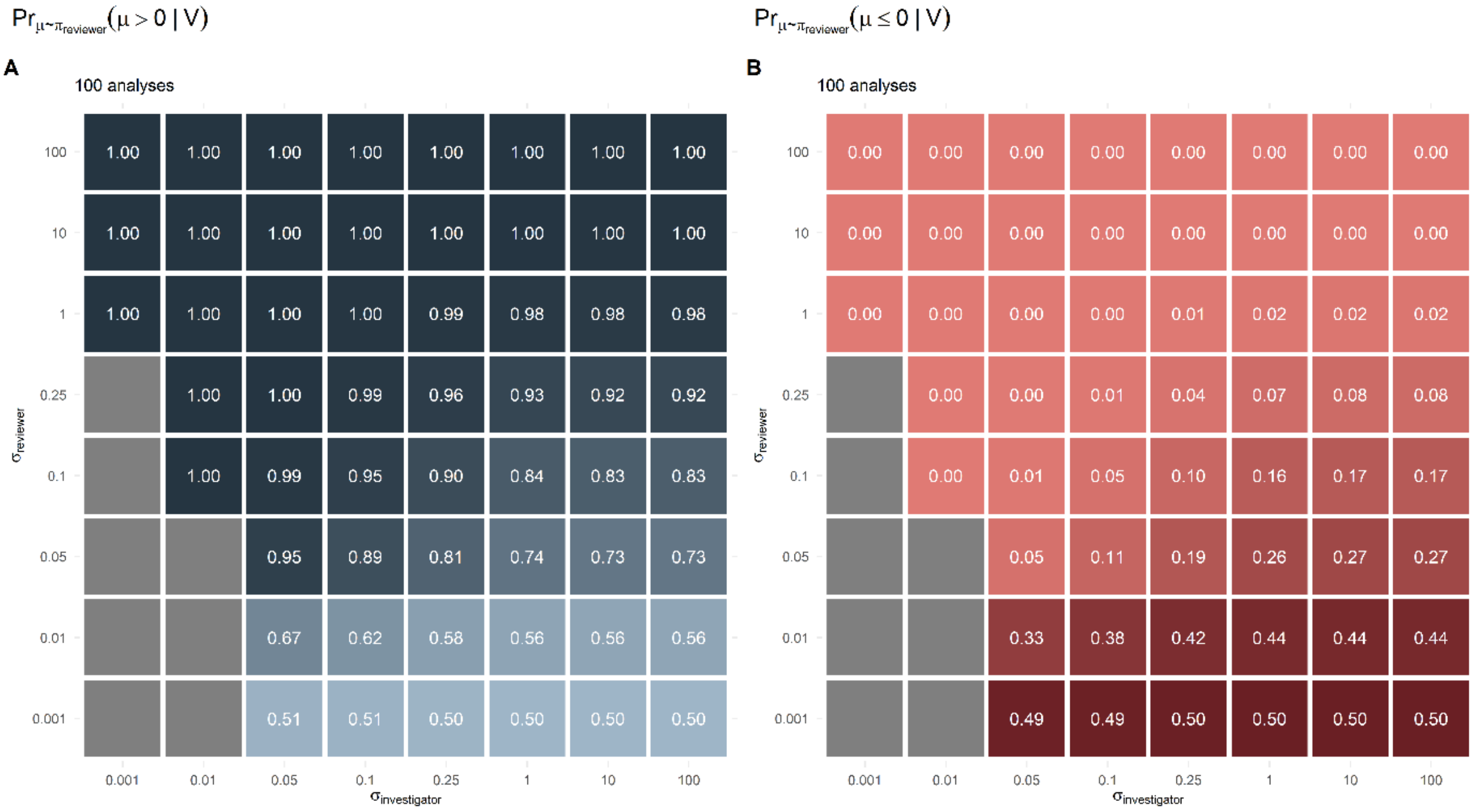

**Figure S35. For 250 analyses.**

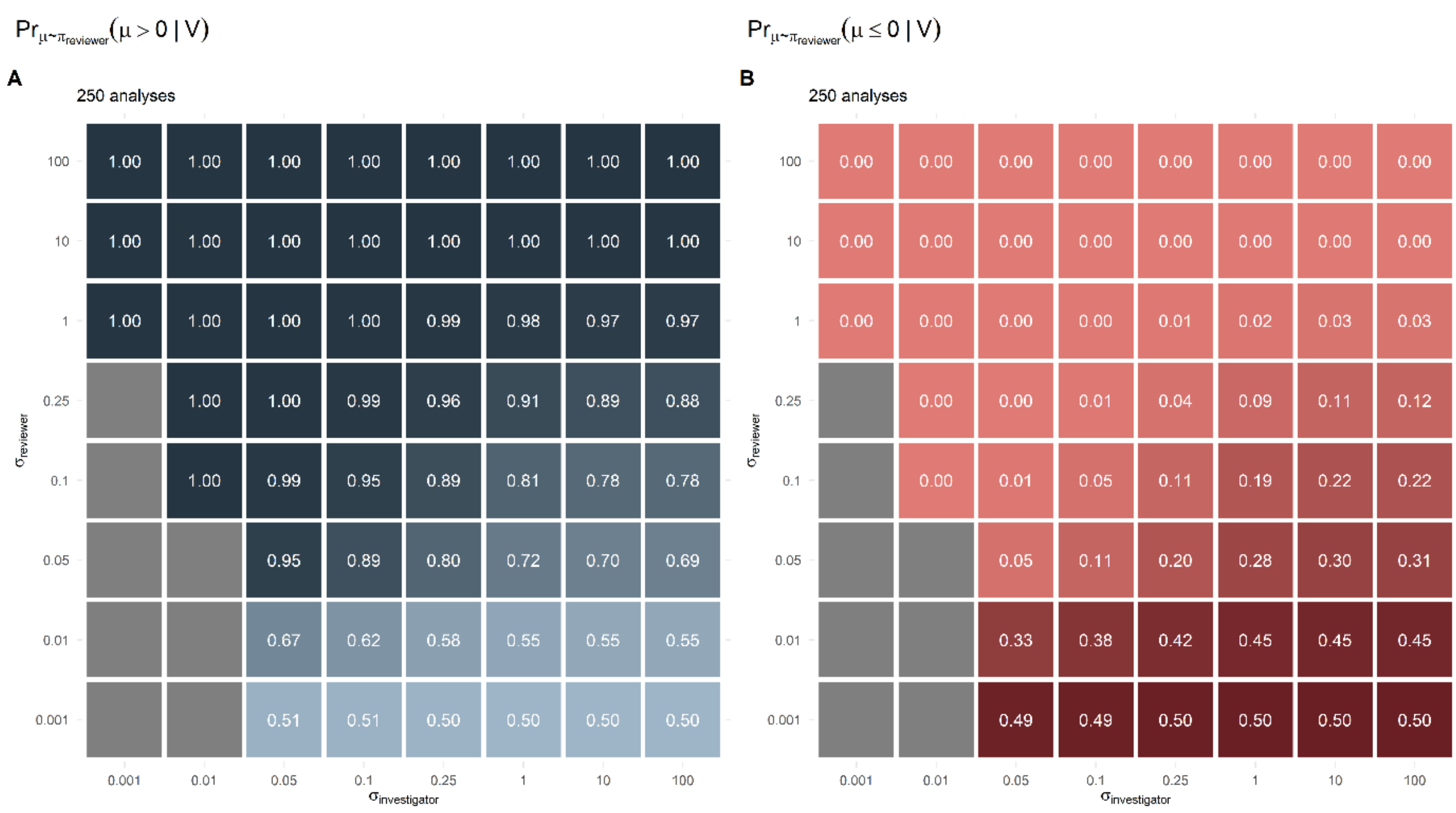

**Figure S36. For 500 analyses.**

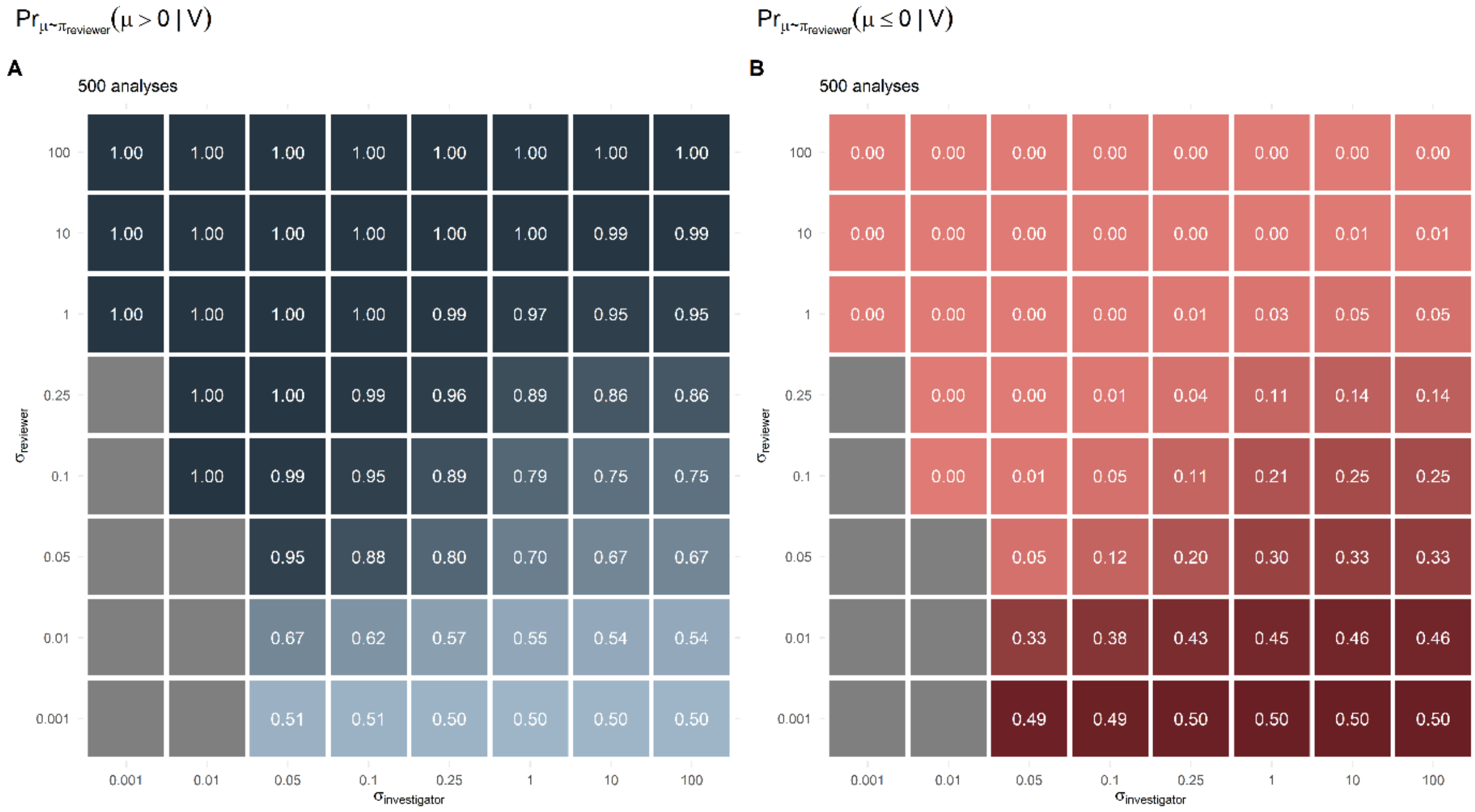

**Table S4. Informative value of a trial concluding efficacy expressed as $pTDO = pTDR/pFDR$, in Bayesian trials with efficacy stopping rules $Pr(\mu > 0|X_{1,\ldots,k_i}) > 0.95, i \in \{1, \ldots, m\}$, futility stopping rules $Pr(\mu > 0|X_{1,\ldots,k_i}) < 0.05, i \in \{1, \ldots, m-1\}$, and no correction for multiplicity of analyses. $\sigma_{investigator}$ is the standard deviation of the prior $\mathcal{N}(0, \sigma^2_{investigator})$ used for the analyses, and $\sigma_{reviewer}$ is the standard deviation of the prior $\mathcal{N}(0, \sigma^2_{reviewer})$ used to compute the operating characteristics of the trial. Empty cells are not computable because no event of efficacy conclusion occurred in the simulations.**

| Number of analyses | $\sigma_{reviewer}$ | $\text{pTDO} = \frac{pTDR}{pFDR} = \frac{Pr_{\mu \sim \pi_{reviewer}}(\mu > 0 \vert V)}{Pr_{\mu \sim \pi_{reviewer}}(\mu \leq 0 \vert V)}$ | | | | | | | |
|---|---|---|---|---|---|---|---|---|---|
| | | $\sigma_{investigator}$ | $\sigma_{investigator}$ | $\sigma_{investigator}$ | $\sigma_{investigator}$ | $\sigma_{investigator}$ | $\sigma_{investigator}$ | $\sigma_{investigator}$ | $\sigma_{investigator}$ |
| | | **0.001** | **0.01** | **0.05** | **0.1** | **0.25** | **1** | **10** | **100** |
| 1 | 0.001 | | | 1.1 | 1.1 | 1.1 | 1.0 | 1.0 | 1.0 |
| 2 | 0.001 | | | 1.1 | 1.1 | 1.0 | 1.0 | 1.0 | 1.0 |
| 5 | 0.001 | | | 1.1 | 1.0 | 1.0 | 1.0 | 1.0 | 1.0 |
| 10 | 0.001 | | | 1.1 | 1.0 | 1.0 | 1.0 | 1.0 | 1.0 |
| 20 | 0.001 | | | 1.1 | 1.0 | 1.0 | 1.0 | 1.0 | 1.0 |
| 50 | 0.001 | | | 1.1 | 1.0 | 1.0 | 1.0 | 1.0 | 1.0 |
| 100 | 0.001 | | | 1.0 | 1.0 | 1.0 | 1.0 | 1.0 | 1.0 |
| 250 | 0.001 | | | 1.0 | 1.0 | 1.0 | 1.0 | 1.0 | 1.0 |
| 500 | 0.001 | | | 1.0 | 1.0 | 1.0 | 1.0 | 1.0 | 1.0 |
| 1 | 0.01 | | | 2.5 | 2.2 | 2.1 | 2.1 | 2.1 | 2.1 |
| 2 | 0.01 | | | 2.3 | 1.9 | 1.8 | 1.8 | 1.8 | 1.8 |
| 5 | 0.01 | | | 2.2 | 1.8 | 1.6 | 1.6 | 1.6 | 1.6 |
| 10 | 0.01 | | | 2.2 | 1.7 | 1.5 | 1.5 | 1.5 | 1.5 |
| 20 | 0.01 | | | 2.1 | 1.7 | 1.5 | 1.4 | 1.4 | 1.4 |
| 50 | 0.01 | | | 2.0 | 1.7 | 1.4 | 1.3 | 1.3 | 1.3 |
| 100 | 0.01 | | | 2.0 | 1.7 | 1.4 | 1.3 | 1.3 | 1.3 |
| 250 | 0.01 | | | 2.0 | 1.6 | 1.4 | 1.2 | 1.2 | 1.2 |
| 500 | 0.01 | | | 2.0 | 1.6 | 1.3 | 1.2 | 1.2 | 1.2 |
| 1 | 0.05 | | | 47 | 24 | 20 | 19 | 19 | 19 |
| 2 | 0.05 | | | 36 | 16 | 12 | 12 | 12 | 12 |
| 5 | 0.05 | | | 29 | 12 | 7.9 | 7.2 | 7.2 | 7.2 |
| 10 | 0.05 | | | 27 | 10 | 6.3 | 5.4 | 5.4 | 5.4 |
| 20 | 0.05 | | | 24 | 9.3 | 5.4 | 4.3 | 4.2 | 4.2 |
| 50 | 0.05 | | | 21 | 8.4 | 4.6 | 3.3 | 3.2 | 3.2 |
| 100 | 0.05 | | | 21 | 8.1 | 4.3 | 2.9 | 2.7 | 2.7 |
| 250 | 0.05 | | | 20 | 7.8 | 4.0 | 2.5 | 2.3 | 2.3 |
| 500 | 0.05 | | | 19 | 7.6 | 3.9 | 2.3 | 2.0 | 2.0 |
| 1 | 0.1 | | ≥ 1000 | 249 | 90 | 70 | 66 | 66 | 66 |
| 2 | 0.1 | | ≥ 1000 | 166 | 52 | 39 | 36 | 35 | 35 |
| 5 | 0.1 | | ≥ 1000 | 142 | 37 | 21 | 18 | 18 | 18 |
| 10 | 0.1 | | ≥ 1000 | 110 | 31 | 15 | 13 | 12 | 12 |
| 20 | 0.1 | | ≥ 1000 | 99 | 27 | 12 | 9.1 | 8.8 | 8.8 |
| 50 | 0.1 | | ≥ 1000 | 82 | 23 | 10 | 6.4 | 6.1 | 6.1 |
| 100 | 0.1 | | ≥ 1000 | 76 | 21 | 9.1 | 5.3 | 4.8 | 4.8 |
| 250 | 0.1 | | ≥ 1000 | 76 | 20 | 8.3 | 4.2 | 3.7 | 3.6 |
| 500 | 0.1 | | ≥ 1000 | 70 | 19 | 7.9 | 3.8 | 3.1 | 3.0 |
| 1 | 0.25 | | ≥ 1000 | 999 | 499 | 332 | 332 | 332 | 332 |
| 2 | 0.25 | | ≥ 1000 | 999 | 249 | 166 | 142 | 142 | 142 |
| 5 | 0.25 | | ≥ 1000 | 999 | 166 | 82 | 70 | 66 | 66 |
| 10 | 0.25 | | ≥ 1000 | 999 | 124 | 55 | 42 | 41 | 41 |
| 20 | 0.25 | | ≥ 1000 | 499 | 110 | 41 | 28 | 26 | 26 |
| 50 | 0.25 | | ≥ 1000 | 499 | 90 | 31 | 17 | 16 | 16 |
| 100 | 0.25 | | ≥ 1000 | 499 | 82 | 27 | 13 | 11 | 11 |
| 250 | 0.25 | | ≥ 1000 | 499 | 76 | 24 | 9.6 | 7.7 | 7.7 |
| 500 | 0.25 | | ≥ 1000 | 332 | 76 | 22 | 8.3 | 6.0 | 6.0 |
| 1 | 1 | ≥ 1000 | ≥ 1000 | ≥ 1000 | 999 | 999 | 999 | 999 | 999 |
| 2 | 1 | ≥ 1000 | ≥ 1000 | ≥ 1000 | 999 | 499 | 499 | 499 | 499 |
| 5 | 1 | ≥ 1000 | ≥ 1000 | ≥ 1000 | 499 | 332 | 249 | 249 | 249 |
| 10 | 1 | ≥ 1000 | ≥ 1000 | ≥ 1000 | 499 | 249 | 199 | 199 | 199 |
| 20 | 1 | ≥ 1000 | ≥ 1000 | ≥ 1000 | 499 | 199 | 124 | 124 | 124 |
| 50 | 1 | ≥ 1000 | ≥ 1000 | 999 | 332 | 142 | 76 | 70 | 70 |
| 100 | 1 | ≥ 1000 | ≥ 1000 | 999 | 332 | 124 | 55 | 47 | 47 |
| 250 | 1 | ≥ 1000 | ≥ 1000 | 999 | 332 | 110 | 39 | 29 | 29 |
| 500 | 1 | ≥ 1000 | ≥ 1000 | 999 | 332 | 99 | 32 | 21 | 21 |
| 1 | 10 | ≥ 1000 | ≥ 1000 | ≥ 1000 | ≥ 1000 | ≥ 1000 | ≥ 1000 | ≥ 1000 | ≥ 1000 |
| 2 | 10 | ≥ 1000 | ≥ 1000 | ≥ 1000 | ≥ 1000 | ≥ 1000 | ≥ 1000 | ≥ 1000 | ≥ 1000 |
| 5 | 10 | ≥ 1000 | ≥ 1000 | ≥ 1000 | ≥ 1000 | ≥ 1000 | ≥ 1000 | ≥ 1000 | ≥ 1000 |
| 10 | 10 | ≥ 1000 | ≥ 1000 | ≥ 1000 | ≥ 1000 | 999 | 999 | 999 | 999 |
| 20 | 10 | ≥ 1000 | ≥ 1000 | ≥ 1000 | ≥ 1000 | 999 | 999 | 999 | 999 |
| 50 | 10 | ≥ 1000 | ≥ 1000 | ≥ 1000 | ≥ 1000 | 999 | 499 | 499 | 499 |
| 100 | 10 | ≥ 1000 | ≥ 1000 | ≥ 1000 | ≥ 1000 | 999 | 499 | 499 | 499 |
| 250 | 10 | ≥ 1000 | ≥ 1000 | ≥ 1000 | ≥ 1000 | 999 | 332 | 249 | 249 |
| 500 | 10 | ≥ 1000 | ≥ 1000 | ≥ 1000 | ≥ 1000 | 999 | 332 | 199 | 199 |
| 1 | 100 | ≥ 1000 | ≥ 1000 | ≥ 1000 | ≥ 1000 | ≥ 1000 | ≥ 1000 | ≥ 1000 | ≥ 1000 |
| 2 | 100 | ≥ 1000 | ≥ 1000 | ≥ 1000 | ≥ 1000 | ≥ 1000 | ≥ 1000 | ≥ 1000 | ≥ 1000 |
| 5 | 100 | ≥ 1000 | ≥ 1000 | ≥ 1000 | ≥ 1000 | ≥ 1000 | ≥ 1000 | ≥ 1000 | ≥ 1000 |
| 10 | 100 | ≥ 1000 | ≥ 1000 | ≥ 1000 | ≥ 1000 | ≥ 1000 | ≥ 1000 | ≥ 1000 | ≥ 1000 |
| 20 | 100 | ≥ 1000 | ≥ 1000 | ≥ 1000 | ≥ 1000 | ≥ 1000 | ≥ 1000 | ≥ 1000 | ≥ 1000 |
| 50 | 100 | ≥ 1000 | ≥ 1000 | ≥ 1000 | ≥ 1000 | ≥ 1000 | ≥ 1000 | ≥ 1000 | ≥ 1000 |
| 100 | 100 | ≥ 1000 | ≥ 1000 | ≥ 1000 | ≥ 1000 | ≥ 1000 | ≥ 1000 | ≥ 1000 | ≥ 1000 |
| 250 | 100 | ≥ 1000 | ≥ 1000 | ≥ 1000 | ≥ 1000 | ≥ 1000 | ≥ 1000 | ≥ 1000 | ≥ 1000 |
| 500 | 100 | ≥ 1000 | ≥ 1000 | ≥ 1000 | ≥ 1000 | ≥ 1000 | ≥ 1000 | ≥ 1000 | ≥ 1000 |

### 2.2.2.2 Futility threshold = 0.10

## *A = Pr(V | μ > 0) and 1 – B = Pr(V | μ ≤ 0)*

**Figure S37 to S45. Probability of concluding efficacy (binary event $V$) conditional on $\mu > 0$ (panel A) and $\mu \leq 0$ (panel B) in Bayesian trials with efficacy stopping rules $Pr(\mu > 0 | X_{1,\ldots,k_i}) > 0.95, i \in \{1, \ldots, m\}$, futility stopping rules $Pr(\mu > 0 | X_{1,\ldots,k_i}) < 0.10, i \in \{1, \ldots, m-1\}$, and no correction for multiplicity of analyses. $\sigma_{investigator}$ is the standard deviation of the prior $\pi_{investigator} = \mathcal{N}(0, \sigma^2_{investigator})$ used for the analyses, and $\sigma_{reviewer}$ is the standard deviation of the prior $\pi_{reviewer} = \mathcal{N}(0, \sigma^2_{reviewer})$ used to compute the operating characteristics of the trial.**

**Figure S37. For one analysis.**

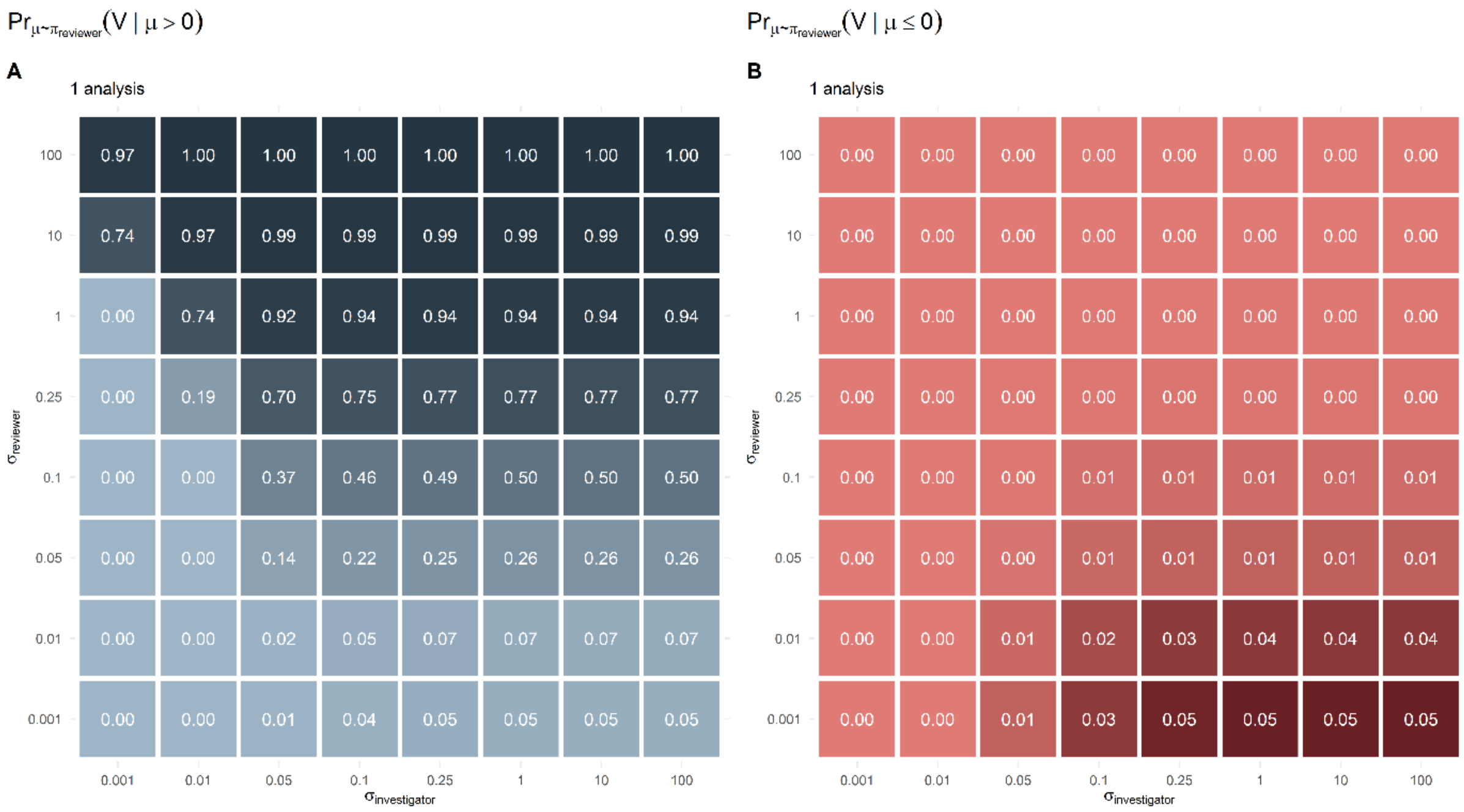

**Figure S38. For two analyses.**

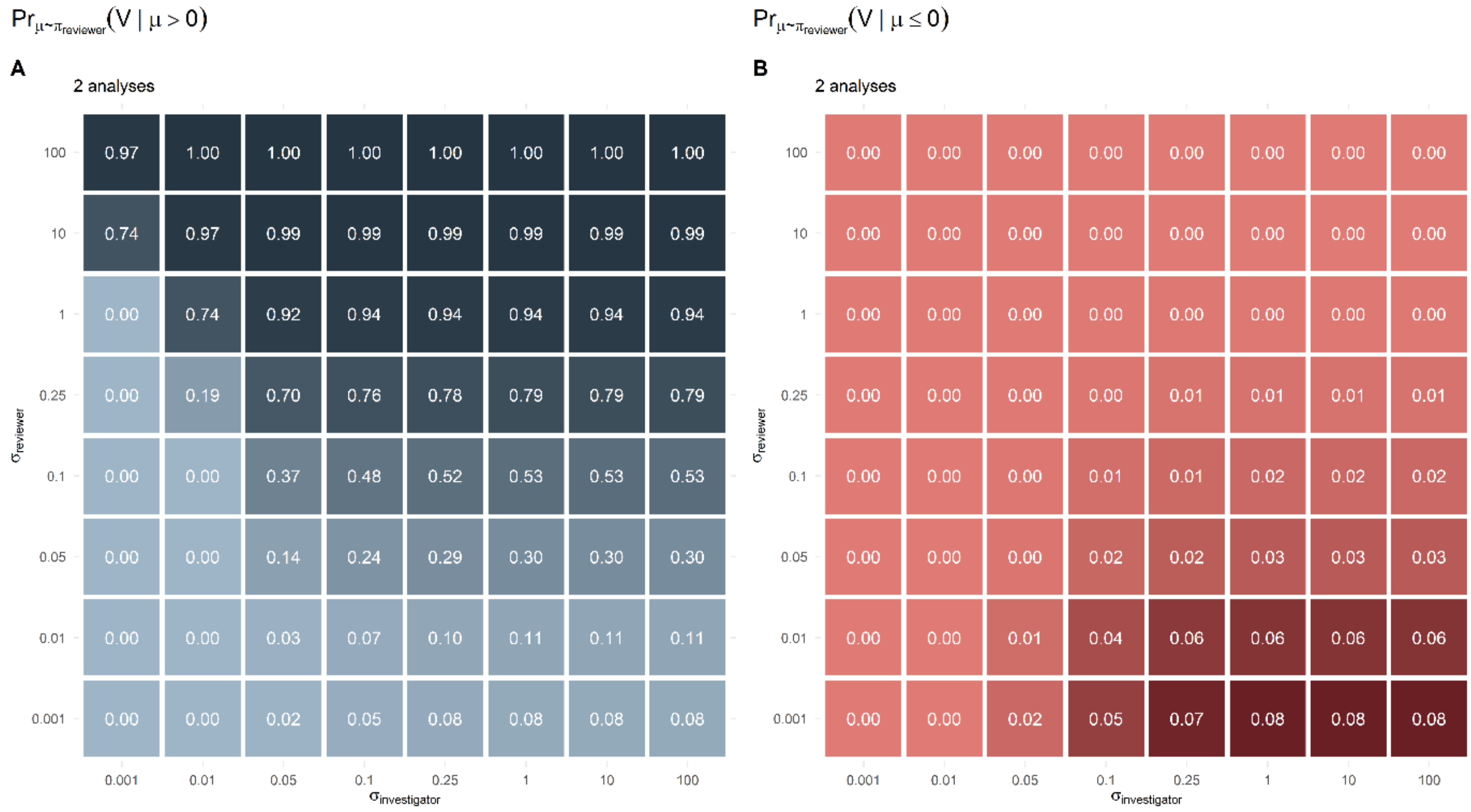

**Figure S39. For five analyses.**

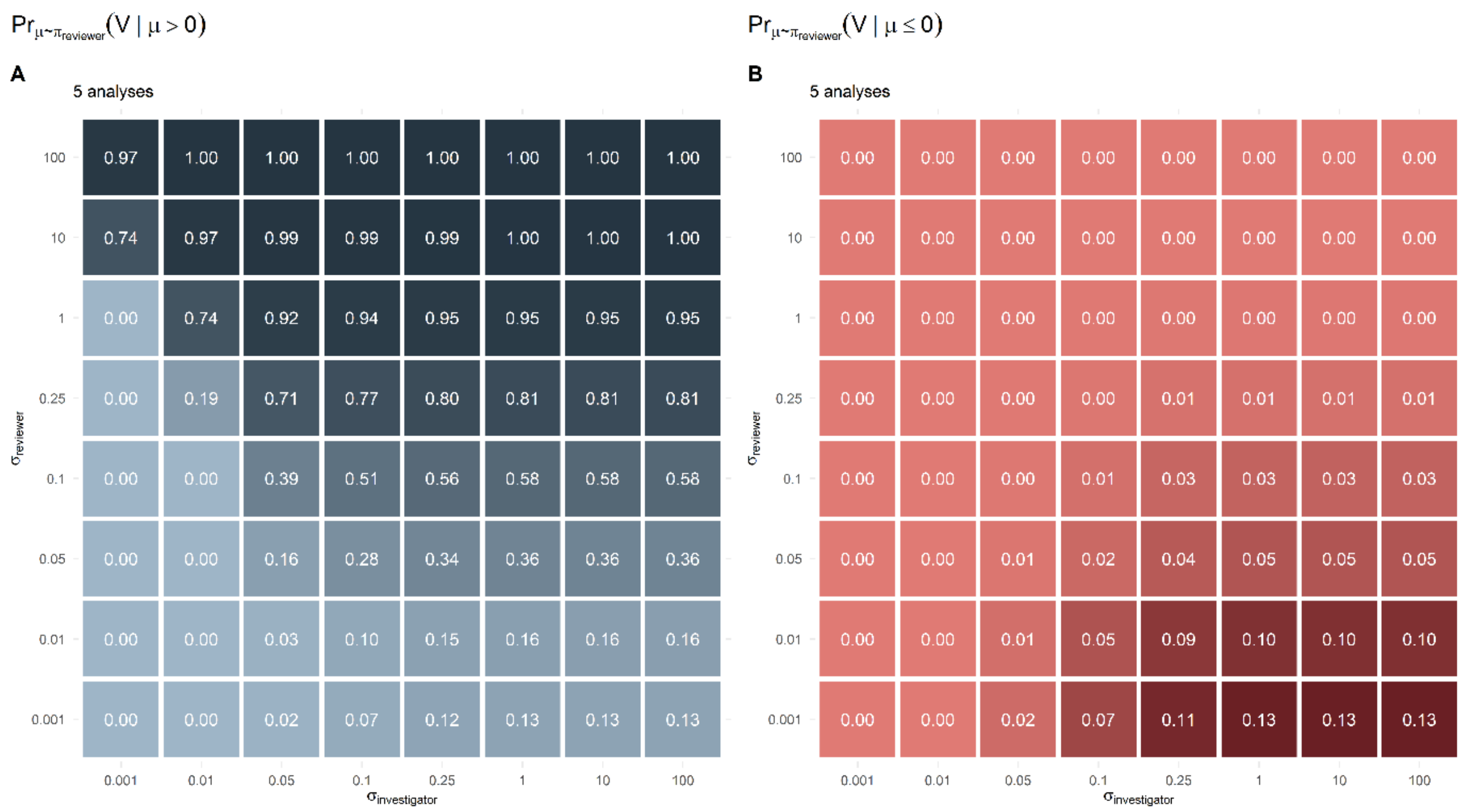

**Figure S40. For 10 analyses.**

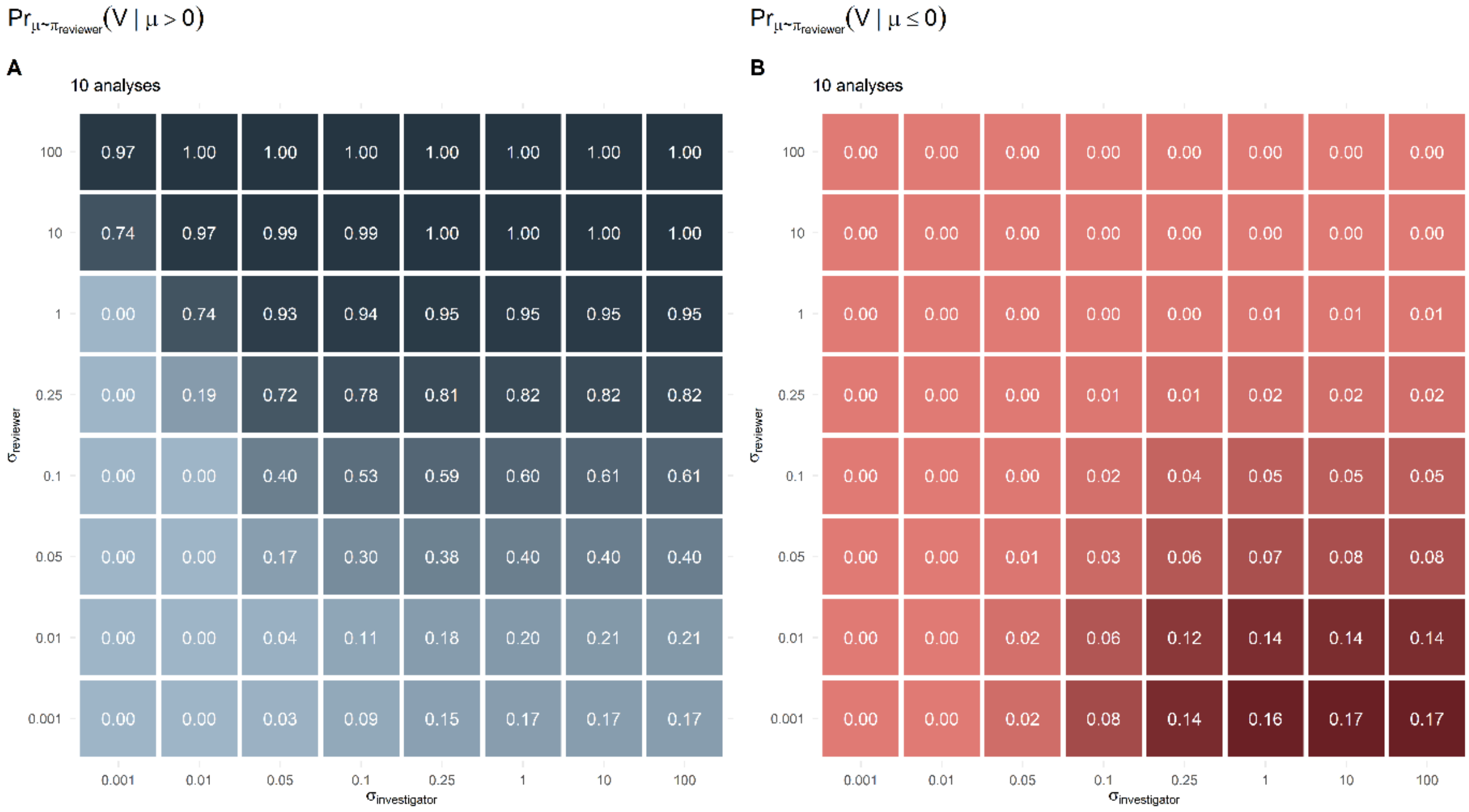

**Figure S41. For 20 analyses.**

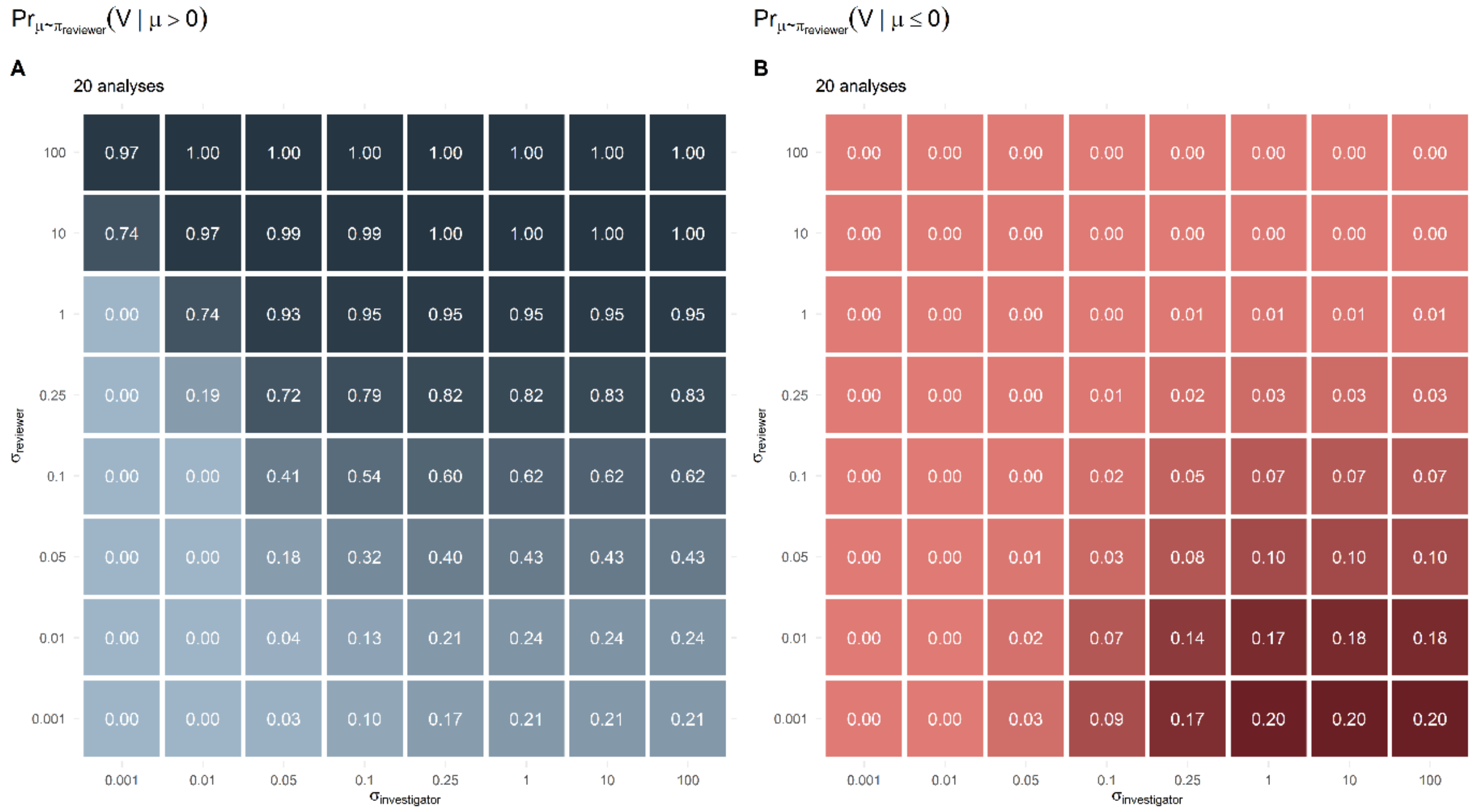

**Figure S42. For 50 analyses.**

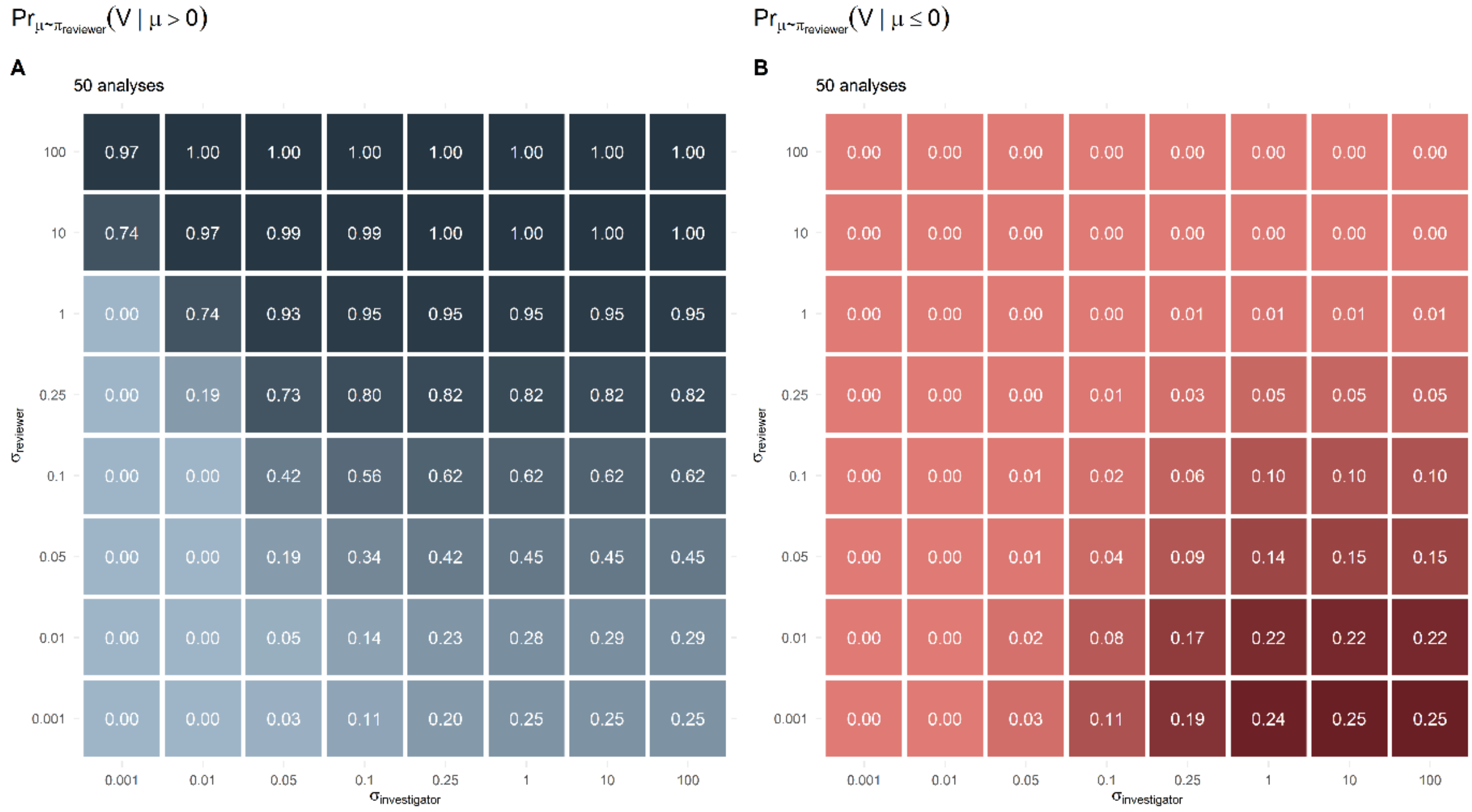

**Figure S43. For 100 analyses.**

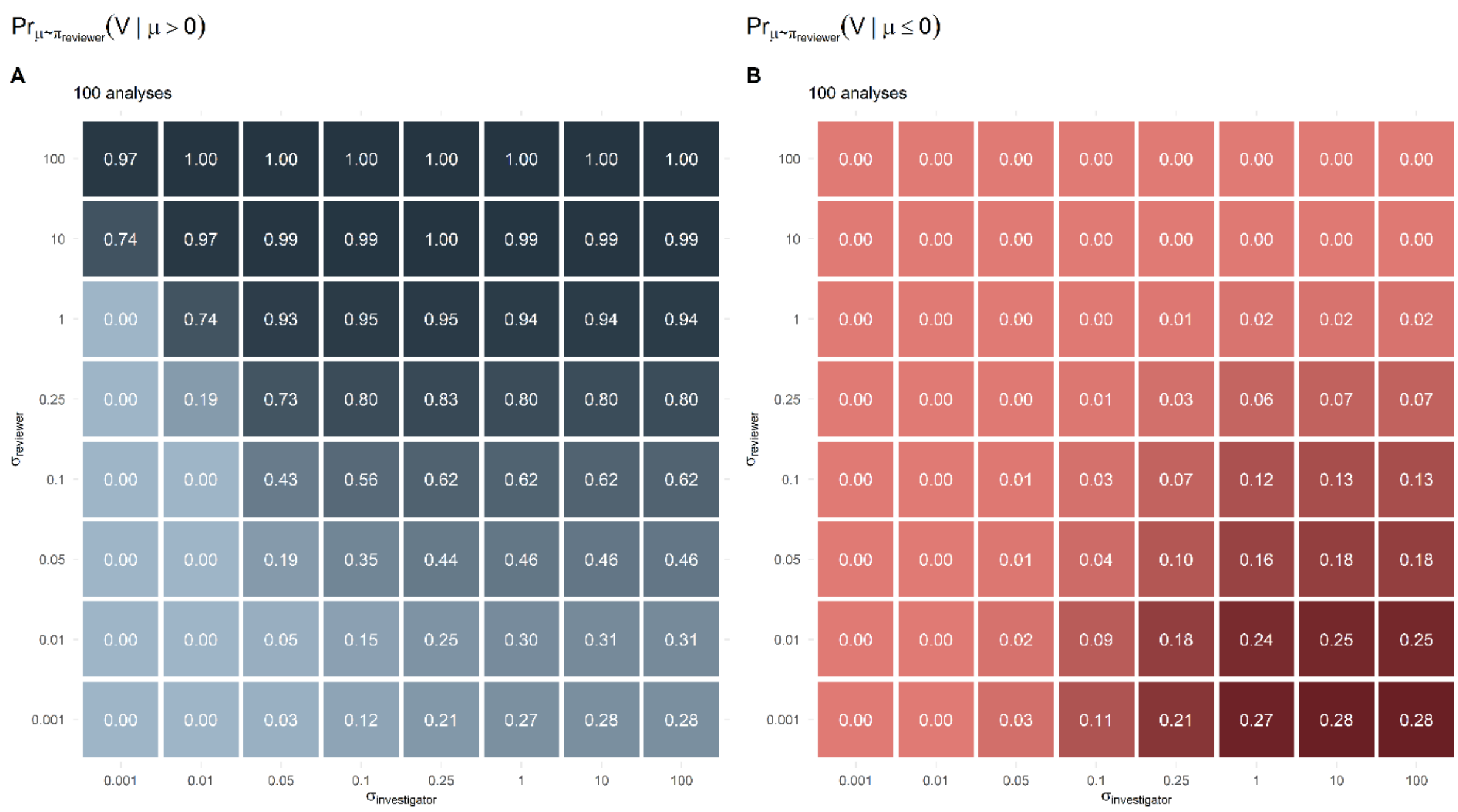

**Figure S44. For 250 analyses.**

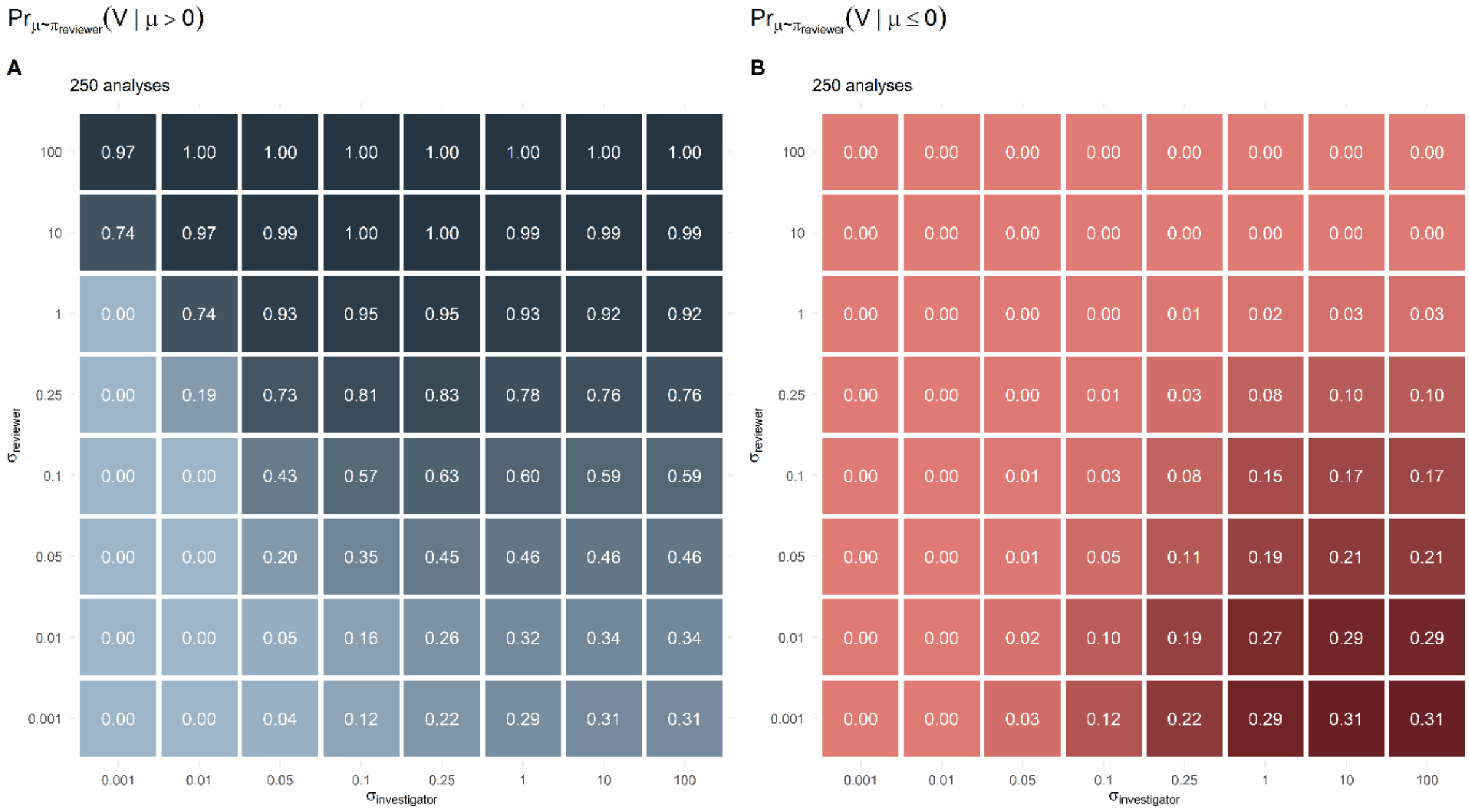

**Figure S45. For 500 analyses.**

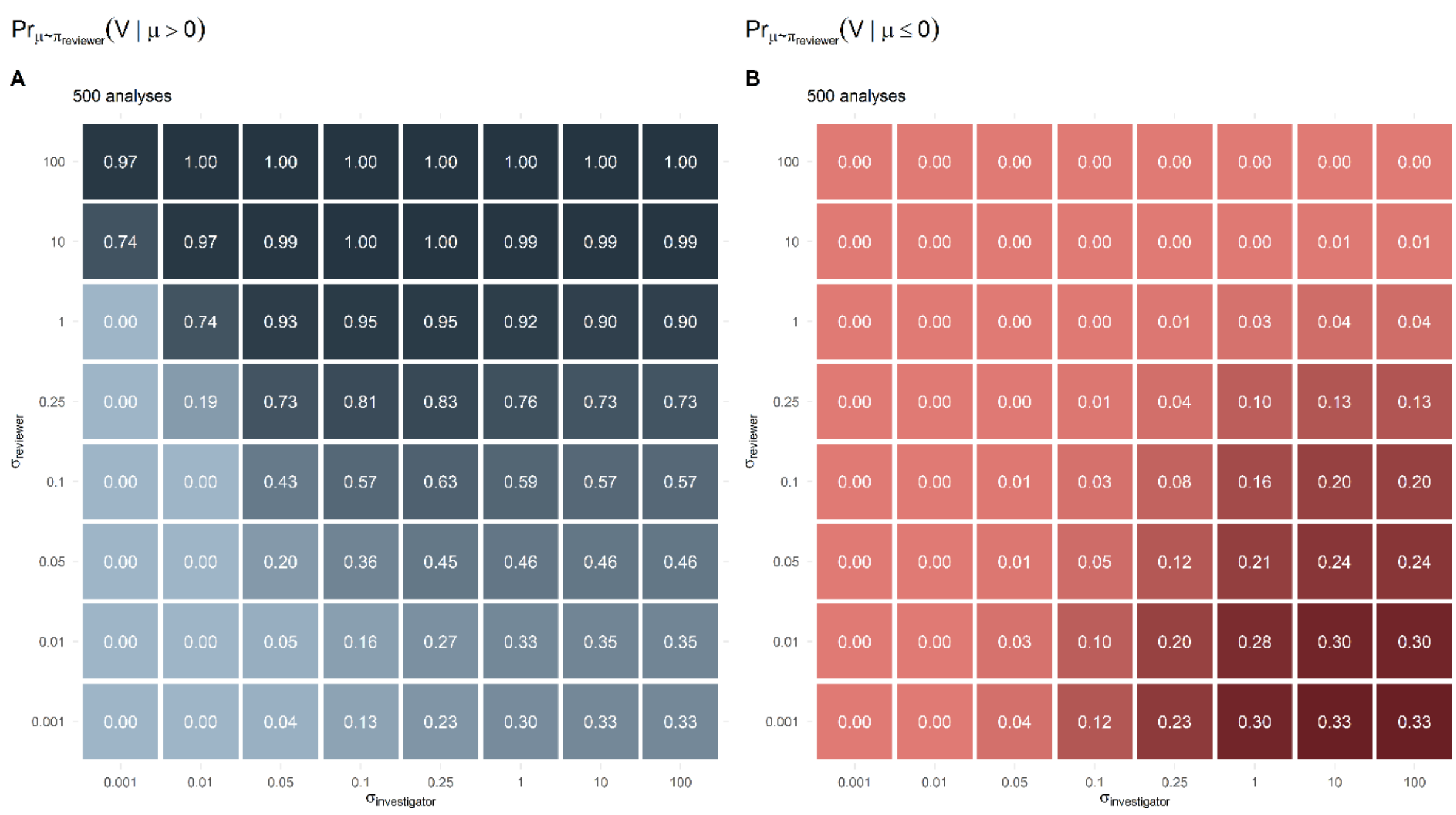

***pTDR = Pr(μ > 0 | V) and pFDR = Pr(μ ≤ 0 | V)***

**Figure S46 to S54. Probability of $\mu > 0$ (panel A) and $\mu \leq 0$ (panel B) conditional on having concluded efficacy (binary event $V$) in Bayesian trials with efficacy stopping rules $Pr(\mu > 0|X_{1,\ldots,k_i}) > 0.95, i \in \{1, \ldots, m\}$, futility stopping rules $Pr(\mu > 0|X_{1,\ldots,k_i}) < 0.10, i \in \{1, \ldots, m-1\}$, and no correction for multiplicity of analyses. $\sigma_{investigator}$ is the standard deviation of the prior $\pi_{investigator} = \mathcal{N}(0, \sigma^2_{investigator})$ used for the analyses, and $\sigma_{reviewer}$ is the standard deviation of the prior $\pi_{reviewer} = \mathcal{N}(0, \sigma^2_{reviewer})$ used to compute the operating characteristics of the trial. Grey cells are not computable because no event of efficacy conclusion occurred in the simulations.**

**Figure S46. For one analysis.**

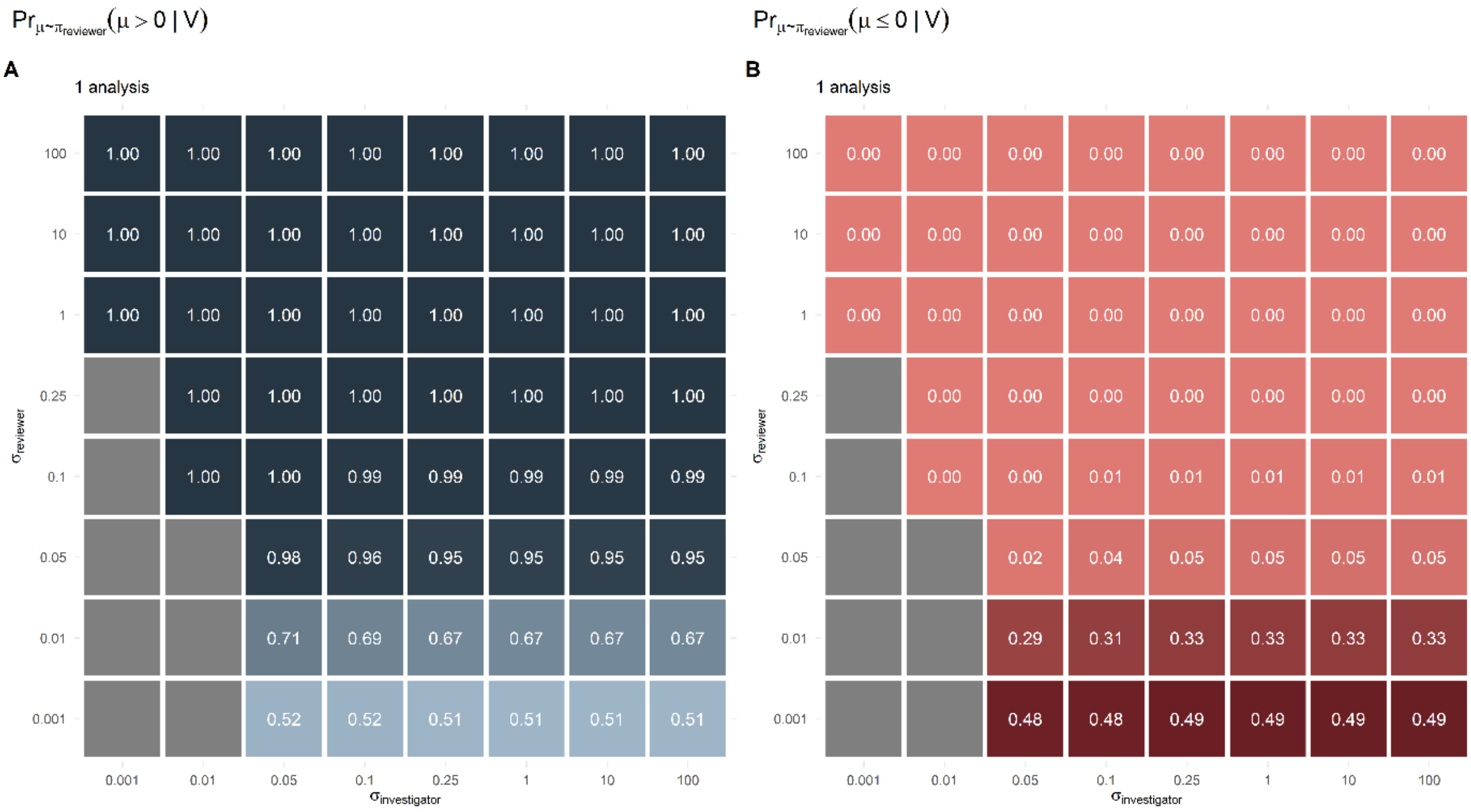

**Figure S47. For two analyses.**

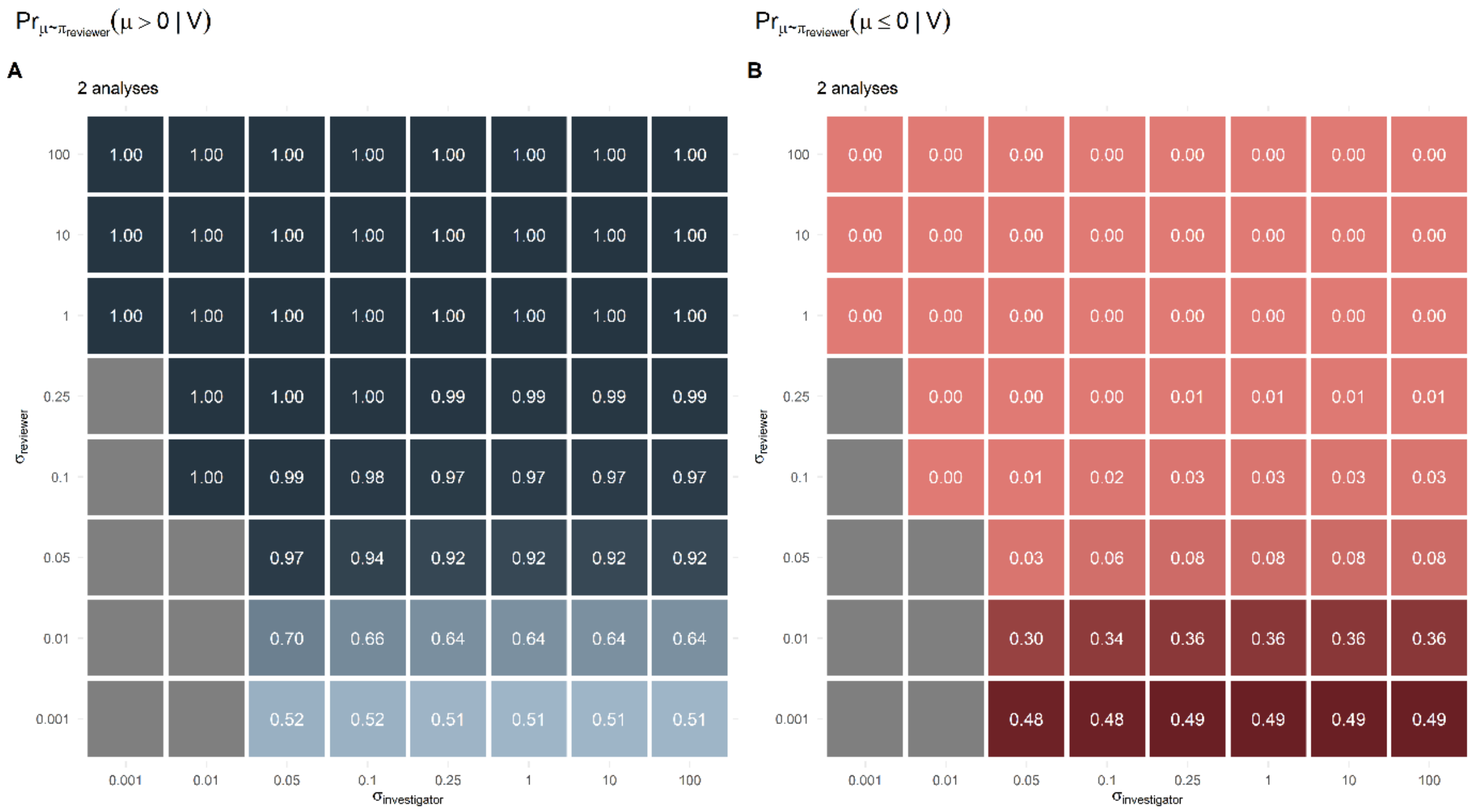

**Figure S48. For five analyses.**

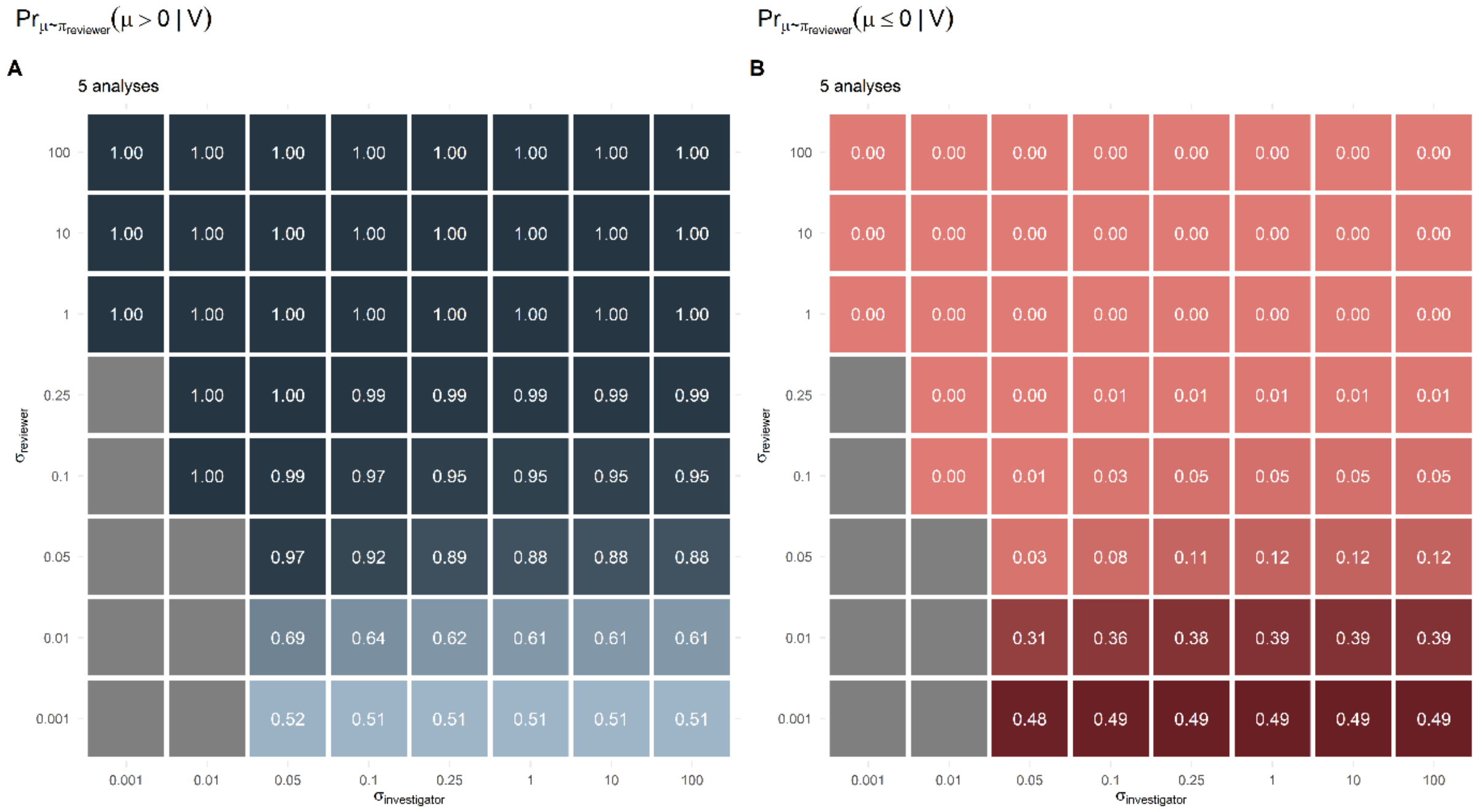

**Figure S49. For 10 analyses.**

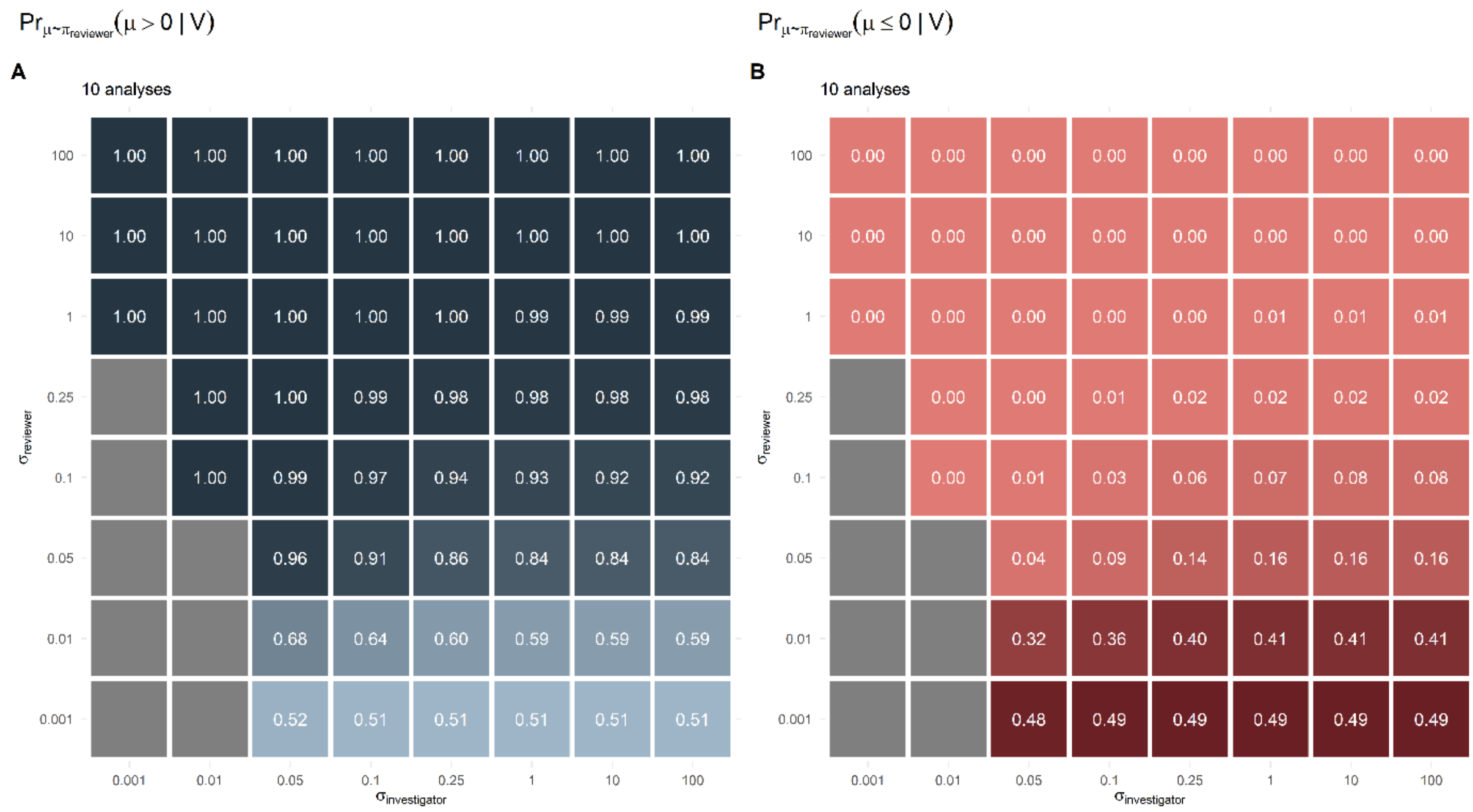

**Figure S50. For 20 analyses.**

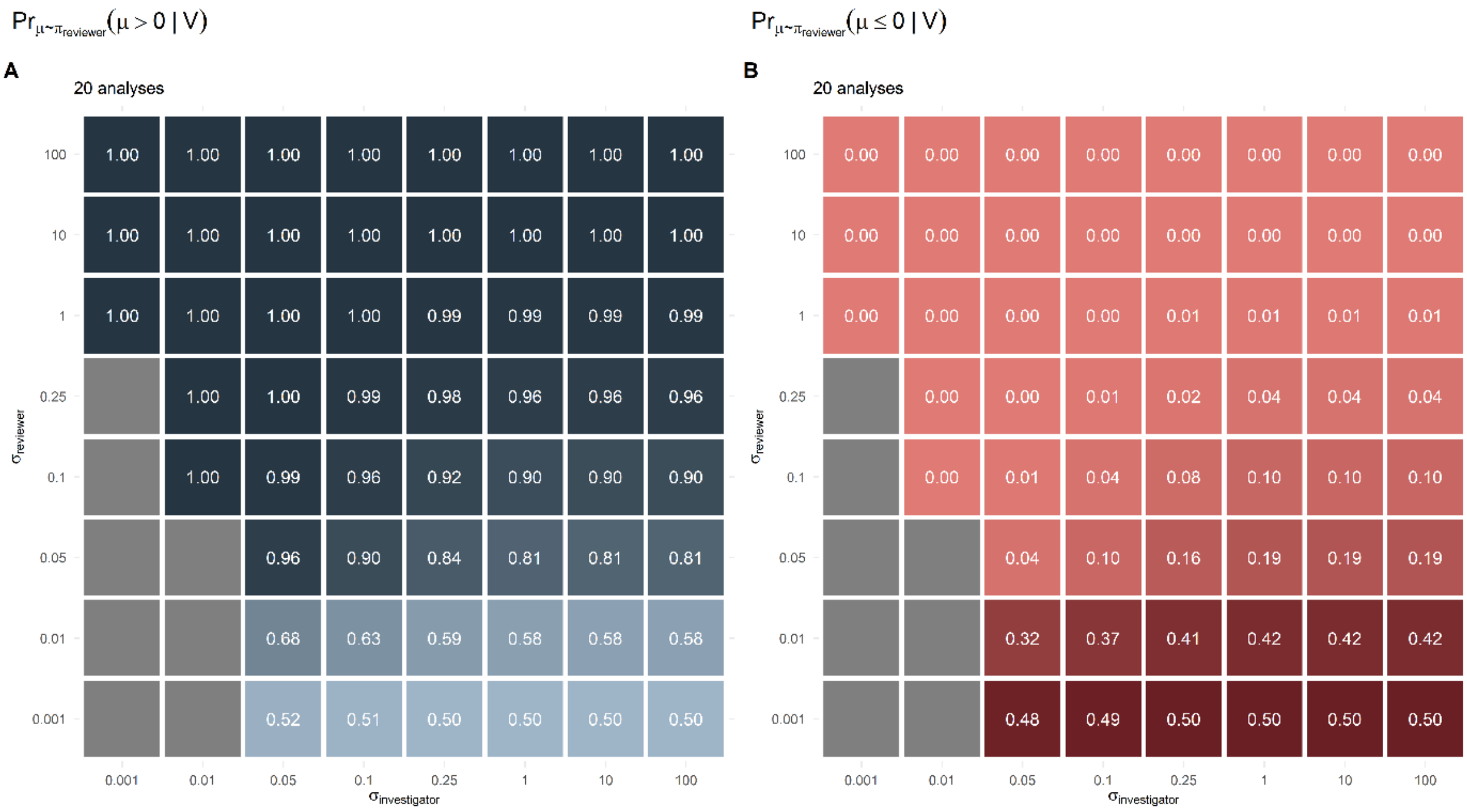

**Figure S51. For 50 analyses.**

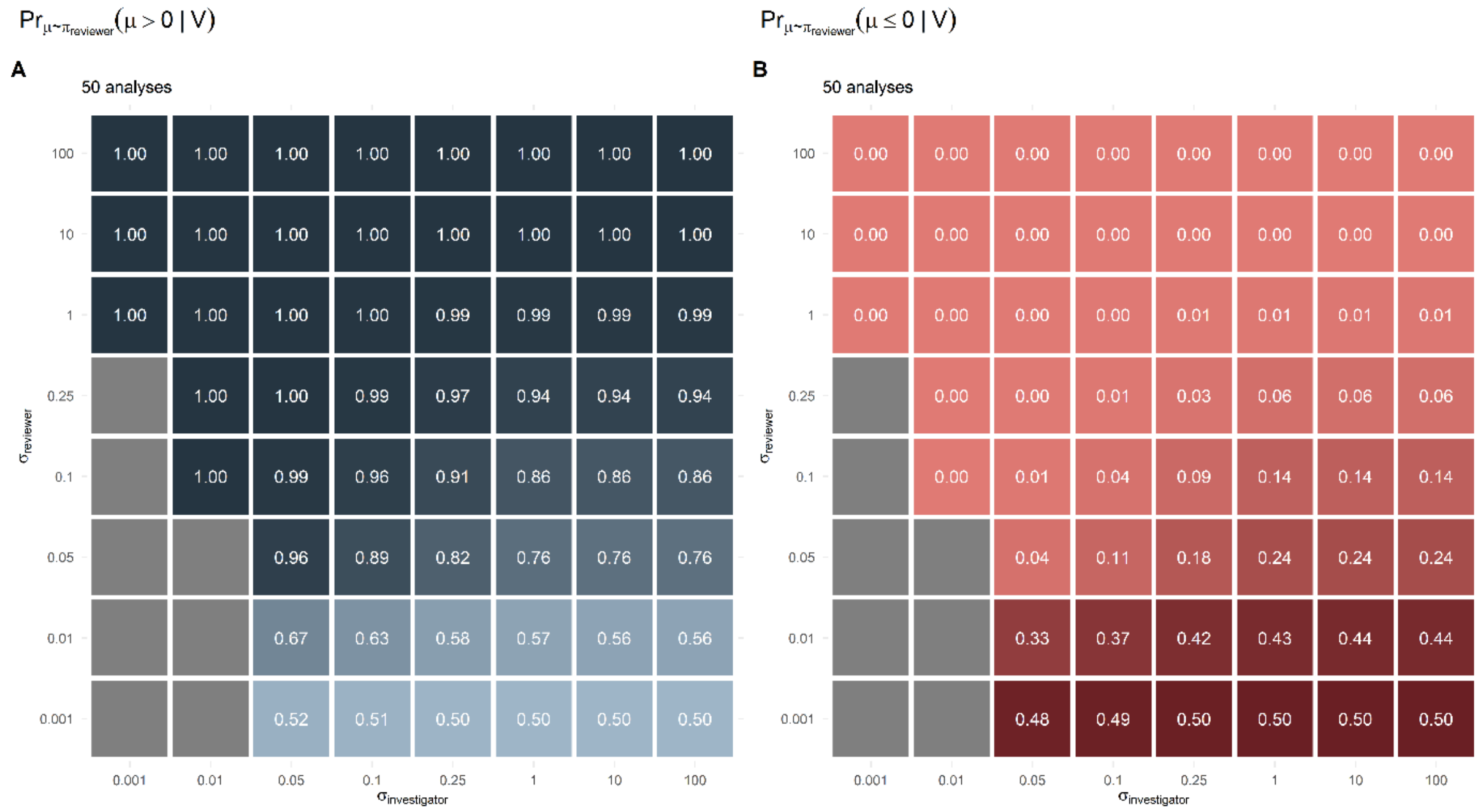

**Figure S52. For 100 analyses.**

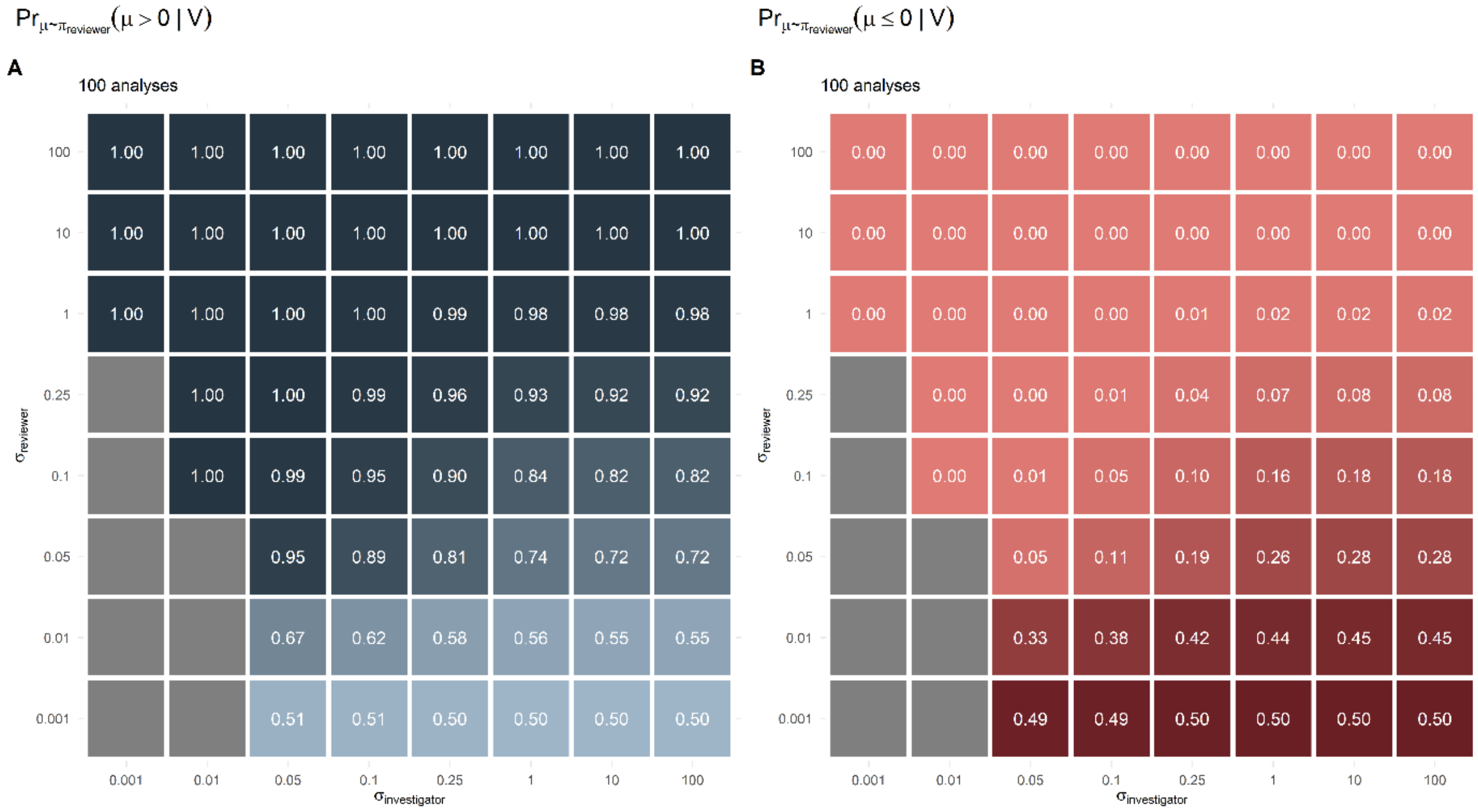

**Figure S53. For 250 analyses.**

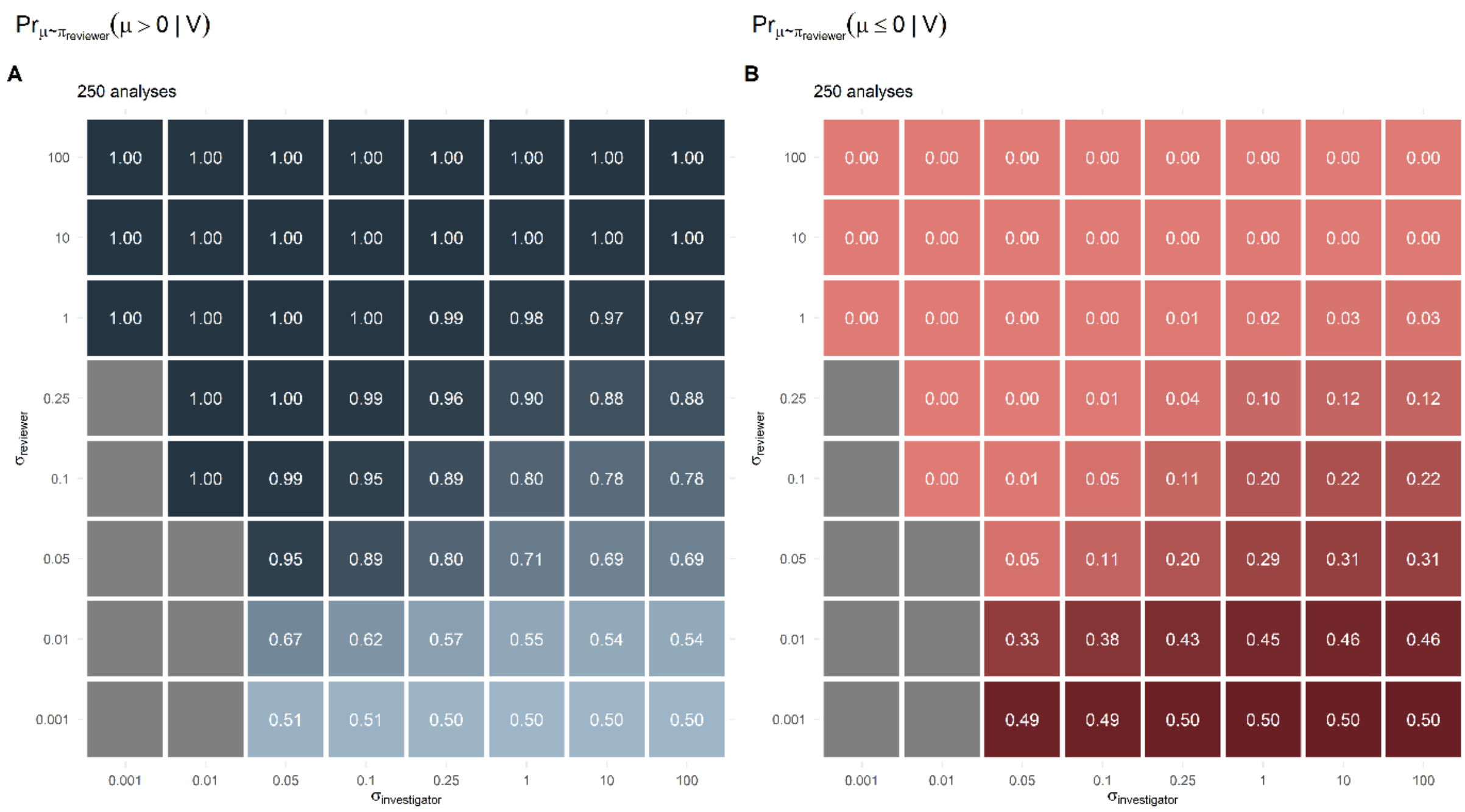

**Figure S54. For 500 analyses.**

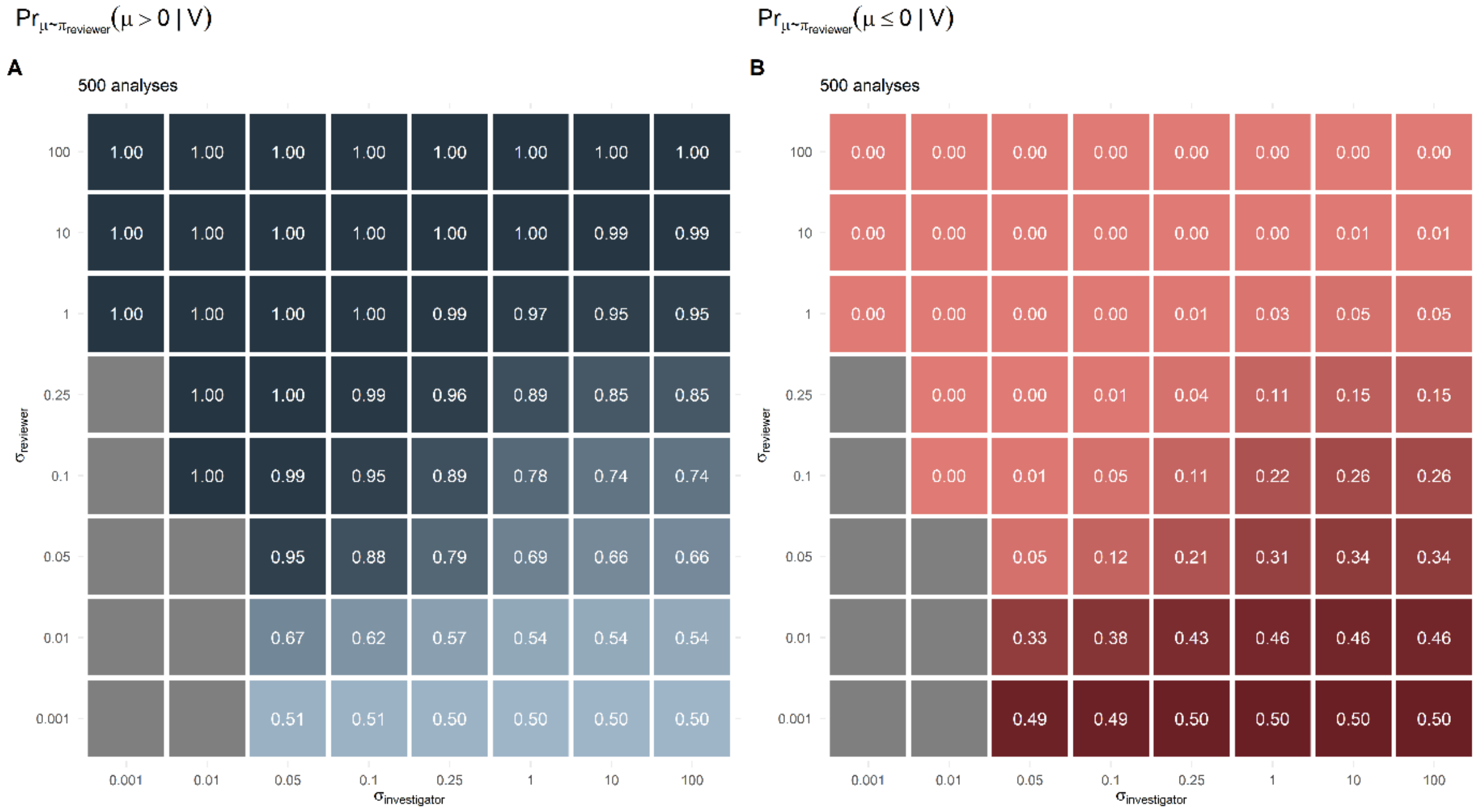

**Table S5. Informative value of a trial concluding efficacy expressed as $pTDO = pTDR/pFDR$, in Bayesian trials with efficacy stopping rules $Pr(\mu > 0|X_{1,\dots,k_i}) > 0.95, i \in \{1, \dots, m-1\}$, futility stopping rules $Pr(\mu > 0|X_{1,\dots,k_i}) < 0.10, i \in \{1, \dots, m\}$, and no correction for multiplicity of analyses. $\sigma_{investigator}$ is the standard deviation of the prior $\mathcal{N}(0, \sigma^2_{investigator})$ used for the analyses, and $\sigma_{reviewer}$ is the standard deviation of the prior $\mathcal{N}(0, \sigma^2_{reviewer})$ used to compute the operating characteristics of the trial. Empty cells are not computable because no event of efficacy conclusion occurred in the simulations.**

| Number of analyses | $\sigma_{reviewer}$ | $\text{pTDO} = \frac{pTDR}{pFDR} = \frac{Pr_{\mu\sim\pi_{reviewer}}(\mu > 0\|V)}{Pr_{\mu\sim\pi_{reviewer}}(\mu \leq 0\|V)}$ | | | | | | | |
|---|---|---|---|---|---|---|---|---|---|
| | | $\sigma_{investigator}$ | $\sigma_{investigator}$ | $\sigma_{investigator}$ | $\sigma_{investigator}$ | $\sigma_{investigator}$ | $\sigma_{investigator}$ | $\sigma_{investigator}$ | $\sigma_{investigator}$ |
| | | **0.001** | **0.01** | **0.05** | **0.1** | **0.25** | **1** | **10** | **100** |
| 1 | 0.001 | | | 1.1 | 1.1 | 1.1 | 1.0 | 1.0 | 1.0 |
| 2 | 0.001 | | | 1.1 | 1.1 | 1.0 | 1.0 | 1.0 | 1.0 |
| 5 | 0.001 | | | 1.1 | 1.0 | 1.0 | 1.0 | 1.0 | 1.0 |
| 10 | 0.001 | | | 1.1 | 1.0 | 1.0 | 1.0 | 1.0 | 1.0 |
| 20 | 0.001 | | | 1.1 | 1.0 | 1.0 | 1.0 | 1.0 | 1.0 |
| 50 | 0.001 | | | 1.1 | 1.0 | 1.0 | 1.0 | 1.0 | 1.0 |
| 100 | 0.001 | | | 1.0 | 1.0 | 1.0 | 1.0 | 1.0 | 1.0 |
| 250 | 0.001 | | | 1.0 | 1.0 | 1.0 | 1.0 | 1.0 | 1.0 |
| 500 | 0.001 | | | 1.0 | 1.0 | 1.0 | 1.0 | 1.0 | 1.0 |
| 1 | 0.01 | | | 2.5 | 2.2 | 2.1 | 2.1 | 2.1 | 2.1 |
| 2 | 0.01 | | | 2.3 | 1.9 | 1.8 | 1.8 | 1.8 | 1.8 |
| 5 | 0.01 | | | 2.2 | 1.8 | 1.6 | 1.6 | 1.6 | 1.6 |
| 10 | 0.01 | | | 2.2 | 1.7 | 1.5 | 1.5 | 1.5 | 1.5 |
| 20 | 0.01 | | | 2.1 | 1.7 | 1.5 | 1.4 | 1.4 | 1.4 |
| 50 | 0.01 | | | 2.0 | 1.7 | 1.4 | 1.3 | 1.3 | 1.3 |
| 100 | 0.01 | | | 2.0 | 1.7 | 1.4 | 1.3 | 1.2 | 1.2 |
| 250 | 0.01 | | | 2.0 | 1.6 | 1.4 | 1.2 | 1.2 | 1.2 |
| 500 | 0.01 | | | 2.0 | 1.6 | 1.3 | 1.2 | 1.2 | 1.2 |
| 1 | 0.05 | | | 47 | 24 | 20 | 19 | 19 | 19 |
| 2 | 0.05 | | | 36 | 16 | 12 | 12 | 12 | 12 |
| 5 | 0.05 | | | 29 | 12 | 7.9 | 7.2 | 7.2 | 7.2 |
| 10 | 0.05 | | | 27 | 10 | 6.2 | 5.4 | 5.3 | 5.3 |
| 20 | 0.05 | | | 24 | 9.3 | 5.3 | 4.2 | 4.2 | 4.2 |
| 50 | 0.05 | | | 21 | 8.4 | 4.5 | 3.2 | 3.1 | 3.1 |
| 100 | 0.05 | | | 21 | 8.1 | 4.2 | 2.8 | 2.6 | 2.6 |
| 250 | 0.05 | | | 20 | 7.8 | 4.0 | 2.4 | 2.2 | 2.2 |
| 500 | 0.05 | | | 19 | 7.6 | 3.8 | 2.2 | 1.9 | 1.9 |
| 1 | 0.1 | | ≥ 1000 | 249 | 90 | 70 | 66 | 66 | 66 |
| 2 | 0.1 | | ≥ 1000 | 166 | 52 | 39 | 36 | 35 | 35 |
| 5 | 0.1 | | ≥ 1000 | 142 | 37 | 21 | 18 | 18 | 18 |
| 10 | 0.1 | | ≥ 1000 | 110 | 31 | 15 | 13 | 12 | 12 |
| 20 | 0.1 | | ≥ 1000 | 99 | 27 | 12 | 9.1 | 8.8 | 8.8 |
| 50 | 0.1 | | ≥ 1000 | 82 | 23 | 10 | 6.3 | 5.9 | 5.9 |
| 100 | 0.1 | | ≥ 1000 | 76 | 21 | 9.0 | 5.1 | 4.7 | 4.7 |
| 250 | 0.1 | | ≥ 1000 | 76 | 20 | 8.2 | 4.1 | 3.5 | 3.5 |
| 500 | 0.1 | | ≥ 1000 | 70 | 19 | 7.8 | 3.6 | 2.9 | 2.8 |
| 1 | 0.25 | | ≥ 1000 | 999 | 499 | 332 | 332 | 332 | 332 |
| 2 | 0.25 | | ≥ 1000 | 999 | 249 | 166 | 142 | 142 | 142 |
| 5 | 0.25 | | ≥ 1000 | 999 | 166 | 82 | 70 | 66 | 66 |
| 10 | 0.25 | | ≥ 1000 | 999 | 124 | 55 | 42 | 41 | 41 |
| 20 | 0.25 | | ≥ 1000 | 499 | 110 | 41 | 27 | 26 | 26 |
| 50 | 0.25 | | ≥ 1000 | 499 | 90 | 31 | 17 | 16 | 16 |
| 100 | 0.25 | | ≥ 1000 | 499 | 82 | 27 | 13 | 11 | 11 |
| 250 | 0.25 | | ≥ 1000 | 499 | 76 | 23 | 9.4 | 7.5 | 7.4 |
| 500 | 0.25 | | ≥ 1000 | 332 | 76 | 22 | 8.0 | 5.8 | 5.7 |
| 1 | 1 | ≥ 1000 | ≥ 1000 | ≥ 1000 | 999 | 999 | 999 | 999 | 999 |
| 2 | 1 | ≥ 1000 | ≥ 1000 | ≥ 1000 | 999 | 499 | 499 | 499 | 499 |
| 5 | 1 | ≥ 1000 | ≥ 1000 | ≥ 1000 | 499 | 332 | 249 | 249 | 249 |
| 10 | 1 | ≥ 1000 | ≥ 1000 | ≥ 1000 | 499 | 249 | 199 | 199 | 199 |
| 20 | 1 | ≥ 1000 | ≥ 1000 | ≥ 1000 | 499 | 199 | 124 | 124 | 124 |
| 50 | 1 | ≥ 1000 | ≥ 1000 | 999 | 332 | 142 | 76 | 70 | 70 |
| 100 | 1 | ≥ 1000 | ≥ 1000 | 999 | 332 | 124 | 55 | 47 | 47 |
| 250 | 1 | ≥ 1000 | ≥ 1000 | 999 | 332 | 110 | 41 | 30 | 29 |
| 500 | 1 | ≥ 1000 | ≥ 1000 | 999 | 332 | 99 | 33 | 21 | 20 |
| 1 | 10 | ≥ 1000 | ≥ 1000 | ≥ 1000 | ≥ 1000 | ≥ 1000 | ≥ 1000 | ≥ 1000 | ≥ 1000 |
| 2 | 10 | ≥ 1000 | ≥ 1000 | ≥ 1000 | ≥ 1000 | ≥ 1000 | ≥ 1000 | ≥ 1000 | ≥ 1000 |
| 5 | 10 | ≥ 1000 | ≥ 1000 | ≥ 1000 | ≥ 1000 | ≥ 1000 | ≥ 1000 | ≥ 1000 | ≥ 1000 |
| 10 | 10 | ≥ 1000 | ≥ 1000 | ≥ 1000 | ≥ 1000 | 999 | 999 | 999 | 999 |
| 20 | 10 | ≥ 1000 | ≥ 1000 | ≥ 1000 | ≥ 1000 | 999 | 999 | 999 | 999 |
| 50 | 10 | ≥ 1000 | ≥ 1000 | ≥ 1000 | ≥ 1000 | 999 | 499 | 499 | 499 |
| 100 | 10 | ≥ 1000 | ≥ 1000 | ≥ 1000 | ≥ 1000 | 999 | 499 | 499 | 499 |
| 250 | 10 | ≥ 1000 | ≥ 1000 | ≥ 1000 | ≥ 1000 | 999 | 332 | 249 | 249 |
| 500 | 10 | ≥ 1000 | ≥ 1000 | ≥ 1000 | ≥ 1000 | 999 | 332 | 199 | 199 |
| 1 | 100 | ≥ 1000 | ≥ 1000 | ≥ 1000 | ≥ 1000 | ≥ 1000 | ≥ 1000 | ≥ 1000 | ≥ 1000 |
| 2 | 100 | ≥ 1000 | ≥ 1000 | ≥ 1000 | ≥ 1000 | ≥ 1000 | ≥ 1000 | ≥ 1000 | ≥ 1000 |
| 5 | 100 | ≥ 1000 | ≥ 1000 | ≥ 1000 | ≥ 1000 | ≥ 1000 | ≥ 1000 | ≥ 1000 | ≥ 1000 |
| 10 | 100 | ≥ 1000 | ≥ 1000 | ≥ 1000 | ≥ 1000 | ≥ 1000 | ≥ 1000 | ≥ 1000 | ≥ 1000 |
| 20 | 100 | ≥ 1000 | ≥ 1000 | ≥ 1000 | ≥ 1000 | ≥ 1000 | ≥ 1000 | ≥ 1000 | ≥ 1000 |
| 50 | 100 | ≥ 1000 | ≥ 1000 | ≥ 1000 | ≥ 1000 | ≥ 1000 | ≥ 1000 | ≥ 1000 | ≥ 1000 |
| 100 | 100 | ≥ 1000 | ≥ 1000 | ≥ 1000 | ≥ 1000 | ≥ 1000 | ≥ 1000 | ≥ 1000 | ≥ 1000 |
| 250 | 100 | ≥ 1000 | ≥ 1000 | ≥ 1000 | ≥ 1000 | ≥ 1000 | ≥ 1000 | ≥ 1000 | ≥ 1000 |
| 500 | 100 | ≥ 1000 | ≥ 1000 | ≥ 1000 | ≥ 1000 | ≥ 1000 | ≥ 1000 | ≥ 1000 | ≥ 1000 |

### 2.2.2.3 *Futility threshold = 0.50*

### *A = Pr(V | μ > 0) and 1 – B = Pr(V | μ ≤ 0)*

**Figure S55 to S63. Probability of concluding efficacy (binary event $V$) conditional on $\mu > 0$ (panel A) and $\mu \leq 0$ (panel B) in Bayesian trials with efficacy stopping rules $Pr(\mu > 0 | X_{1,\ldots,k_i}) > 0.95, i \in \{1, \ldots, m\}$, futility stopping rules $Pr(\mu > 0 | X_{1,\ldots,k_i}) < 0.50, i \in \{1, \ldots, m-1\}$, and no correction for multiplicity of analyses. $\sigma_{investigator}$ is the standard deviation of the prior $\pi_{investigator} = \mathcal{N}(0, \sigma^2_{investigator})$ used for the analyses, and $\sigma_{reviewer}$ is the standard deviation of the prior $\pi_{reviewer} = \mathcal{N}(0, \sigma^2_{reviewer})$ used to compute the operating characteristics of the trial.**

**Figure S55. For one analysis.**

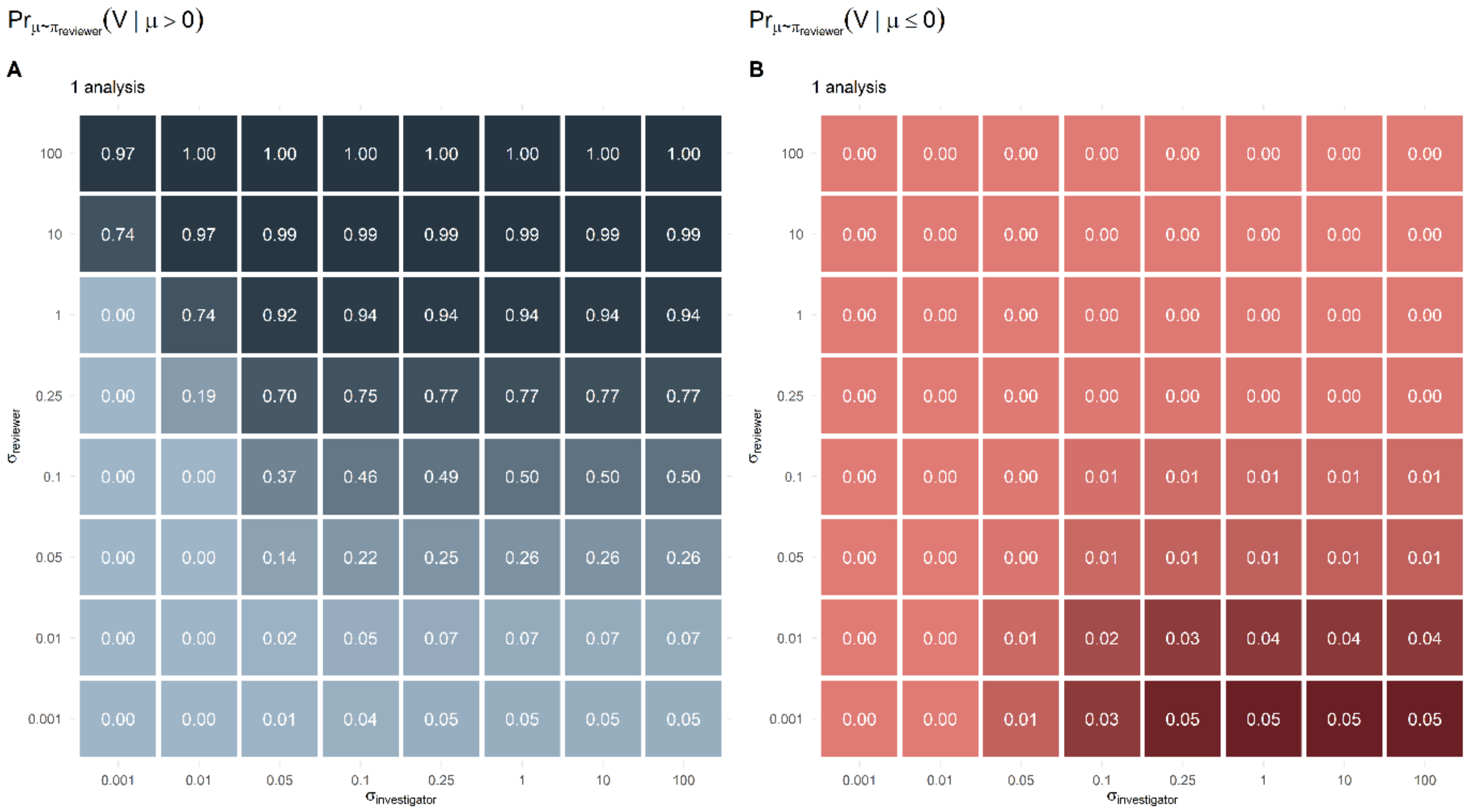

**Figure S56. For two analyses.**

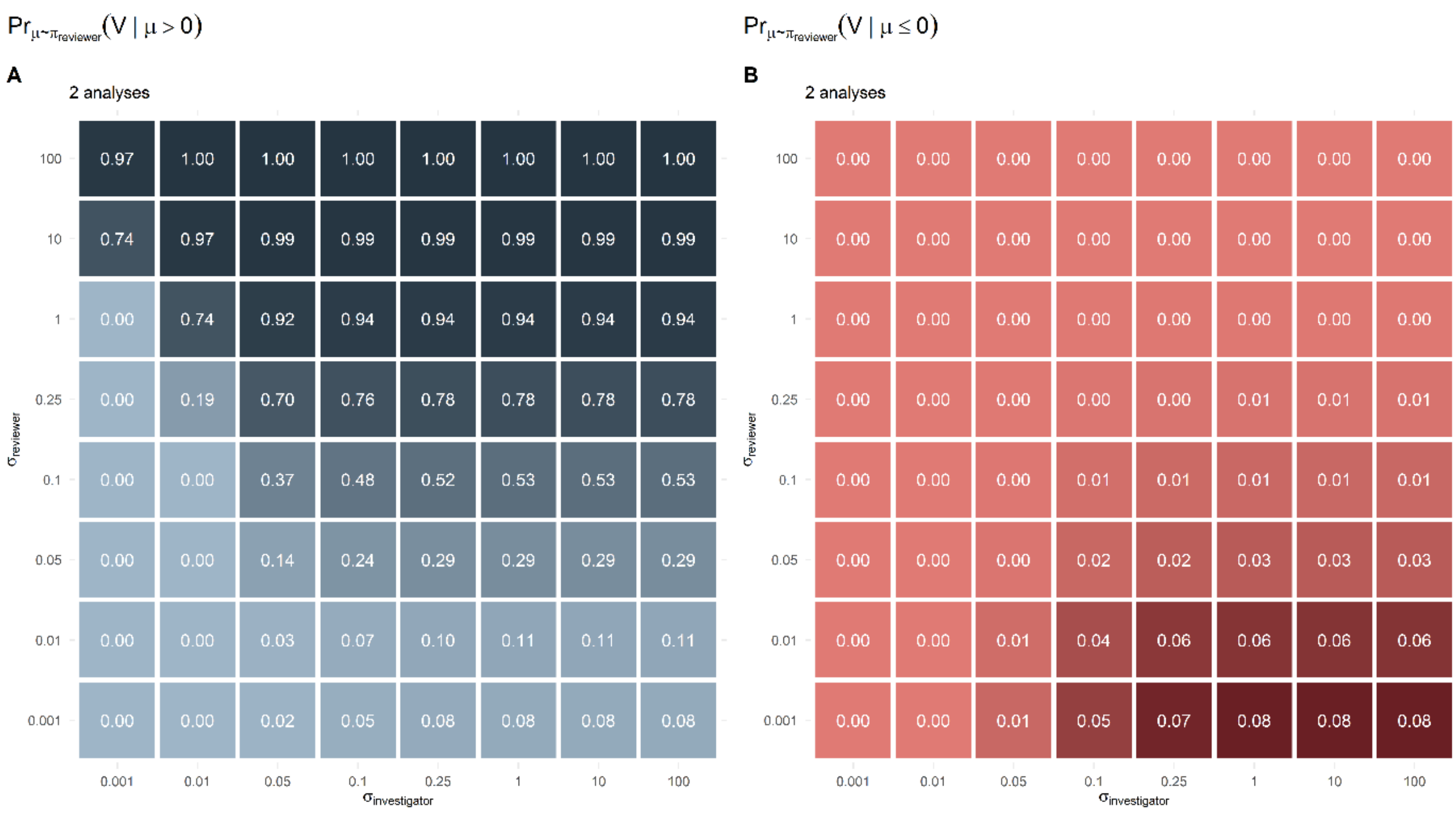

**Figure S57. For five analyses.**

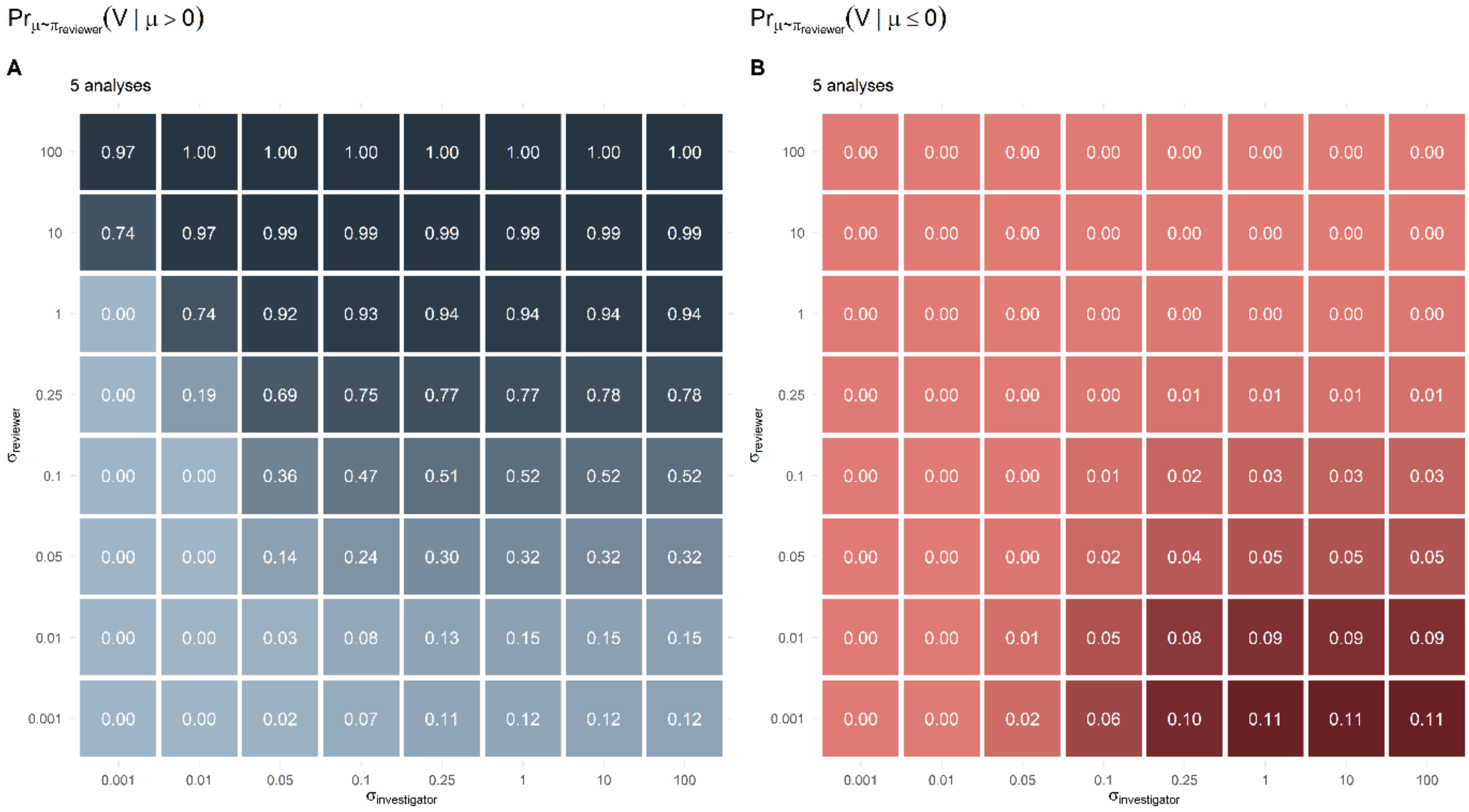

**Figure S58. For 10 analyses.**

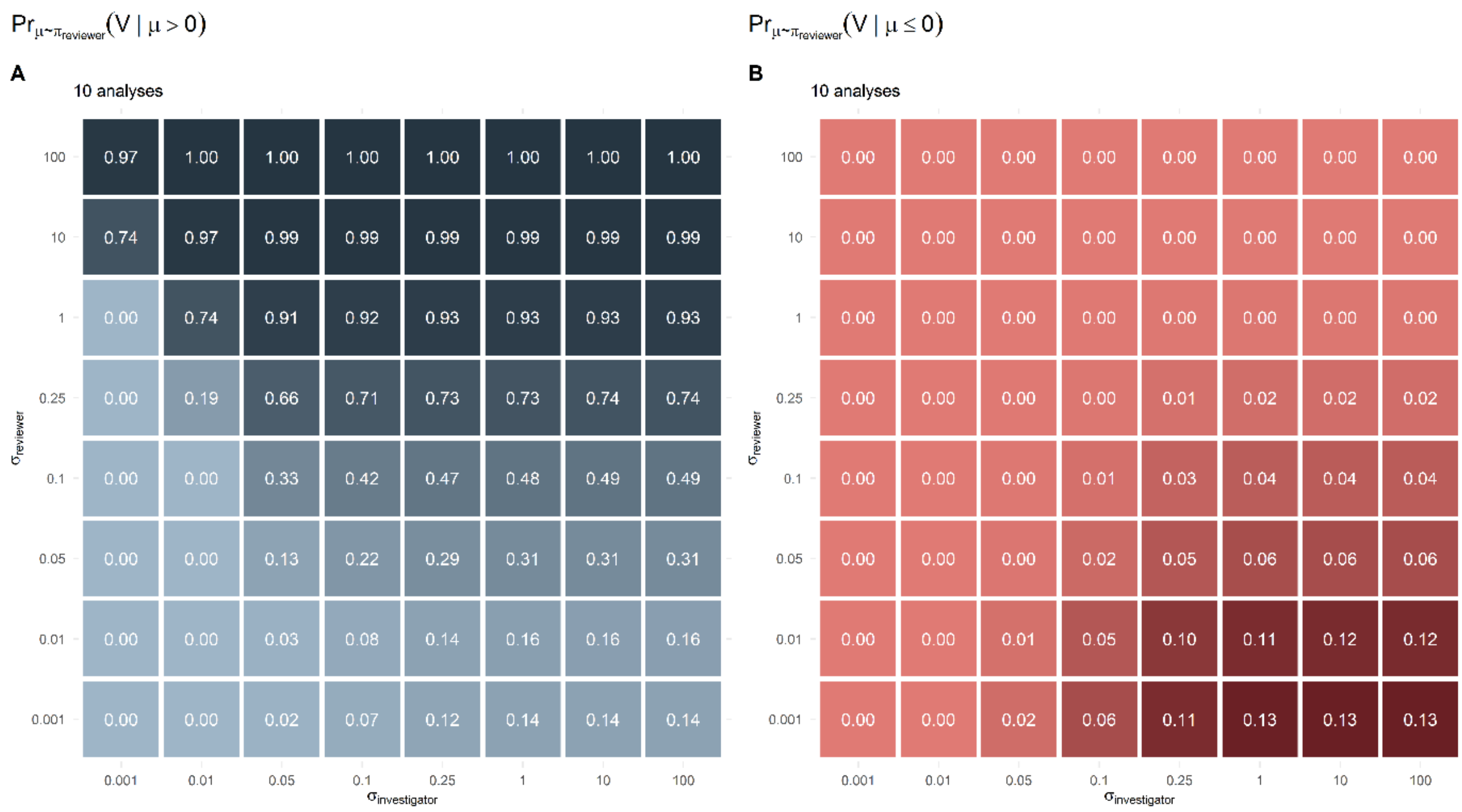

**Figure S59. For 20 analyses.**

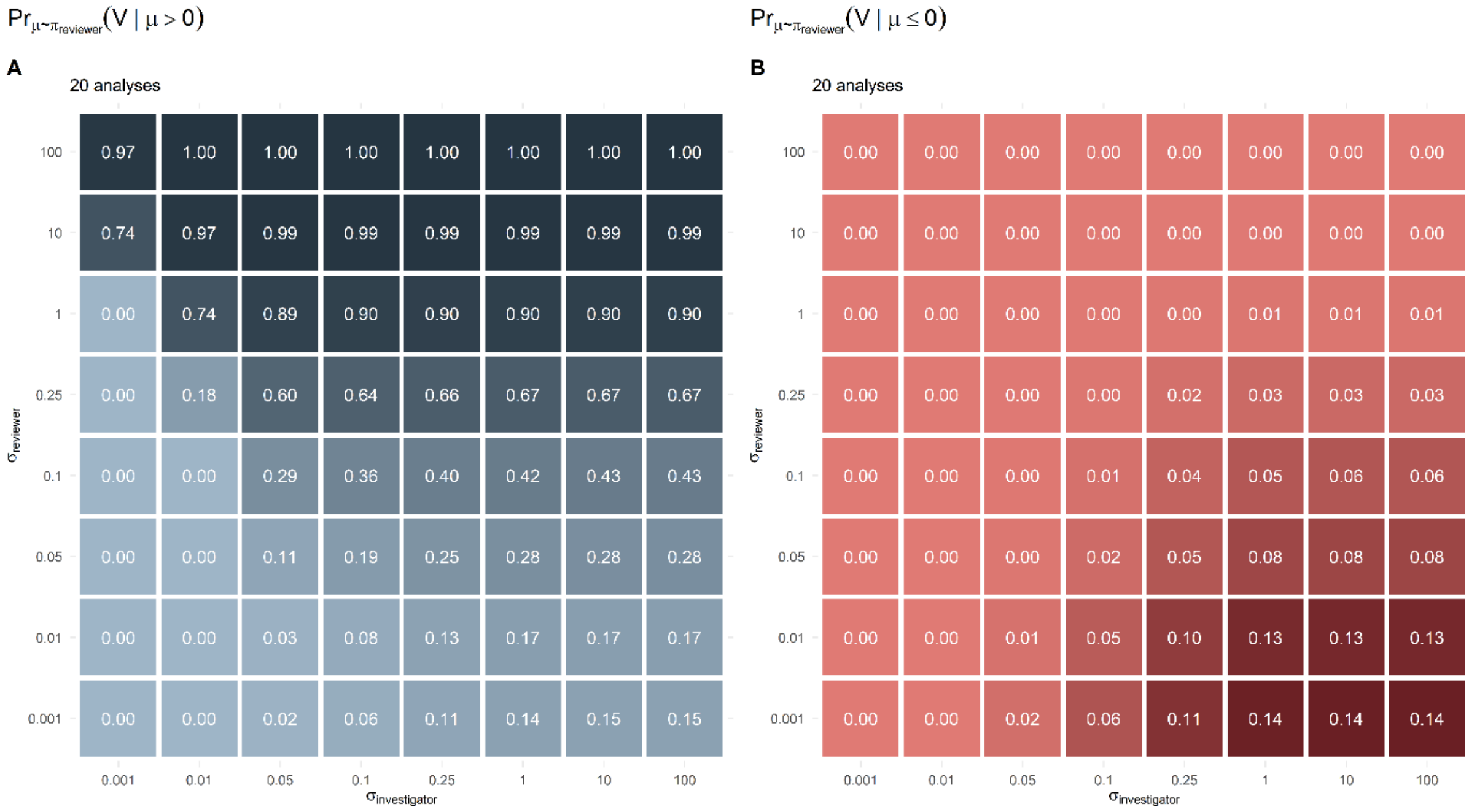

**Figure S60. For 50 analyses.**

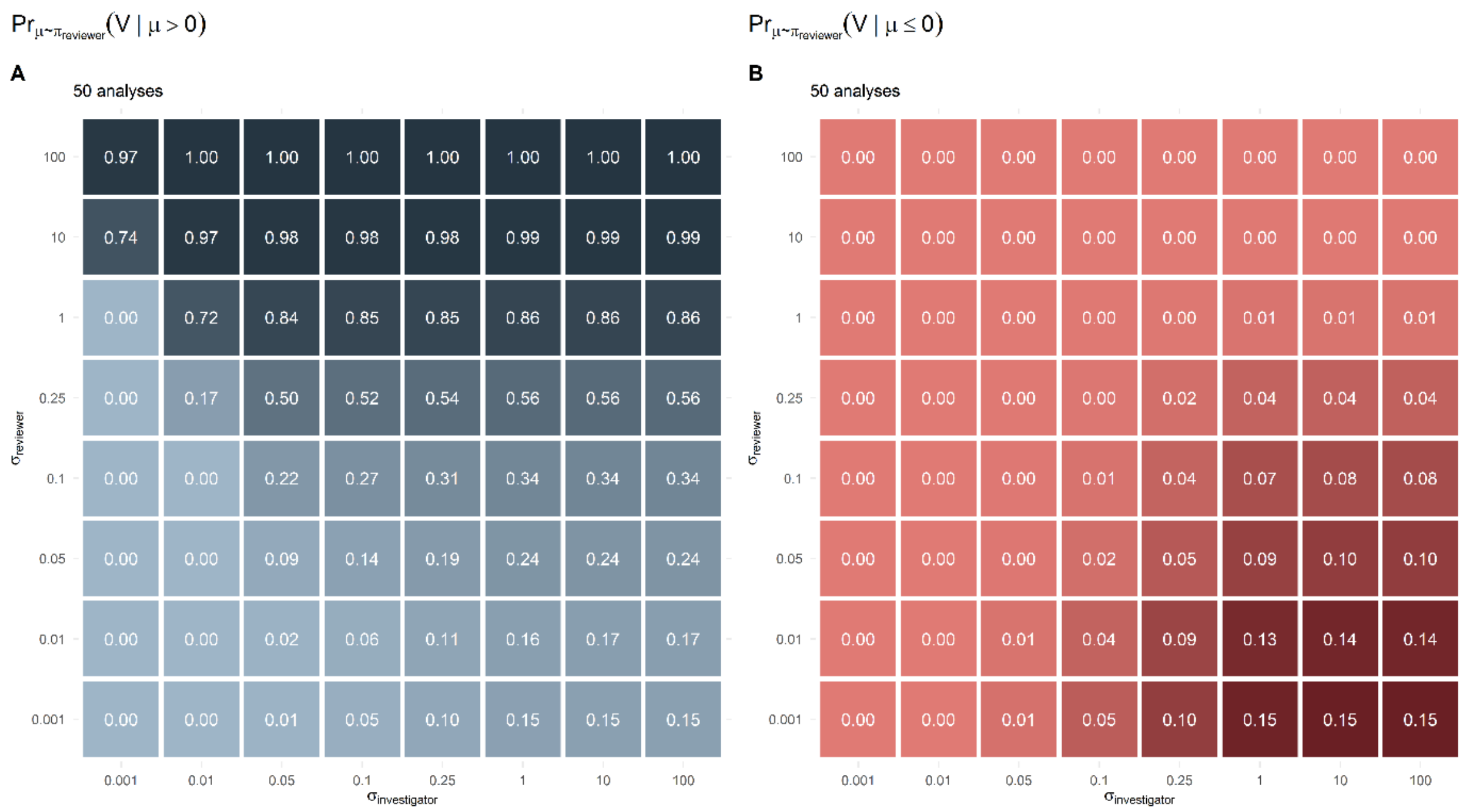

**Figure S61. For 100 analyses.**

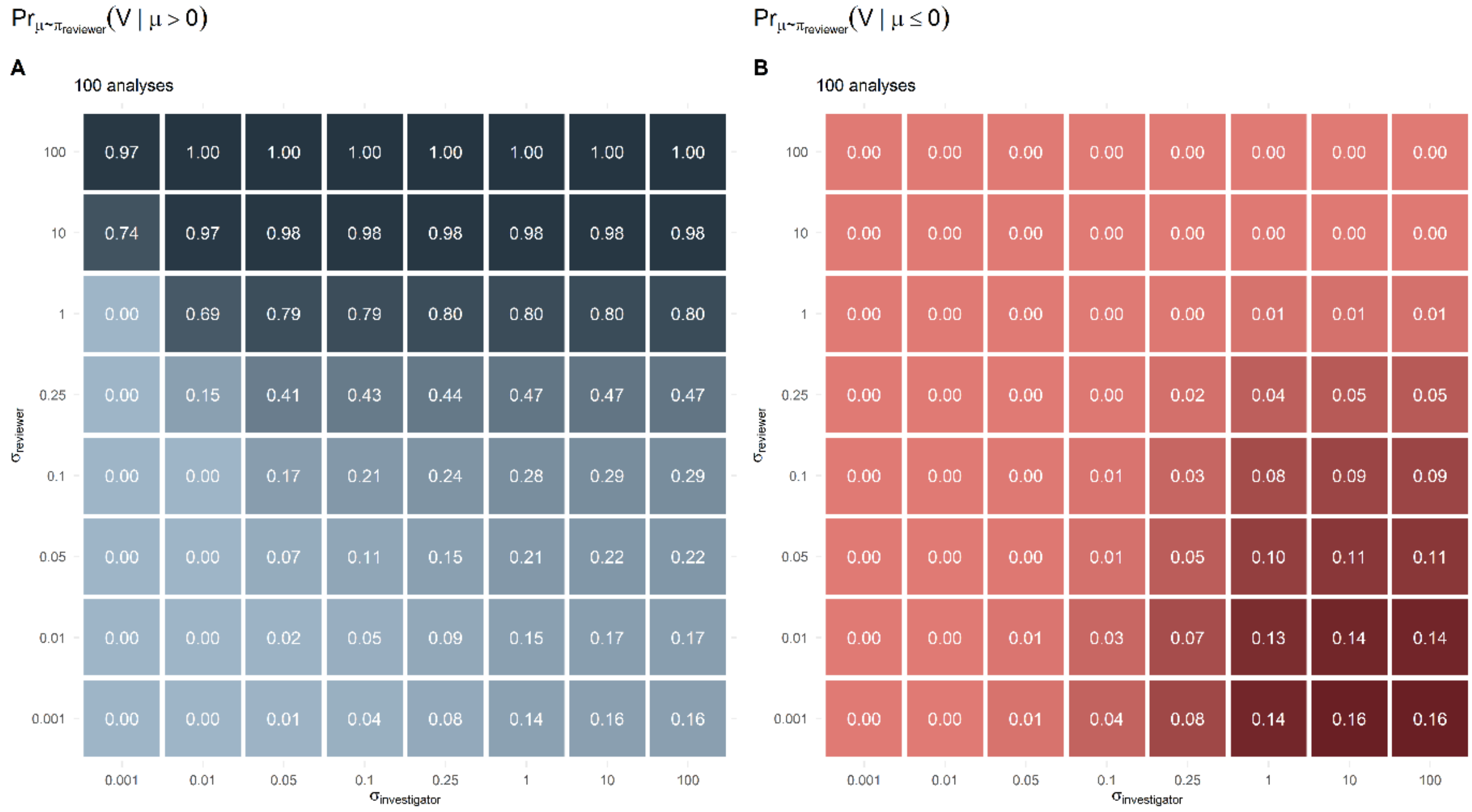

**Figure S62. For 250 analyses.**

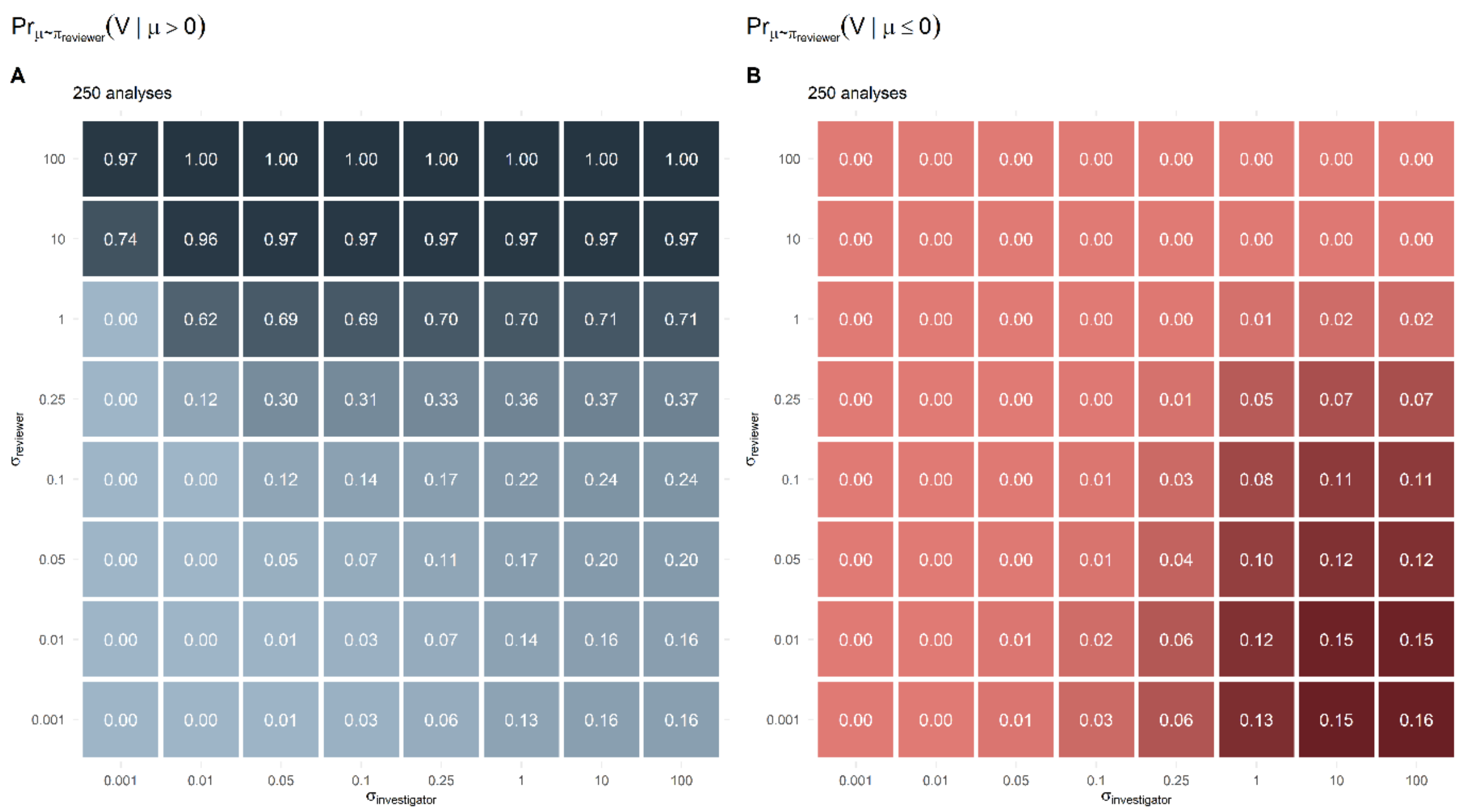

**Figure S63. For 500 analyses.**

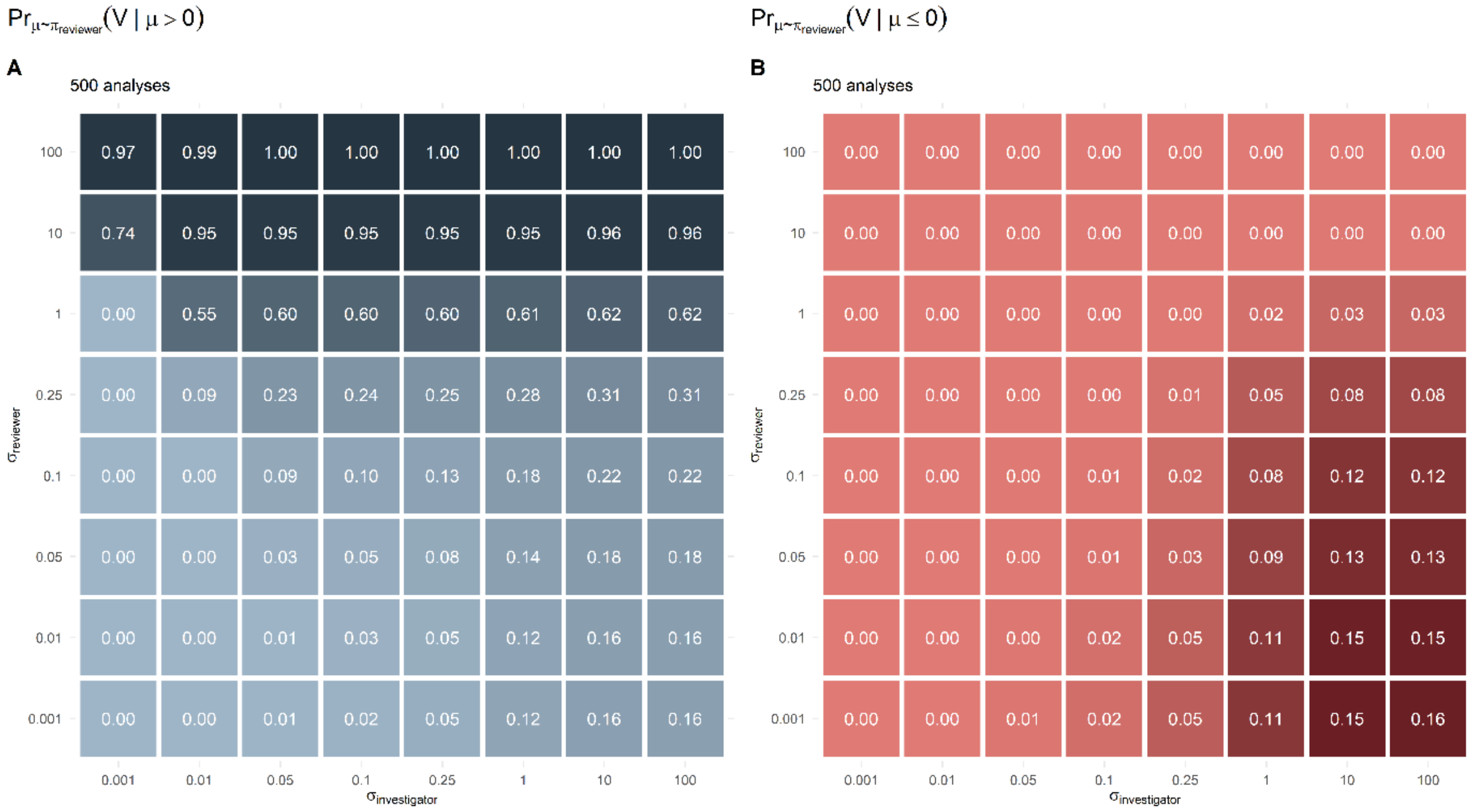

***pTDR = Pr(μ > 0 | V) and pFDR = Pr(μ ≤ 0 | V)***

**Figure S64 to S72. Probability of $\mu > 0$ (panel A) and $\mu \leq 0$ (panel B) conditional on having concluded efficacy (binary event $V$) in Bayesian trials with efficacy stopping rules $Pr(\mu > 0|X_{1,\dots,k_i}) > 0.95, i \in \{1,\dots,m\}$, futility stopping rules $Pr(\mu > 0|X_{1,\dots,k_i}) < 0.50, i \in \{1,\dots,m-1\}$, and no correction for multiplicity of analyses. $\sigma_{investigator}$ is the standard deviation of the prior $\pi_{investigator} = \mathcal{N}(0, \sigma^2_{investigator})$ used for the analyses, and $\sigma_{reviewer}$ is the standard deviation of the prior $\pi_{reviewer} = \mathcal{N}(0, \sigma^2_{reviewer})$ used to compute the operating characteristics of the trial. Grey cells are not computable because no event of efficacy conclusion occurred in the simulations.**

**Figure S64. For one analysis.**

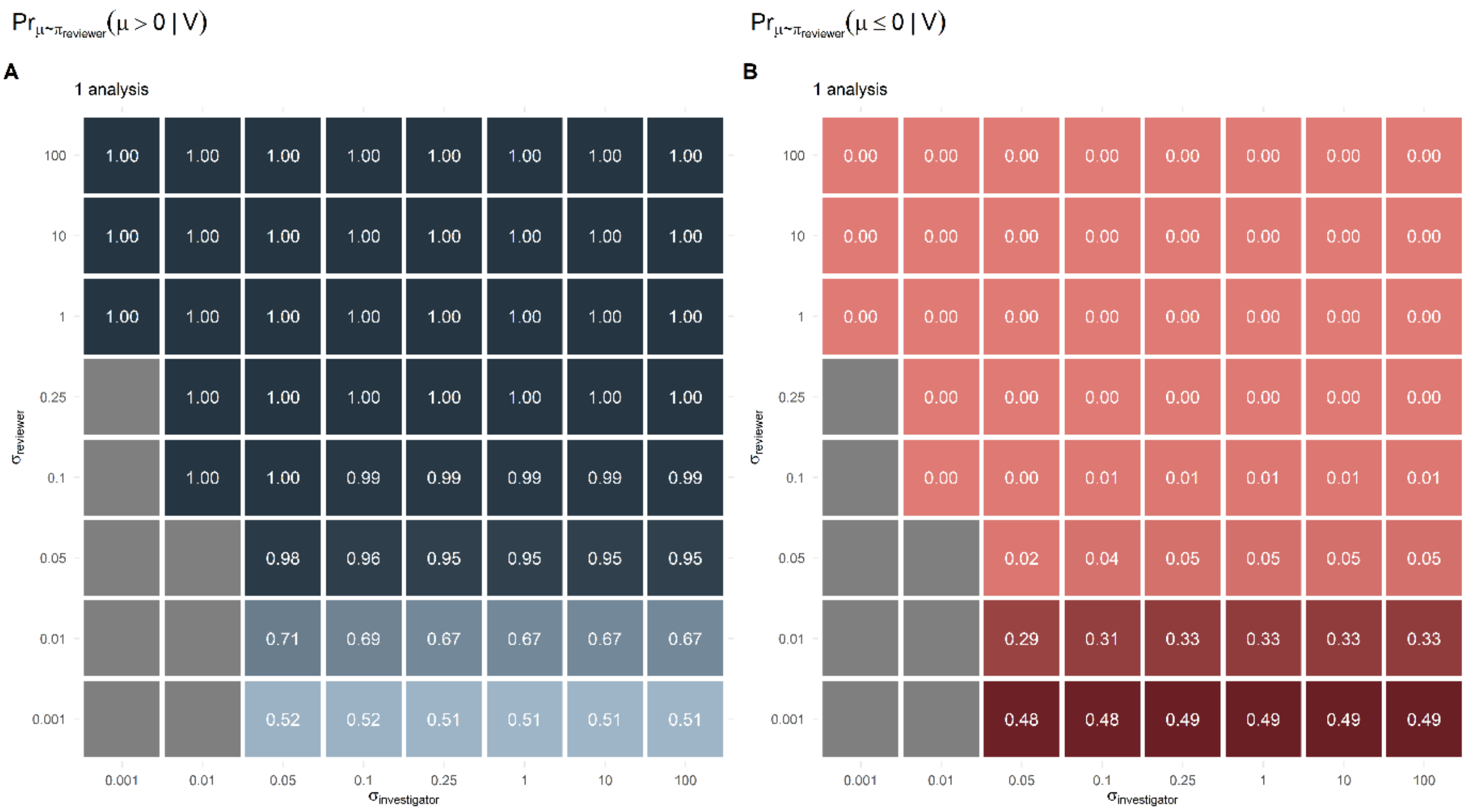

**Figure S65. For two analyses.**

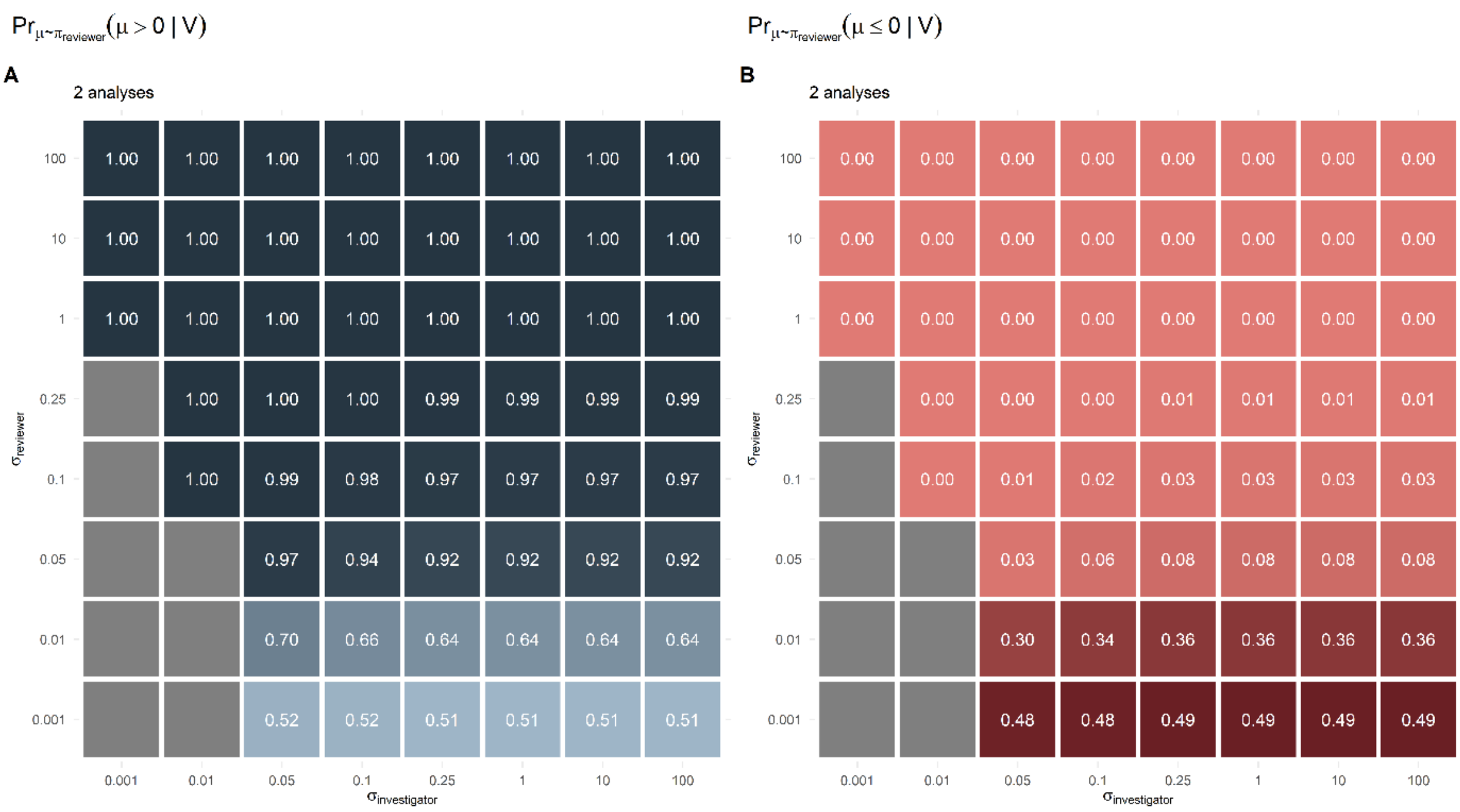

**Figure S66. For five analyses.**

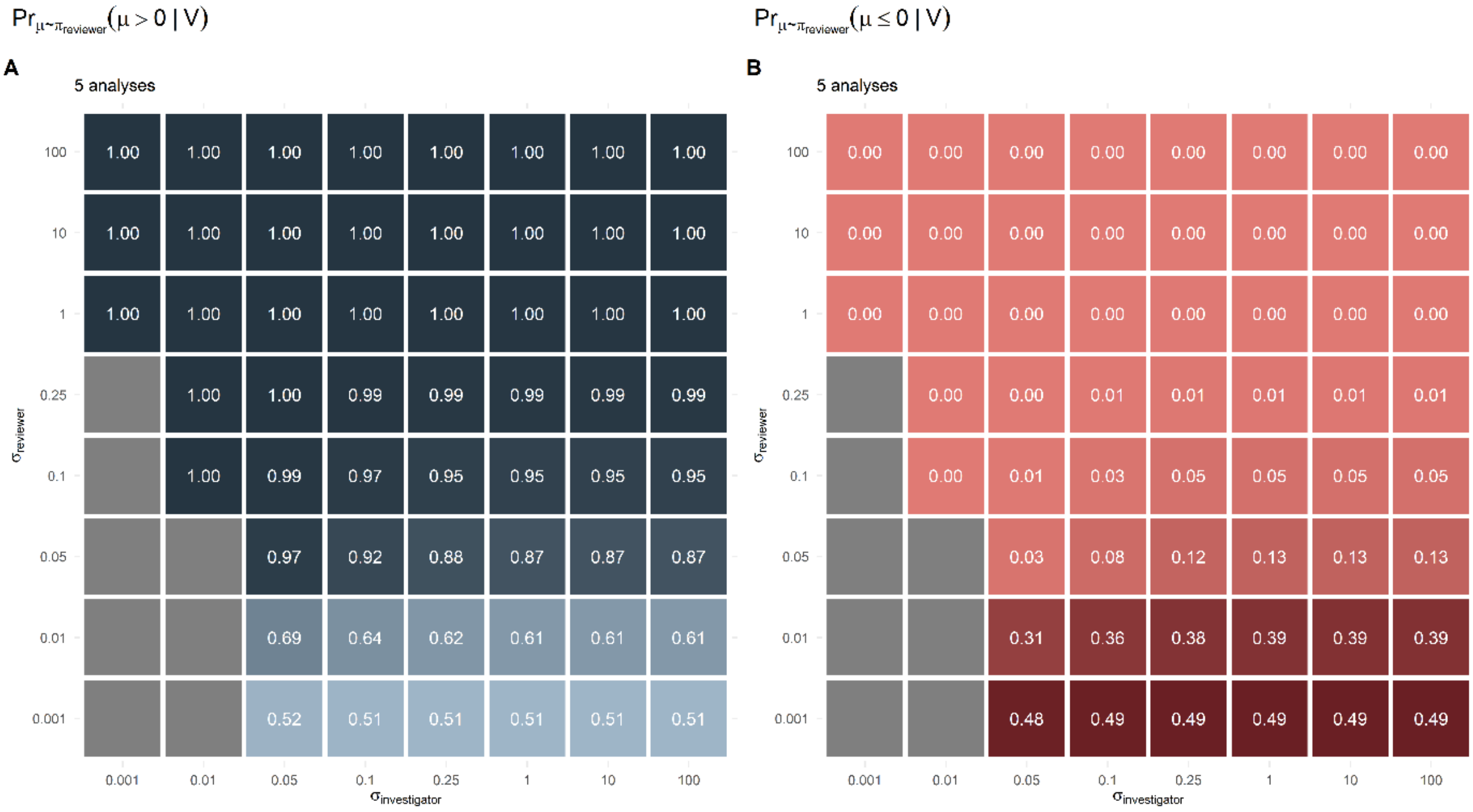

**Figure S67. For 10 analyses.**

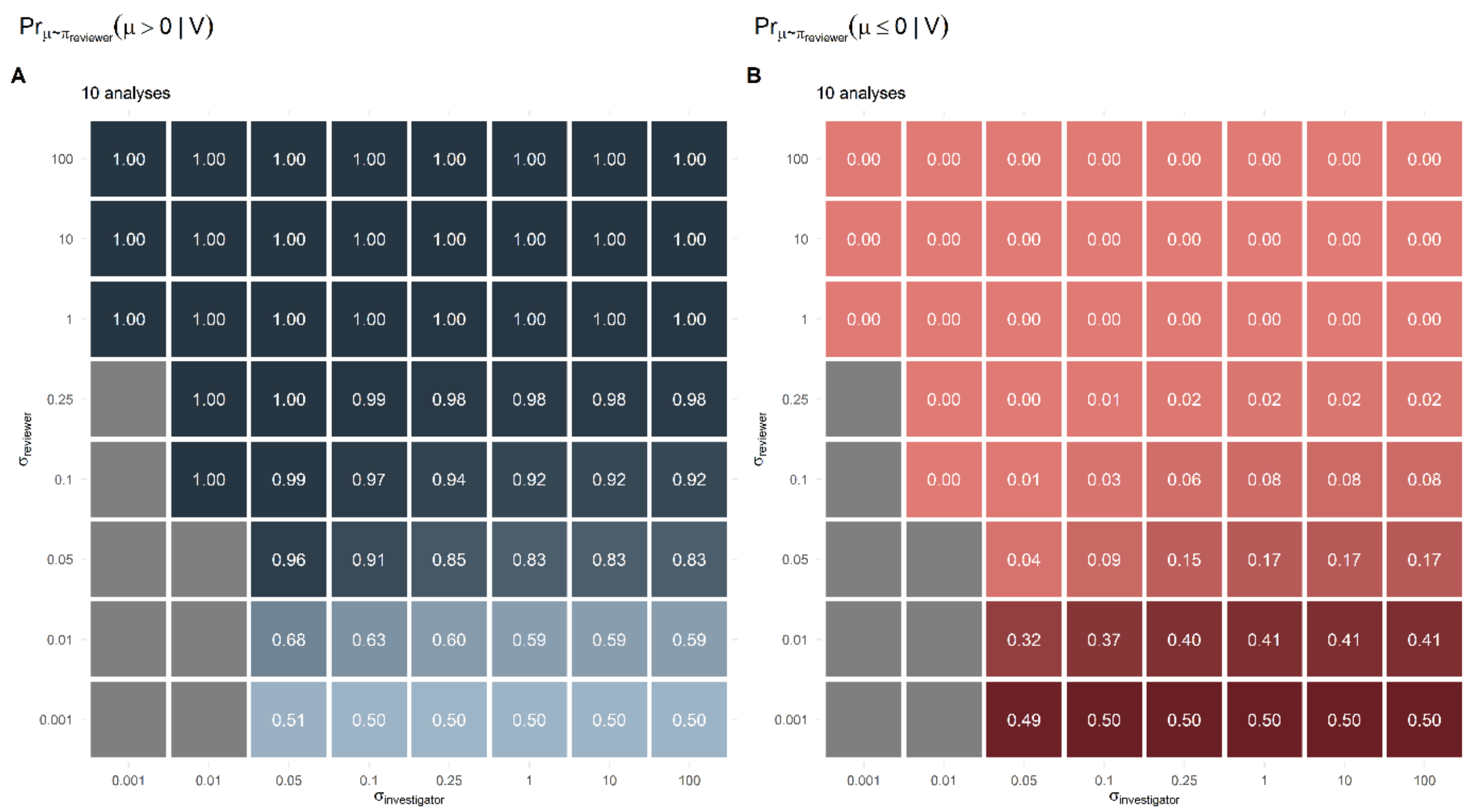

**Figure S68. For 20 analyses.**

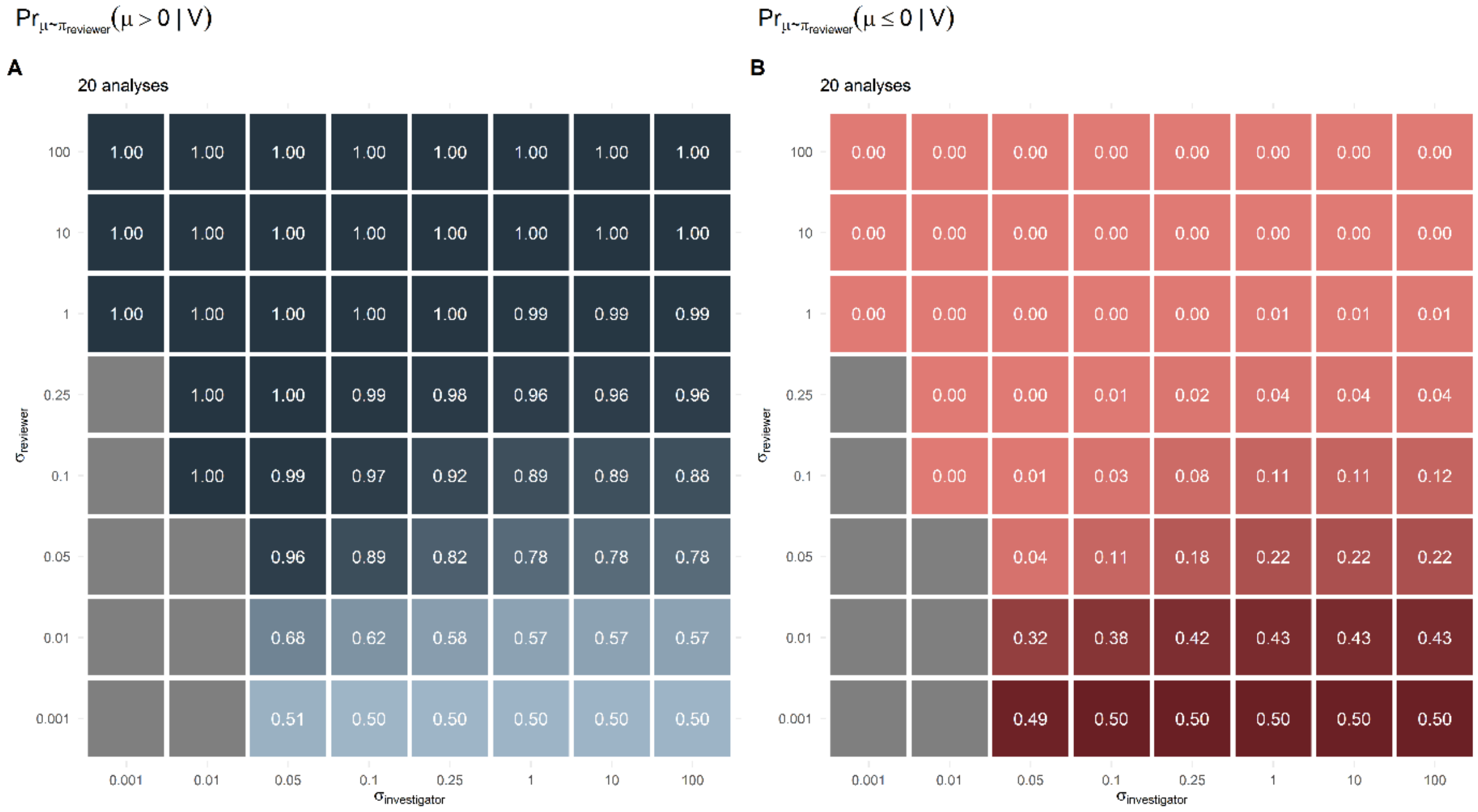

**Figure S69. For 50 analyses.**

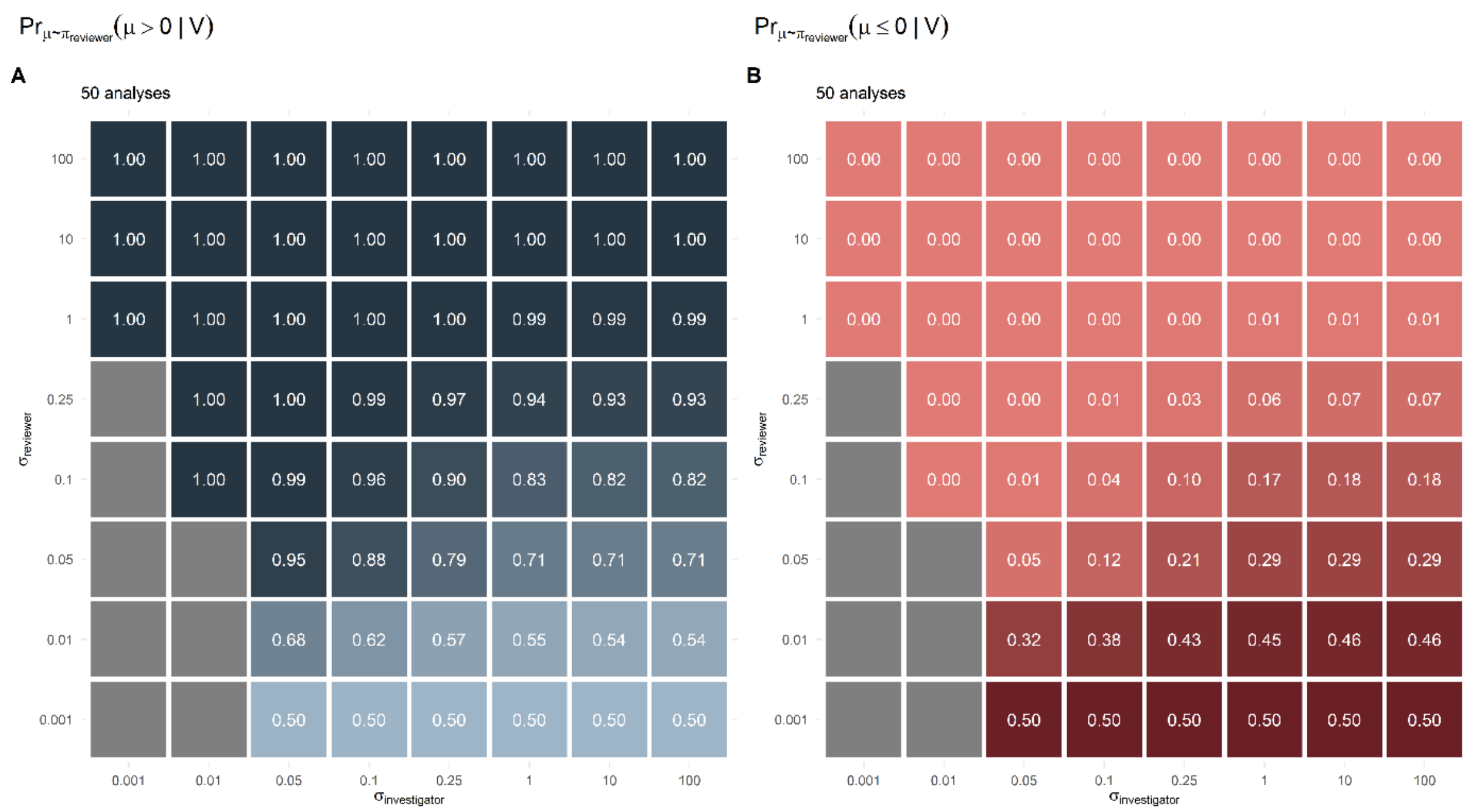

**Figure S70. For 100 analyses.**

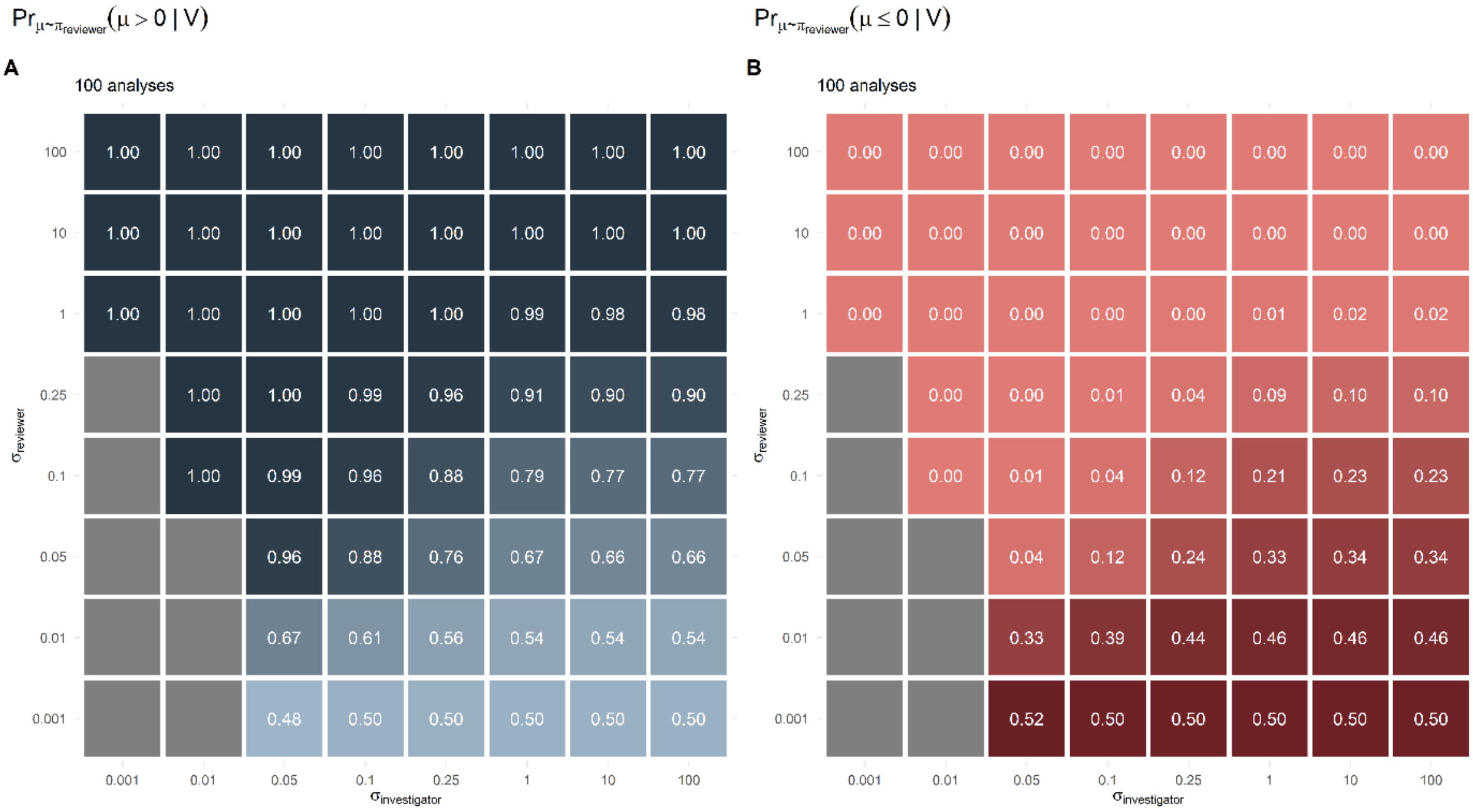

**Figure S71. For 250 analyses.**

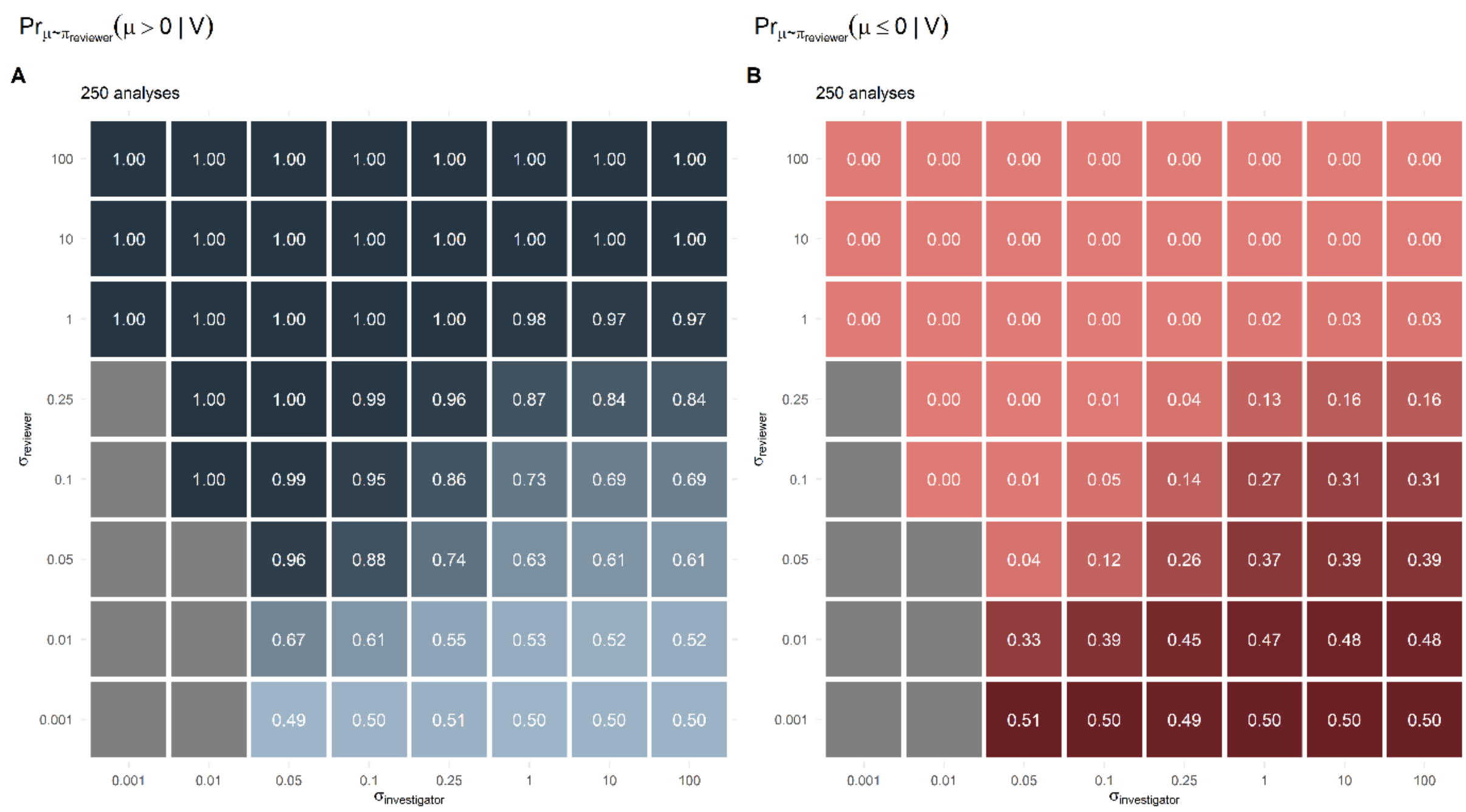

**Figure S72. For 500 analyses.**

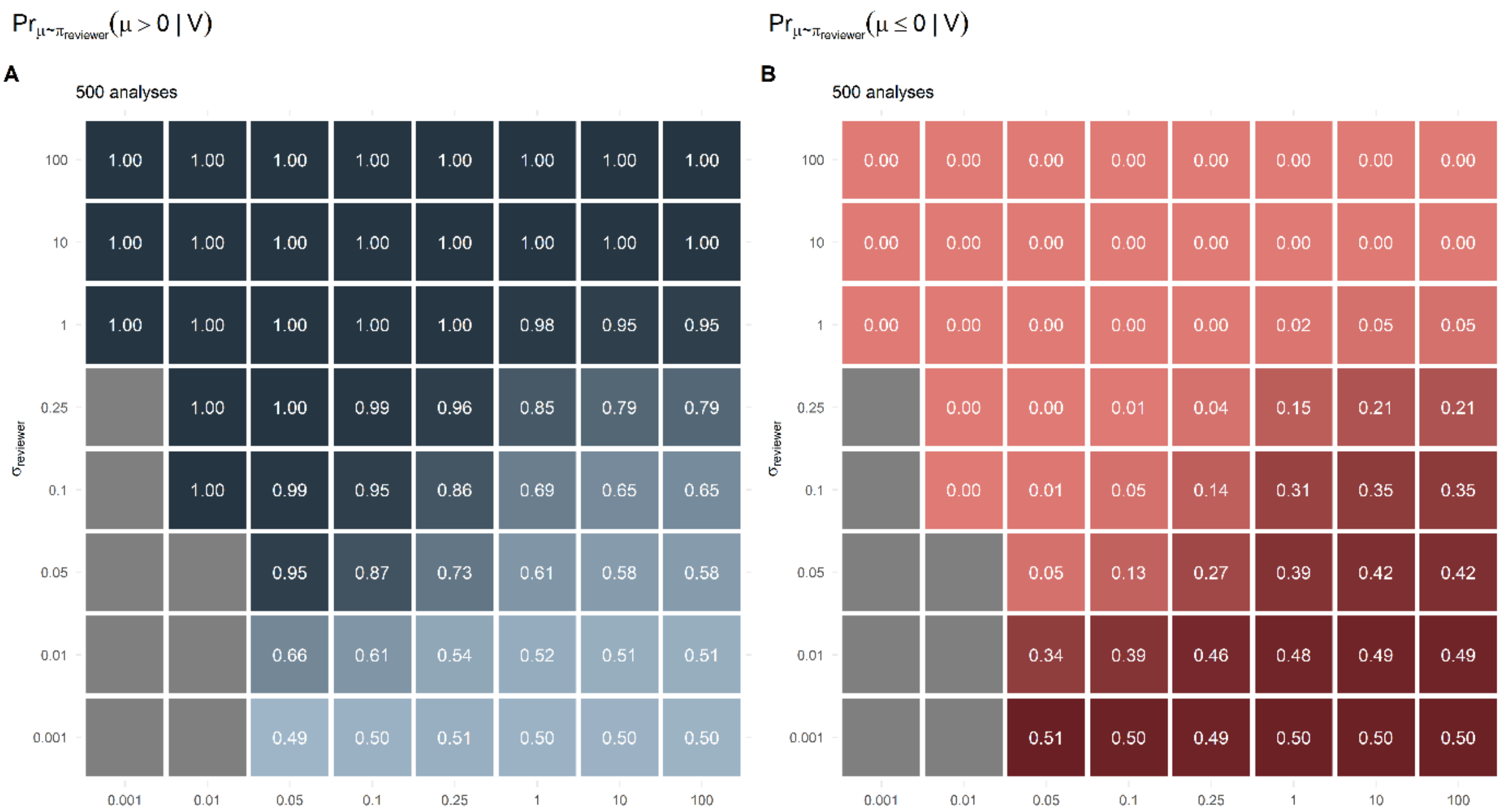

**Table S6. Informative value of a trial concluding efficacy expressed as $pTDO = pTDR/pFDR$, in Bayesian trials with efficacy stopping rules $Pr(\mu > 0|X_{1,\dots,k_i}) > 0.95, i \in \{1,\dots,m-1\}$, futility stopping rules $Pr(\mu > 0|X_{1,\dots,k_i}) < 0.50, i \in \{1,\dots,m\}$, and no correction for multiplicity of analyses. $\sigma_{investigator}$ is the standard deviation of the prior $\mathcal{N}(0, \sigma^2_{investigator})$ used for the analyses, and $\sigma_{reviewer}$ is the standard deviation of the prior $\mathcal{N}(0, \sigma^2_{reviewer})$ used to compute the operating characteristics of the trial. Empty cells are not computable because no event of efficacy conclusion occurred in the simulations.**

| Number of analyses | $\sigma_{reviewer}$ | $\text{pTDO} = \frac{pTDR}{pFDR} = \frac{Pr_{\mu\sim\pi_{reviewer}}(\mu > 0\|V)}{Pr_{\mu\sim\pi_{reviewer}}(\mu \leq 0\|V)}$ | | | | | | | |
|---|---|---|---|---|---|---|---|---|---|
| | | $\sigma_{investigator}$ | $\sigma_{investigator}$ | $\sigma_{investigator}$ | $\sigma_{investigator}$ | $\sigma_{investigator}$ | $\sigma_{investigator}$ | $\sigma_{investigator}$ | $\sigma_{investigator}$ |
| | | **0.001** | **0.01** | **0.05** | **0.1** | **0.25** | **1** | **10** | **100** |
| 1 | 0.001 | | | 1.1 | 1.1 | 1.1 | 1.0 | 1.0 | 1.0 |
| 2 | 0.001 | | | 1.1 | 1.1 | 1.0 | 1.0 | 1.0 | 1.0 |
| 5 | 0.001 | | | 1.1 | 1.0 | 1.0 | 1.0 | 1.0 | 1.0 |
| 10 | 0.001 | | | 1.1 | 1.0 | 1.0 | 1.0 | 1.0 | 1.0 |
| 20 | 0.001 | | | 1.1 | 1.0 | 1.0 | 1.0 | 1.0 | 1.0 |
| 50 | 0.001 | | | 1.0 | 1.0 | 1.0 | 1.0 | 1.0 | 1.0 |
| 100 | 0.001 | | | 0.9 | 1.0 | 1.0 | 1.0 | 1.0 | 1.0 |
| 250 | 0.001 | | | 1.0 | 1.0 | 1.0 | 1.0 | 1.0 | 1.0 |
| 500 | 0.001 | | | 1.0 | 1.0 | 1.0 | 1.0 | 1.0 | 1.0 |
| 1 | 0.01 | | | 2.5 | 2.2 | 2.1 | 2.1 | 2.1 | 2.1 |
| 2 | 0.01 | | | 2.3 | 1.9 | 1.8 | 1.8 | 1.8 | 1.8 |
| 5 | 0.01 | | | 2.2 | 1.8 | 1.6 | 1.6 | 1.6 | 1.6 |
| 10 | 0.01 | | | 2.1 | 1.7 | 1.5 | 1.4 | 1.4 | 1.4 |
| 20 | 0.01 | | | 2.1 | 1.7 | 1.4 | 1.3 | 1.3 | 1.3 |
| 50 | 0.01 | | | 2.1 | 1.6 | 1.3 | 1.2 | 1.2 | 1.2 |
| 100 | 0.01 | | | 2.1 | 1.6 | 1.3 | 1.2 | 1.2 | 1.2 |
| 250 | 0.01 | | | 2.0 | 1.6 | 1.2 | 1.1 | 1.1 | 1.1 |
| 500 | 0.01 | | | 1.9 | 1.5 | 1.2 | 1.1 | 1.1 | 1.1 |
| 1 | 0.05 | | | 47 | 24 | 20 | 19 | 19 | 19 |
| 2 | 0.05 | | | 36 | 16 | 12 | 12 | 12 | 12 |
| 5 | 0.05 | | | 29 | 12 | 7.6 | 6.9 | 6.9 | 6.9 |
| 10 | 0.05 | | | 27 | 9.9 | 5.8 | 4.8 | 4.8 | 4.8 |
| 20 | 0.05 | | | 23 | 8.4 | 4.6 | 3.5 | 3.5 | 3.5 |
| 50 | 0.05 | | | 21 | 7.7 | 3.7 | 2.5 | 2.4 | 2.4 |
| 100 | 0.05 | | | 22 | 7.3 | 3.2 | 2.1 | 1.9 | 1.9 |
| 250 | 0.05 | | | 22 | 7.2 | 2.9 | 1.7 | 1.6 | 1.6 |
| 500 | 0.05 | | | 21 | 6.8 | 2.6 | 1.5 | 1.4 | 1.4 |
| 1 | 0.1 | | ≥ 1000 | 249 | 90 | 70 | 66 | 66 | 66 |
| 2 | 0.1 | | ≥ 1000 | 166 | 52 | 39 | 36 | 35 | 35 |
| 5 | 0.1 | | ≥ 1000 | 142 | 37 | 21 | 18 | 18 | 18 |
| 10 | 0.1 | | ≥ 1000 | 110 | 31 | 15 | 12 | 12 | 12 |
| 20 | 0.1 | | ≥ 1000 | 99 | 28 | 11 | 8.0 | 7.7 | 7.7 |
| 50 | 0.1 | | ≥ 1000 | 82 | 24 | 8.6 | 4.8 | 4.6 | 4.6 |
| 100 | 0.1 | | ≥ 1000 | 82 | 23 | 7.4 | 3.7 | 3.3 | 3.3 |
| 250 | 0.1 | | ≥ 1000 | 76 | 21 | 6.3 | 2.6 | 2.3 | 2.3 |
| 500 | 0.1 | | ≥ 1000 | 70 | 20 | 5.9 | 2.3 | 1.8 | 1.8 |
| 1 | 0.25 | | ≥ 1000 | 999 | 499 | 332 | 332 | 332 | 332 |
| 2 | 0.25 | | ≥ 1000 | 999 | 249 | 166 | 142 | 142 | 142 |
| 5 | 0.25 | | ≥ 1000 | 999 | 166 | 82 | 70 | 70 | 70 |
| 10 | 0.25 | | ≥ 1000 | 999 | 142 | 58 | 42 | 42 | 42 |
| 20 | 0.25 | | ≥ 1000 | 999 | 142 | 42 | 27 | 26 | 26 |
| 50 | 0.25 | | ≥ 1000 | 499 | 124 | 32 | 15 | 14 | 14 |
| 100 | 0.25 | | ≥ 1000 | 499 | 124 | 28 | 10 | 9.1 | 9.1 |
| 250 | 0.25 | | ≥ 1000 | 999 | 124 | 24 | 6.9 | 5.3 | 5.2 |
| 500 | 0.25 | | ≥ 1000 | 499 | 124 | 22 | 5.5 | 3.7 | 3.7 |
| 1 | 1 | ≥ 1000 | ≥ 1000 | ≥ 1000 | 999 | 999 | 999 | 999 | 999 |
| 2 | 1 | ≥ 1000 | ≥ 1000 | ≥ 1000 | 999 | 999 | 499 | 499 | 499 |
| 5 | 1 | ≥ 1000 | ≥ 1000 | ≥ 1000 | 999 | 332 | 332 | 332 | 332 |
| 10 | 1 | ≥ 1000 | ≥ 1000 | ≥ 1000 | 999 | 249 | 199 | 199 | 199 |
| 20 | 1 | ≥ 1000 | ≥ 1000 | ≥ 1000 | 999 | 249 | 142 | 142 | 142 |
| 50 | 1 | ≥ 1000 | ≥ 1000 | ≥ 1000 | 999 | 249 | 90 | 82 | 82 |
| 100 | 1 | ≥ 1000 | ≥ 1000 | ≥ 1000 | 999 | 249 | 66 | 55 | 55 |
| 250 | 1 | ≥ 1000 | ≥ 1000 | ≥ 1000 | 999 | 249 | 49 | 32 | 32 |
| 500 | 1 | ≥ 1000 | ≥ 1000 | ≥ 1000 | 999 | 249 | 39 | 20 | 20 |
| 1 | 10 | ≥ 1000 | ≥ 1000 | ≥ 1000 | ≥ 1000 | ≥ 1000 | ≥ 1000 | ≥ 1000 | ≥ 1000 |
| 2 | 10 | ≥ 1000 | ≥ 1000 | ≥ 1000 | ≥ 1000 | ≥ 1000 | ≥ 1000 | ≥ 1000 | ≥ 1000 |
| 5 | 10 | ≥ 1000 | ≥ 1000 | ≥ 1000 | ≥ 1000 | ≥ 1000 | ≥ 1000 | ≥ 1000 | ≥ 1000 |
| 10 | 10 | ≥ 1000 | ≥ 1000 | ≥ 1000 | ≥ 1000 | ≥ 1000 | 999 | 999 | 999 |
| 20 | 10 | ≥ 1000 | ≥ 1000 | ≥ 1000 | ≥ 1000 | 999 | 999 | 999 | 999 |
| 50 | 10 | ≥ 1000 | ≥ 1000 | ≥ 1000 | ≥ 1000 | 999 | 999 | 999 | 999 |
| 100 | 10 | ≥ 1000 | ≥ 1000 | ≥ 1000 | ≥ 1000 | 999 | 499 | 499 | 499 |
| 250 | 10 | ≥ 1000 | ≥ 1000 | ≥ 1000 | ≥ 1000 | ≥ 1000 | 499 | 332 | 332 |
| 500 | 10 | ≥ 1000 | ≥ 1000 | ≥ 1000 | ≥ 1000 | ≥ 1000 | 499 | 249 | 249 |
| 1 | 100 | ≥ 1000 | ≥ 1000 | ≥ 1000 | ≥ 1000 | ≥ 1000 | ≥ 1000 | ≥ 1000 | ≥ 1000 |
| 2 | 100 | ≥ 1000 | ≥ 1000 | ≥ 1000 | ≥ 1000 | ≥ 1000 | ≥ 1000 | ≥ 1000 | ≥ 1000 |
| 5 | 100 | ≥ 1000 | ≥ 1000 | ≥ 1000 | ≥ 1000 | ≥ 1000 | ≥ 1000 | ≥ 1000 | ≥ 1000 |
| 10 | 100 | ≥ 1000 | ≥ 1000 | ≥ 1000 | ≥ 1000 | ≥ 1000 | ≥ 1000 | ≥ 1000 | ≥ 1000 |
| 20 | 100 | ≥ 1000 | ≥ 1000 | ≥ 1000 | ≥ 1000 | ≥ 1000 | ≥ 1000 | ≥ 1000 | ≥ 1000 |
| 50 | 100 | ≥ 1000 | ≥ 1000 | ≥ 1000 | ≥ 1000 | ≥ 1000 | ≥ 1000 | ≥ 1000 | ≥ 1000 |
| 100 | 100 | ≥ 1000 | ≥ 1000 | ≥ 1000 | ≥ 1000 | ≥ 1000 | ≥ 1000 | ≥ 1000 | ≥ 1000 |
| 250 | 100 | ≥ 1000 | ≥ 1000 | ≥ 1000 | ≥ 1000 | ≥ 1000 | ≥ 1000 | ≥ 1000 | ≥ 1000 |
| 500 | 100 | ≥ 1000 | ≥ 1000 | ≥ 1000 | ≥ 1000 | ≥ 1000 | ≥ 1000 | ≥ 1000 | ≥ 1000 |

## 2.3 Bayesian correction for multiplicity of analyses

**Figure S73. Bayesian operating characteristics $pTDR = Pr_{\mu \sim N(0,\sigma^2_{reviewer})}(\mu > 0 \mid V)$ and $pFDR = Pr_{\mu \sim N(0,\sigma^2_{reviewer})}(\mu \leq 0 \mid V)$ computed under various $\sigma_{reviewer}$. The efficacy thresholds $1 - e_i$ for the decision rule $Pr(\mu > 0 \mid X_{1,\ldots,k_i}) > 1 - e_i$, and the maximum sample size were calibrated according to the number of analyses to achieve $Pr_{\mu \sim N(0,0.25^2)}(V \mid \mu \leq 0) = 0.05$ and $Pr_{\mu \sim N(0,0.25^2)}(V \mid \mu > 0) = 0.80$.**

A. $Pr_{\mu \sim \pi_{reviewer}}(\mu > 0 \mid V)$

| $\sigma_{reviewer}$ | | | | | | | |
|---|---|---|---|---|---|---|---|
| 100 | 1.00 | 1.00 | 1.00 | 1.00 | 1.00 | 1.00 | 1.00 |
| 10 | 1.00 | 1.00 | 1.00 | 1.00 | 1.00 | 1.00 | 1.00 |
| 1 | 0.99 | 0.99 | 0.99 | 0.99 | 0.99 | 0.99 | 0.99 |
| 0.25 | 0.94 | 0.94 | 0.94 | 0.94 | 0.94 | 0.94 | 0.94 |
| 0.1 | 0.84 | 0.84 | 0.84 | 0.85 | 0.84 | 0.85 | 0.85 |
| 0.05 | 0.73 | 0.73 | 0.73 | 0.74 | 0.73 | 0.74 | 0.74 |
| 0.01 | 0.55 | 0.55 | 0.55 | 0.55 | 0.55 | 0.55 | 0.55 |
| 0.001 | 0.50 | 0.50 | 0.50 | 0.50 | 0.50 | 0.50 | 0.50 |
| Number of analyses | 1 | 2 | 5 | 10 | 20 | 50 | 100 |
| $\sigma_{investigator}$ | 0.25 | 0.25 | 0.25 | 0.25 | 0.25 | 0.25 | 0.25 |
| Efficacy threshold | 0.701 | 0.787 | 0.86 | 0.89 | 0.904 | 0.919 | 0.925 |
| Maximum sample size | 121 | 145 | 178 | 199 | 207 | 220 | 225 |

B. $Pr_{\mu \sim \pi_{reviewer}}(\mu \leq 0 \mid V)$

| $\sigma_{reviewer}$ | | | | | | | |
|---|---|---|---|---|---|---|---|
| 100 | 0.00 | 0.00 | 0.00 | 0.00 | 0.00 | 0.00 | 0.00 |
| 10 | 0.00 | 0.00 | 0.00 | 0.00 | 0.00 | 0.00 | 0.00 |
| 1 | 0.01 | 0.01 | 0.01 | 0.01 | 0.01 | 0.01 | 0.01 |
| 0.25 | 0.06 | 0.06 | 0.06 | 0.06 | 0.06 | 0.06 | 0.06 |
| 0.1 | 0.16 | 0.16 | 0.16 | 0.15 | 0.16 | 0.15 | 0.15 |
| 0.05 | 0.27 | 0.27 | 0.27 | 0.26 | 0.27 | 0.26 | 0.26 |
| 0.01 | 0.45 | 0.45 | 0.45 | 0.45 | 0.45 | 0.45 | 0.45 |
| 0.001 | 0.50 | 0.50 | 0.50 | 0.50 | 0.50 | 0.50 | 0.50 |
| | 1 | 2 | 5 | 10 | 20 | 50 | 100 |
| | 0.25 | 0.25 | 0.25 | 0.25 | 0.25 | 0.25 | 0.25 |
| | 0.701 | 0.787 | 0.86 | 0.89 | 0.904 | 0.919 | 0.925 |
| | 121 | 145 | 178 | 199 | 207 | 220 | 225 |

## 2.4 Posterior probability when concluding efficacy

**Figures S74 to S78. Median posterior probability $Pr_{\mu\sim\pi_{investigator}}(\mu > 0 \mid X_{1,\dots,k_i})$ in trials having concluded efficacy with efficacy stopping rule $Pr_{\mu\sim\pi_{investigator}}(\mu > 0 \mid X_{1,\dots,k_i}) > 0.95$, no futility stopping rules, and no correction for multiplicity of analyses. $\sigma_{investigator}$ is the standard deviation of the prior $\mathcal{N}(0, \sigma^2_{investigator})$ used for the analyses, and $\sigma_{reviewer}$ is the standard deviation of the prior $\mathcal{N}(0, \sigma^2_{reviewer})$ used to simulate trials with $X \sim \mathcal{N}(\mu, 1)$.**

**Figure S74. According to the number of analyses, for $\sigma_{reviewer} = \sigma_{investigator}$**

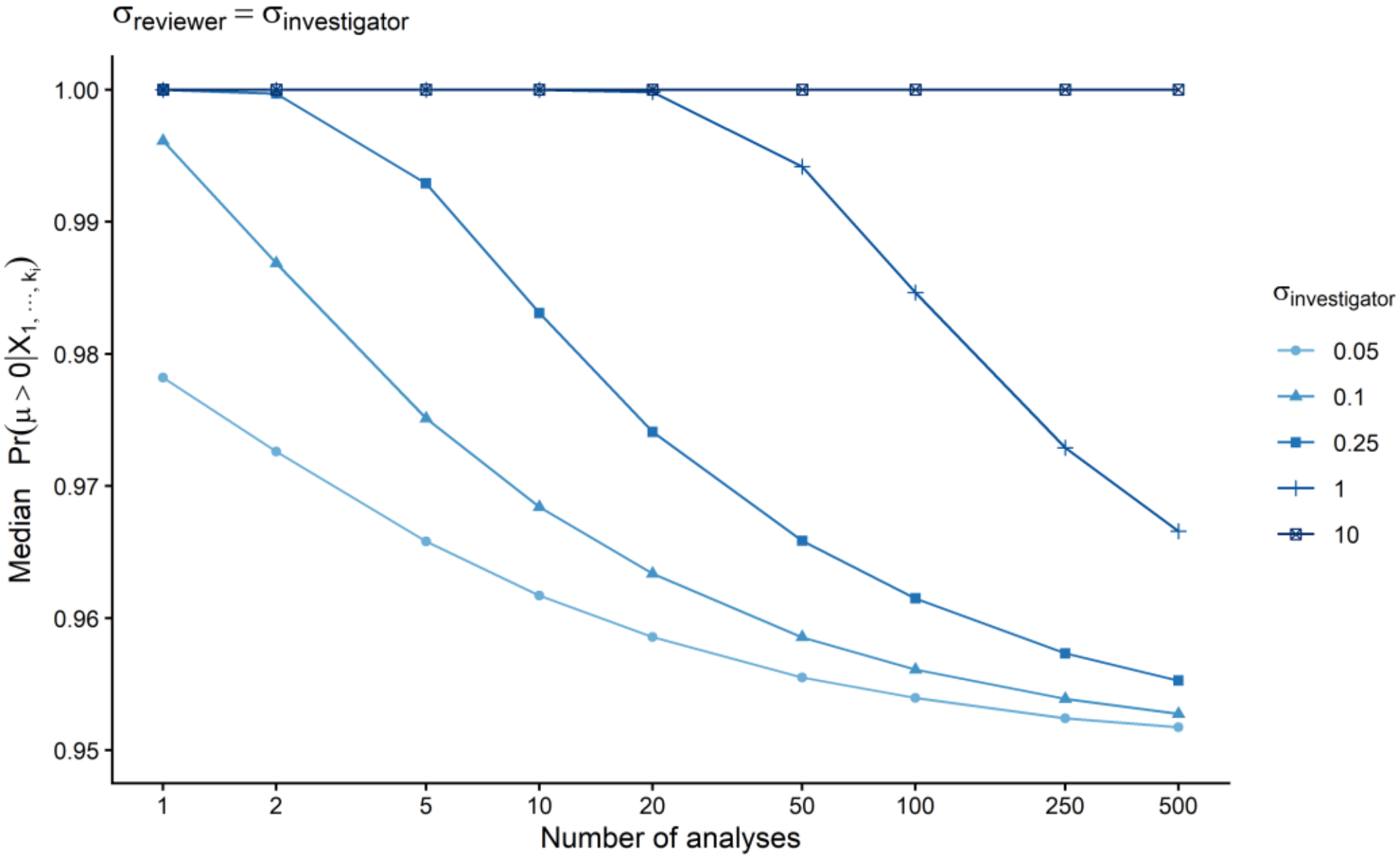

**Figure S75. According to the number of analyses, for $\sigma_{reviewer} = 0$ $(\mu = 0)$**

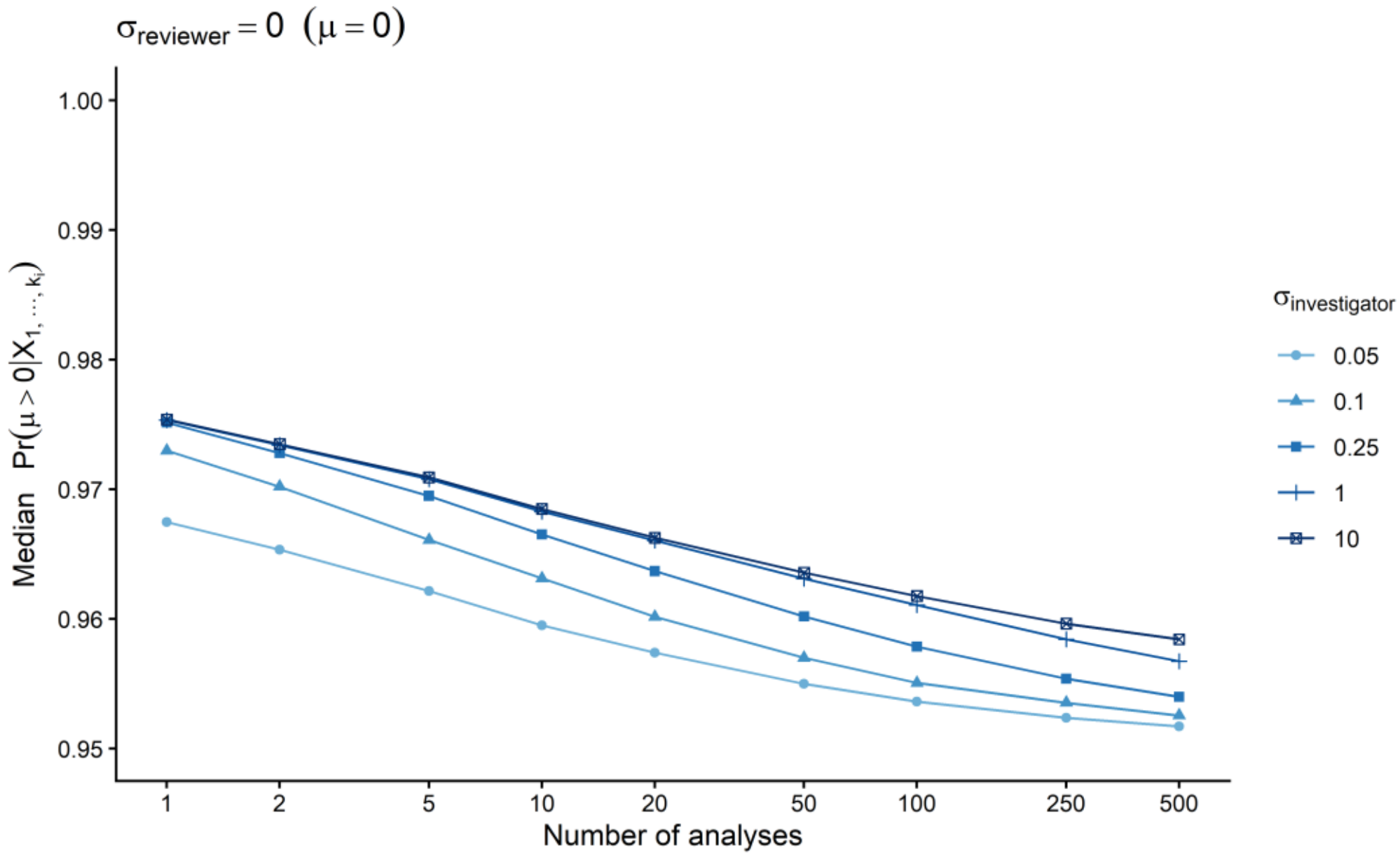

**Figure S76. According to $\sigma_{reviewer}$ and $\sigma_{investigator}$, for 500 analyses**

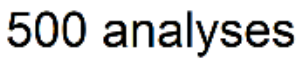

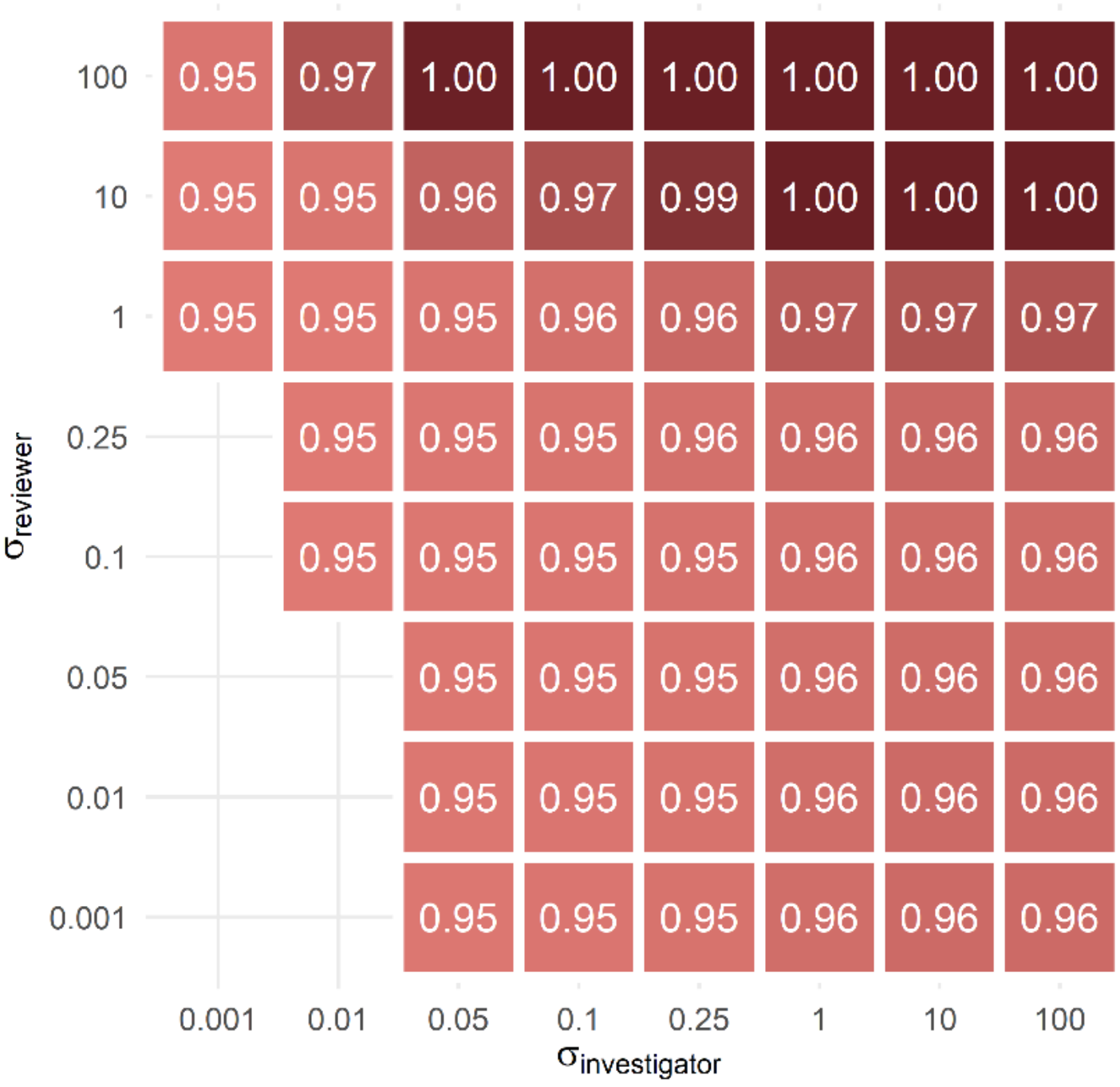

**Figure S77. According to $\sigma_{reviewer}$ and $\sigma_{investigator}$, for 10 analyses**

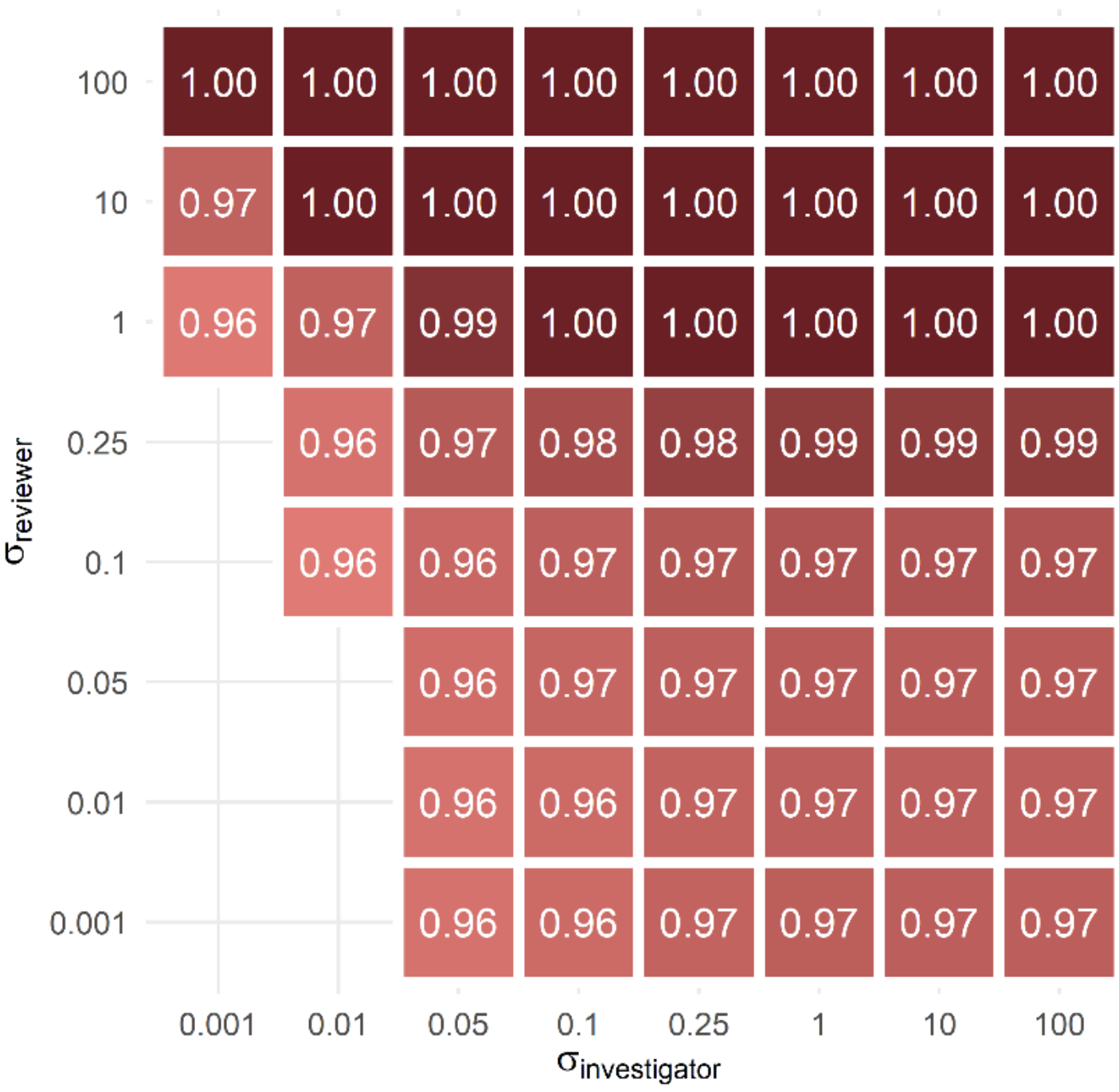

**Figure S78. According to $\sigma_{reviewer}$ and $\sigma_{investigator}$, for 1 analysis**

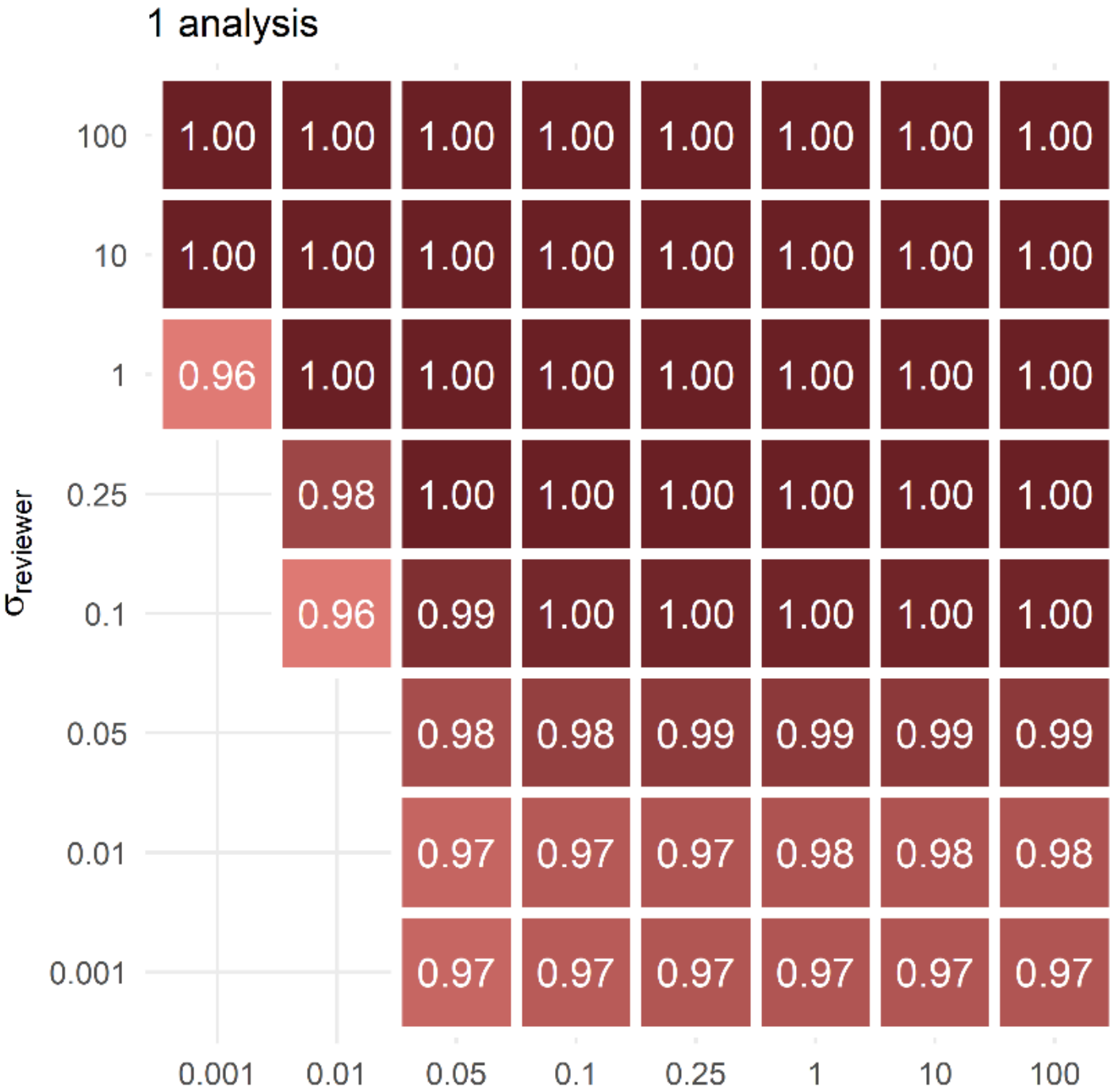